\documentclass[11pt,twoside,openright]{report}
\usepackage{cite}
\usepackage[a4paper,margin=2.5cm]{geometry}
\usepackage{epsfig}
\usepackage{tabularx}
\usepackage{picins}
\usepackage{here}
\usepackage{afterpage}
\usepackage{array}
\usepackage{pifont}
\usepackage{multirow}
\usepackage{hhline}
\usepackage{rotating}
\usepackage{euscript}
\usepackage{oldgerm}
\usepackage{calc}
\usepackage{fancyhdr}
\usepackage{amsmath,amssymb,amsthm}
\usepackage{floatflt}
\usepackage{caption, subfigure, fancybox}
\usepackage{placeins}   % for 'flushing' floats in each section

\usepackage{longtable}
\usepackage{subfigure}
\usepackage{ccaption} % figures spreading multiple pages
\usepackage{My_physics}

\usepackage{hyperref}
\hypersetup{pdfstartview={Fit -32768},
  pdfpagemode=UseOutlines,
  plainpages=true,
  bookmarksnumbered=true
  }
\hypersetup{bookmarksopenlevel=2}

\usepackage{ifthen}

\begin{document}

{\begin{titlepage}
\topmargin = 0pt

\vspace{1.5cm}
\begin{center}
\vspace{0.4cm}

\huge The discovery potential of the \\ $\chinonn$ in mSUGRA 
	in the $\tau$-channel \\ at high $\tan \beta$ at the LHC 

\end{center}
\vspace{0.4cm}
\begin{center}
\includegraphics[width=8cm]{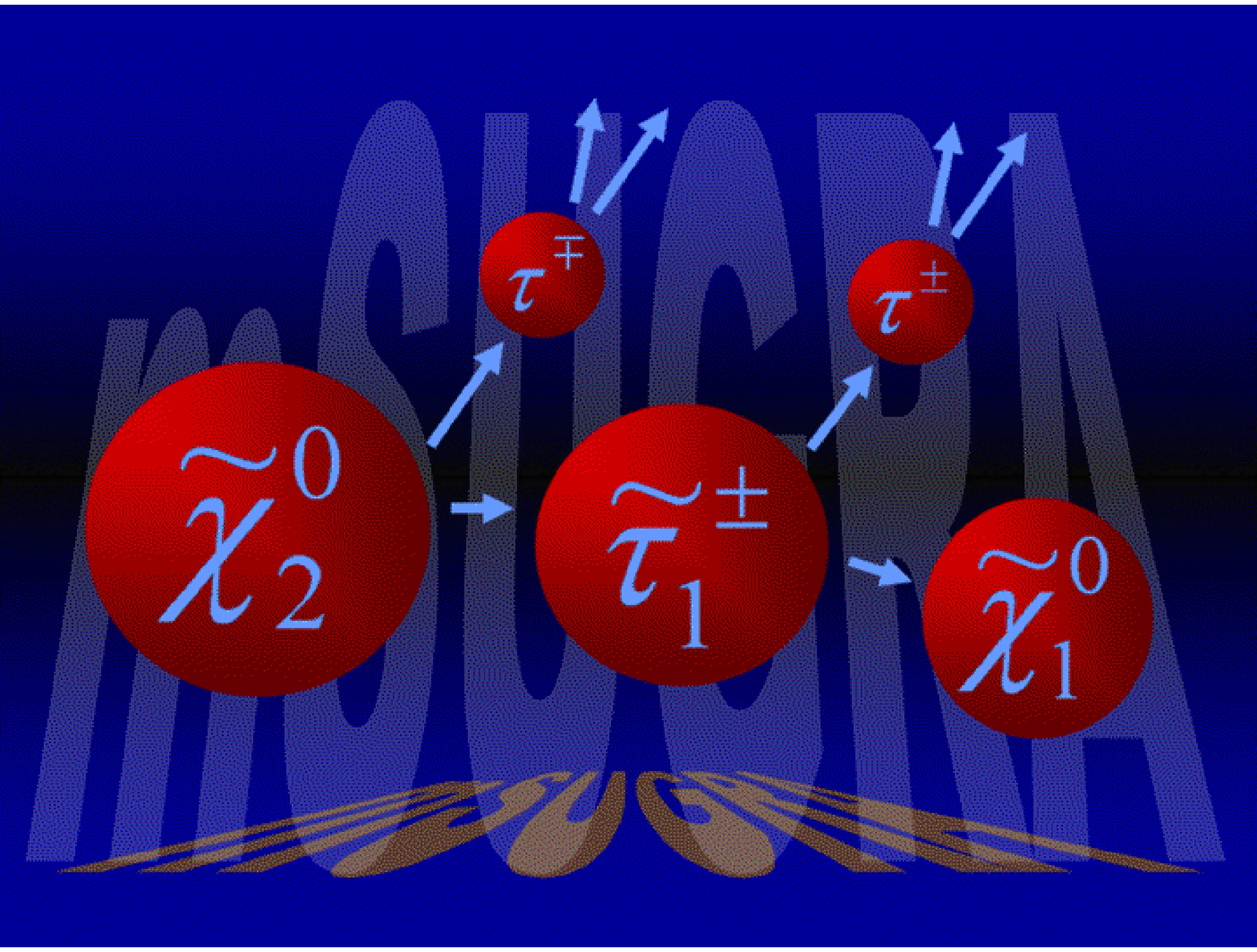}
\end{center}

\begin{center}

\Large{$\qquad$ } \\

\Large{Diploma thesis} \\
\vspace{0.75cm}
\Large{Florian Heinemann} \\
\vspace{0.5cm}
\normalsize{ETH Zurich, Switzerland} \\
\vspace{1.5cm}
\normalsize{performed at CERN} \\
\vspace{0.25cm}
\normalsize{supervised by} \\
\vspace{0.75cm}
\Large{Dr. Andr\'{e} Holzner,} \normalsize{ETHZ} \\
\vspace{0.15cm}
\Large{Dr. Luc Pape,} \normalsize{CERN} \\
\vspace{0.15cm}
\Large{Prof. Dr. Felicitas Pauss,} \normalsize{ETHZ} \\
\vspace{1.55cm}

\large{ November 2003 to March 2004} \\
\end{center}

\end{titlepage}

}{}

\pagestyle{plain}

\renewcommand{\thepage}{\roman{page}}
\setcounter{page}{1}

\tableofcontents
\newpage
\clearpage

\renewcommand{\thepage}{\roman{page}}

%\include{DTtitle}
%\topmargin = 0.25in
%----------------------------------------------------------------------	
{
 \begin{abstract} One of the goals of the Large Hadron Collider (LHC)
is to search for supersymmetric particles predicted by the minimal
Supergravity Model (mSUGRA). In previous studies the discovery
potential for the next-to-lightest neutralino ($\chinonn$) has been
investigated in many areas of its parameter space. It has been pointed
out that the leptonic decay of the $\chinonn$ offers kinematic
properties which can be well used for mass reconstructions of
sparticles. However, in the parameter space where taus dominate the
leptonic decay, the $\chinonn$ discovery is especially difficult since
a full tau reconstruction will not be possible at LHC. In several
recent publications it has been indicated that studies are necessary
in that region.

This study analyses the possibility of using kinematic endpoints for a
sparticle mass reconstruction at $\tan \beta = 35$, $A_{0} = 0$ and
$\mu > 0$. A region of 25 points around the benchmark point I' with
$m_{0} = 181$ and $m_{1/2} = 350$ is chosen where the cascade decay
$\chinonn \rightarrow \tau^{\pm} + \tilde{\tau}^{\mp}
\rightarrow \tau^{\pm} + \tau^{\mp} + \chinon$ has a branching ratio of
over $95\%$. In this cascade decay at least three particles, the
$\chinon$ and two $\nu_{\tau}$, escape the detector. Therefore, a
precise mass reconstruction in the investigated region using the
kinematic limits is not possible. However, this study shows that some
invariant mass endpoints can be well estimated, using a $m_{0}$ and
$m_{1/2}$ independent correlation between the distribution maximum and
the kinematic limit. Furthermore, a linear fit at the tail of the
invariant mass distribution gives in most of the cases a good estimate
as well. In addition, a method is described how to measure two
endpoints within the single rho with one quark invariant mass
distribution. The influence of the mixing between $\tilde{b}_{1}$
($\tilde{t}_{1}$) and the heavier $\tilde{b}_{2}$ ($\tilde{t}_{2}$) on
endpoint measurements is also discussed.

The simulations are performed with PYTHIA 6.220 in combination with
ISASUGRA 7.69.  The results are at Monte Carlo level, and give the
minimal systematic and statistical uncertainties on measurements of 13
kinematic endpoints. For the mass reconstruction of sparticles with a
$30\ fb^{-1}$ data sample the minimal uncertainties of $20.4\gev$
(7.9\%) for the $\chinonn$, $20.5\gev$ (14.8\%) for the $\chinon$,
$21.2\gev$ (14.0\%) for the $\tilde{\tau}^{\pm}_{1}$, $28.2\gev$
(3.6\%) for the $\tilde{d}_{1}$ and $\tilde{s}_{1}$, $25.7\gev$
(3.6\%) for the $\tilde{b}_{2}$ and $28.8\gev$ (3.9\%) for the
$\tilde{t}_{2}$ have been calculated.
\end{abstract}

 \cleardoublepage
}{}

%\tableofcontents\markboth{}{}
%\include{DTabstract}
%----------------------------------------------------------------------

\renewcommand{\thepage}{\arabic{page}}
\setcounter{page}{1}

%----------------------------------------------------------------------
% set up the page style (headers and footers)
\pagestyle{fancy}
\fancyhf{}
\renewcommand{\chaptermark}[1]{\markboth{\sf\chaptername\ \thechapter:\ #1}{}}
\renewcommand{\sectionmark}[1]{\markright{\sf\thesection\ #1}}
\cfoot{\rm\thepage}    %       le numero
\fancyhead[LO]{\rightmark}      %       a droite
\fancyhead[RE]{\leftmark}       %       a gauche
%----------------------------------------------------------------------

 \chapter{Introduction}
 \markboth{Introduction}{Introduction} Today, the Standard Model
\cite{SM} of electroweak and strong interactions describes nature at
the smallest scales accessible in high energy physics.  Both theories
are based on the same fundamental principle, the local gauge
invariance. Through precision measurements at LEP and SLC the
electroweak theory has been checked very thoroughly, and no obvious
deviation or inconsistency has been observed.  The only particle of
that theory that still has to be discovered is the Higgs boson.  It is
necessary for explaining the existence of massive fundamental
particles in nature. Quantum Chromodynamic (QCD), describing the
strong interactions within the Standard Model, is tested very well at
HERA. Only for gravitation, it has not been possible yet to construct a
theory based on quantum mechanics which can be tested in experiments.

However, there are a lot of open questions which have to be
solved. The origin of mass, the reason for the huge matter antimatter
asymmetry in the universe and the nature of cold dark matter are not
known until now. The Standard Model has a large number of parameters
which - instead of arising from a more fundamental theory - have to be
measured. Furthermore, there are theoretical motivations for physics
beyond the Standard Model.  One is the hierarchy problem concerning
the quadratically divergent fermion loop corrections to the Higgs
mass.  Additionally, - for reasons of concinnity - theorists want to
unify the gauge couplings.  J.C.  Maxwell working out his famous
equations as well as S.  Glashow, S.  Weinberg and A.  Salam - all
three responsible for the electroweak unification - have shown that a
unification of theories can be very successful.

In order to answer some of these questions, one possibility is to
extend the Standard Model Poincar\'e algebra to a supersymmetric
graduated Lie algebra which leads to a larger particle spectrum. The
model with the minimal particle content is the Minimal Supersymmetric
Standard Model (MSSM). It is an important prediction of this theory
that every particle described by the Standard Model has a
supersymmetric partner, a sparticle: Each fermion has a bosonic
and each boson has a fermionic superpartner.  However, none of the
predicted sparticles has been discovered yet.  If our universe really
was supersymmetric directly at the Big Bang, Supersymmetry had
to be broken at a very short time after it.  Otherwise, every particle
would have the same mass as its superpartner, which is definitely
excluded by experiments. The MSSM solves the hierarchy problem,
provided there are sparticles in the $1 - 10\tev$ mass region.  It
makes a unification of the gauge couplings possible, and in some cases
even predicts a candidate for cold dark matter.  Furthermore, the MSSM
gives strong constraints for the lightest Higgs mass which there has
to be lower than $130\gev$~\cite{lightHiggs}.

The mechanism for breaking Supersymmetry is not specified in the
general MSSM.  The breaking through gravity is one possibility, and is
given by the minimal Supergravity model (mSUGRA)~\cite{mSUGRA}. This
theory is very appealing because it reduces the 124-dimensional
parameter space of the MSSM to five dimensions.  Since in mSUGRA an
ideal candidate for cold dark matter can be predicted, also
experimentalists use this model.

The start of the Large Hadron Collider (LHC) at CERN near Geneva is
foreseen in the year 2007.  In addition to searches for the Higgs
boson, one of the main goals is to test the predictions of prospective
theories.  If the MSSM describes nature, it is the aim to produce
enough supersymmetric particles such that the two detectors CMS and
ATLAS are able to discover them. After the discovery, it is important
to measure their properties since these properties are necessary for
distinguishing between different ways of Supersymmetry breaking.  It
is thus essential to determine the best methods for sparticle searches
within the different models before the start of the LHC.
%----------------study-----------------------------------------------

\bigskip
A short introduction into the principles of Supersymmetry is given in
chapter~\ref{chap:Supersymmetry}. There the focus is set on the
Minimal Supersymmetric Standard Model (MSSM), since it provides the
simplest supersymmetric extension of the Standard Model being
consistent with experiments.  Furthermore, the mSUGRA model is
introduced being one possible realisation of the MSSM and the
framework in which this study has been carried out. The Monte Carlo
generators used in this work are presented in chapter~\ref{chap:MCG}.
In chapter~\ref{chap:Experiments}, three important experiments
concerning this study are presented. Part of the mSUGRA parameter
space has been excluded due to limits on sparticle masses with the
Large Electron Positron collider (LEP) at CERN.  The Wilkinson
Microwave Anisotropy Probe (WMAP), a NASA satellite, provides strong
constraints for the cold dark matter density in the universe. In
mSUGRA the $\chinon$ is the lightest supersymmetric particle. Assuming
the $\chinon$ to be the candidate for cold dark matter, the parameter
space can be further substantially reduced.  In the future, the Large
Hadron Collider (LHC) will be an ideal machine to search for
Supersymmetry.  Chapter~\ref{chap:Kinematics} gives an introduction to
the basic kinematics used in this analysis.  The formulae and
configurations for the kinematic endpoints are explained.
Chapter~\ref{chap:Searches} is the main chapter where the analysis
methods and the results for $\chinonn$ searches at high $\tan \beta$
are presented. Many surveys have been made in several areas of the
mSUGRA parameter space which are accessible at the LHC. It has been
shown that the leptonic decay of the next-to-lightest neutralino,
\begin{equation}
\chinonn \rightarrow l^{\pm}  +
\tilde{l}^{\mp}  \rightarrow l^{\pm}  +   l^{\mp} + \chinon \ ,\nonumber
\end{equation}
has a useful kinematic feature: The dilepton invariant mass spectrum
has a sharp edge near the kinematic upper limit.  This feature was
first discussed in \cite{FirstEP}.  If the decay into electrons and
muons has a sufficient branching ratio, the endpoint can be measured
very precisely, and therefore can be used for sparticle mass
reconstructions.  In some regions of the parameter space, however, the
decay into taus dominates.  At the benchmark point I' with $\tan \beta
= 35$, $A_{0} = 0$, $\mu > 0$, $m_{0} = 181$ and $m_{1/2} = 350$
\cite{wmappara} $96\%$ of all $\chinonn$ decays are into taus. Therefore,
methods introduced for the dileptonic decay of the $\chinonn$ are
applied to the tau channel.  Since the endpoint is always smeared out
due to the presence of neutrinos in the final states, the existence of
a correlation between the different endpoints and the distribution
maxima has been investigated. For distributions where the ratio
between both values shows a linear dependence, the measurement
precision can be improved. The single rho with the associated quark
channel offers two endpoints which can be well estimated in most of
the cases without knowing whether the corresponding tau comes from the
slepton or the $\chinonn$. Additionally, it turned out that by using
more endpoints than sparticle masses the determination of the
sparticle masses can be improved significantly. The outlook in chapter
\ref{chap:Outlook} summarises problems and ideas which have not been
treated in this work.
\label{chap:introduction}
 \newpage
 \cleardoublepage

 \chapter{Supersymmetry}\label{chap:Supersymmetry}
 The Standard Model successfully passed all particle physics experiments
in the last 30 years.  Especially the LEP and SLC experiments have
tested its predictions with high precision.  Indeed, there are
experiments like the measurements of the anomalous magnetic moment of
the muon in Brookhaven \cite{g2} and the WMAP measuring the cold dark
matter density in our universe (see section \ref{sec:wmap}) which may
give a hint that physics beyond the Standard Model exists. But they do
not exclude it. Thus, the motivation for introducing Supersymmetry is
based on theoretical aspects.

In nature it is often possible to explain its principles and make
predictions using symmetries.  Since the discovery of the Pauli
Principle, particles are divided in two groups, the fermions whose spin
is an odd multiple of $\frac{1}{2}$ and the bosons with integer spin.
A supersymmetric operation on the representation of a particle
belonging to one group transforms it to a member of the other
group. Thus, the spin changes but the mass and the couplings to other
particles remain the same.  The electron with spin $\frac{1}{2}$ and a
mass of $0.511 \mev$ would have a bosonic partner with spin 0 and the
same mass in a supersymmetric world.  However, no supersymmetric
partner has been discovered until now. It is therefore certain that - if
Supersymmetry exists at a high energy scale - it is broken at the
electroweak scale which is at the moment accessible for experiments.

\section{MSSM}
Particles and their interactions are described by Quantum Field
Theories.  The fundamental principles of nature are described through
particles and their interactions whereas also the interactions are
treated as particles in such theories.  A quantum field is a complex
n-dimensional function $\psi(x)$ representing one or more particles.
$\psi(x)$ has the property that $|\psi(x)|^2$ gives the probability to
find that particle at the space-time point $x$.  Since only
$|\psi(x)|^2$ is measurable, the phase of $\psi(x)$ is a free
parameter. It is natural to assume that at any point in the universe
the phase can be chosen differently without changing the particle's
physical state:
\begin{equation}
  \label{eq:example-u1-gauge-transformation}
  \psi(x) \ra e^{i\theta(x)}\psi(x).
\end{equation}
This is the basis of all gauge theories like the Standard Model and
its minimal supersymmetric extension. The direct consequence is that
gauge bosons are responsible for the interactions between fermions.
In the Standard Model the electroweak interactions result from a local
$SU(2)_{L} \times U(1)_{Y}$ symmetry and the strong interactions from
a local $SU(3)$ symmetry. The interactions within the MSSM are based
on the same symmetries.
%-----------------------------------------------------------------

The MSSM is minimal in the sense that it is the simplest possible
supersymmetric extension of the Standard Model with a minimal particle
content. The MSSM has an additional second Higgs doublet, describes
more than twice as many particles than the Standard Model and solves
the following problems:
\begin{enumerate}
\item One of the main problems of the Standard Model is the
\emph{Hierarchy Problem}. The radiative corrections to the
Higgs mass including the fermion loops at one-loop-level has a
unphysical quadratic divergence. Indeed, this divergence can be
canceled by a mass counter term which, however, has to be fine tuned
at each order in perturbation theory. In the MSSM, two additional
scalar particles $S$, provided they are in the region of $1-10\tev$,
are responsible for the cancellation of that divergence. The remaining
divergence is then only logarithmic, multiplied with ($m_{S}^{2}
- m_{f}^{2}$) being zero in the supersymmetric limit.
\item Within the Standard Model, it is not possible to achieve a
\emph{unification of gauge couplings} - at any scale - which reduces the
number of open parameters, and is therefore favoured by many
theorists. The MSSM provides the possibility of gauge coupling
unification at a scale of about $10^{16}\gev$.
\item In some models within the MSSM there exists a \emph{candidate
for cold dark matter}. Experiments measuring the rotation velocities
of galaxies have shown that it is very likely that there is more than
just visible matter in our universe. One component of the so called
dark matter seems to be cold, meaning that it consists of particles
moving at a non-relativistic speed. Indeed, the Standard Model axion can
be responsible for cold dark matter. But, it has not been discovered
until now, and it is not clear whether it can explain the measured cold
dark matter density sufficiently
\cite{axion}. If mSUGRA is the model describing the Supersymmetry
breaking in nature and if the lightest supersymmetric particle is a
neutralino, it would be an excellent candidate for cold dark matter.
\item In the MSSM the mechanism of \emph{electroweak symmetry
breaking} occurs radiatively at the level of perturbation theory,
without the need for any new strong interaction. Thus, in contrast to
the SM the MSSM gives an explanation for the Higgs mechanism.
\end{enumerate}
However, the MSSM gives neither a hint to the source of CP
violation nor is it able to make predictions for any particle
mass. Furthermore, the MSSM still is a theory without quantum
gravity.  String theories may solve some of these problems, although a
phenomenologically viable string theory has yet to be constructed
\cite{MSSMworks}.

There are two different kinds of superfields in the
Minimal Supersymmetric Standard Model at the GUT-scale and above:
\begin{enumerate}
\item The \emph{Chiral Scalar Superfields} consisting of a complex
scalar field,\ $S$, and a 2-component Majorana fermion field,\ $\zeta$.
The SM quarks and leptons and the MSSM higgsinos at the$\tev$-scale are
connected with $\zeta$ and its partners, the squarks, sleptons and Higgses,
with $S$.
\item The \emph{Massless Vector Superfields} consisting of a gauge
field $F^{A}_{\mu\nu}$ and a 2-component Majorana fermion 
field,\ $\lambda_{A}$. The field $F^{A}_{\mu\nu}$ is connected with
the Standard Model gauge bosons and $\lambda_{A}$ with its
Superpartners, the gauginos, at the$\tev$-scale.
\end{enumerate}
These fields respect the same $SU(3) \times SU(2)_{L} \times U(1)_{Y}$
gauge symmetries as do the Standard Model fields. In order to confirm
or confute the predictions of the MSSM it is first necessary to find
the new particles meaning to find the mass peaks of the mass
eigenstates (Column \emph{Mass} in Table \ref{tab:MSSMparticles}). In
the MSSM every supersymmetric particle has the same gauge couplings
than its Standard Model partner.  Thus, the second step would be to
measure the interaction couplings and the spin of the new particles.
\begin{table}[H]
\begin{center}
\begin{tabular*}{\textwidth}{@{\extracolsep{\fill}}|c|c|c||c|c|c|} \hline\hline
\multicolumn{6}{|c|}{\emph{MSSM Particles}}\\\hline\hline
\multicolumn{3}{|c}{Extended Standard
Model} & \multicolumn{3}{c|}{Supersymmetric Partners}\\\hline
\emph{Interaction} & \emph{Mass} & \emph{Spin} & 
\emph{Interaction} &\emph{Mass} &\emph{Spin} \\\hline 
$e^{\pm}_{L/R}, \mu^{\pm}_{L/R}, \tau^{\pm}_{L/R}$ & 
$e^{\pm}, \mu^{\pm}, \tau^{\pm}$ & $1/2$ & 
$\tilde{e}^{\pm}_{L/R},\tilde{ \mu}^{\pm}_{L/R}, \tilde{\tau}^{\pm}_{L/R}$ & 
$\tilde{e}^{\pm}_{1/2}, \tilde{\mu}^{\pm}_{1/2}, \tilde{\tau}^{\pm}_{1/2}$ & $0$\\

$\nu_{e}, \nu_{\mu}, \nu_{\tau}$ &
$\nu_{e}, \nu_{\mu},\nu_{\tau}$ & $1/2$ & 
$\tilde{\nu}_{e}, \tilde{\nu}_{\mu},\tilde{\nu}_{\tau}$ &
$\tilde{\nu}_{e}, \tilde{\nu}_{\mu},\tilde{\nu}_{\tau}$ & $0$\\\hline

$u^{'}_{L/R},c^{'}_{L/R},t^{'}_{L/R}$ &
$u,c,t$ & $1/2$ & 
$\tilde{u}^{'}_{L/R},\tilde{c}^{'}_{L/R},\tilde{t}^{'}_{L/R}$ & 
$\tilde{u}_{1/2},\tilde{c}_{1/2},\tilde{t}_{1/2}$ & $0$\\

$d^{'}_{L/R},s^{'}_{L/R},b^{'}_{L/R}$ &
$d,s,b$ & $1/2$ & 
$\tilde{d}^{'}_{L/R},\tilde{s}^{'}_{L/R},\tilde{b}^{'}_{L/R}$ &
$\tilde{d}_{1/2},\tilde{s}_{1/2},\tilde{b}_{1/2}$ & $0$\\\hline

$B^{0},W^{0}$ &
$\gamma, Z^{0}$ & $1$ &
$\tilde{B}^{0},\tilde{W}^{0}$ &
 & $1/2$ \\

$H^{0}_{u}, H^{0}_{d}$ &
$h, H, A^{0}, A^{\pm}$ & $0$ &
$\tilde{H}^{0}_{u}, \tilde{H}^{0}_{d}$ &
\raisebox{1.1ex}[0cm][0cm]{$\chinon,\chinonn,\chinonnn,\chinonnnn$} & $1/2$ \\\hline

$W^{\pm}$ &
$W^{\pm}$ & $1$ &
$\tilde{W}^{\pm}$ &
 & $1/2$ \\

$H^{\pm}$ &
$H^{\pm}$ & $1$ &
$\tilde{H}^{\pm}$ &
\raisebox{1.1ex}[0cm][0cm]{$\chionepm,\chitwopm$} & $1/2$ \\\hline

$g$ &
$g$ & $1$ &
$\tilde{g}$ &
$\tilde{g}$ & $1/2$\\\hline\hline
\end{tabular*}
\caption{This is the particle spectrum of the MSSM. The extended
Standard Model has instead of only one an additional second Higgs
doublet. The labels ``L'' and ``R'' denote the left- and right-handed
electroweak eigenstates. The mass eigenstates $x_{1/2}$ are defined as
$x_{1/2} = x_{L/R}\cdot \cos \theta \pm x_{R/L}\cdot \sin
\theta$. Since $\theta$ is close to zero for electrons, muons, up, down, charmed and
strange quarks, the mass eigenstates are mostly denoted with ``L/R''
instead of ``1/2'', in the literature.}\label{tab:MSSMparticles}
\end{center}
\end{table}

 \section{Superpotential and R-parity}\label{sec:RParity} In the SM, the
effects due to the masses of quarks and leptons are parametrised by
the Yukawa potential. The Yukawa couplings in the MSSM are specified
by means of a function called superpotential. The general
superpotential of the MSSM contains terms where the baryon and the
lepton numbers are violated.  These terms have significant
experimental restrictions especially from measurements of the proton
decay and the double beta decay. There are two ways to make the theory
describing the data. First, the involved parameters can be tuned such
that they are in accordance with experimental results.  Alternatively,
by introducing a new symmetry this problem is solved.  This symmetry
and its corresponding transformation is well explained in
\cite{MSSMworks} and \cite{SusyPrimer}.  It leads to a multiplicative
quantum number, the R-parity, which by convention is $1$ for all
Standard Model particles and $-1$ for their superpartners. It is
given by
\begin{equation}
  \label{eq:RParity}
   R = {(-1)}^{3(B-L) + 2s}
\end{equation}
for a particle with baryon number B, lepton number  L and spin s. This
symmetry has an important influence on the MSSM phenomenology since it
requires  that supersymmetric particles always  are  produced in pairs
and  that no supersymmetric   particle can decay into solely  Standard
Model particles. In particular, the lightest supersymmetric particle
is stable.

\section{mSUGRA} 
\begin{figure}[H]
  \begin{center} \includegraphics[width=12cm]{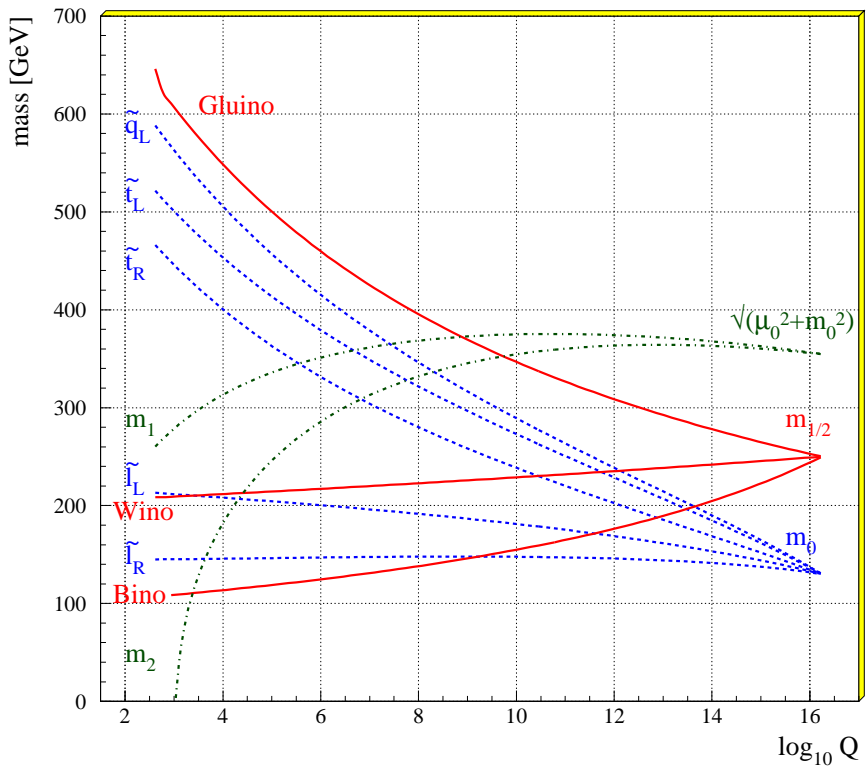}
  \end{center} \caption{Evolution of the particle masses in mSUGRA
  through the Renormalisation Group equations (RNG) as a function of
  the energy scale. The mass spectrum is reduced to three distinct
  values at the GUT scale, whereas the particle spectrum at the
  experimentally accessible scale is split. In mSUGRA the gluino is
  always the heaviest gaugino.}  \label{mSUGRA}
\end{figure}

In mSUGRA it is assumed that the three interactions are unified at the
Grand Unification Theory (GUT) scale ($10^{16}\gev$) and that
Supersymmetry is broken due to gravity.  The beauty of mSUGRA lies in
the fact that this model needs only four more parameters and the sign
of a fifth additional parameter with respect to the known free
parameters in the Standard Model.  If mSUGRA describes the breaking of
the Supersymmetry correctly, all scalar particles like sfermions and
Higgs bosons have a common mass $m_{0}$ at the GUT scale.  Also the
gaugino masses $M_{1}$, $M_{2}$ and $M_{3}$, corresponding to the
U(1), SU(2) and SU(3) gauge symmetries, unify to a common mass
$m_{1/2}$ at the GUT scale which can be seen in figure \ref{mSUGRA}.
All trilinear coupling parameters $A_{ijk}$ have the same value
$A_{0}$ at $M_{GUT}$, and the ratio of the vacuum expectation values of
the two Higgs doublets is given by $\tan \beta$.  The magnitude of the
Higgsino mass parameter $\mu$ is given by the four mSUGRA parameters
but its sign is the fifth open parameter.  In mSUGRA with conserved
R-parity (see section~\ref{sec:RParity}) the lightest supersymmetric
particle (LSP) has to be stable.  In some parts of the mSUGRA
parameter space the $\tilde{\tau}_{1}^{\pm}$ is the LSP.  However, a
charged stable supersymmetric particle is cosmologically excluded. In
other parts, the lightest neutralino is the LSP. This neutral weakly
interacting particle is an ideal candidate for cold dark matter. If
the latter is a concentration of lightest neutralinos, the WMAP data
(see section~\ref{sec:wmap}), giving constraints on the cold dark
matter density in the universe, can be used to restrict the mSUGRA
parameter space. Very often, this restriction is shown in the $m_{0} -
m_{1/2}$ plane for different values of $\tan \beta$, $A_{0}$ and
sign($\mu$).

One of the challenges of the LHC is to decide whether mSUGRA is the
right theory for the description of nature. If this is the case, the
next step will be to determine the values of the five parameters.

 \newpage
 \cleardoublepage
 
\chapter{Monte Carlo generators}\label{chap:MCG}
 At LHC, protons - which are made of quarks and gluons - will be
accelerated to $7\tev$, and will collide at different interaction
regions. These collisions will produce leptons together with quarks
and gluons, both fragmenting into jets, in the final state. In most of
the cases it is not possible to calculate analytically the Quantum
Field theory based production cross sections and 4-momenta of
particles produced in such collisions.  Therefore, Monte Carlo
generators, generally used for statistical models in science, are a
useful tool for simulations in particle physics.  In order to estimate
the sensitivity of analysis methods, which will be utilised after
having collected the first real data, simulations of proton-proton
collisions at the LHC are necessary. Here, the first step is to
generate quark-quark, quark-gluon and gluon-gluon collisions with a
Monte Carlo generator. The particle distribution functions which
describe how the quarks and gluons are distributed within the protons
are measured at HERA and by other experiments. The choice of the
special Monte Carlo generator depends on the physics model for which
sensitivity studies are made.  The output at this level are the
precise 4-momenta of all particles in such an event. Then detector
effects and effects due to event reconstruction have to be included.

\section{ISASUGRA} 
In this study ISASUGRA~7.69 is used which is part of
ISAJET~7.69~\cite{isajet}. It calculates numerically the masses, the
branching ratios and the weak-scale mass parameters within the mSUGRA
model. The particle spectrum calculated by ISASUGRA can be used as an
input for PYTHIA~\cite{pythia}. The cross sections are evaluated by
ISAJET or PYTHIA. ISASUGRA takes the following values as input: The
choice of the mSUGRA parameters and the top mass. The top mass is set
to $175\gev$ in this study which is the same value used by most of the
recent publications of mSUGRA studies. Presently, the updated
value is $174.3 \pm 5.1\gev$~\cite{topmass}. Furthermore, $A_{0}$ is
set to zero, $\mu$ is positive and the value for $\tan \beta$ is
35. The values for $m_{0}$ and $m_{1/2}$ and the motivation for the
parameter choice are given in chapter~\ref{chap:Searches}.

 \section{PYTHIA} PYTHIA is a program for the Monte Carlo generation of
high energy physics events arising from collisions between elementary
particles such as $e^{+}$, $e^{-}$, $p$ and $\bar{p}$ in various
combinations.  It is based on theories and models for a number of
physics processes, including hard and soft interactions, parton
distributions, initial and final state parton showers, multiple
interactions, fragmentation and decay. The version used in this study
is the corrected (See appendix \ref{sec:bug}) version of PYTHIA
6.220. In PYTHIA it is possible to include new particles, cross
sections, processes, and also to specify the decays of particles. In
order to save CPU time, the taus in this study have been forced to
decay solely into charged rhos. The production cross section
calculated by PYTHIA did not change, with respect to the cross section
with all tau channels included, due to that adjustment (See appendix
\ref{sec:cs}). Therefore, in order to normalise all distributions to
30 $fb^{-1}$, the PYTHIA 6.220 branching ratio of 24.9 $\%$ for
$\tau^{\pm} \ra \rho^{\pm} \nu_{\tau}$ has to be included. In this
study an event is defined as one produced $\chinonn$ decaying into two
$\tau$. Thus, the number of produced events has been multiplied by
the squared branching ratio.

 \newpage
 \cleardoublepage

\chapter{Experiments}\label{chap:Experiments}
 \section{Large Electron Positron Collider}\label{sec:LEP} The
operation of the Large Electron Positron (LEP) collider at CERN
started in August 1989. With a circumference of 27 km, LEP was the
largest accelerator in the world. It stopped in November 2000, but the
analysis of data is still going on. LEP accelerated electrons and
positrons - in opposite directions in a vacuum pipe inside a retaining
ring of magnets - which collided head-on at different interaction
points. The four LEP experiments were called ALEPH, DELPHI, L3 and
OPAL.

\begin{figure}[H]
  \begin{center} \includegraphics[width = \textwidth]{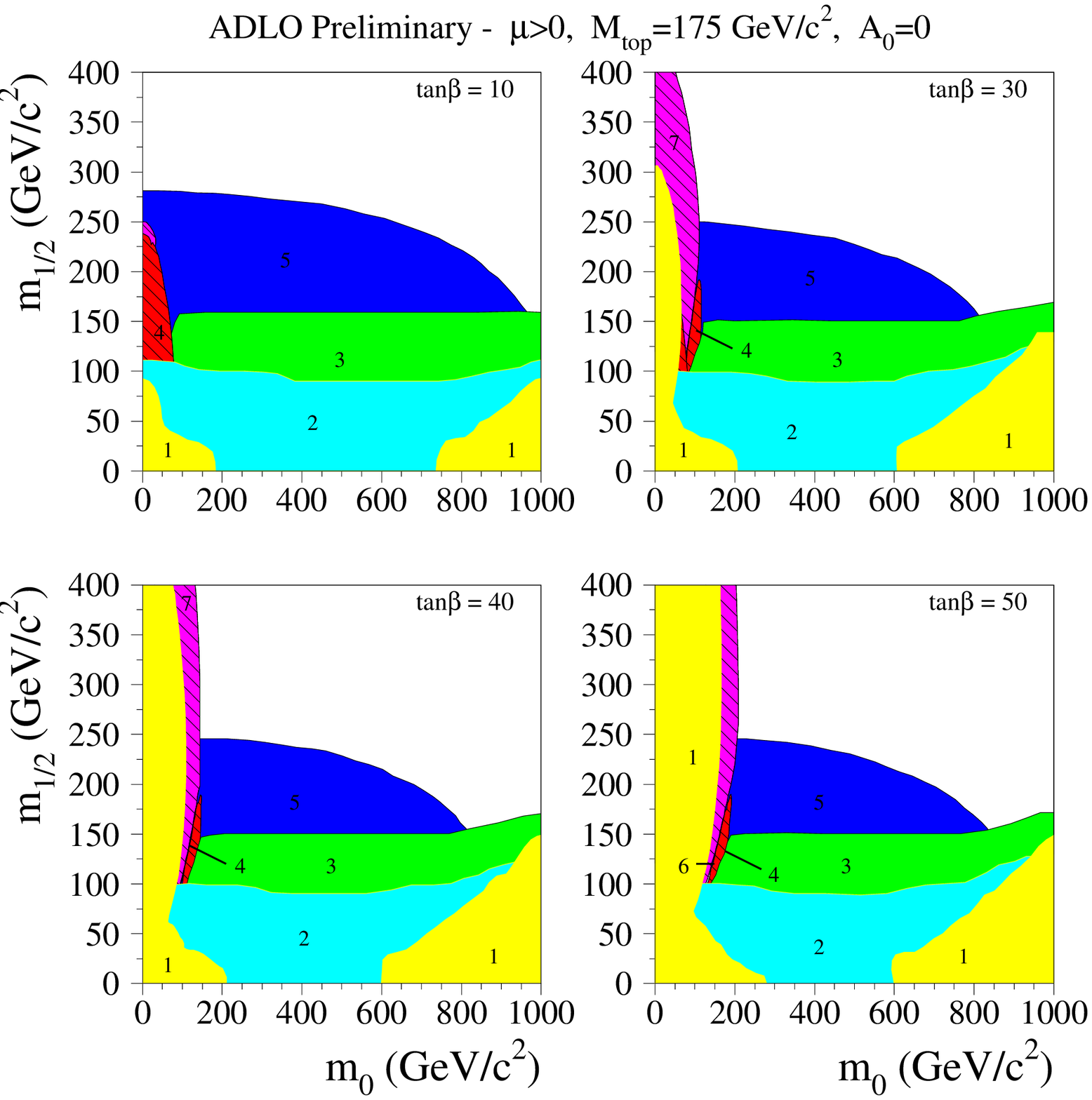}
  \end{center} \caption{The excluded regions in the
  $m_{0}$-$m_{1/2}$-plane through measurements with LEP for $\mu > 0$
  and $A_{0} = 0$. Yellow (1): no mSUGRA solution: no EWSB or
  tachyonic particles; Light blue (2): regions inconsistent with the
  measurement of the electroweak parameters at LEP1; Green (3):
  regions excluded by chargino searches; Red (4): regions excluded by
  selectron or stau standard searches; Dark Blue (5): regions excluded
  by the search for hZ; Brown (6): regions excluded by the neutralino
  stau cascade searches.  Magenta (7): regions excluded by the search
  for heavy stable charged particles applied to staus \cite{lepsusy}.}
  \label{LEPsusy1}
\end{figure}

\begin{figure}[H]
  \begin{center} \includegraphics[width = \textwidth]{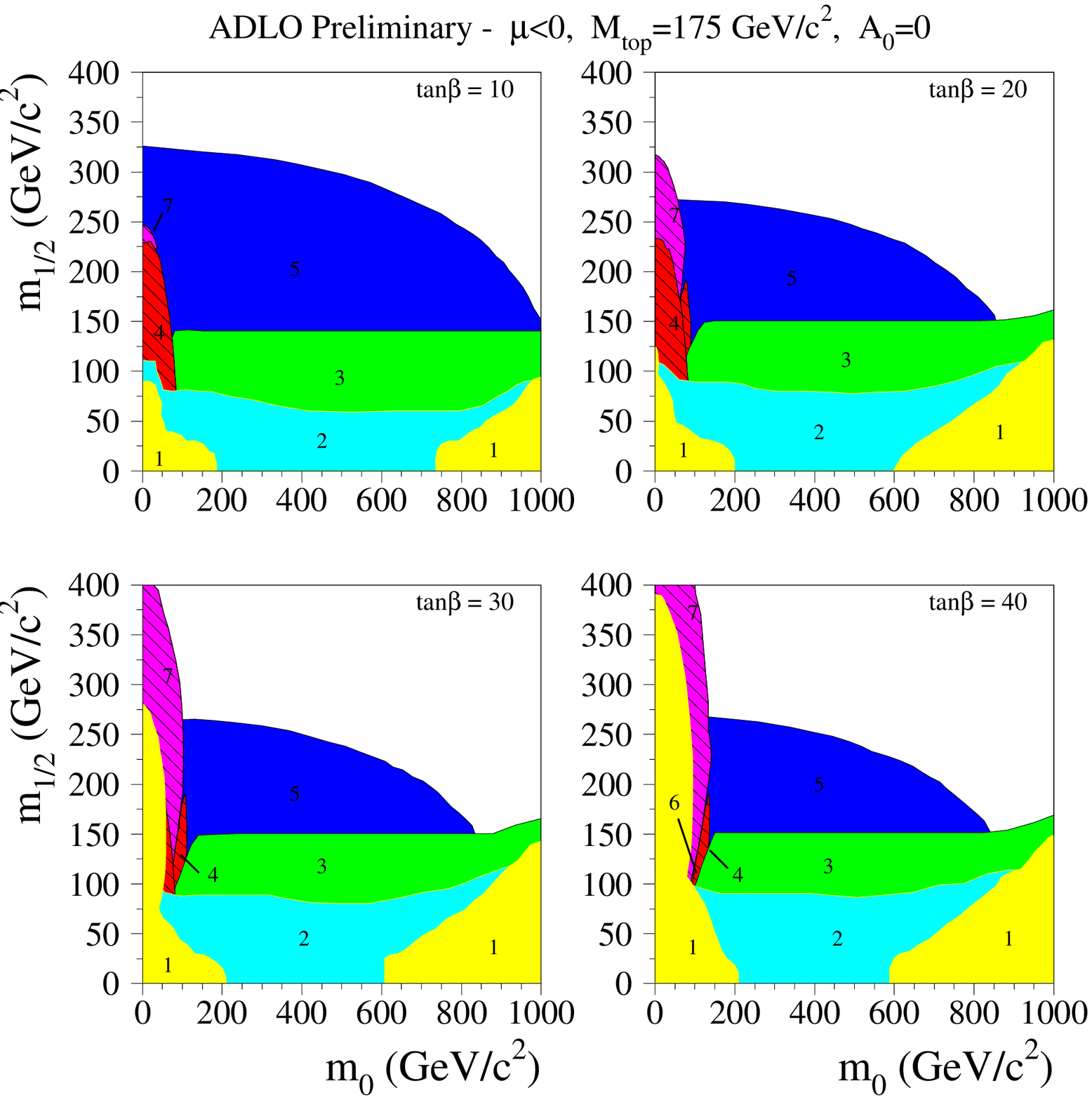}
  \end{center} \caption{The excluded regions in the
  $m_{0}$-$m_{1/2}$-plane through measurements with LEP for $\mu < 0$
  and $A_{0} = 0$. Yellow (1): no mSUGRA solution: no EWSB or
  tachyonic particles; Light blue (2): regions inconsistent with the
  measurement of the electroweak parameters at LEP1; Green (3):
  regions excluded by chargino searches; Red (4): regions excluded by
  selectron or stau standard searches; Dark Blue (5): regions excluded
  by the search for hZ; Brown (6): regions excluded by the neutralino
  stau cascade searches.  Magenta (7): regions excluded by the search
  for heavy stable charged particles applied to staus \cite{lepsusy}.}
  \label{LEPsusy2}
\end{figure}

The MSSM predicts a light Higgs with $M_{h^{0}} \le 130\gev$, and with
LEP a lower constraint of $M_{h^{0}} \ge 114\gev$ was obtained. In
some parts of the mSUGRA parameter space, especially for low values of
$m_{0}$ and $m_{1/2}$, LEP should have produced enough supersymmetric
particles necessary for a discovery. However, there was no evidence
for Supersymmetry at LEP. Thus, a value for $\tan \beta$ lower than
2.4 has been excluded \cite{lepsusy2}. The LEP Susy Working Group has
obtained preliminary results which are shown in figure~\ref{LEPsusy1}
and \ref{LEPsusy2}. These results are based on
\begin{enumerate}
\item a combination of $E_{CM} = 183-208\gev$ ALEPH, DELPHI, L3 and OPAL searches
for pair production of $\tilde{e}$ in the $\tilde{e} \ra e \chinon$
channel,
\item a combination of  $E_{CM} = 183-208\gev$ ALEPH, DELPHI , L3 and OPAL
searches for pair production of $\tilde{\tau}$ in the $\tilde{\tau}
\ra \tau \chinon$ channel,
\item a combination of the $E_{CM} =  205-208\gev$ ALEPH, L3 and OPAL searches
for pair production of $\chionepm$,
\item the $E_{CM} \le 202 \gev$ ALEPH searches for pair production of
heavy stable charged particles,
\item the DELPHI and ALEPH searches for neutralinos decaying into
$\tilde{\tau}$ and $\tau$,
\item a combination of $E_{CM} < 209 \gev$ ALEPH, DELPHI , L3 and OPAL searches
for the lightest scalar neutral Higgs in the $e e \ra h Z$ reaction.
\end{enumerate}


\section{Wilkinson Microwave Anisotropy Probe}\label{sec:wmap}
\begin{figure}[H]
  \begin{center}
    \includegraphics[angle=270,width = \textwidth]{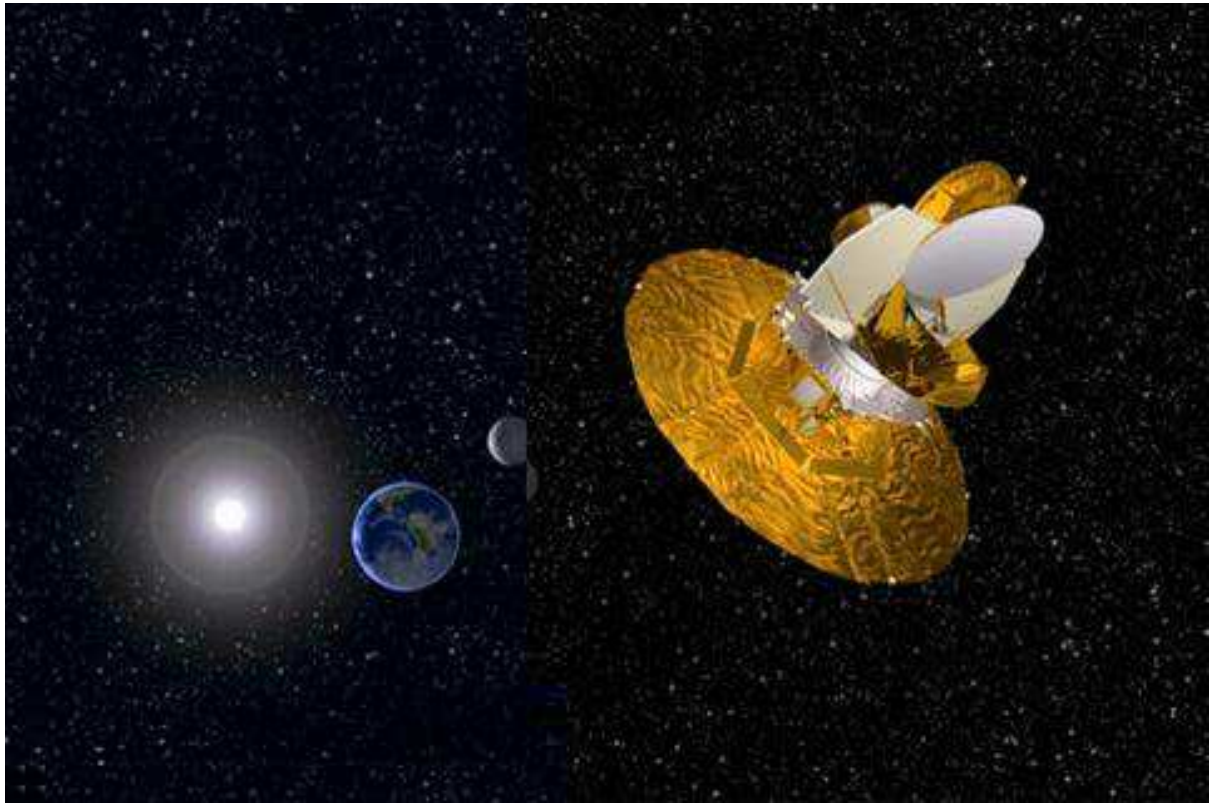}
  \end{center}
  \caption{A view of the WMAP satellite in the universe \cite{wmappic}.}
  \label{wmappicture}
\end{figure}
The cosmic microwave background (CMB) radiation is the radiant heat
left over from the Big Bang.  It was first observed in 1965 by Arno
Penzias and Robert Wilson at the Bell Telephone Laboratories in Murray
Hill, New Jersey \cite{CMB_PW}.  The properties of the radiation give
information about the physical conditions in the early universe, and
can therefore be used for a better understanding of the formation of
cosmic structures such as galaxies and dark matter.  As a consequence,
assuming that cold dark matter is made of neutralinos, the CMB
radiation data leads to restrictions on the allowed mSUGRA parameters.
Thus, this data is important for making constraints on the mSUGRA
parameter space.

If the $\chinon$ were abundantly produced directly after the Big Bang
they would have had mostly annihilated to the relic density which has
been measured with WMAP. In neutralino relic density calculations for
mSUGRA there is a 'bulk' region of relatively low values of $m_{1/2}$
and $m_{0}$, where neutralinos annihilate dominantly through t-channel
slepton exchange ($\chinon\chinon \ra l\ \bar{l}$). For a large
neutralino mass, the relic density is generally large. However,
neutralino-stau coannihilation ($\chinon\tilde{\tau}
\ra \tau V$) can reduce the relic density to satisfy the cosmologically
favored relic density bounds. The region where this is possible
($M_{\tilde{\tau}} - M_{\chinon} \ll M_{\chinon}$) has a highly
fine-tuned relic density since a small variation in $m_{0}$ would lead
to a large change in $\Omega_{LSP} h^{2}$.  Furthermore, there is a
region where rapid annihilation through s-channel Higgs boson
resonances ($\chinon\chinon \ra H
\ra b\bar{b},\tau\bar{\tau}$) can reduce the neutralino relic
density to the measured region. Finally, in the focus point region at
large $m_{0}$ - where the $\chinon$ is Higgsino like - the relic
density is achieved mainly through the $\chinon\chinon \ra W^{+}W^{-}$
and the $\chinon\chinon \ra ZZ$ channels.
\begin{figure}[H]
  \begin{center} \includegraphics[angle=270,width=\textwidth]
  {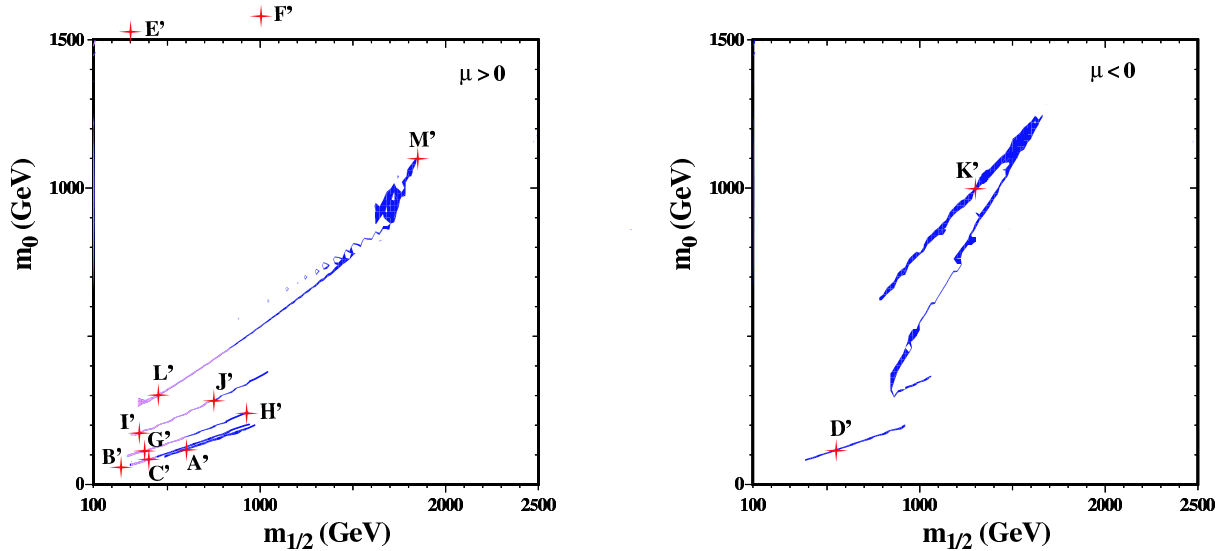} \end{center} \caption{The shaded strips
  display the regions of the ($m_{1/2}$,$ m_{0}$) plane that are
  compatible with $0.094 < \Omega_{LSP} h^{2} < 0.129$ at the
  2-$\sigma$ level where $\Omega_{LSP} h^{2}$ is the normalised
  density. These areas are in the bulk (B',C',G',I',L'),
  coannihilation tail (A',D',H',J') and rapid-annihilation funnel
  (K',M') regions. Left plot: $\mu > 0$ and $\tan
  \beta$\ = 5 (A'), 10 (B',C'), 20 (G',H'), 35 (I',J') and 50
  (L',M'). Right plot: $\mu < 0$ and $\tan \beta$\ = 10
  (D') and 35 (K'). The parts of these WMAP lines for $\mu > 0$
  compatible with the $g_{\mu}$-2 measurements at BNL \cite{bnl} at
  the 2-$\sigma$ level have lighter (pink) shading \cite{wmappara}.}
  \label{wmapconstraints}
\end{figure}
The Wilkinson Microwave Anisotropy Probe (WMAP), shown in
figure~\ref{wmappicture}, is a NASA satellite, and presently the
best experiment to measure the CMB radiation and its spatial
fluctuations. Launched in 2001, WMAP is now able to resolve small
temperature fluctuations of the order of only millionths of a degree.
Instead of measuring the absolute temperature values of the CMB
radiation WMAP measures the temperature differences between two
different points in the sky, taking advantage of the higher accuracy
of differential measurements.  Figure \ref{wmapconstraints} shows that
the region in the two-dimensional $m_{1/2}$-$ m_{0}$ plane consistent
with experimental data is reduced to nearly one dimensional strips due
to the WMAP measurements. Therefore, the WMAP experiment gives a hint
where we have to focus the searches for mSUGRA.

\section{Large Hadron Collider}
\begin{figure}[h]
  \begin{center} 
	\includegraphics[angle=270,width=\textwidth]{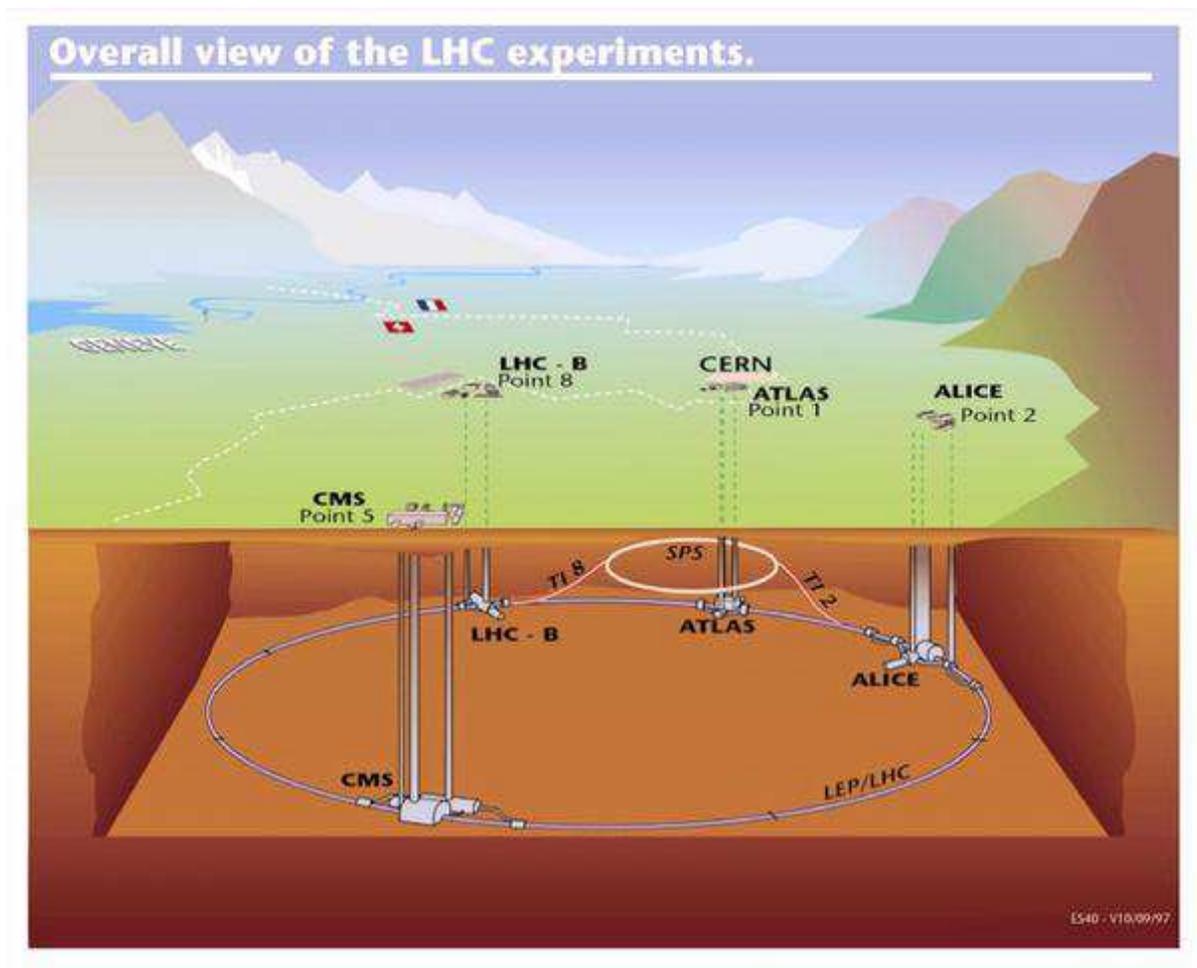}
  \end{center} \caption{A schematic view of the LHC at CERN near
  Geneva \cite{lhcpic}.}  \label{lhc}
\end{figure}

This study concerns physical processes that might happen at the Large
Hadron Collider (LHC). The construction of the LHC has been approved
by the CERN's Council in December 1994. The LHC, shown in figure
\ref{lhc}, is the successor of the Large Electron Positron collider
(LEP), and will be installed in the 27 kilometre circumference
LEP tunnel being about 100 metre underground.
The LHC will operate at a center-of-mass energy $\sqrt{s}$ of $14\tev$
being the highest ever reached in the world.  The design luminosity is
$2.3 \cdot {10}^{34}\ {cm}^{-2}{s}^{-1}$.  Furthermore, lead-ion
collisions are planned with a total $\sqrt{s}$ of $1148\tev$ and a
luminosity of $1.0 \cdot {10}^{27}\ {cm}^{-2}{s}^{-1}$.  The
accelerator consists of two interleaved synchrotron rings, whose main
elements are 1232 superconducting NbTi dipole and 392 quadrupole
magnets operating in superfluid helium at a temperature of \mbox{1.9
K}.  The dipole magnets guide the particles along the ring while the
quadrupole magnets are focusing the particle bunches.  The nominal
field for LHC magnets to handle $7\tev$ beams is $8.34$ Tesla, and the
goal is to achieve $9.0$ Tesla
\cite{CernCourier}. Two $450\gev$ beams  coming from  the Super Proton
Synchrotron (SPS) will be injected into the LHC, and then
accelerated with superconducting cavities, 8 cavities per ring,
operating at 2 MV and 400 MHz.

\begin{figure}[h]
  \begin{center} \includegraphics[width=\textwidth]{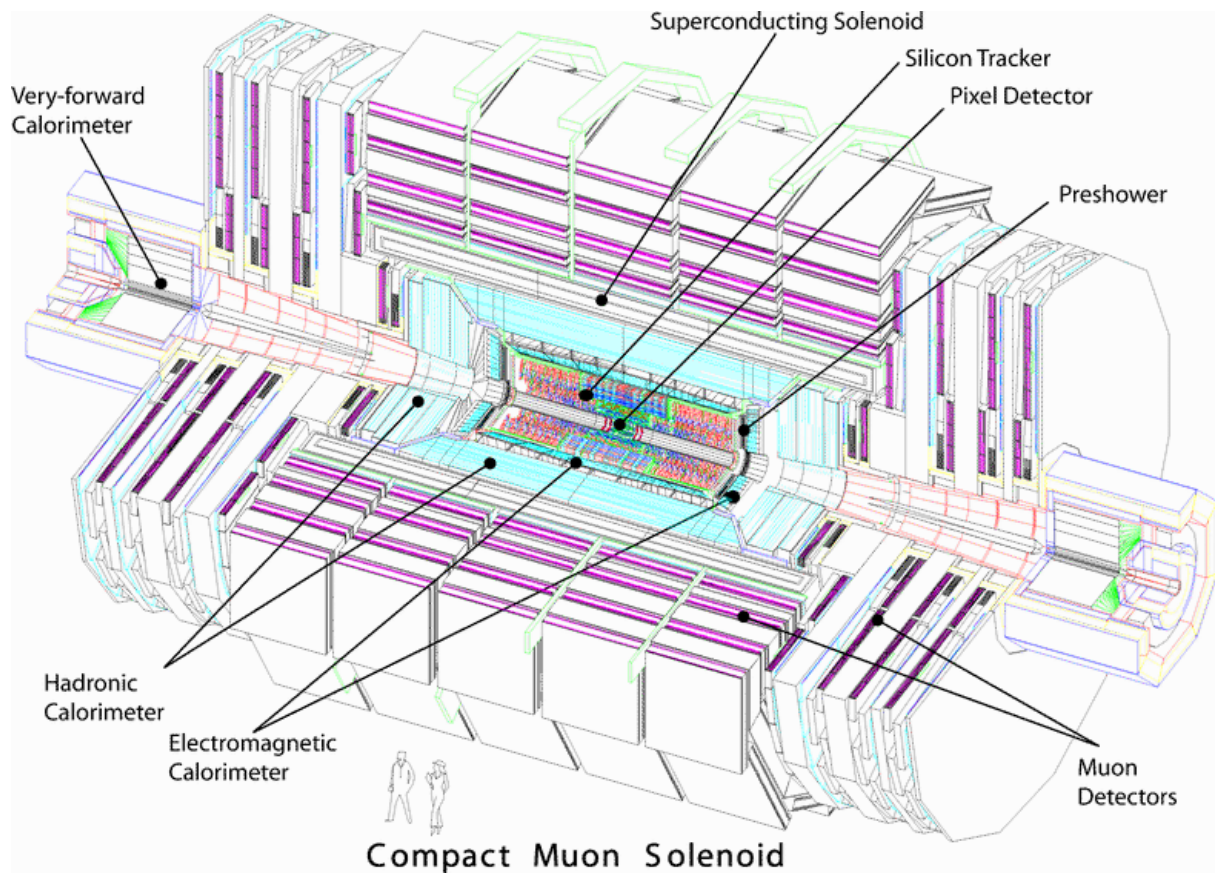}
  \end{center} \caption{A schematic view of the CMS detector. The
  pixel detector is closest to the interaction vertex. The second
  innermost detector is the silicon tracker followed by the crystal
  electromagnetic calorimeter and the hadronic calorimeter. Outside of
  the superconducting solenoid which generates a magnetic field of 4
  Tesla there are the muon chambers. The total weight is 12500 T which
  is nearly twice as much as the weight of the Tour Eiffel.} 
  \label{cms}
\end{figure}

The experimental program of the LHC will include the dedicated
heavy-ion detector ALICE (A Large Ion Collider Experiment), the
specialised B-physics spectrometer LHCb and the general purpose
detectors CMS (Compact Muon Solenoid) and ATLAS (A Toroidal LHC
ApparatuS).  The physics goals of CMS (Figure \ref{cms}) and ATLAS
include searches for the Higgs boson, which is predicted by the
Standard Model, as well as for supersymmetric particles which appear
in Supersymmetry theories like mSUGRA or the Gauge Mediated Susy
Breaking (GMSB) scenario. If the latter particles exist they will be
produced mainly through squark or gluino production at LHC. These
processes can lead to long decay chains which allow to give
constraints for all involved sparticle masses.

 \newpage
 \cleardoublepage
 
\chapter{\texorpdfstring{Kinematics for $\chinonn$
searches}{Kinematics for next-to-lightest neutralino searches}}\label{chap:Kinematics}
 At the Large Hadron Collider the colliding parton's center of mass and
exact center of mass energy is not known. Therefore, it is sensible to
concentrate on Lorentz invariant quantities like the invariant
mass. The invariant mass of n particles with energy $E_{i}$ and
3-momentum $p_{i}$ is frame independent, and given by
\begin{eqnarray}
M^{2} &=& \left( \sum_{i = 1}^{n} E_{i} \right)^{2} - \left( \sum_{i =
1}^{n} p_{i} \right)^{2} 
\label{im_n}
\end{eqnarray}
The lowest limit of $M$ is in any case the sum of all n particle rest
frame masses. If there is no external force the invariant mass of a
system of particles does not change, even if some particles decay or
annihilate. In the following calculations the speed of light $c$ and
Planck's constant $\hbar$ are set to one.

Invariant mass distributions carry information about the particles at
the end of a decay chain as well as about the decaying particles. For
this study, the kinematic formulae and configurations presented in
this chapter are relevant. In general it is necessary to distinguish
between massive and massless particles. If the energy of a particle
is a few magnitudes higher than its mass, it is a good estimation to
treat it as massless. Therefore, in the following calculations
\cite{pape} all Standard Model quarks and leptons - except the top
quark in section~\ref{sec:top} - are chosen to be massless. In the Monte
Carlo simulations, however, the correct masses are taken into account.
 
\section{Sequential two-body decays}\label{sec:twobody}
 Only two-body decays have to be taken into account for the processes
 in this study.  In the two-body decay of a particle $M$ into $m_1$
 and $m_2$, the energy, $E_i^*$, and 3-momentum, $p_i^*$, in the rest
 frame of $M$, which is denoted with a `` * '', are given by
\begin{eqnarray}
 M = E_1^* + E_2^* \;,\; 
 |p_1^*| = |p_2^*| = p^*.
\label{eq.gen.Epcons}
\end{eqnarray}
Squaring the first equation gives
\begin{eqnarray*}
 2 E_1^* E_2^* = M^2 - m_1^2 - m_2^2 -2p^{*2}.
\end{eqnarray*}
Squaring again and extracting $p^*$ leads to
\begin{eqnarray*}
 p^{*2} &=& \frac{1}{4M^2}[(M^2 - m_1^2 - m_2^2)^2 - 4 m_1^2 m_2^2] \nonumber \\
        &=& \frac{1}{4M^2}[M^4 - 2M^2(m_1^2+m_2^2) + (m_1^2-m_2^2)^2] \nonumber \\
        &=& \frac{1}{4M^2}[M^2 - (m_1+m_2)^2][M^2 - (m_1-m_2)^2].
\end{eqnarray*}
The energy becomes
\begin{eqnarray*}
 E_1^{*2} &=& p^{*2} + m_1^2 \nonumber \\
          &=& \frac{1}{4M^2}[M^4 - 2M^2(m_1^2+m_2^2) + (m_1^2-m_2^2)^2
	      + 4M^2 m_1^2] \nonumber \\
	  &=& \frac{1}{4M^2}[M^2 + m_1^2 - m_2^2]^2.
\end{eqnarray*}
In summary:
\begin{eqnarray}
 E_1^* &=& \frac{M^2 + m_1^2 - m_2^2}{2 M} \label{eq.gen.Ener} \\
 |p_1^*| = |p_2^*| &=&
 \frac{\left[ \left( M^2-(m_1+m_2)^2\right)\left( M^2-(m_1-m_2)^2\right)
       \right]^{1/2}}
      {2M}
\label{eq.gen.Mom}
\end{eqnarray}
The formulae are considerably simplified if one of the decay particles
is massless.  Assuming that $m_2 = 0$ one obtains:
\begin{eqnarray}
 E_1^* = \frac{M^2 + m_1^2}{2 M} \; , \;
 E_2^* = \frac{M^2 - m_1^2}{2 M} \; , \;
 |p_1^*| = |p_2^*| = \frac{M^2 - m_1^2}{2 M}.
\label{eq.gen.1massless}
\end{eqnarray}
The transformation from the rest frame of $M$ to the lab system is
defined by two kinematic variables of $M$, the velocity $\beta =
q/E$ and $\gamma = E/M$, where $q$ and $E$ are the lab frame momentum
and energy of $M$.  The Lorentz transformation to the lab system is
most easily decomposed into a longitudinal and a transverse part:
\begin{eqnarray}
 E_i   &=& \gamma \left[ E^*_i + \beta p^* cos \theta^*_i \right] \nonumber \\
 p_{Ti} &=& p_{Ti}^* = p^* sin \theta^*_i                           \nonumber \\
 p_{Li} &=& \gamma \left[ \beta E^*_i + p^* cos \theta^*_i \right]
\label{eq.gen.Lorenz}
\end{eqnarray}
with $\theta^*_2 = \theta^*_1 + \pi$.  The maximum and minimum values
of $E_i$ and $p_{Li}$ are obtained for $cos\theta^*_i = +1$ and $-1$
respectively. 

\begin{figure}[H]
\begin{center}
\includegraphics[width=\textwidth/2]{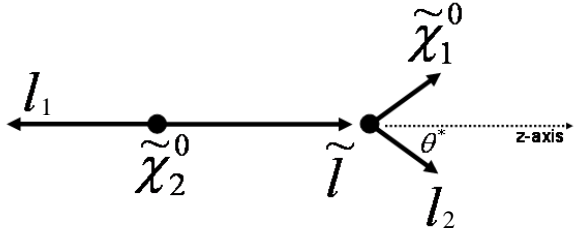}
\caption{Kinematics of the $\chi_2^{0}$ decay via a slepton $\tilde{l}$.}
\label{fig.ino21.decsferm}
\end{center}
\end{figure}

The investigated leptonic $\chinonn$ decay chain in the analysis
presented here is
\begin{eqnarray}
 \chinonn \rightarrow \tilde{l}^{\pm} + l^{\mp} \; , \;
\tilde{l}^{\pm} \rightarrow \chinon + l^{\pm}.
\end{eqnarray}
The calculations are valid for taus as well as for other leptons $l$,
which are assumed to be massless. Moreover, maxima and minima of
effective masses correspond to collinear configurations, where the
particles are emitted along or opposite to the direction of the
Lorentz boost.  In this case,
\begin{eqnarray}
 E_{l2}   = \gamma E^*_{l2} \left( 1 \pm \beta \right). 
\label{eq.onestepmassless.Lorenz}
\end{eqnarray}
As the Lorentz transformation parameters are
\begin{eqnarray}
  \beta = \frac{M_{\chinonn}^2 - M_{\tilde{l}}^2}{M_{\chinonn}^2 + M_{\tilde{l}}^2} \; , \;
  \gamma = \frac{M_{\chinonn}^2 + M_{\tilde{l}}^2}{2 M_{\chinonn} M_{\tilde{l}}},
\label{eq.onestepmassless.betagamma}
\end{eqnarray}
this yields to
\begin{eqnarray}
  \gamma \left( 1 \pm \beta \right) = \frac{1}{2 M_{\chinonn} M_{\tilde{l}}}
  \left[ (M_{\chinonn}^2 + M_{\tilde{l}}^2) \pm (M_{\chinonn}^2 -
  M_{\tilde{l}}^2) \right]
\end{eqnarray}
so that the maximum energies associated with the boost direction and
minimum energies associated with the opposite-to-the-boost direction
are:
\begin{eqnarray}
 E_{l2}^{max}   = \frac{M_{\chinonn}}{M_{\tilde{l}}} E^*_{l2} \; , \;
 E_{l2}^{min}   = \frac{M_{\tilde{l}}}{M_{\chinonn}} E^*_{l2}.
\label{eq.onestepmassless.Emaxmin}
\end{eqnarray}
Similarly, for the momentum with the sign being measured in the boost
direction it is
\begin{eqnarray}
 p_{l2}   = \gamma E^*_{l2} \left( \beta \pm 1 \right) 
\end{eqnarray}
leading to 
\begin{eqnarray}
 p_{l2}^{max}   = E_{l2}^{max} \; , \;
 p_{l2}^{min}   = - E_{l2}^{min}.
\label{eq.onestepmassless.Pmaxmin}
\end{eqnarray}
These expressions considerably simplify the later calculations. In the
following the kinematic limit for both leptons is calculated. The
effective mass can be computed in the rest frame of ${\chinonn}$. The
z-axis is chosen along the direction of the slepton, as depicted on
figure \ref{fig.ino21.decsferm}.
The dileptonic effective mass is given by
\begin{eqnarray}
 M_{ll}^2 &=& (E_{l1} + E_{l2})^2 - (-p_{l1}+p_{Ll2})^2 - (0 + p_{T{l2}})^2\\\nonumber
          &=& 2 E_{l1} E_{l2} + 2 p_{l1} p_{Ll2} = 2 E_{l1} (E_{l2} + p_{Ll2})
\label{eq.onestepmassless.Meff}
\end{eqnarray}
where the indices $L$ and $T$ denote the longitudinal and transverse
components.  As particle $l1$ is massless, the energy and momentum of
$l1$ and $\tilde{l}$ are given by the expressions
(\ref{eq.gen.1massless}) and similarly for particle $l2$ in the rest
frame of $\tilde{l}$.
\begin{eqnarray}
 E_{l2}^* = |p_{l2}^*| = \frac{M_{\tilde{l}}^2 - M_{\chinon}^2}{2 M_{\tilde{l}}}  
\end{eqnarray}
The Lorentz transformation of the massless particle $l2$ 
into the rest frame of $\chinonn$ becomes
\begin{eqnarray}
 E_{l2}    &=& \gamma E_{l2}^* ( 1 + \beta cos \theta^* )  \nonumber \\
 p_{Ll2} &=& \gamma E_{l2}^* ( \beta + cos \theta^* )
\end{eqnarray}
from which 
\begin{eqnarray}
 E_{l2} + p_{Ll2} = \gamma E_{l2}^* ( 1 + \beta )(1 + cos \theta^* ) \nonumber \\
 = \frac{M_{\chinonn}^2}{2 M_{\tilde{l}}} (M_{\tilde{l}}^2 - M_{\chinon}^2 ) (1 + cos \theta^* ).
\end{eqnarray}
The effective mass then takes the simple expression
\begin{eqnarray}
 M_{ll}^2 &=& 2 \frac{M_{\chinonn}^2 - M_{\tilde{l}}^2}{2 M_{\chinonn}} 
            \frac{M_{\chinonn}^2}{2 M_{\tilde{l}}} (M_{\tilde{l}}^2 - M_{\chinon}^2 ) (1 + cos \theta^* ) \nonumber \\
          &=& \frac{(M_{\chinonn}^2 - M_{\tilde{l}}^2)(M_{\tilde{l}}^2 - M_{\chinon}^2)}{2 M_{\tilde{l}}^2} (1 + cos \theta^* )
\label{eq.XY.Mffcost}
\end{eqnarray}
with the upper endpoint of the mass distribution given by
\begin{eqnarray}
 M_{ll}^{max} = M_{\chinonn} \sqrt{\oneMrat{M_{\tilde{l}}}{M_{\chinonn}}\oneMrat{M_{\chinon}}{M_{\tilde{l}}}}.
\label{eq.XY.Mffmax}
\end{eqnarray}
This corresponds to the configuration where the two leptons are emitted
back-to-back in the rest frame of the $\chinonn$. Since the
$\tilde{l}$ is a particle with spin 0 it should decay
isotropically. Thus, formula (\ref{eq.XY.Mffcost}) shows that the
distribution in $M_{ll}$ increases linearly with $M_{ll}$ leading to a
sharp edge at the kinematic limit given by formula (\ref{eq.XY.Mffmax}).

The following section shows a summary of all formulae for the
endpoints which are used in the analysis and the related
configurations. They are calculated in the same way as described above.

\section{Formulae and kinematic configurations}\label{sec.seqmassless.summary}
Here the endpoints available for the decay chain $\tilde{q} \rightarrow q
+ \chinonn$ , $ \chinonn \rightarrow l1 + \tilde{l}$ , $\tilde{l}
\rightarrow l2 + \chinon$ are summarized.

\begin{enumerate}

\item $\mathbf{M_{ll}^{max}}$ \\
\begin{equation}
  M^{max}_{ll} = M_{\chinonn} \sqrt{\oneMrat{M_{\tilde{l}}}{M_{\chinonn}}\oneMrat{M_{\chinon}}{M_{\tilde{l}}}}\label{eq:llmax}
\end{equation}
\begin{figure}[H]
 \begin{center}
     \includegraphics[width=\textwidth/2]{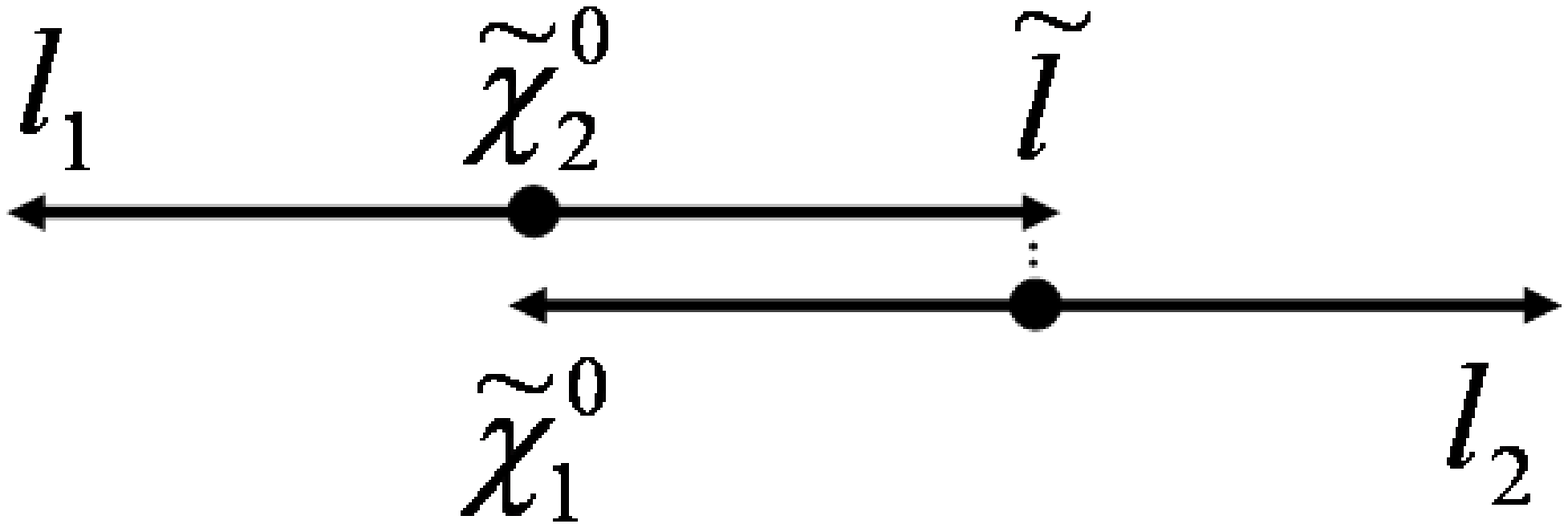} 
  \end{center}
\end{figure}

\item $\mathbf{M_{l1q}^{max}}$ (1st lepton)
\begin{equation}
  M^{max}_{l_1q} = M_{\tilde{q}} \sqrt{\oneMrat{M_{\chinonn}}{M_{\tilde{q}}}\oneMrat{M_{\tilde{l}}}{M_{\chinonn}}}\label{eq:l1qmax}
\end{equation}
\begin{figure}[H]
  \begin{center}
     \includegraphics[width=\textwidth/2]{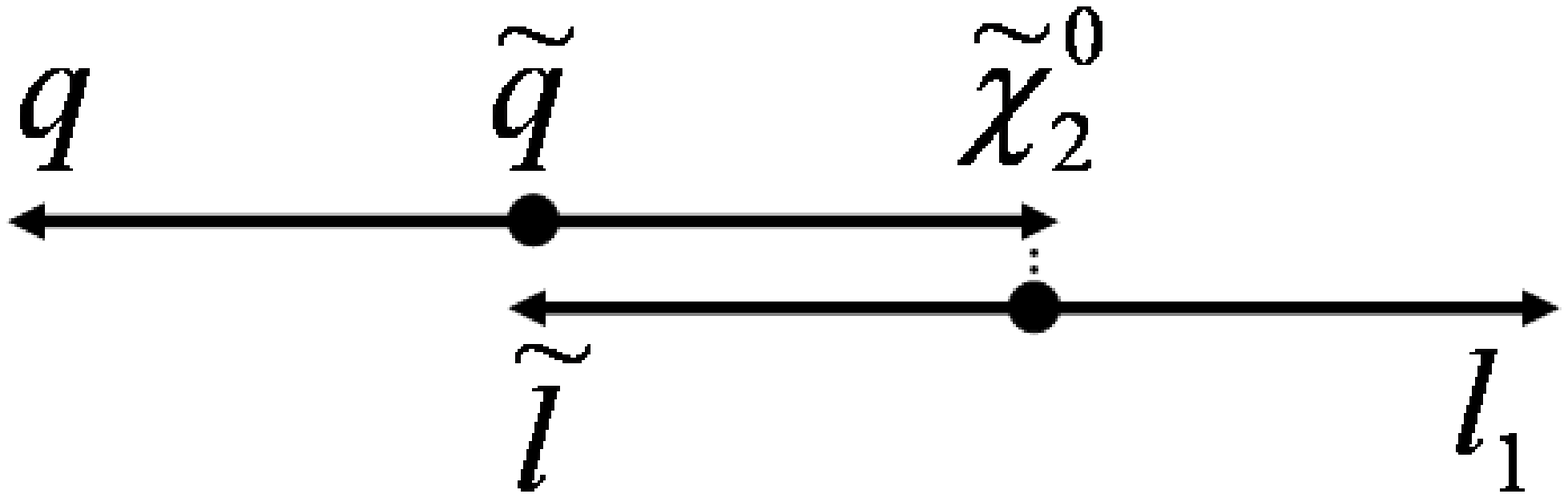} 
  \end{center}
\end{figure}

\item $\mathbf{M_{l2q}^{max}}$ (2nd lepton)
\begin{equation}
 M^{max}_{l_2q} = M_{\tilde{q}} \sqrt{\oneMrat{M_{\chinonn}}{M_{\tilde{q}}}\oneMrat{M_{\chinon}}{M_{\tilde{l}}}}\label{eq:l2qmax}
\end{equation}
\begin{figure}[H]
  \begin{center}
     \includegraphics[width=\textwidth/2]{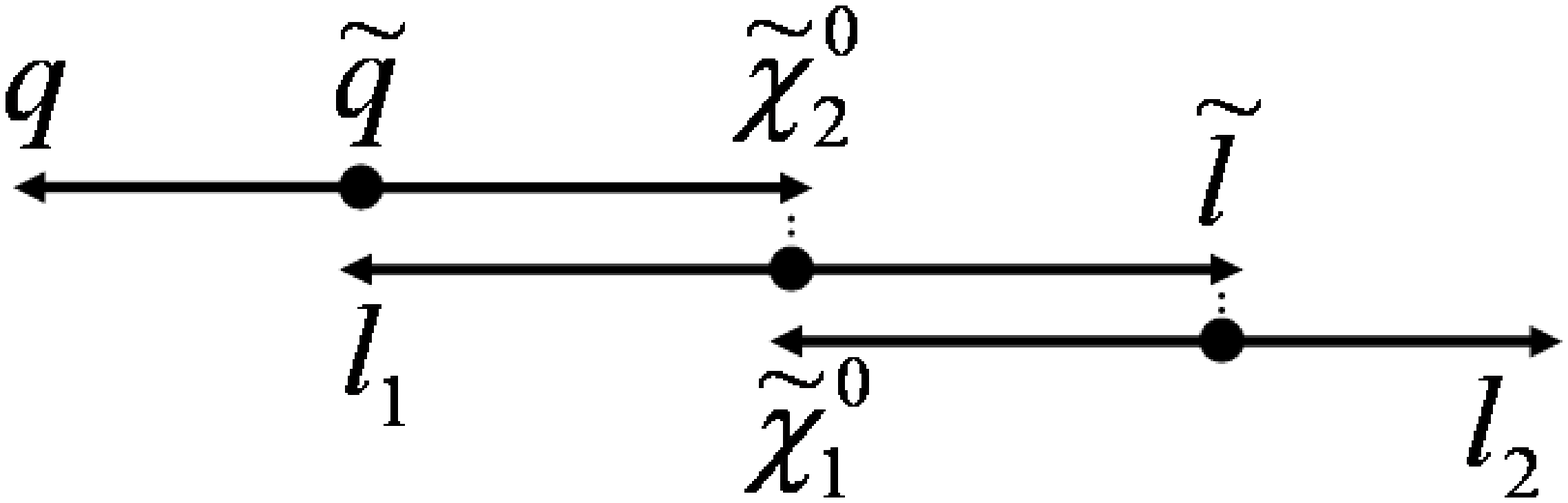}
  \end{center}
\end{figure}

\item $\mathbf{M_{llq}^{max}}$, for $M_{\chinonn}^2 < M_{\tilde{q}} M_{\chinon}$
\begin{equation}
M^{max}_{llq} = M_{\tilde{q}} \sqrt{\oneMrat{M_{\chinonn}}{M_{\tilde{q}}}\oneMrat{M_{\chinon}}{M_{\chinonn}}}\label{eq:llqmax}
\end{equation}
\begin{figure}[H]
  \begin{center}
     \includegraphics[width=\textwidth/2]{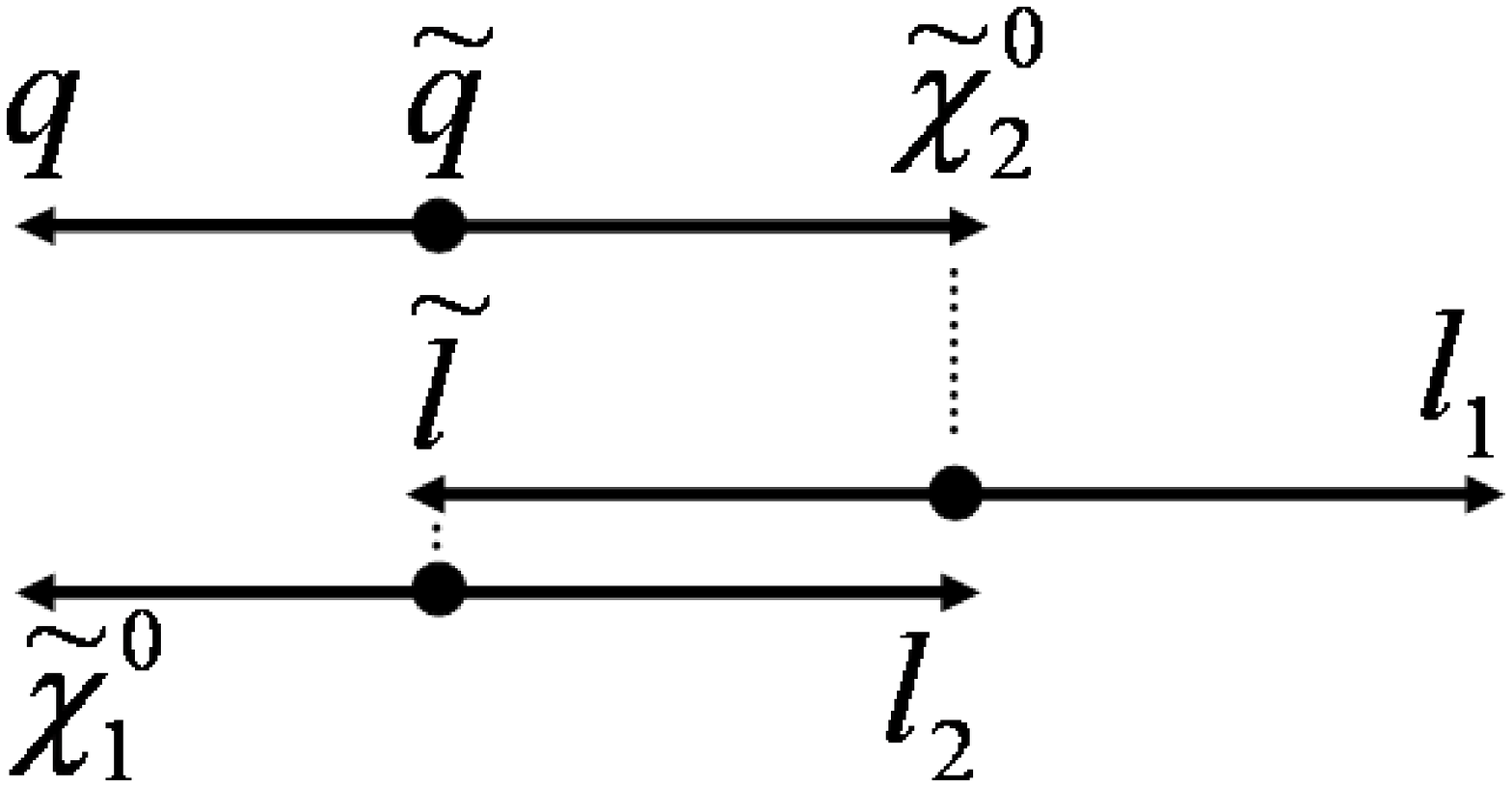}
  \end{center}
\end{figure}

\item $\mathbf{(M_{l1q} + M_{l2q})^{max}}$ 
\begin{equation}
 M^{max}_{l_1q + l_2q} = M^{max}_{l_1q} +
\frac{M_{\tilde{l}}}{M_{\chinonn}} M^{max}_{l_2q} \label{eq:l1ql2qmax}
\end{equation}
\begin{figure}[H]
  \begin{center}
     \includegraphics[width=\textwidth/2]{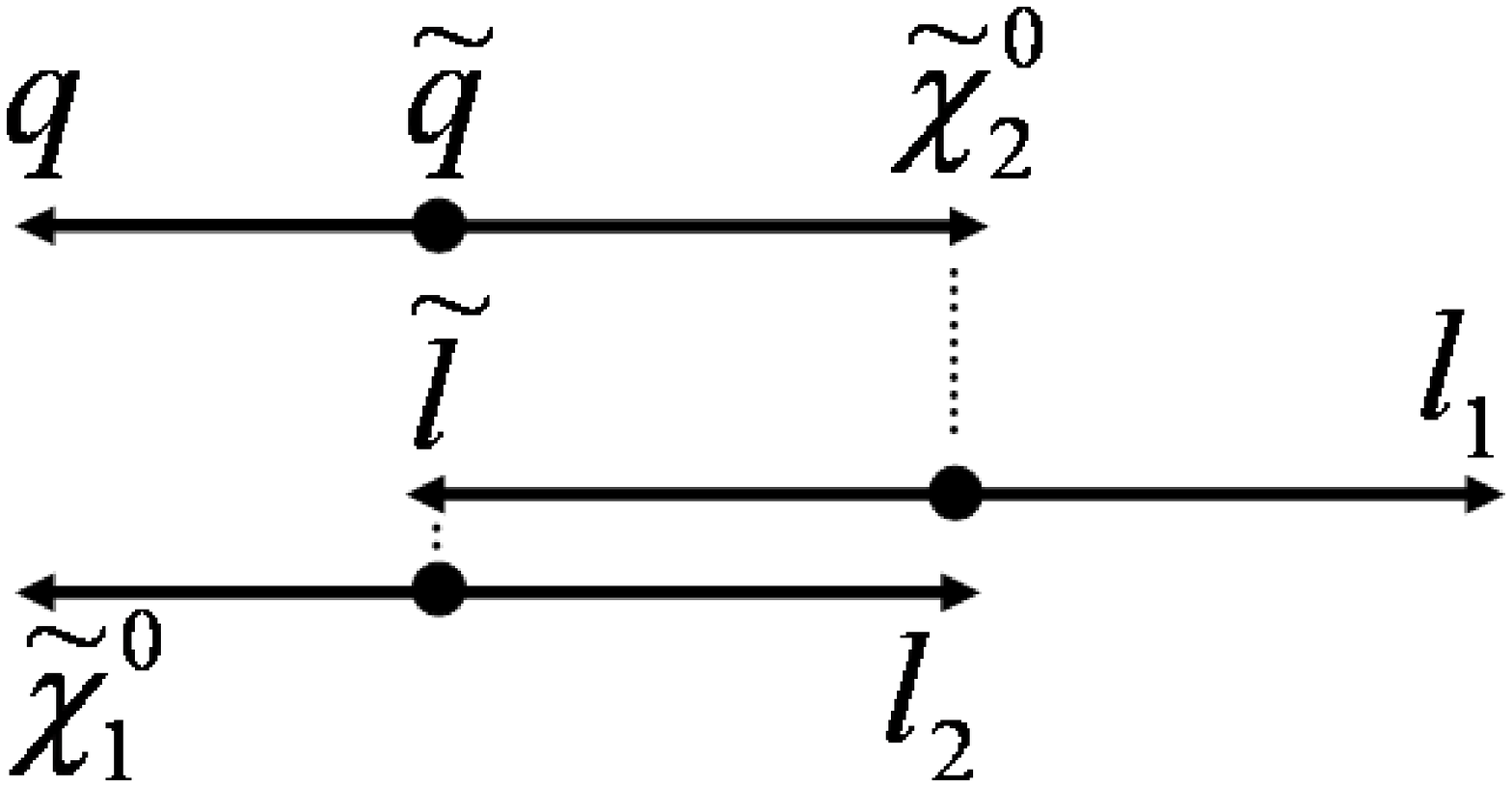}
  \end{center}
\end{figure}
\end{enumerate}
Four of these five endpoints are in principle necessary to determine
all involved masses, provided that several consecutive decay channels
are open (long decay chains). However, it will not always be possible
to measure them precisely. Therefore, it will be sensible to use all
available endpoints.

\section{Endpoints for top quark events}\label{sec:top}
The mass of the top quark is very large and therefore not
negligible. Thus, not only upper endpoints but also lower endpoints
arise in the invariant mass distributions. In the analysis presented
in section \ref{ssec:llqtop} to \ref{sec:Ml1t_Ml2t} the top quark is
involved. The corresponding decay chain is
\begin{eqnarray}
\sTop \ra t + \chinonn \ra t + \tilde{l} + l \ra t + l + l + \chinon.
\end{eqnarray}
With
\begin{eqnarray}
 E_t = \frac{M_{\tilde{t}}^2 + m_t^2 - M_{\chinonn}^2}{2 M_{\tilde{t}}} & , &
 p_t = \sqrt{E_t^2 - m_t^2} \\
 E_{\chinonn} = \frac{M_{\tilde{t}}^2 + M_{\chinonn}^2 - m_t^2}{2
M_{\tilde{t}}} & , &
 p_{\chinonn} = \sqrt{E_{\chinonn}^2 - M_{\chinonn}^2} 
\end{eqnarray}
the endpoint formulae are the following:
\begin{eqnarray}
 (M_{l1t}^{max})^2 &=& m_t^2 + M_{\tilde{t}}^2 
   (1 - \frac{M_{\chinonn}^2}{M_{\tilde{t}}^2} + \frac{m_t^2}{M_{\tilde{t}}^2})
   ( 1 - \frac{M_{\tilde{l}}^2}{M_{\chinonn}^2}) \frac{E_{\chinonn} +
p_{\chinonn}}{M_{\tilde{t}}}\\\nonumber 
&&\times \frac{1}{2}
   \left( 1 + \sqrt{1 - (\frac{m_t}{E_t})^2}\right)\\
%---------------------------------------------------------------------
 (M_{l1t}^{min})^2 &=& m_t^2 + M_{\tilde{t}}^2 
   (1 - \frac{M_{\chinonn}^2}{M_{\tilde{t}}^2} + \frac{m_t^2}{M_{\tilde{t}}^2})
   ( 1 - \frac{M_{\tilde{l}}^2}{M_{\chinonn}^2}) \frac{E_{\chinonn} -
p_{\chinonn}}{M_{\tilde{t}}}\\\nonumber
&&\times \frac{1}{2}
   \left( 1 - \sqrt{1 - (\frac{m_t}{E_t})^2}\right)\\
%---------------------------------------------------------------------
 (M_{l2t}^{max})^2 &=& m_t^2 + M_{\tilde{t}}^2 
   (1 - \frac{M_{\chinonn}^2}{M_{\tilde{t}}^2} + \frac{m_t^2}{M_{\tilde{t}}^2})
   ( 1 - \frac{M_{\chinon}^2}{M_{\tilde{l}}^2}) \frac{E_{\chinonn} +
p_{\chinonn}}{M_{\tilde{t}}}\\\nonumber
&&\times \frac{1}{2}
   \left( 1 + \sqrt{1 - (\frac{m_t}{E_t})^2}\right)\\
%---------------------------------------------------------------------
 (M_{l2t}^{min})^2 &=& m_t^2 + M_{\tilde{t}}^2 
   (1 - \frac{M_{\chinonn}^2}{M_{\tilde{t}}^2} + \frac{m_t^2}{M_{\tilde{t}}^2})
   \frac{M_{\tilde{l}}^2}{M_{\chinonn}^2} ( 1 -
\frac{M_{\chinon}^2}{M_{\tilde{l}}^2}) 
   \frac{E_{\chinonn} - p_{\chinonn}}{M_{\tilde{t}}}\\\nonumber
&&\times \frac{1}{2}
   \left( 1 - \sqrt{1 - (\frac{m_t}{E_t})^2}\right)\\
%----------------------------------------------------------------------
 (M_{llt}^{max})^2 &=& m_t^2 + M_{\tilde{t}}^2 
   (1 - \frac{M_{\chinonn}^2}{M_{\tilde{t}}^2} + \frac{m_t^2}{M_{\tilde{t}}^2})
   ( 1 - \frac{M_{\chinon}^2}{M_{\chinonn}^2}) \frac{E_{\chinonn} +
p_{\chinonn}}{M_{\tilde{t}}}\\\nonumber
&&\times \frac{1}{2}\label{Mlltmax}
   \left( 1 + \sqrt{1 - (\frac{m_t}{E_t})^2}\right)\\
%----------------------------------------------------------------------
 (M_{llt}^{min})^2 &=& m_t^2 + M_{\tilde{t}}^2 
   (1 - \frac{M_{\chinonn}^2}{M_{\tilde{t}}^2} + \frac{m_t^2}{M_{\tilde{t}}^2})
   ( 1 - \frac{M_{\chinon}^2}{M_{\chinonn}^2}) \frac{E_{\chinonn} -
p_{\chinonn}}{M_{\tilde{t}}}\\\nonumber
&&\times \frac{1}{2}\label{Mlltmin}
   \left( 1 - \sqrt{1 - (\frac{m_t}{E_t})^2}\right)
\end{eqnarray}
The configurations for the upper endpoints here are the same as for
the massless case described in section
\ref{sec.seqmassless.summary}. The lower endpoints are realised with a
configuration where all detected particles go into the same direction.
		
 \newpage
 \cleardoublepage

\chapter{\texorpdfstring{Searches for the $\chinonn$}{Searches for the
next-to-lightest neutralino}}\label{chap:Searches}
 Since mSUGRA is a Supersymmetry breaking scenario within the minimal
supersymmetric extension of the Standard Model favoured by many
theorists, it is of high importance to develop analysis methods for
this model. Although the measurements at LEP (see
section~\ref{sec:LEP}) and with the WMAP satellite (see
section~\ref{sec:wmap}) give a strong constraint on its parameter
space there is still a huge number of open possibilities with very
different phenomenologies. Many studies for mSUGRA have been performed
in the past years. Most of them concentrate on methods for models with
$\tan \beta$ lower than $10$. Actually, there is no reason for
preferring low $\tan \beta$-models. It has been pointed out in several
recent publications that due to the different phenomenology at
different $\tan \beta$ values, studies for high $\tan \beta$ values
are very important.

\begin{figure}[H]
  \begin{center}
  \includegraphics[width=\textwidth/3]{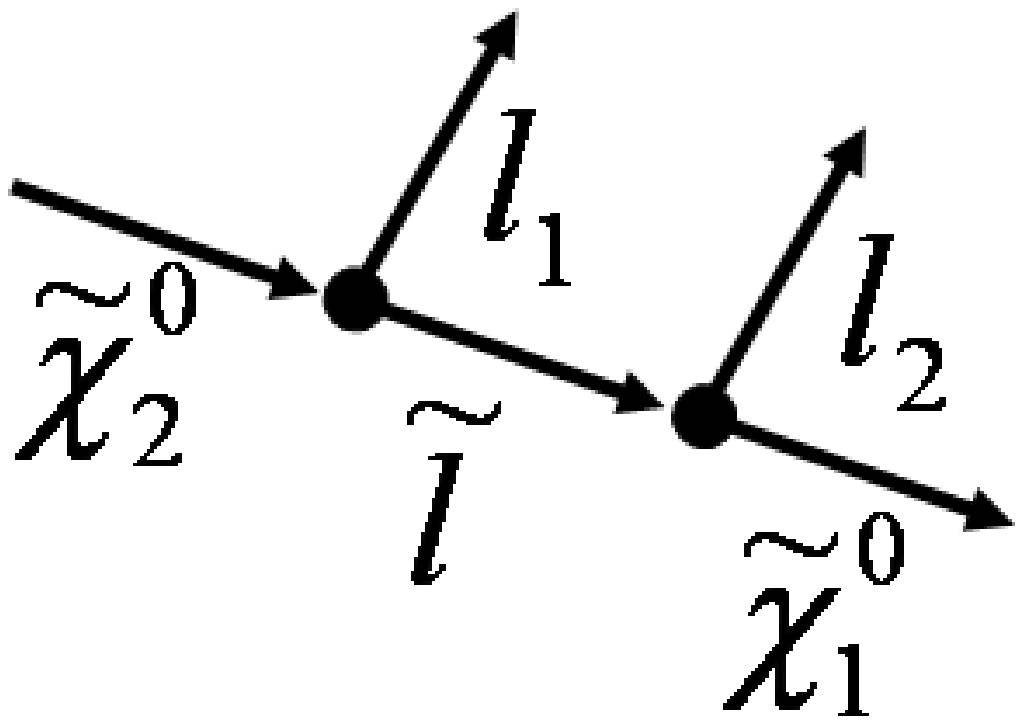}
\caption{Leptonic cascade decay of $\chinonn$.}\label{LCD_N2}		
  \end{center}
\end{figure}

Since there is no evidence for a heavy charged stable particle in the
universe and assuming R-parity conservation, the lightest
supersymmetric particle in mSUGRA is the stable
neutralino~$\chinon$. A neutral weakly interacting stable particle
escapes the detector, thus, a direct mass reconstruction is
not possible. However, it is useful to search for the next-to-lightest
neutralino, the $\chinonn$, which decays in detectable particles
within the detector. It is mainly produced in gluino and squark
decays. The focus here is on the leptonic cascade decay channel shown
in figure~\ref{LCD_N2}.

It dominates in areas of the parameter space where $M_{\chinonn}$ is
larger than $M_{\tilde{l}}$. Results from studies for
$\chinonn$-searches in mSUGRA at low $\tan
\beta$~\cite{N2Studies} are presented very briefly in the following. Then the
results of the Monte Carlo analysis at $\tan \beta = 35$ performed in
this work are discussed. The latter shows that in principle there are
ways to reconstruct the involved sparticle masses in that area of the
parameter space.
%----------------------------------------------------------------------
\section{\texorpdfstring{Searches at low $\tan \beta$}{Searches at low
tan beta}}
\begin{figure}[H]
  \begin{center}
  \includegraphics[angle=270,width=\textwidth]{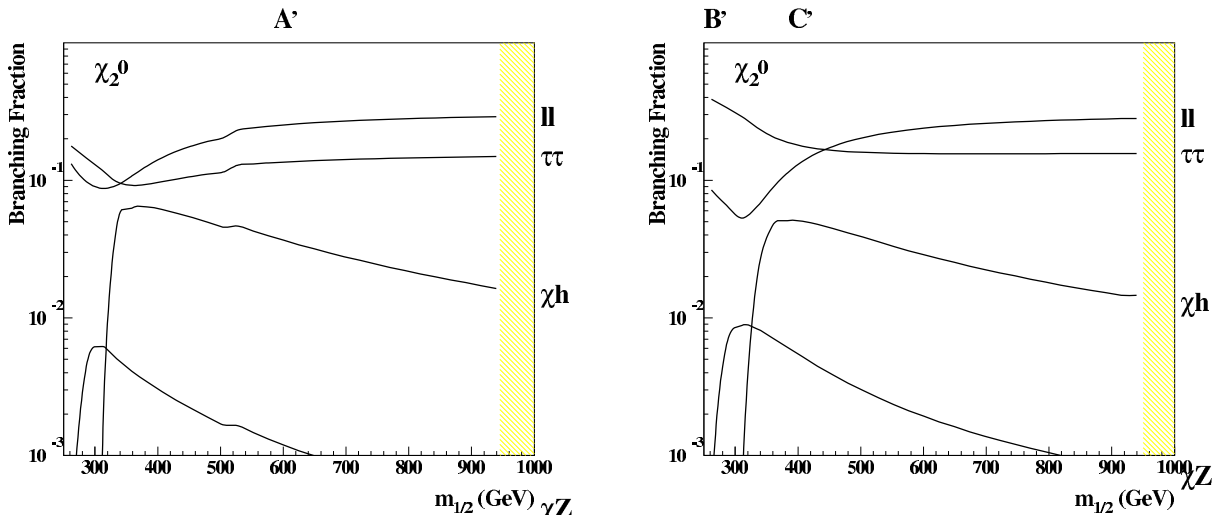}
  \caption{Dominant branching ratios along the WMAP line for the
  $\chinonn$ as a function of $m_{1/2}$ at $\tan \beta = 5$ (left) and
  $\tan \beta = 10$ (right). $\mu$ is positive and $A_{0}$ is set to
  zero. The locations of the updated benchmark points
  A'($m_{0}=107$,$m_{1/2}=600$), B'(57,250) and C'(80,400) are
  indicated \cite{wmappara}. Here leptons l are electrons and muons
  whereas taus are shown separately.} \label{br5_10} \end{center}
\end{figure}

\begin{figure}[H]
\begin{center}
  \includegraphics[width=\textwidth]{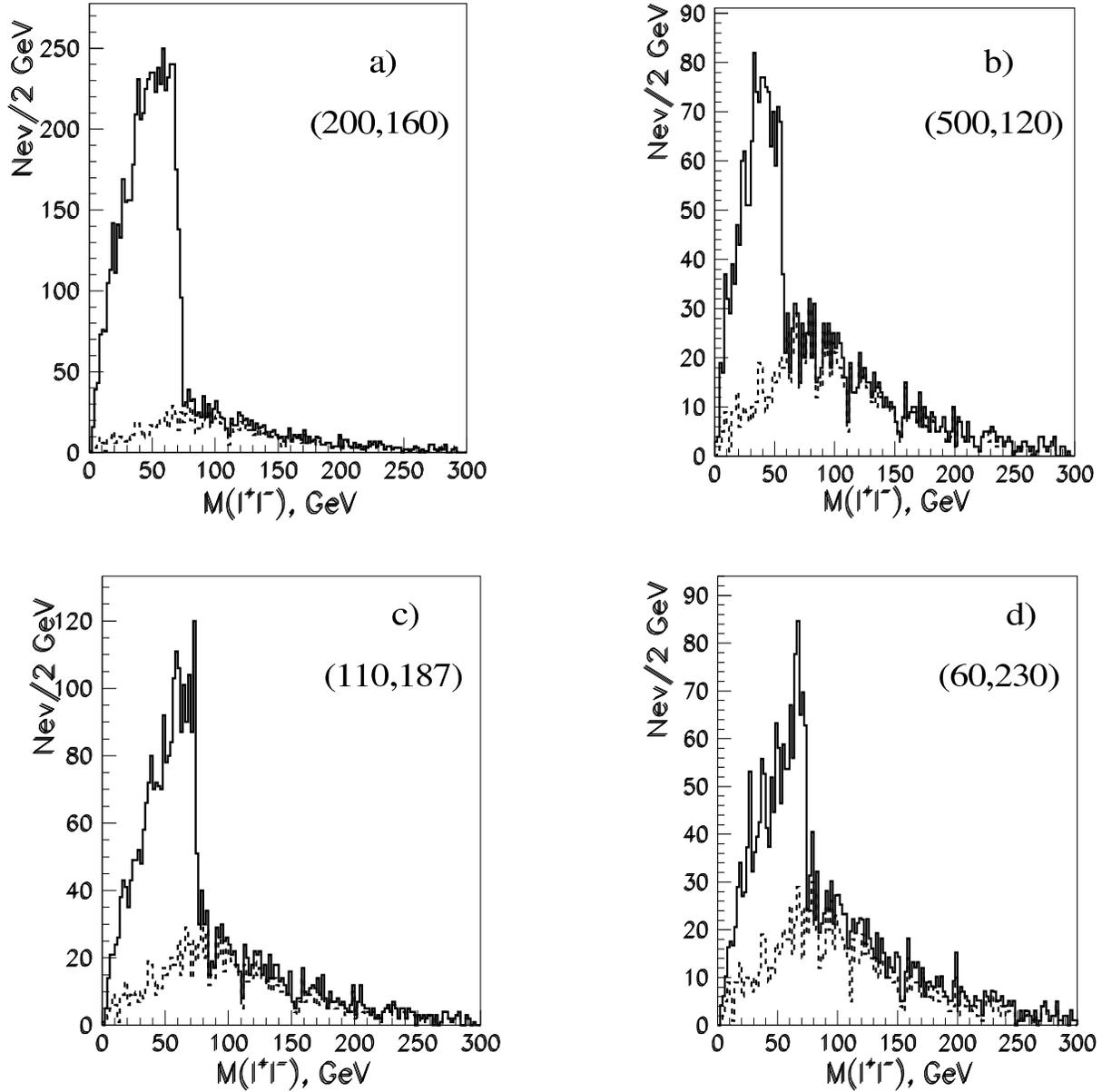}
\caption{The solid line shows the invariant mass distribution of two
same-flavour, opposite-sign leptons of $\chinonn$ decays at various
($m_{0}$,$m_{1/2}$) points for $\tan \beta = 2$ and $\mu < 0$ with
$L_{int}~=~10^{3}\ pb^{-1}$. The Standard Model background is given by
the dashed line.}\label{lowtbresults}
\end{center}
\end{figure}

Leptonic decays of the $\chinonn$ have a useful kinematic
characteristic. The dilepton invariant mass spectrum has a sharp edge
near the kinematic upper limit which is given by formula
(\ref{eq:llmax}). Thus, constraints on the involved sparticle masses
which are independent of the underlying model can be obtained by
measuring the endpoint of this distribution. The most straightforward
signature for selecting the $\chinonn$ decays is provided by the
topology with two same-flavour opposite-sign leptons accompanied by
large missing transverse energy which comes from the $\chinon$. It is
the advantage of low $\tan
\beta$ models that many of the charged leptons coming from the $\chinonn$ are
electrons and muons which can be fully
reconstructed. Figure~\ref{br5_10} shows two representative examples
of $\chinonn$ branching ratios as a function of $m_{1/2}$ within the
WMAP constrained parameter space.
In the two opposite-sign leptons + $E_{T}^{miss}$ + $(jets)$ channel
the largest Standard Model background is $t\bar{t}$ production, with
$t \ra b W$, and both $\W$s decaying into leptons, or one of the
leptons coming from a $\W$ decay and the other from the $b$-decay of
the same $t$-quark. Defining the significance $\sigma$ as $\sigma =
(N_{EV} - N_{B})/\sqrt{N_{EV}}$ it has been shown that the
observability of the kinematic endpoint varies from $\sigma = 77$ at
$m_{0} = 200$ and $m_{1/2} = 160$ to $\sigma = 27$ at $m_{0} = 60$ and
$m_{1/2} = 230$ for an integrated luminosity $L_{int} = 10^{3}\
pb^{-1}$~\cite{N2Studies}. For this luminosity, the appearance of the
edges in the distributions is already sufficiently pronounced in a
significant part of the $m_{0}$-$m_{1/2}$-plane. The position of the
kinematic edge can be measured with a precision of about
$0.5\gev$. Figure~\ref{lowtbresults} shows different realisations of
such an edge very clearly.
With increasing $m_{0}$ and $m_{1/2}$ the cross sections are
decreasing and therefore a higher integrated luminosity is needed. In
order to perform a full mass reconstruction of the sparticles further
analysis methods are necessary which are described to some extent in
\cite{N2Studies}.
%----------------------------------------------------------------------
\section{\texorpdfstring{Searches at $\tan \beta = 35$}{Searches at
tan beta = 35}} 
\begin{figure}[H]
\begin{center}
  \includegraphics[angle=270,width = \textwidth/3*2]{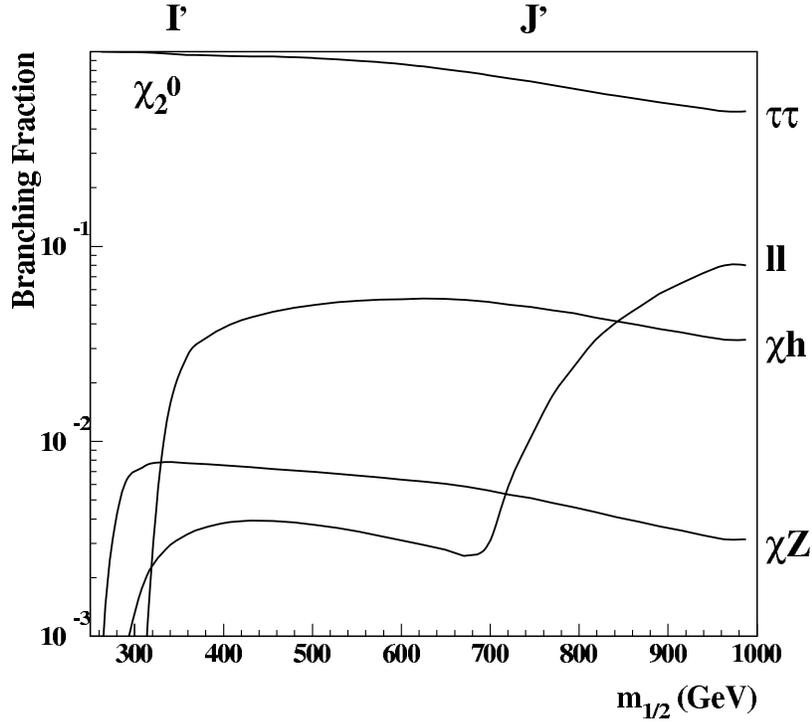}
  \caption{Dominant branching ratios along the WMAP line for
  $\chinonn$ as a function of $m_{1/2}$ at $\tan \beta = 35$, $\mu >
  0$ and $A_{0} = 0$. The location of updated benchmark points
  I'($m_{0}=181$,$m_{1/2}=350$) and J'(299,750) are
  indicated~\cite{wmappara}. Here leptons l are electrons and muons
  whereas taus are shown separately.}\label{br35} \end{center}
\end{figure}

The kinematic characteristic of the leptonic
$\chinonn$ decay does not change with increasing $\tan
\beta$. However, the branching ratios do. Simulations made with
\mbox{ISASUGRA} show that the decay into taus became more and more
important with increasing $\tan \beta$. Figure~\ref{br35} illustrates
the $\chinonn$ branching ratios, evaluated at $\tan \beta = 35$.
Neither with CMS nor with ATLAS will it be possible to fully
reconstruct taus since they decay into at least one $\nu_{\tau}$ which
escapes the detector. Therefore, the kinematic endpoint in the
dilepton invariant mass distribution will in general be smeared
out. It is the aim of the following study to investigate the
$\chinonn$ discovery potential in this region of the mSUGRA parameter
space. Furthermore, the goal is to use the $\chinonn$ discovery for
sparticle mass reconstructions.
%----------------------------------------
\subsection{Choice of parameters}
At the moment it is not known whether the MSSM with mSUGRA describing
the Supersymmetry breaking is the right extension of the Standard
Model. However, there are experimental (see section~\ref{sec:LEP} and
\ref{sec:wmap}) and theoretical constraints on the mSUGRA parameter
space. For the development of analysis methods it is necessary to
choose parameters in a characteristic region which is not yet excluded
by any experiment and which is accessible for future experiments like
the Large Hadron Collider.
\begin{figure}[H]
  \begin{center}
  \includegraphics[width=\textwidth]{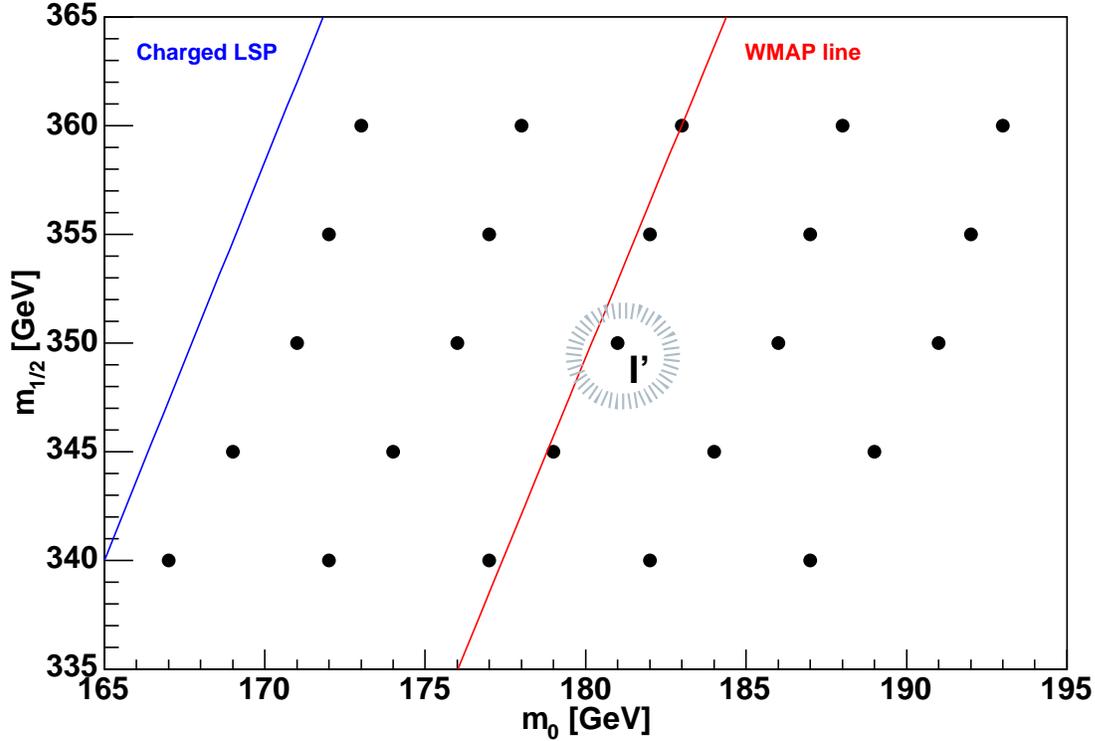}
  \caption{Points for the analysis in the $m_{0}$-$m_{1/2}$ plane in
  an area around the benchmark point I'. The upper left side is
  excluded due to cosmological constraints.}  \label{25Points}
  \end{center}
\end{figure}
The parameters in this analysis are chosen with the following motivation:
\begin{enumerate}
\item $\tan \beta = 35$: There is no experimental limit which excludes
high $\tan \beta$ models. It is still necessary to develop methods for
the $\chinonn$ discovery at LHC assuming these models. It is especially
necessary to determine whether it is possible to discover the
$\chinonn$ in the $\tau$-channel which is favoured at $\tan \beta =
35$.
\item $m_{0} \in [167,193], m_{1/2} \in [340,360]$ around I' in figure
\ref{br35}: In that region the branching ratio BR($\chinonn \ra
\tilde{\tau}^{\pm}_{1} \tau^{\mp}$) is with over $95\%$ especially
high. For the analysis 25 points in that region are chosen. In order
to consider the WMAP results and the cosmologically excluded region
the points are not equidistant which can be seen in
figure~\ref{25Points}.
\item $A_{0} = 0$: A non zero value would only be necessary for the
electroweak symmetry breaking if the top mass was lower than the
measured value. However, it would lead to different sparticle masses.
\item $\mu > 0$: There is no special reason to choose $\mu > 0$, but
the results are not very sensitive to that choice \cite{pape}.
\end{enumerate} 
The masses of the sparticles relevant for this study in the chosen
area of the mSUGRA parameter space are given in table~\ref{tab:masses}.
\begin{table}[H]
\begin{center}
\begin{tabular*}{\textwidth}{@{\extracolsep{\fill}}|c|rrrrr|} \hline\hline
\multicolumn{6}{|c|}{\emph{Masses in$\gev$ for} ($m_{0}$,$m_{1/2}$)}\\\hline\hline
\emph{Particle} & (167,340) & (187,340) & (181,350) & (173,360) & (193,360)\\\hline\hline 
$\tilde{g}$
& 812.06 & 812.49 & 834.16 & 855.61 & 856.23 \\\hline

$\tilde{d}_{1}/ \tilde{s}_{1}$
& 759.81 & 764.82 & 782.61 & 800.09 & 804.65 \\\hline

$\tilde{d}_{2}/ \tilde{s}_{2}$
& 730.15 & 735.04 & 752.02 & 768.81 & 773.64 \\\hline

$\tilde{u}_{1}/ \tilde{c}_{1}$
& 755.24 & 760.29 & 778.17 & 795.75 & 800.34 \\\hline

$\tilde{u}_{2}/ \tilde{c}_{2}$
& 731.79 & 736.67 & 753.80 & 770.73 & 775.06 \\\hline

$\tilde{b}_{1}$
& 657.94 & 661.48 & 678.14 & 694.20 & 697.50 \\\hline

$\tilde{b}_{2}$
& 704.72 & 708.19 & 724.93 & 740.96 & 744.21 \\\hline

$\tilde{t}_{1}$
& 565.70 & 568.46 & 583.27 & 598.06 & 600.47 \\\hline

$\tilde{t}_{2}$
& 729.25 & 731.74 & 747.90 & 763.69 & 766.14 \\\hline

$\chinonn$
& 257.10 & 257.23 & 265.52 & 273.74 & 273.92 \\\hline

$\chinon$
& 133.93 & 133.99 & 138.11 & 142.28 & 142.41 \\\hline

$\tilde{\tau}_{1}$
& 135.74 & 154.52 & 150.14 & 144.24 & 162.56 \\\hline\hline
\end{tabular*}
\caption{Sparticle masses in$\gev$ calculated with ISASUGRA
7.69. $A_{0}$ is set to zero, $\mu$ is positive and $\tan
\beta$ is 35.}\label{tab:masses}
\end{center}
\end{table}
%----------------------------------------
\subsection{\texorpdfstring{Production, decay and selection of
$\chinonn$}{Production, decay and selection of the next-to-lightest
neutralino}}\label{sec:prod_decay_sel} There are different
possibilities of producing $\chinonn$, and there are different
channels in which the $\chinonn$ can decay. The four $\chinonn$ decays
with the highest branching ratios are shown in
table~\ref{tab:chi2dec}. Table~\ref{tab:prod} shows eight processes
with the highest cross sections which either produce a neutralino
directly, or indirectly in decays of squarks and gluinos.
\begin{figure}[H]
  \begin{center}
      	\begin{tabular}{c@{\qquad}c}
  	\includegraphics[width=\textwidth/11*5]{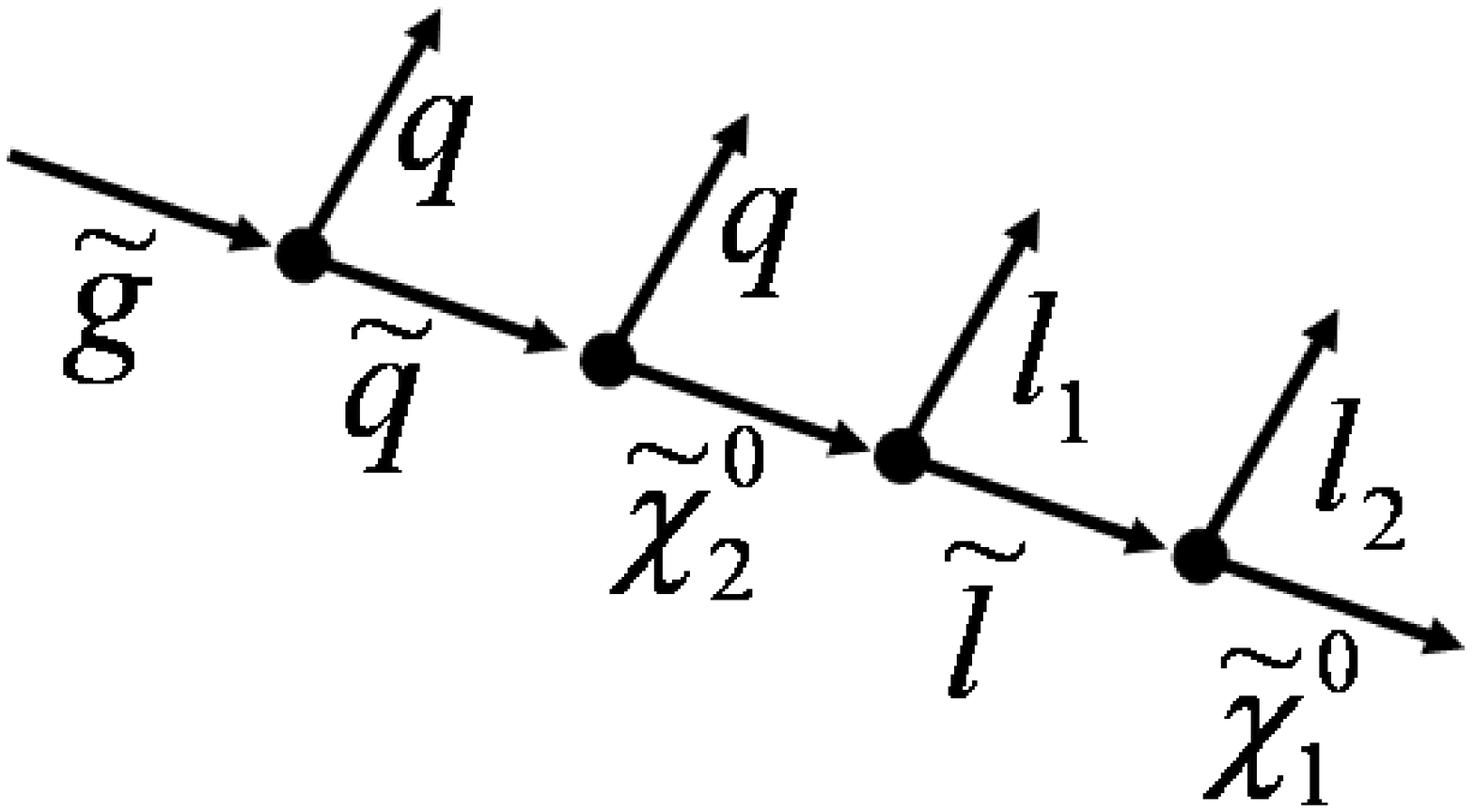}&
  	\includegraphics[width=\textwidth/11*4]{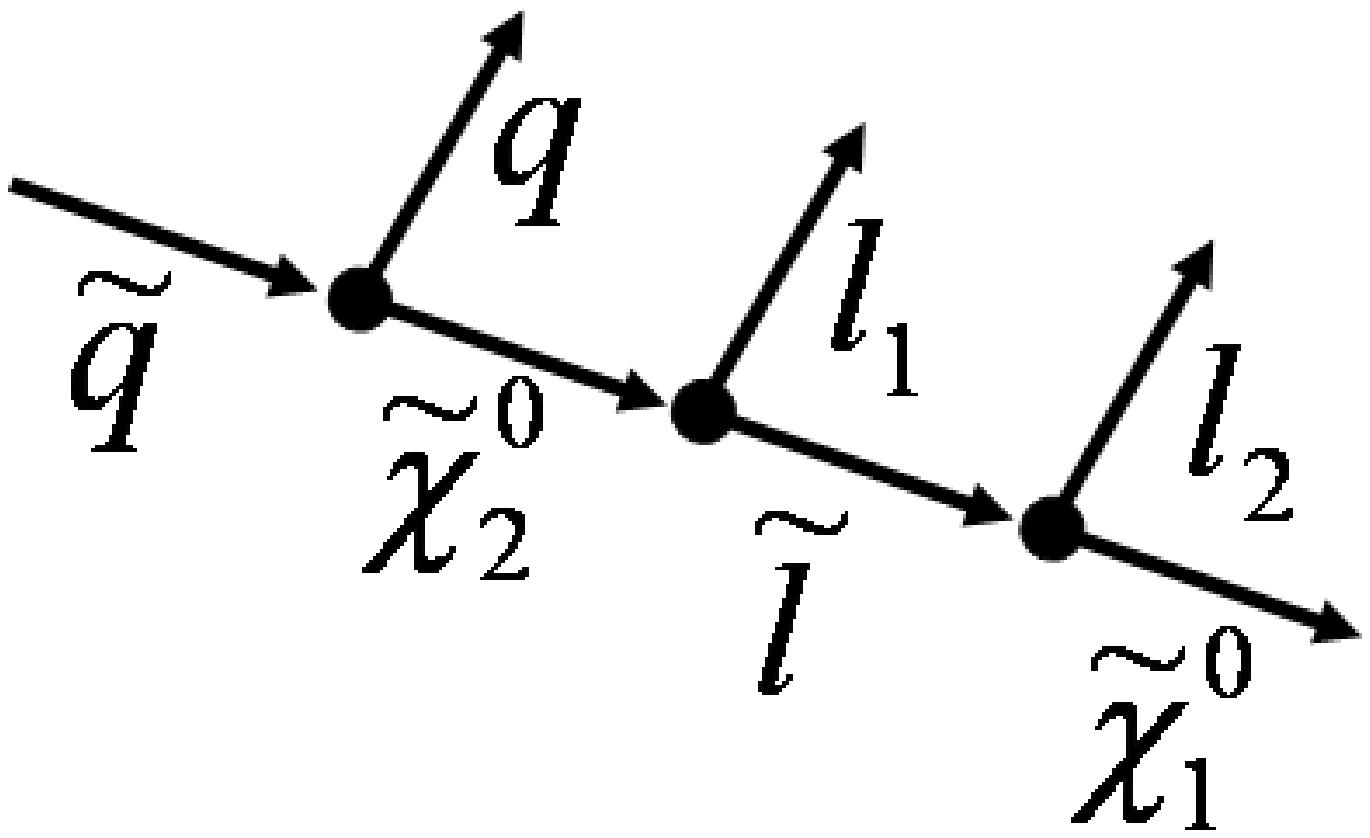}	
	\end{tabular}
	\caption{Cascade decay of $\tilde{g}$ and $\tilde{q}$.}\label{CD_GL_SQ}
  \end{center}
\end{figure}
The latter processes are used in this study since there are one or two
additional quarks in the decay chain which carry information about the
involved sparticle masses. They can be combined with both leptons in
order to get more kinematic endpoints. The processes are shown in the
two diagrams in figure~\ref{CD_GL_SQ}. Events with directly produced
$\chinonn$ can only be used for finding the dilepton kinematic
endpoint. The cross sections are calculated using PYTHIA~6.220 in
combination with ISASUGRA~7.69 at the given ($m_{0}$,$m_{1/2}$)
values. The reliability of these cross sections is described in
appendix~\ref{sec:cs}. For the indirect production, the squark
branching ratios of the decay into the $\chinonn$ - shown in
table~\ref{tab:sqtochi2} - are important. The left handed squarks show
much larger branching ratios for the decay into wino-like $\chinonn$.
The gluinos will in most of the cases first decay into a squark and a
quark and then through the squarks produce
$\chinonn$. Table~\ref{tab:gluino} shows that a significant amount of
left squarks is produced through gluino production. Additionally, it
can be seen that the sbottom channel is favoured in the gluino decay.

In this study the analysed data samples correspond to 100~000
generated events in which the tau decay is specified. Depending on the
values for $m_{0}$ and $m_{1/2}$ this leads to between 24278 and 25379
selected events. The event selection in this study is at the Monte
Carlo level. In the real experiment it will be an additional challenge
to select and reconstruct the taus and the associated quarks. Here an
event is defined as follows:
\begin{enumerate}
\item Two opposite-sign charged rho mesons (see
section~\ref{taudecay}) which come from two opposite-sign tau
leptons. These tau leptons must come from the cascade decay of the
$\chinonn$ which is produced through squarks. At this level the exact
origin of each particle is known. In order to take the experimental
reconstruction of the rho mesons into account, the distribution of the
rho mass is included into the simulation. In the real experiment two
opposite-sign isolated tau-like jets have to be searched with a
$\pi^{0}$ and a $\pi^{\pm}$ in the detector having an invariant mass
around the $\rho^{\pm}$ mass.
\item The quarks used in the analysis come only from squarks which
decay into a $\chinonn$. The flavour is known and also the precise
4-momentum of the quark is used for the analysis. It is mostly called
``associated'' quark in this study. In the experiment at least one
additional jet is needed.
\item No cut on the $E_{T}^{miss}$ has been
applied. However, in the real experiment $E_{T}^{miss}$ should
required to be greater than 100 $\gev$. This is the value chosen in
most of the low $\tan \beta$ studies. It has to be investigated
whether this limit is also a good value for high $\tan \beta$.
\end{enumerate} 
\begin{table}[H]
\begin{center}
\begin{tabular*}{\textwidth}{@{\extracolsep{\fill}}|l|rrrrr|} \hline\hline
\multicolumn{6}{|c|}{\emph{Branching Ratio in \% for} ($m_{0}$,$m_{1/2}$)}\\\hline\hline
\emph{Decay} & (167,340) & (187,340) & (181,350) & (173,360) & (193,360)\\\hline\hline 
$\chinonn \ra \tilde{\tau}_1^{\pm} + \tau^{\mp}$
& 96.8 & 96.2 & 96.0 & 96.0 & 95.2 \\\hline

$\chinonn \ra \chinon  + h^0$
& 2.1 & 2.6 & 2.8 & 2.9 & 3.6\\\hline

$\chinonn \ra \chinon  + Z^0$
& 0.7 & 0.9 & 0.8 & 0.7 & 0.9\\\hline

$\chinonn \ra \tilde{l}_R^{\pm} + \l^{\mp}$
& 0.4 & 0.2 & 0.3 & 0.4 & 0.3\\\hline\hline
\end{tabular*}
\caption{The largest branching ratios of the $\chinonn$ for the chosen
mSUGRA parameters. $A_{0}$ is set to zero, $\mu$ is positive and $\tan
\beta$ is 35. The letter $l$ denotes electrons and
muons.}\label{tab:chi2dec}
\end{center}
\end{table}
\begin{table}[H]
%\hspace{-10mm}
\begin{center}
\begin{tabular*}{\textwidth}{@{\extracolsep{\fill}}|lcl|rrrrr|} \hline\hline
\multicolumn{8}{|c|}{\emph{Cross Section (pb) for} ($m_{0}$,$m_{1/2}$)}\\\hline\hline
\multicolumn{3}{|l|}{\emph{Subprocess}}&(167,340)&(187,340)&(181,350)&(173,360)&(193,360)\\\hline\hline
$g\ + g\ $ &$\ra$& $\tilde{g} + \tilde{g}$ 
& 0.636 & 0.634 & 0.526 & 0.434 & 0.435\\\hline
$q_j +g\ $ &$\ra$& $\tilde{q}_{jL} + \tilde{g}$ 
& 1.865 & 1.824 & 1.578 & 1.366 & 1.338\\
$q_j +g\ $ &$\ra$& $\tilde{q}_{jR} + \tilde{g}$ 
& 2.030 & 1.987 & 1.718 & 1.490 & 1.470\\\hline
$q\ + \bar{q}\ $ &$\ra$& $\chinonn + \chionepm$ 
& 0.271 & 0.276 & 0.241 & 0.214 & 0.217\\
$q_i + q_j $ &$\ra$& $\tilde{q}_{iL} + \tilde{q}_{jL}$ 
& 0.502 & 0.496 & 0.443 & 0.399 & 0.393\\
$q_i + q_j $ &$\ra$& $\tilde{q}_{iR} + \tilde{q}_{jR}$ 
& 0.585 & 0.575 & 0.515 & 0.466 & 0.459\\
$q_i + q_j $ &$\ra$& $(\tilde{q}_{iL} + \tilde{q}_{jR})$
& 0.467 & 0.454 & 0.398 & 0.365 & 0.356\\
 &+&$(\tilde{q}_{iR} + \tilde{q}_{jL})$ & & & & &\\
$q_i + \bar{q}_j $ &$\ra$& $(\tilde{q}_{iL} + \tilde{\bar{q}}_{jR})$
& 0.518 & 0.509 & 0.444 & 0.398 & 0.384\\
 &+&$(\tilde{q}_{iR} + \tilde{\bar{q}}_{jL})$ & & & & &\\\hline\hline
\end{tabular*}
\caption{Most important processes for direct and indirect (in cascade
decays of $\tilde{g}$ and $\tilde{q}$) $\chinonn$
production at LHC. $A_{0}$ is set to zero, $\mu$ is positive and $\tan
\beta$ is 35.}\label{tab:prod}
\end{center}
\end{table}
\begin{table}[H]
\begin{center}
\begin{tabular*}{\textwidth}{@{\extracolsep{\fill}}|l|rrrrr|} \hline\hline
\multicolumn{6}{|c|}{\emph{Branching Ratio in \% for} ($m_{0}$,$m_{1/2}$)}\\\hline\hline
\emph{Decay} & (167,340) & (187,340) & (181,350) & (173,360) & (193,360)\\\hline\hline
$\tilde{d}_L \ra \chinonn + d$ 
& 31.1 & 31.0 & 31.1 & 31.2 & 31.2 \\
$\tilde{d}_R \ra \chinonn + d$ 
& 0.3 & 0.3 & 0.3 & 0.2 & 0.2 \\\hline
$\tilde{u}_L \ra \chinonn + u$ 
& 31.5 & 31.5 & 31.6 & 31.6 & 31.6 \\
$\tilde{u}_R \ra \chinonn + u$ 
& 0.3 & 0.3 & 0.3 & 0.2 & 0.2 \\\hline
$\tilde{s}_L \ra \chinonn + s$ 
& 31.1 & 31.0 & 31.1 & 31.2 & 31.2 \\
$\tilde{s}_R \ra \chinonn + s$ 
& 0.3 & 0.3 & 0.3 & 0.2 & 0.2 \\\hline
$\tilde{c}_L \ra \chinonn + c$ 
& 31.5 & 31.5 & 31.6 & 31.6 & 31.6 \\
$\tilde{c}_R \ra \chinonn + c$ 
& 0.3 & 0.3 & 0.3 & 0.2 & 0.2 \\\hline
$\tilde{b}_1 \ra \chinonn + b$ 
& 29.3 & 29.0 & 28.5 & 28.1 & 27.9 \\
$\tilde{b}_2 \ra \chinonn + b$ 
& 7.3 & 7.0 & 7.0 & 7.1 & 6.9 \\\hline
$\tilde{t}_1 \ra \chinonn + t$ 
& 14.3 & 14.3 & 14.3 & 14.4 & 14.3 \\
$\tilde{t}_2 \ra \chinonn + t$ 
& 9.1 & 9.2 & 9.3 & 9.4 & 9.5 \\\hline\hline
\end{tabular*}
\caption{Squark branching ratios for the decay into
$\chinonn$. $A_{0}$ is set to zero, $\mu$ is positive and $\tan
\beta$ is 35.}\label{tab:sqtochi2}
\end{center}
\end{table}
\begin{table}[H]
\begin{center}
\begin{tabular*}{\textwidth}{@{\extracolsep{\fill}}|l|rrrrr|} \hline\hline
\multicolumn{6}{|c|}{\emph{Branching Ratio in \% for} ($m_{0}$,$m_{1/2}$)}\\\hline\hline
\emph{Decay} & (167,340) & (187,340) & (181,350) & (173,360) & (193,360)\\\hline\hline 
$\tilde{g} \ra \chinonn + t + \bar{t}$ & 0.01 & 0.01 & 0.01 & 0.01 &
0.01\\\hline
$\tilde{g} \ra \tilde{d}_L + d$
& 3.5 & 3.2 & 3.3 & 3.5 & 3.2 \\
$\tilde{g} \ra \tilde{d}_R + d$
& 8.1 & 7.9 & 8.0 & 8.0 & 7.8 \\\hline
$\tilde{g} \ra \tilde{u}_L + u$
& 4.1 & 3.8 & 3.9 & 4.0 & 3.8 \\
$\tilde{g} \ra \tilde{u}_R + u$
& 7.8 & 7.6 & 7.7 & 7.7 & 7.5 \\\hline
$\tilde{g} \ra \tilde{s}_L + s$
& 3.5 & 3.2 & 3.3 & 3.5 & 3.2 \\
$\tilde{g} \ra \tilde{s}_R + s$
& 8.1 & 7.9 & 8.0 & 8.0 & 7.8 \\\hline
$\tilde{g} \ra \tilde{c}_L + c$
& 4.1 & 3.8 & 3.9 & 4.0 & 3.8 \\
$\tilde{g} \ra \tilde{c}_R + c$
& 7.8 & 7.6 & 7.7 & 7.7 & 7.5 \\\hline
$\tilde{g} \ra \tilde{b}_1 + b$
& 24.8 & 25.8 & 24.8 & 24.0 & 24.8 \\
$\tilde{g} \ra \tilde{b}_2 + b$
& 13.9 & 14.2 & 14.0 & 13.8 & 14.1 \\\hline
$\tilde{g} \ra \tilde{t}_1 + t$
& 14.3 & 15.0 & 15.5 & 15.8 & 16.5 \\
$\tilde{g} \ra \tilde{t}_2 + t$
& 0.0 & 0.0 & 0.0 & 0.0 & 0.0 \\\hline\hline
\end{tabular*}
\caption{Gluino branching ratios for the given
($m_{0}$,$m_{1/2}$) values. $A_{0}$ is set to zero, $\mu$ is positive
and $\tan \beta$ is 35. For the chosen parameters the $\tilde{g}$
cannot decay into $\tilde{t}_2 + t$ because $M_{\tilde{g}} <
M_{\tilde{t}_2} + M_{t}$.}\label{tab:gluino}
\end{center}
\end{table}
%----------------------------------------
\subsection{Methods to find the endpoint}\label{sec:methods}
Due to the presence of neutrinos in the investigated $\chinonn$ decay,
the distribution of the invariant mass of all visible $\chinonn$ decay
products is fundamentally different from the case where the $\chinonn$
decays into a $\chinon$ and a pair of electrons or muons. The
neutrinos smear out the kinematic endpoint, which is shown in
figure~\ref{InvMassll181350}. Therefore, it is necessary to define a
method how to find the kinematic endpoints in the measured invariant
mass ($M_{INV}$) distributions. In the following two ``simple'' types
are presented. ``Simple'' means here that they do in principle not
depend on the model. However, the included corrections for the
improvement of the measurement and the obtained systematic
uncertainties are based on mSUGRA Monte Carlo simulations.

\subsubsection{Linear fit}
First, there is the possibility to perform a linear fit $f(M_{INV})$
to the invariant mass distribution near the kinematic limit. The
invariant mass range in which the fit is performed needs to be chosen
carefully in order to obtain a sensible fit. Certainly, the region is
between the bin with the highest content and the end of the tail. Good
results have been achieved using an area starting at the bin with $50
\%$ and ending at the bin with $5\%$ of the maximum bin
content. Definitely, this definition is only meaningful if no
background is included. The measured endpoint $Ep^{M}_{i}$ for each
point $i$ of the $N = 25$ investigated points in the parameter space
is then defined as the intercept point of the fit line and the
abscissa. $A_{i}$ is a free parameter which is not needed for the
endpoint measurement.
\begin{eqnarray}
f(M_{INV}) &=& A_{i} \cdot \left( M_{INV} - B_{i} \right)\\
Ep^{M}_{i} &=& B_{i}
\end{eqnarray}
The mean statistical uncertainty on $Ep^{M}$ is 
\begin{eqnarray}
\delta_{stat} Ep^{M} = \frac{\sum_{i = 1}^{N} \delta B_{i}}{N}
\end{eqnarray}
with $\delta B_{i}$ being the statistical uncertainty of $B_{i}$.
The systematic shift $S$ of the endpoint measurement is given by the mean
difference of the theoretical endpoint $Ep^{T}_{i}$ and $Ep^{M}_{i}$,
which includes a systematic and a statistical uncertainty. 
\begin{eqnarray}
S &=& \frac{\sum_{i = 1}^{N} Ep^{T}_{i} - Ep^{M}_{i}}{N}\label{S}\\
\delta S &=& \sqrt{\frac{\sum_{i = 1}^{N} \left( Ep^{T}_{i} -
Ep^{M}_{i} - S \right)^{2}}{N}}\label{DS}
\end{eqnarray}
The corrected measured endpoint is then
\begin{eqnarray}
Ep^{M,c}_{i} &=& Ep^{M}_{i} + S.
\end{eqnarray}
The pure systematic uncertainty on the endpoint measurement is then
\begin{eqnarray}
\delta_{syst} Ep^{M}_{i} = \sqrt{\left
( \delta S \right)^{2} - \left( \delta_{stat} Ep^{M}_{i} \right)^{2} }
\end{eqnarray}
Certainly, the calculated $S$ can and will change when a different
region in the mSUGRA parameter space is chosen.

\subsubsection{Gaussian fit}
Second, the shape of the distribution can be investigated. In the
chosen region the neutrinos smear out the visible invariant mass
distribution near the kinematic limit. As a result the maximum of the
invariant mass distribution is at a mass much lower than the
endpoint. The measurable maximum $G_{i}$ is here defined as the
maximum of an iterated gaussian fit. The maximum position is
determined as follows:
\begin{enumerate}
\item The bin with the maximal bin content is the centre for the
first gaussian fit. Unless otherwise noted for the investigated
invariant mass distributions a sensible area for the gaussian fit is
a symmetric area around the centre of $25 \%$ of the centre value.
\item  The resulting gaussian maximum is then the input for the centre
of the second fit.
\item It turned out that three iterations are enough for a stable
maximum. The last (out of three) gaussian maximum is taken for $G_{i}$.
\end{enumerate}
$G_{i}$ has a statistical uncertainty $\delta G_{i}$. If there is a
correlation between the maximum and the endpoint which is independent
of $m_{0}$ and $m_{1/2}$ over a large area, then it is possible to get
the endpoint value out of the maximum measurement. In the following it is
assumed that there exists a linear dependence between $G_{i}$ and
$Ep^{T}_{i}$. A linear least squares fit of $Ep^{T}_{i}$ as a function
of $G_{i} \pm \delta G_{i}$ with the fit function $f(G) = C \cdot G +
D$ gives the values for this dependence. The measured endpoint
$Ep^{M}_{i}$ with its mean statistical uncertainty $\delta_{stat}
Ep^{M}$ is then
\begin{eqnarray}
Ep^{T}_{i} &=& C \cdot G_{i} + D,\label{eq:gaussfit}\\
\delta_{stat} Ep^{M} &=& C \cdot\frac{\sum_{i = 1}^{N} \delta
G_{i}}{N}.\label{eq:deltagaussfit} 
\end{eqnarray}
The corrected endpoint, the systematic shift with its uncertainty and
the pure systematic uncertainty are equally defined as in the case of
the linear fit. Since the gaussian maximum should be zero if
$Ep^{T}_{i}$ is zero it is interesting to check whether $D$ is close
to zero or not. This value gives a hint for the stability of the
assumed correlation on further changes in $m_{0}$ and $m_{1/2}$. In
the following analysis therefore also the slope $C$ and the resulting
uncertainties on the endpoint measurement are given if $D$ is forced
to zero. It is obvious that the gaussian method relies on Monte Carlo
simulations in order to find $C$ and $D$ whereas the measurements of a
sharp edge at low $\tan \beta$ are model independent.
%----------------------------------------

The difference between $Ep^{T}_{i}$ and $Ep^{M}_{i} + S$ in the linear
and in the gaussian case may have in some cases a $m_{0}$ and
$m_{1/2}$ dependence. This is discussed for each invariant mass
distribution separately. For the analysis the binning and the choice
of the fit area have been optimised such that they are most
universally valid within the investigated area of the mSUGRA parameter
space.

 \subsection{Selection of the leptonic decay}\label{taudecay}
\begin{figure}[H]
  \begin{center} \includegraphics[width=\textwidth/6*5]
  {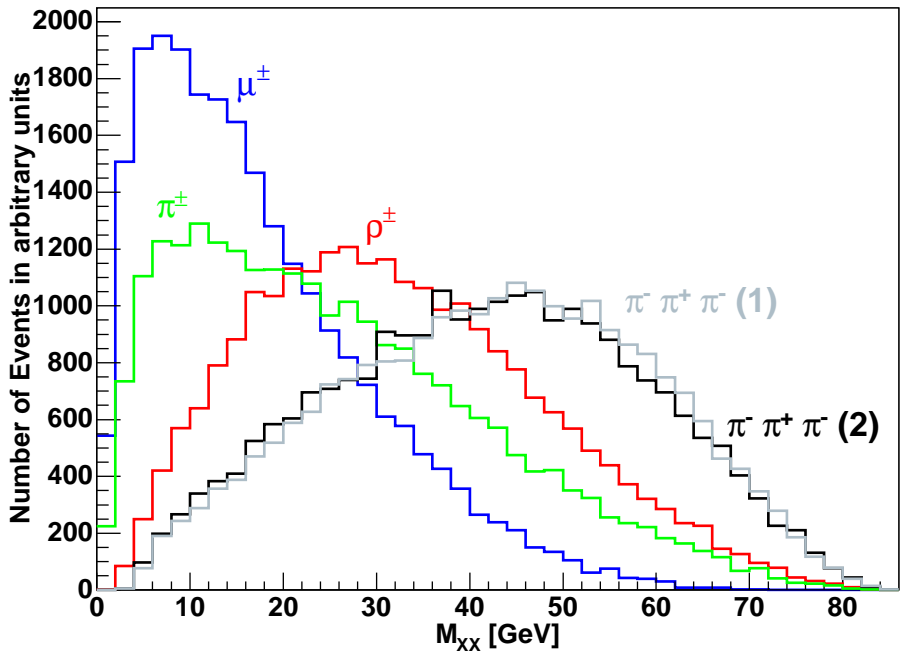} \caption{The red line shows the visible
  dilepton invariant mass distribution in the $\rho^{\pm}$-channel,
  the green line in the $\pi^{\pm}$-channel and the blue line in the
  $\mu^{\pm}$-channel for $m_{0} = 181$ and $m_{1/2} = 350$. The gray
  (1) and the black (2) line show the distribution of the 3-prong
  channel. The first includes the $\tau^{\pm} \ra \pi^{\pm} \rho^{0}
  \nu_{\tau}$ and the direct $\tau^{\pm} \ra \pi^{\pm} \pi^{\mp}
  \pi^{\pm} \nu_{\tau}$ decay whereas in the second only the latter is
  included.}  \label{InvMassll181350} \end{center}
\end{figure}
Unlike electrons and muons, the tau leptons cannot be fully
reconstructed, since they decay in at least one neutrino within the
detector. However, it is the aim to find the invariant mass upper
limit of both tau leptons in the leptonic $\chinonn$ decay. The
kinematic endpoint of the visible dilepton invariant mass
distribution is the same as the limit of the true invariant mass
because neutrinos are massless.  However, its shape is strongly
dependent on the chosen visible particles. The branching ratios for
the tau decays in detectable particles are shown in
table~\ref{tauBR}. For the $\tau^{\pm} \ra \rho^{\pm}
\nu_{\tau}$ channel, the visible mass of the $\chinonn$ decay
products tends to be higher than for the other channels, except the
3-prong, which can be seen in figure~\ref{InvMassll181350}.
There the neutrinos get less energy than in the other decay chains
because of the relatively large rho mass. The branching ratio of
$25.41\%$, compared to the $14.57\%$ of the 3-prong channel, is a
further advantage of this channel. Therefore, in the following
analysis the $\tau^{\pm} \ra \rho^{\pm} \nu_{\tau}$ channel is chosen.
The $\rho^{\pm}$ used here are at Monte Carlo level with a mean mass
of $766\mev$ and a width of about $126\mev$ given by PYTHIA.
\begin{table}[H]
\begin{center}
\begin{tabular}{ll}
$\tau^{\pm} \ra \rho^{\pm} \nu_{\tau}$  &(25.41\%)\\\nonumber
$\tau^{\pm} \ra \mu^{\pm} \nu_{\mu} \nu_{\tau}$ &(17.36\%)\\\nonumber
$\tau^{\pm} \ra e^{\pm} \nu_{e} \nu_{\tau}$ &(17.84\%)\\\nonumber
$\tau^{\pm} \ra h^{-} h^{-}  h^{+} \geq 0$\ neutrals\ $\nu_{\tau}$
 &(14.57\%)\\\nonumber \ \ \ \ (3-prong,\ ex.\ $K^{0}_{S} \ra \pi^{+}
\pi^{-}$)\\\nonumber
$\tau^{\pm} \ra \pi^{\pm} \nu_{\tau}$ &(11.06\%)\\\nonumber
Other  &(13.76\%)\\\nonumber
\end{tabular}
\caption{The branching ratios of the tau lepton~\cite{topmass}.}\label{tauBR}
\end{center}
\end{table}
%----------------------------------------
\subsection{\texorpdfstring{The kinematic limit of
$M_{\rho^{\pm}\rho^{\mp}}$}{The kinematic limit of
M(rr)}}\label{sec:Mll} 
\begin{figure}[H]
 \begin{center}
    \shadowbox{\includegraphics[width=\textwidth/2]{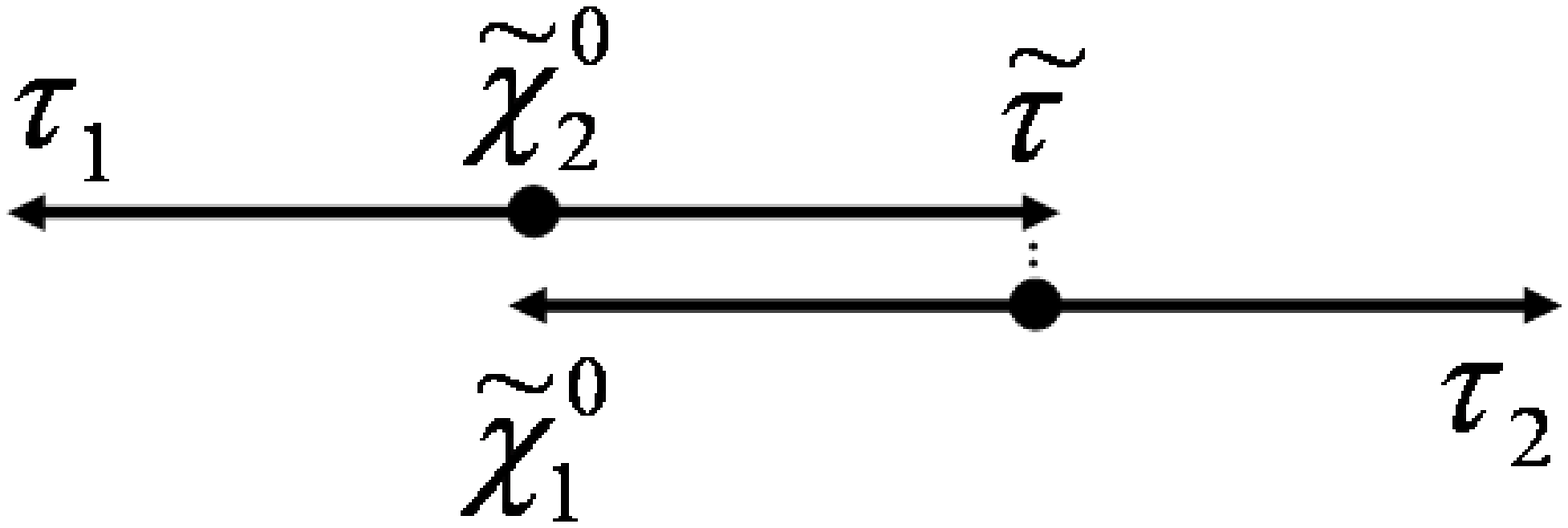}}
  \end{center}
\end{figure}
\begin{figure}[H]
  \begin{center} \includegraphics[height=\textheight/2 - 4cm]
  {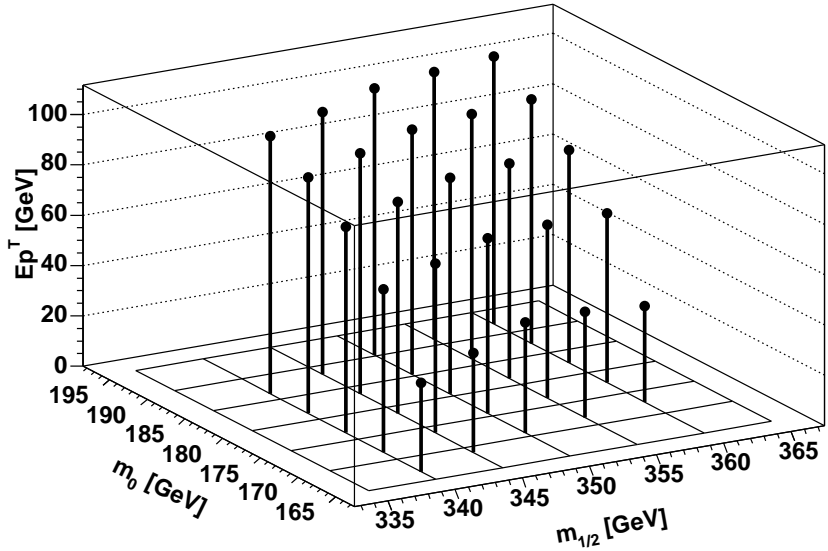} \caption{Theoretical kinematic endpoints in$\gev$
  for the invariant mass of both opposite-sign rhos coming from the
  $\chinonn$ and the $\tilde{\tau}$ respectively. The endpoint masses
  decrease significantly for decreasing values of $m_{0}$.}
  \label{Mll_TEP} \end{center}
\end{figure}

The mass of the $\tilde{\tau}^{\pm}$ is quite similar to that of the
$\chinon$ for small values of $m_{0}$ in the chosen mSUGRA parameter
space. Formula~(\ref{eq:llmax}) shows that in this case the endpoint
is close to zero. Therefore, the theoretical values for the dilepton
kinematic limit given in figure~\ref{Mll_TEP} have a strong $m_{0}$
and $m_{1/2}$ dependence. In the decay chain of the $\chinonn$ at high
$\tan \beta$ there will in about 6\% of the cases be two opposite-sign
rho mesons which can be used for the endpoint measurement.  The shape
of the invariant mass spectrum of both rho mesons shows that there is
no sharp edge near the kinematic limit since it is smeared out due to
the neutrinos. However, there are two possibilities to measure the
endpoints (see section~\ref{sec:methods}). In the following analysis
the whole data samples with about 25000 events are used to determine
the corrections and systematic uncertainties for both methods.

\begin{figure}[H]
  \begin{center} \includegraphics[height=\textheight/2 - 4cm]
  {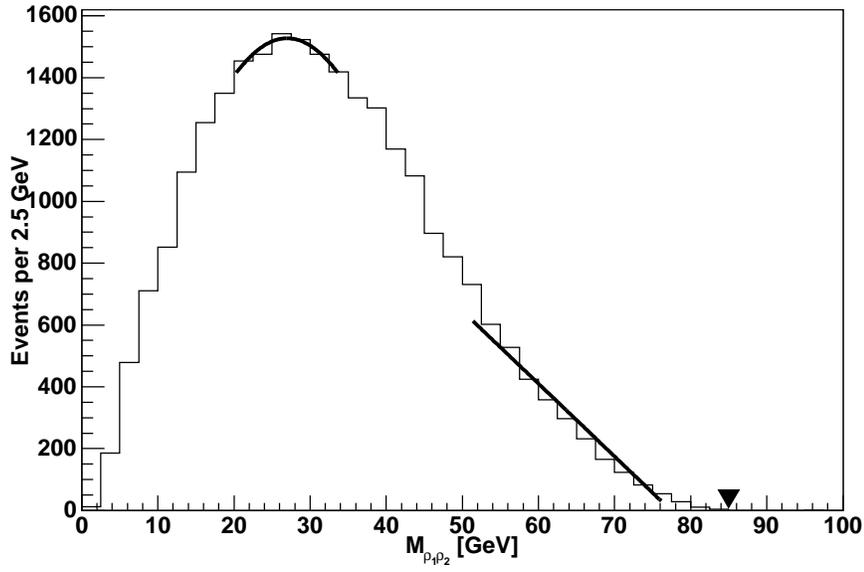} \caption{Example for the linear and the
  gaussian fit at $m_{0} = 181$ and $m_{1/2} = 350$. In this
  distribution 25071 events are represented. The triangle shows the
  theoretical kinematic endpoint at $85.8\gev$.} 
  \label{Mll_Ex_181_350} \end{center}
\end{figure}

\begin{figure}[H]
  \begin{center} \includegraphics[height=\textheight/2 - 4cm]
  {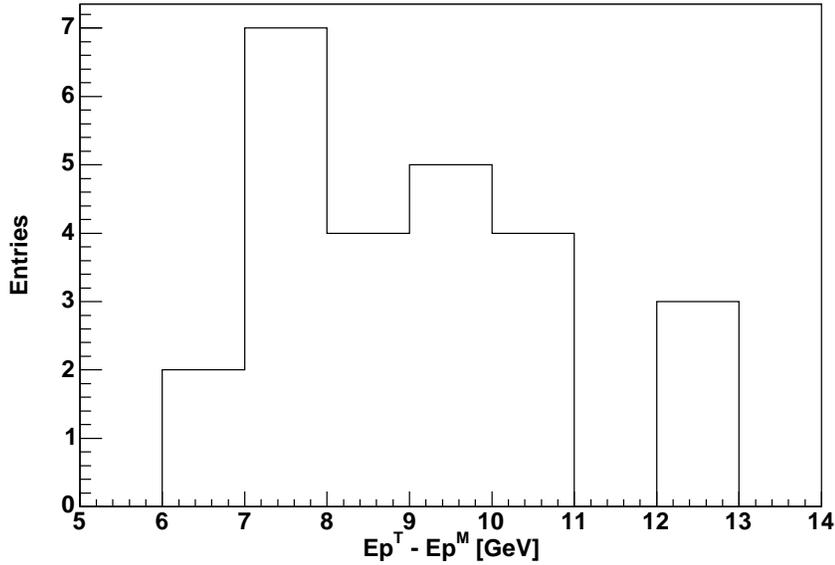} \caption{The difference of the
  theoretical and the measured $M_{\rho^{\pm}\rho^{\mp}}$ endpoint
  in$\gev$ obtained with the linear fit at the 25 investigated points
  in the $m_{0}$-$m_{1/2}$ plane. It shows a mean value of $9.1\gev$
  with a root-mean-square deviation of $1.8\gev$.} 
  \label{Mll_Linear_TEP-MEP} \end{center}
\end{figure}

\begin{figure}[H]
  \begin{center} \includegraphics[height=\textheight/2 - 4cm]
  {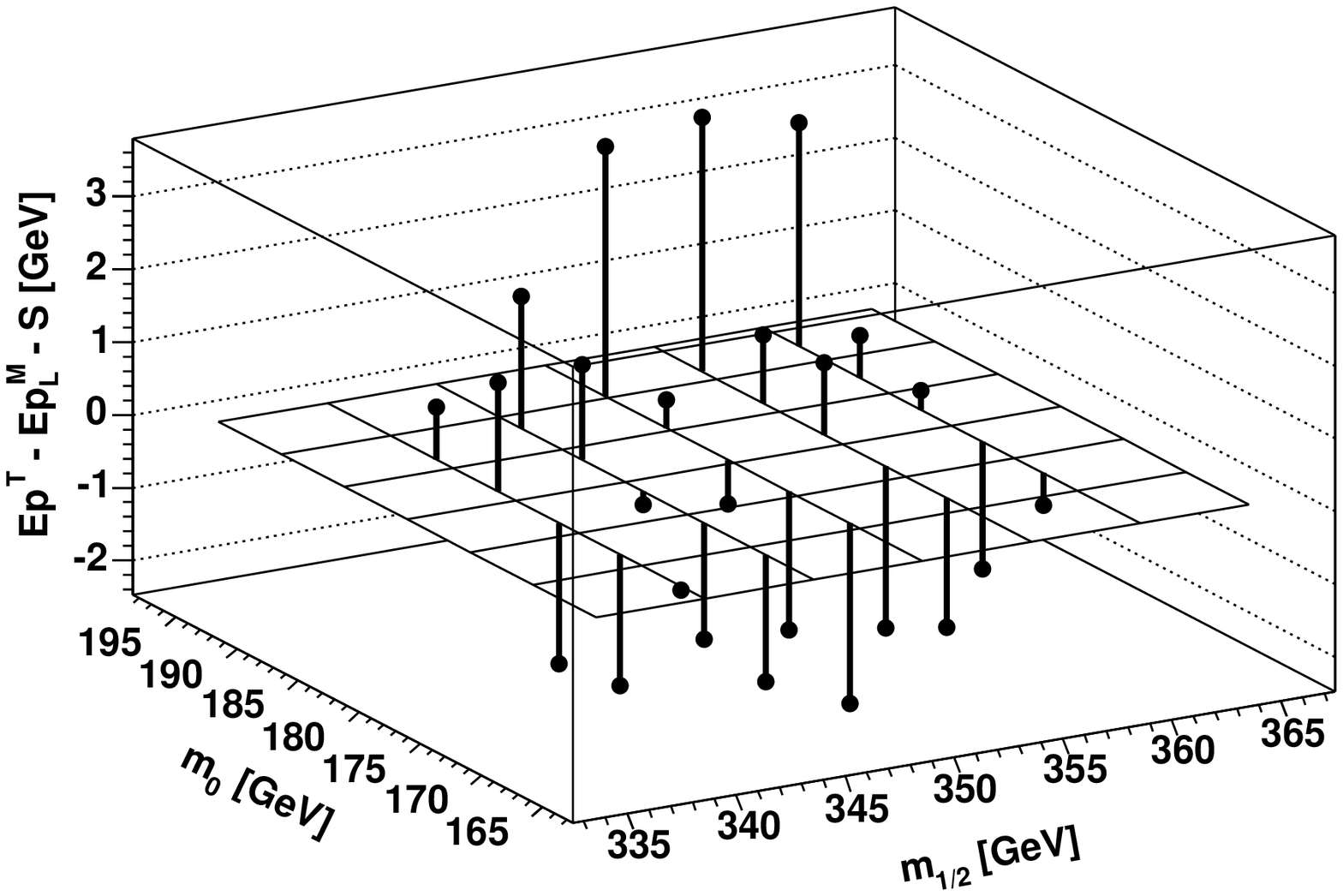} \caption{The theoretical
  $M_{\rho^{\pm}\rho^{\mp}}$ endpoint minus the measured endpoint
  in$\gev$ obtained with the linear fit after applying the constant
  correction of $9.1\gev$. The values decrease significantly for
  decreasing values of $m_{0}$.} \label{Mll_Linear_TEP-MEP-Cor}
  \end{center}
\end{figure}

%--linear--
First, it is possible to perform a linear fit to the tail of the
distribution. An example is given in figure~\ref{Mll_Ex_181_350}. The
measured endpoint mass is too small for every point in the
investigated region. The mean statistical uncertainty $\delta_{stat}
Ep^{M}$ on this measurement for these large data samples is $0.2\gev$
and therefore negligible. The mean correction $S$ is $9.0\gev$ which
can be seen in figure~\ref{Mll_Linear_TEP-MEP}. Indeed, the
uncertainty $\delta S = 1.8\gev$ on this correction is small, but as
can be seen in figure~\ref{Mll_Linear_TEP-MEP-Cor} the real needed
correction should increase with increasing $m_{0}$. The pure mean
systematic uncertainty $\delta_{syst} Ep^{M}$ is $1.7\gev$.

\begin{figure}[H]
  \begin{center} \includegraphics[height=\textheight/2 - 4cm]
  {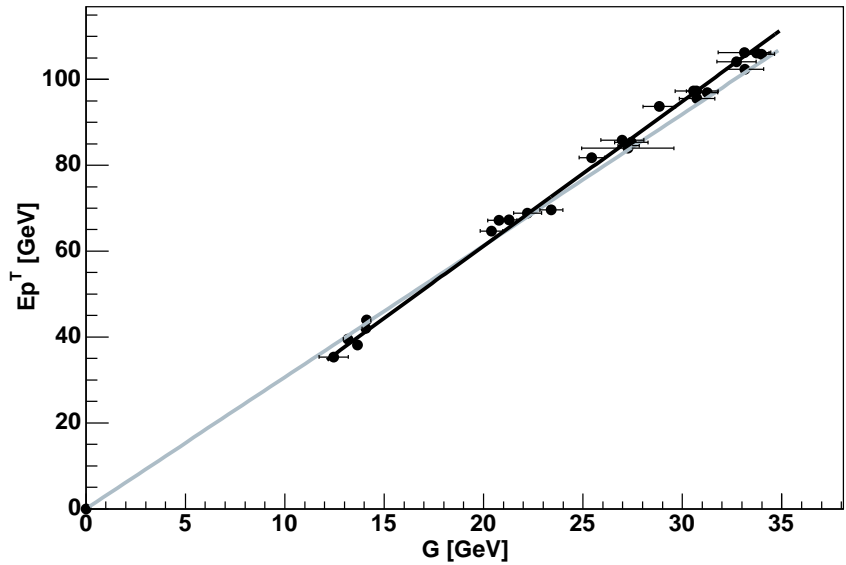} 
\caption{The theoretical dilepton kinematic endpoints as a function
of the measured gaussian maximum for all 25 investigated points in the
$m_{0}$-$m_{1/2}$ plane. The gray line is forced to go through the
origin whereas the black has an optimised ordinate value.}
\label{Mll_Gauss_TE_vs_MM_Fit} \end{center}
\end{figure}

\begin{figure}[H]
  \begin{center} \includegraphics[height=\textheight/2 - 4cm]
  {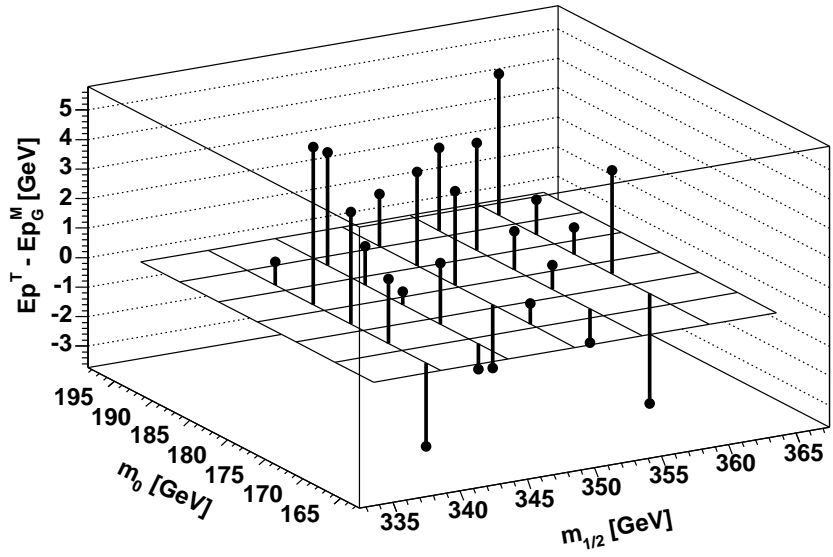} \caption{The difference of the
  theoretical and the measured $M_{\rho^{\pm}\rho^{\mp}}$ endpoint
  in$\gev$ obtained with the gaussian fixed origin method. The
  measured values are on average lower by $1.5\gev$.} 
  \label{Mll_GaussFZ_TEP-MEP} \end{center}
\end{figure}

\begin{figure}[H]
  \begin{center} \includegraphics[height=\textheight/2 - 4cm]
  {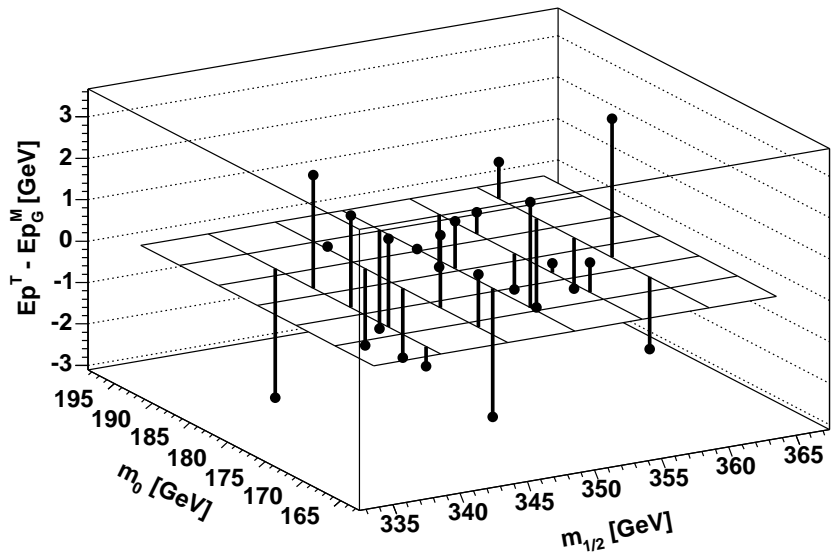} \caption{The difference of the
  theoretical and the measured $M_{\rho^{\pm}\rho^{\mp}}$ endpoint
  in$\gev$ obtained with the gaussian non-fixed origin method. } 
  \label{Mll_GaussNFZ_TEP-MEP} \end{center}
\end{figure}

%--gaussian--
The second method can be applied for this distribution since the
gaussian fit is stable for all 25 points and the assumed linear
dependence between $G_{i}$ and $Ep^{T}_{i}$ is sensible (see
figure~\ref{Mll_Gauss_TE_vs_MM_Fit}). If the linear fit of
$Ep^{T}_{i}$ as a function of $G_{i} \pm \delta G_{i}$ is forced to go
through the origin, the resulting slope $C$ is 3.1. The statistical
uncertainty $\delta_{stat} Ep^{M}$ on the endpoint measurement has
then a mean value of $2.2\gev$. However, in this case a systematic
shift $S$ of $1.5 \pm 2.3\gev$ is needed in the whole region to
improve the endpoint measurement (see
figure~\ref{Mll_GaussFZ_TEP-MEP}). If the fit is not forced to
intersect the origin the value $C$ changes to 3.4, no systematic
shift $S$ is needed and $\delta S$ decreases to $1.9\gev$.  The fit
here intersects the ordinate at $D = -6.1\gev$.  The mean statistical
uncertainty $\delta_{stat} Ep^{M}$ on the endpoint measurement
increases slightly to $2.4\gev$. The mean pure systematic
uncertainty is not defined here because $\delta S$ is dominated by the
statistical uncertainty in the data samples. The difference of the
theoretical endpoint $Ep^{T}_{i}$ and $Ep^{M}_{i}$ shown in
figure~\ref{Mll_GaussNFZ_TEP-MEP} has no significant $m_{0}$ and
$m_{1/2}$ dependence.
\begin{table}[H]
\begin{center}
\begin{tabular}{|l|rrr|} \hline
\emph{Value} & Linear & Gaussian FO & Gaussian NFO\\\hline\hline
$\delta_{stat} Ep^{M}$ & 0.2 & 2.2 & 2.4\\
$S$ & 9.1 & 1.5 & 0\\
$\delta S$ & 1.8 & 2.3 & 1.9\\\hline
\end{tabular}
\caption{A summary of the calculated mean uncertainties and systematic
shifts in$\gev$. $\delta_{stat} Ep^{M}$ is the statistical uncertainty
on the endpoint measurement, $S$ is the mean difference of the
measured and the theoretical endpoint with the root-mean-square
deviation $\delta S$ defined in formula~\ref{S} and \ref{DS}. FO
denotes the results with fixed origin and NFO with non-fixed origin in
the $G_{i}$-$Ep^{T}_{i}$ plane.}\label{sum:Mll}
\end{center}
\end{table}

\pagebreak
%-----------------------------------------------------------------
\subsection{Selection of quarks}
\begin{figure}[H]
  \begin{center} \includegraphics[width=\textwidth]
  {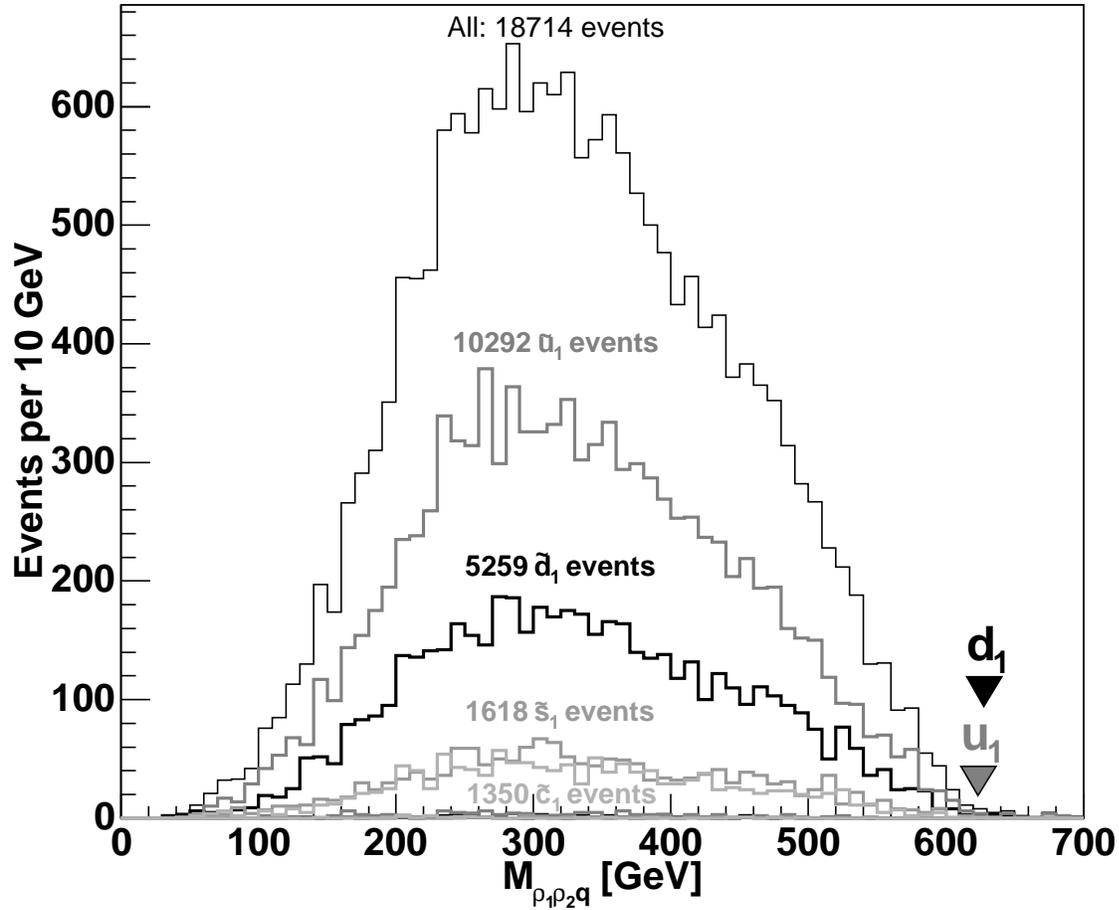} \caption{The invariant mass distributions for
  two opposite-sign rhos with the associated light quark $q$
  ($\tilde{q} \ra q + \chinonn$). The gray triangle shows the
  theoretical endpoint for the $\tilde{u}_{1}$ and $\tilde{c}_{1}$
  distribution at $624\gev$ and the black triangle that for the
  $\tilde{d}_{1}$ and $\tilde{s}_{1}$ distribution at $628\gev$. The
  $\tilde{u}_{1}$ distribution clearly dominates, but due to the small
  difference of both theoretical endpoints the effect is small, too.} 
  \label{Light_Mixing} \end{center}
\end{figure}

In order to get more equations for the reconstruction of the sparticle
masses it is necessary to include the quark which comes from the
process $\tilde{q}_i \ra \chinonn + q_i$. Since the squark masses are
flavour dependent the kinematic limits are as well. In the detector it
may in principle be possible to distinguish up, down, charmed and
strange quarks from bottom and top quarks. The top quark decays mainly
into a bottom quark and a $\W$ boson. Therefore, by using the b-tagging
it is in principle possible to divide the processes into two groups,
light and heavy quarks. The upper kinematic limits of the invariant
mass distributions with quarks depend on the most heavy
squark. In the investigated region the heaviest squarks among the
$\tilde{u}$, $\tilde{d}$, $\tilde{c}$ and $\tilde{s}$ are the left
handed $\tilde{d}_{1}$ and $\tilde{s}_{1}$. Therefore only this mass
can be reconstructed. The kinematic limits for events with
$\tilde{d}_{2}$, $\tilde{s}_{2}$, $\tilde{u}_{1,2}$ and
$\tilde{c}_{1,2}$ are hidden within the measured distributions.
However, due to the proton structure the $\tilde{u}_{1}$ dominates the
distributions as can be seen in figure~\ref{Light_Mixing}.  The heavy
quark distributions are strongly dominated by the $\tilde{b}_{1}$
($\tilde{t}_{1}$), however, $\tilde{b}_{2}$ ($\tilde{t}_{2}$) are
heavier and determine the value for the upper kinematic limit. This
problem is discussed in more detail in section~\ref{sec:Mllb}. In this
study bottom and top quarks are treated separately in order to
understand their different contributions to the invariant mass
distributions. This knowledge can then be used for more realistic
simulations. The following results, which include information from
quarks, are strongly dependent on the precision of the quark energy
reconstruction. For the analysis quarks are used at the parton
level. Indeed, this is a very crude approach but, nevertheless, the
methods developed here may also be valid for the real experiment.
%----------------------------------------
\subsection{\texorpdfstring{The kinematic limit of $M_{\rho^{\pm}\rho^{\mp}q}$ for
light quarks}{The kinematic limit of M(rrq) for light
quarks}}\label{sec:Mllq} 
\begin{figure}[H]
  \begin{center}
    \shadowbox{\includegraphics[width=\textwidth/2]{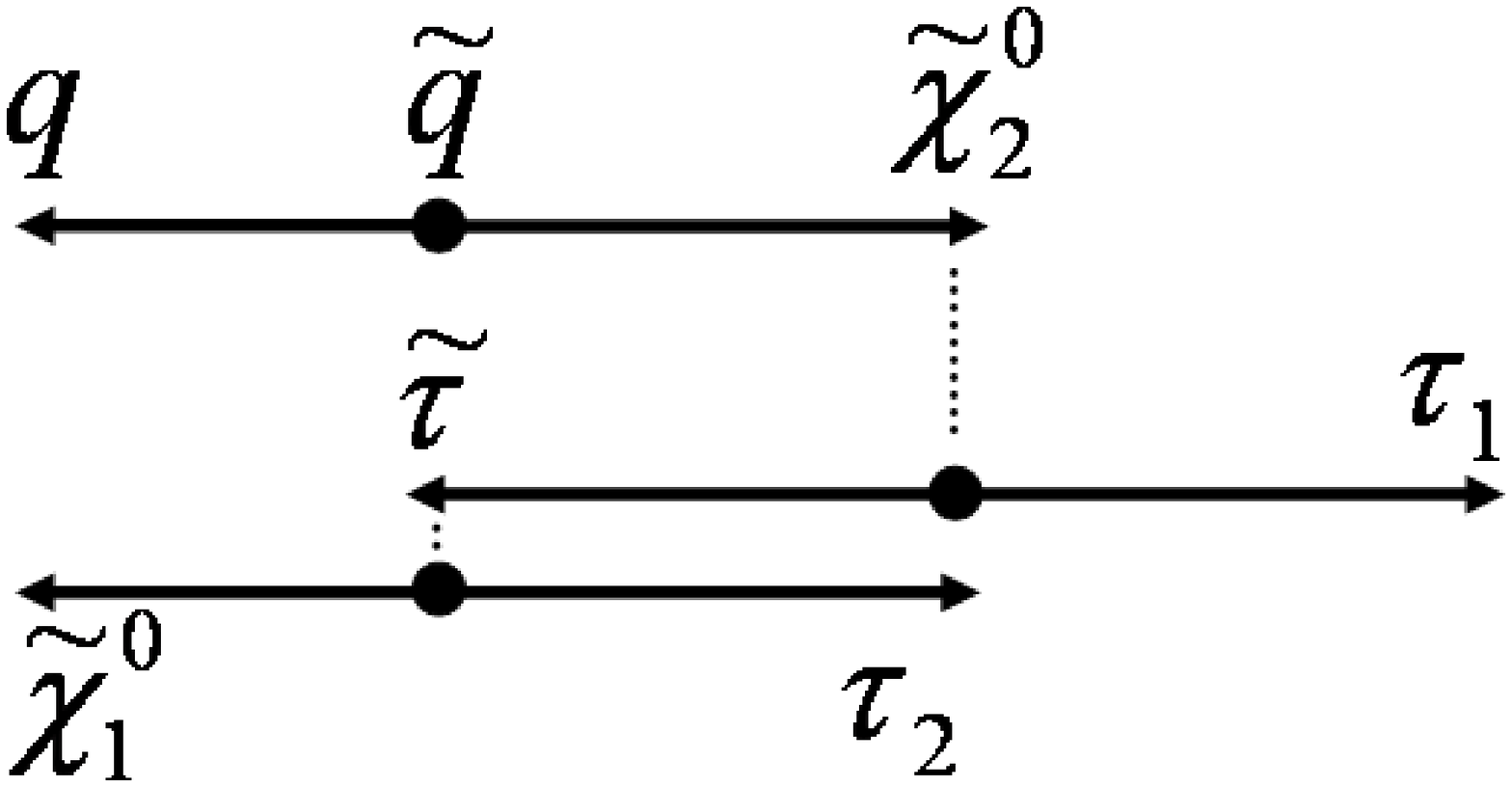}}
  \end{center}
\end{figure}
\begin{figure}[H]
  \begin{center} \includegraphics[height=\textheight/2 - 4cm]
  {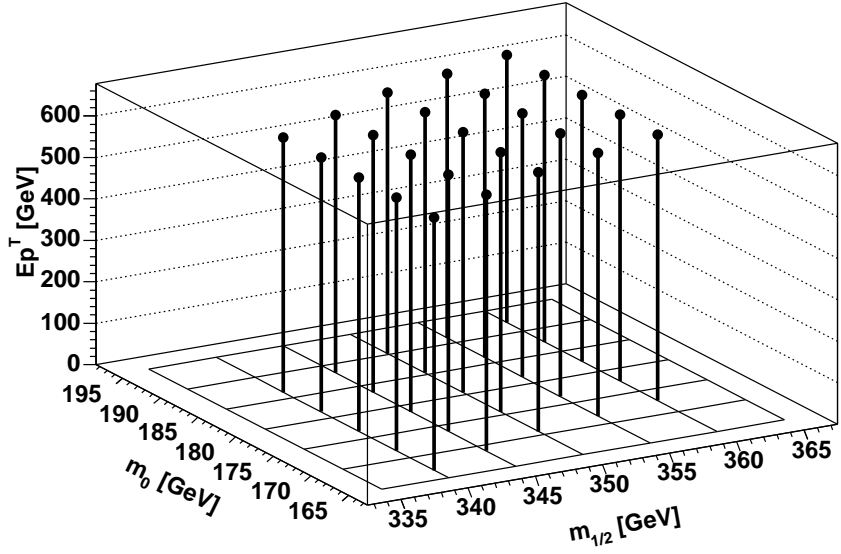} \caption{Theoretical kinematic endpoints in$\gev$
  for the invariant mass of both opposite-sign rhos, coming from the
  $\chinonn$ and the $\tilde{\tau}$ respectively, and the associated
  quark from $\tilde{s}_1,\tilde{d}_1 \ra s,d + \chinonn$ in the
  investigated area.}  \label{Mllq_TEP} \end{center}
\end{figure}
\begin{figure}[H]
  \begin{center} \includegraphics[height=\textheight/2 - 4cm]
  {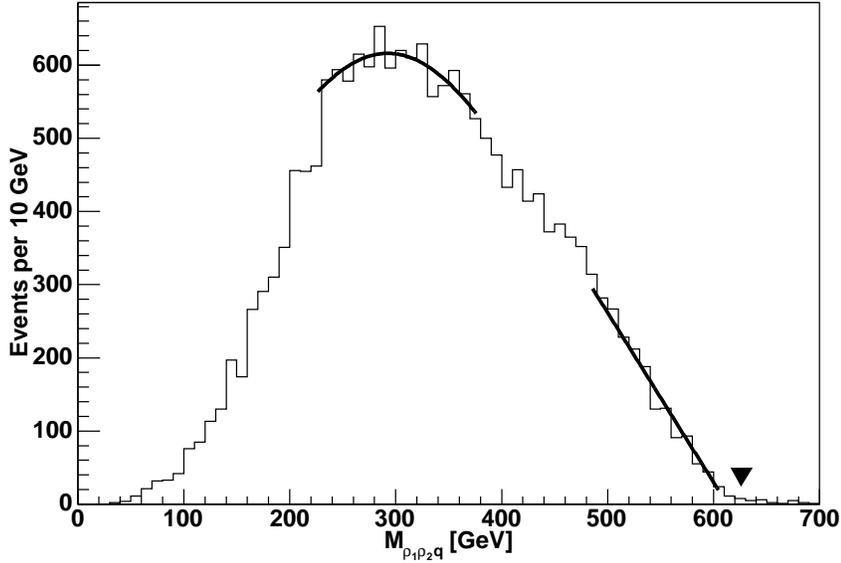} \caption{Example for the linear and the
  gaussian fit at $m_{0} = 181$ and $m_{1/2} = 350$. The invariant
  mass distribution is based on 18714 light quark events which are
  defined in section~\ref{sec:prod_decay_sel}. The triangle shows the
  theoretical kinematic endpoint of $M_{\rho^{\pm}\rho^{\mp}q}$ at
  $627.8\gev$.} \label{Mllq_Ex_181_350} \end{center}
\end{figure}
\begin{figure}[H]
  \begin{center} \includegraphics[height=\textheight/2 - 4cm]
  {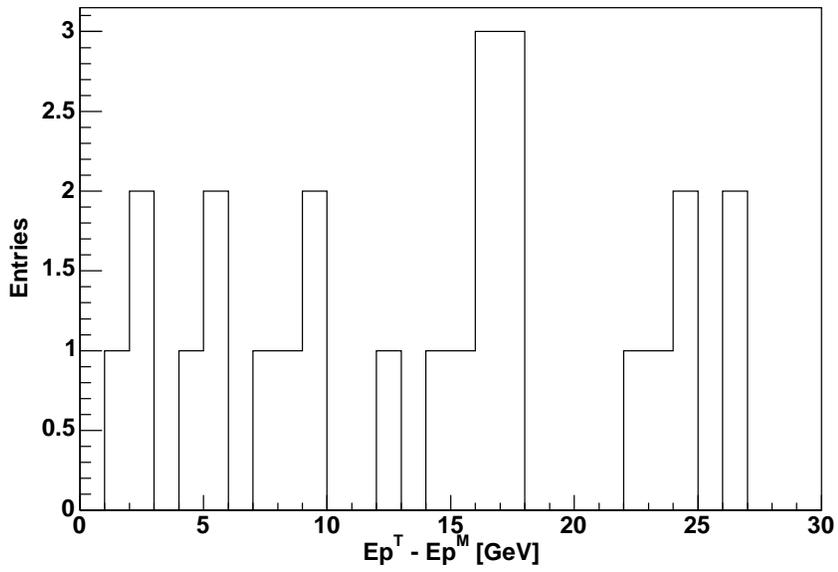} \caption{The theoretical
  $M_{\rho^{\pm}\rho^{\mp}q}$ endpoint minus the measured endpoint
  in$\gev$ for all 25 points. The mean value is $14.0\gev$ with a
  root-mean-square deviation of $7.8\gev$.} 
  \label{Mllq_Linear_TEP-MEP} \end{center}
\end{figure}
\begin{figure}[H]
  \begin{center} \includegraphics[height=\textheight/2 - 4cm]
  {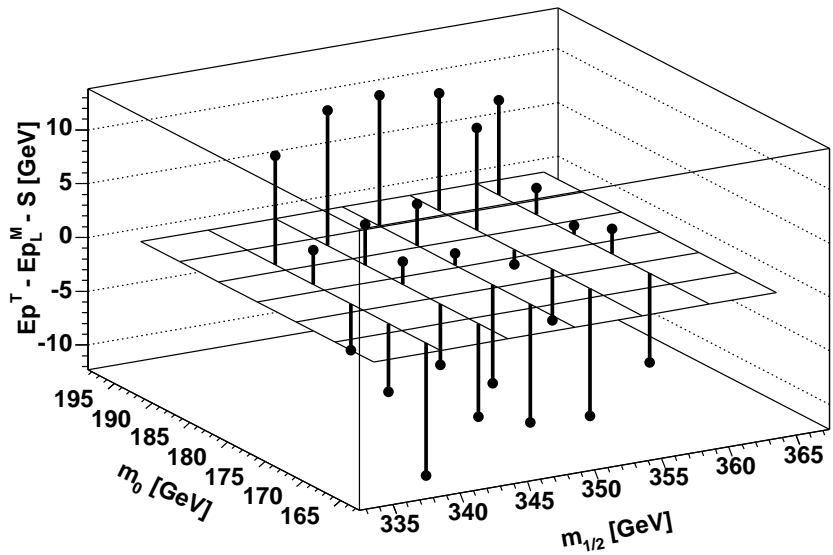} \caption{The theoretical endpoint
  minus the measured endpoint in$\gev$ after applying a constant
  correction of $14.0\gev$.} \label{Mllq_Linear_TEP-MEP-Cor}
  \end{center}
\end{figure}
\begin{figure}[H]
  \begin{center} \includegraphics[height=\textheight/2 - 4cm]
  {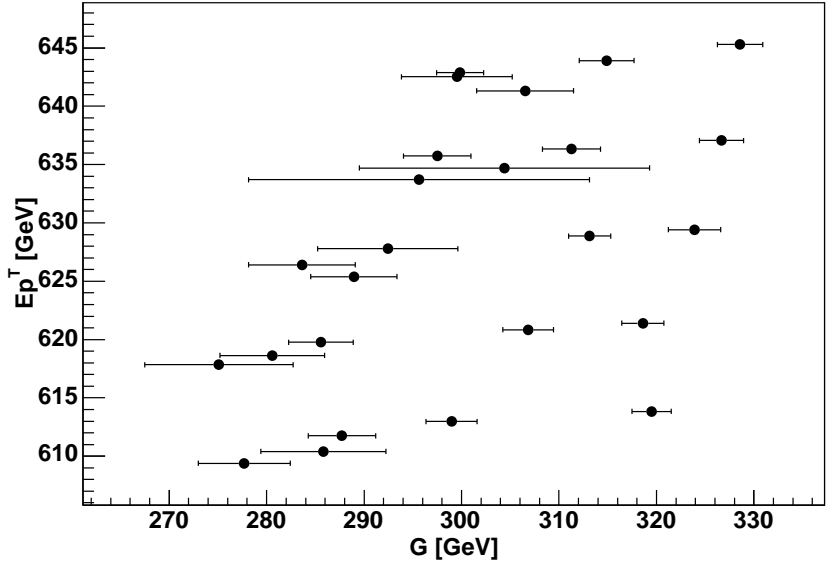} 
\caption{The theoretical kinematic dilepton
  endpoints as a function of the measured gaussian maximum for all 25
  investigated points in the $m_{0}$-$m_{1/2}$ plane. A unique mapping
  of the measured maximum to the theoretical endpoint can not be
  established since several theoretical endpoint values are associated
  with the same measured maximum. Additionally, some distributions are
  flat around the maximum which leads to large uncertainties.}
  \label{Mllq_Gauss_TE_vs_MM_Fit} \end{center}
\end{figure}

The invariant mass distribution of the quark
which comes from the decay $\tilde{q} \ra q + \chinonn$ and the same
two rho mesons used in section~\ref{sec:Mll} carries further
information which can be used for the reconstruction of sparticle
masses. Figure~\ref{Mllq_TEP} shows that the theoretical values of the
kinematic limits have a very small $m_{0}$ dependence and a slightly
stronger $m_{1/2}$ dependence in that area compared to the dilepton
case. The reason is that the last factor in formula~(\ref{eq:llqmax})
in that region is never close to zero in contrast to that in
formula~(\ref{eq:llmax}). The $m_{0}$ dependence is only caused by the
change in the squark mass whereas the $m_{1/2}$ dependence takes also
the changes of both neutralino masses into account.  For the following
analysis data samples of about 18000 events have been used. They
include the same events as used in the previous section with the
constraint that the associated quarks have the flavour up, down,
charmed or strange.
%--linear--
An estimation of the kinematic limit can be achieved with a linear
fit. An example is given in figure~\ref{Mllq_Ex_181_350}.  The
measured value is too low for every point in the parameter space. As
can be seen in figure~\ref{Mllq_Linear_TEP-MEP} and
figure~\ref{Mllq_Linear_TEP-MEP-Cor} the actually needed correction is
not constant within the investigated area. The gaussian method cannot
be applied to this case, since there is no linear dependence between
the maximum and the endpoint - which can be
seen in figure~\ref{Mllq_Gauss_TE_vs_MM_Fit}.

\begin{table}[H]
\begin{center}
\begin{tabular}{|l|r|} \hline
\emph{Value} & Linear \\\hline\hline
$\delta_{stat} Ep^{M}$ & 1.7\\
$S$ & 14.0\\
$\delta S$ & 7.8\\\hline
\end{tabular}
\caption{A summary of the calculated correction and the mean
uncertainties in$\gev$ for the endpoint measurement in the
distribution of $M_{\rho^{\pm}\rho^{\mp}q}$.}\label{sum:Mllq}
\end{center}
\end{table}

%--------------------------------------------------------------------
\subsection{\texorpdfstring{Two kinematic limits in $M_{\rho^{\pm}q}$ for light
quarks}{Two kinematic limits in M(rq) for light
quarks}}\label{sec:Mlq} 
\begin{figure}[H]
  \begin{center}
    \shadowbox{\begin{tabular}{c}
	\includegraphics[width=\textwidth/2]{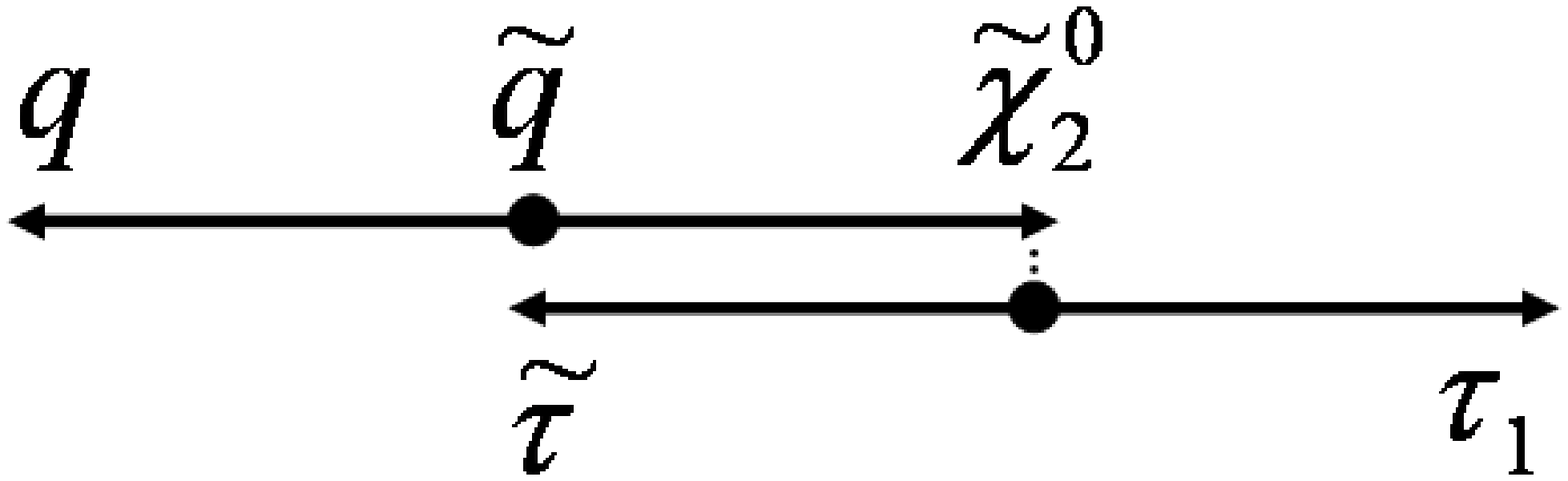}\\
    \includegraphics[width=\textwidth/2]{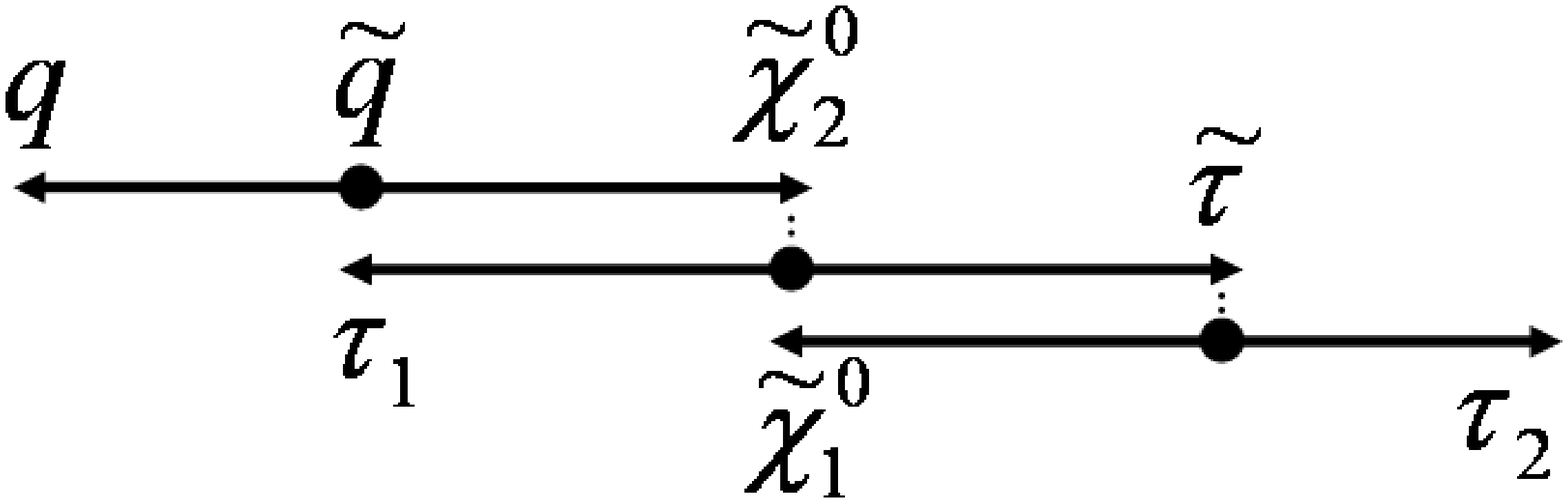}
	\end{tabular}}
  \end{center}
\end{figure}
As described in section~\ref{sec:twobody} there are two different
leptons in the investigated decay chain. The first comes directly from
the $\chinonn$ and the second from the slepton. Both leptons have
different kinematic properties so that the endpoint of the invariant
mass of first lepton with the light quark is not the same as that with
the second lepton. However, in the experiment it will not be possible
to know for each event which is the first lepton and which the
second. In the following it is described how, nevertheless, both
endpoints can be separated if the values are not too close to each
other. For the latter constraint it is necessary to compare
formula~(\ref{eq:l1qmax}) with formula~(\ref{eq:l2qmax}). In the
chosen area of $m_{0}$ and $m_{1/2}$ the $\tilde{\tau}$ mass is much
closer to the $\chinon$ mass than to the $\chinonn$ mass. Thus, the
second endpoint is much lower than the first one and it will be
possible to distinguish both endpoints. First, the method for finding
the second endpoint will be described as its outcome is useful for
measuring the first endpoint.

\begin{figure}[H]
  \begin{center} \includegraphics[width=\textwidth]
  {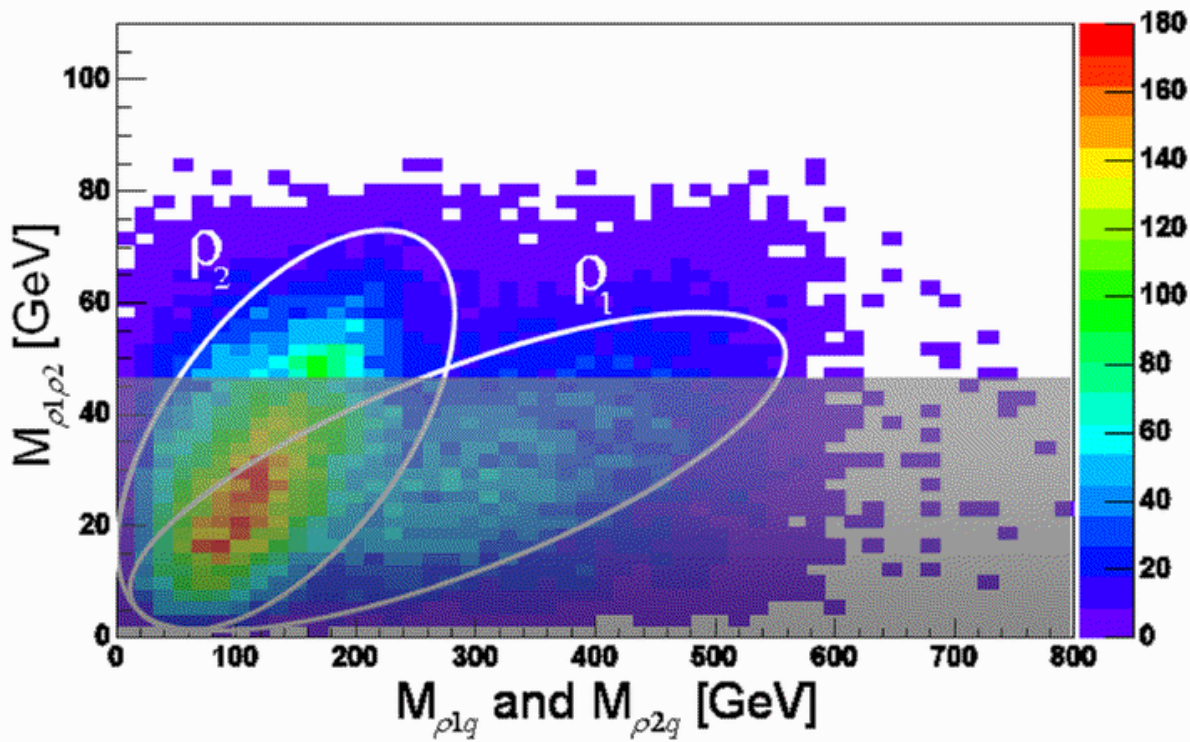} \caption{The invariant mass of both
  opposite-sign rho mesons versus the invariant mass of a single rho
  with the associated quark in$\gev$. It is an example for the point
  I'. The right ellipse denotes the events with the first rho whereas
  the left ellipse includes the second rho events. The lower
  $M_{\rho^{\pm}q}$ endpoint is determined from the events with
  $M_{\rho^{\pm}\rho^{\mp}} > 46\gev$ (unshaded region).}
  \label{R1R2_vs_R1R2Jet} \end{center}
\end{figure}

\begin{figure}[H]
  \begin{center} \includegraphics[height=\textheight/2 - 4cm]
  {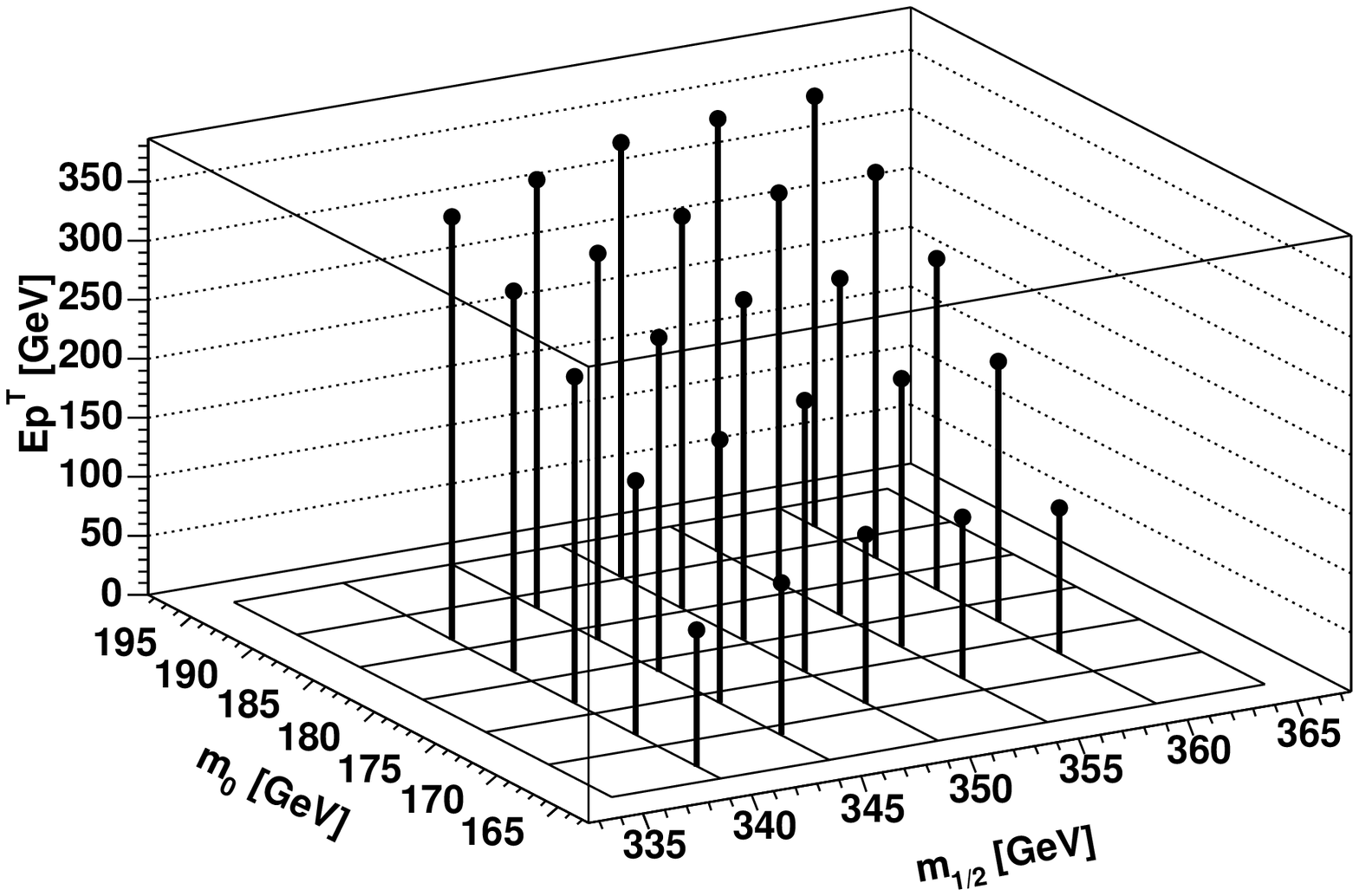} \caption{Theoretical kinematic endpoints in$\gev$
  for the invariant mass of the second rhos coming from the
  $\tilde{\tau}^{\pm}$ and the associated light quark coming from
  the squark in the $\tilde{s}_1,\tilde{d}_1 \ra s,d + \chinonn$
  decay. The endpoints decrease significantly for decreasing values of
  $m_{0}$.}  \label{Ml2q_TEP} \end{center}
\end{figure}

\begin{figure}[H]
  \begin{center} \includegraphics[height=\textheight/2 - 4cm]
  {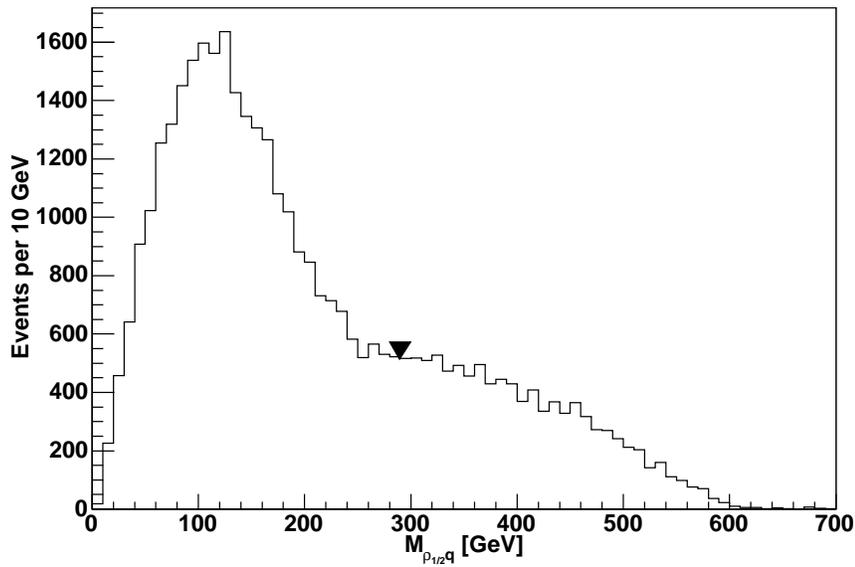} \caption{Example for the invariant mass
  $M_{\rho^{\pm}q}$ distribution at $m_{0} = 181$ and $m_{1/2} = 350$
  before the cut on $M_{\rho^{\pm}\rho^{\mp}}$. In this distribution
  37428 events are represented. The triangle shows the theoretical
  kinematic endpoint of $M_{\rho^{\pm}_{2}q}$ at $288.0\gev$.} 
  \label{Ml2q_Ex_181_350_NC} \end{center}
\end{figure}

\begin{figure}[H]
  \begin{center} \includegraphics[height=\textheight/2 - 4cm]
  {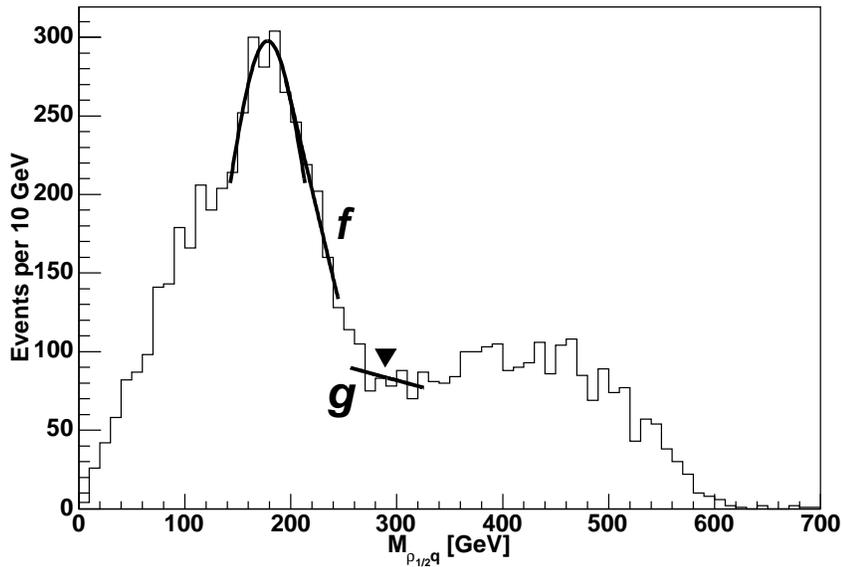} \caption{Example for the two linear fits
  $f$ and $g$ and the gaussian fit at $m_{0} = 181$ and $m_{1/2} =
  350$. In this distribution 6904 events after the cut (55\%) on
  $M_{\rho^{\pm}\rho^{\mp}}$ are represented. The triangle shows the
  theoretical kinematic endpoint of $M_{\rho^{\pm}_{2}q}$ at
  $288.0\gev$.} \label{Ml2q_Ex_181_350} \end{center}
\end{figure}

\begin{figure}[H]
  \begin{center} \includegraphics[height=\textheight/2 - 4cm]
  {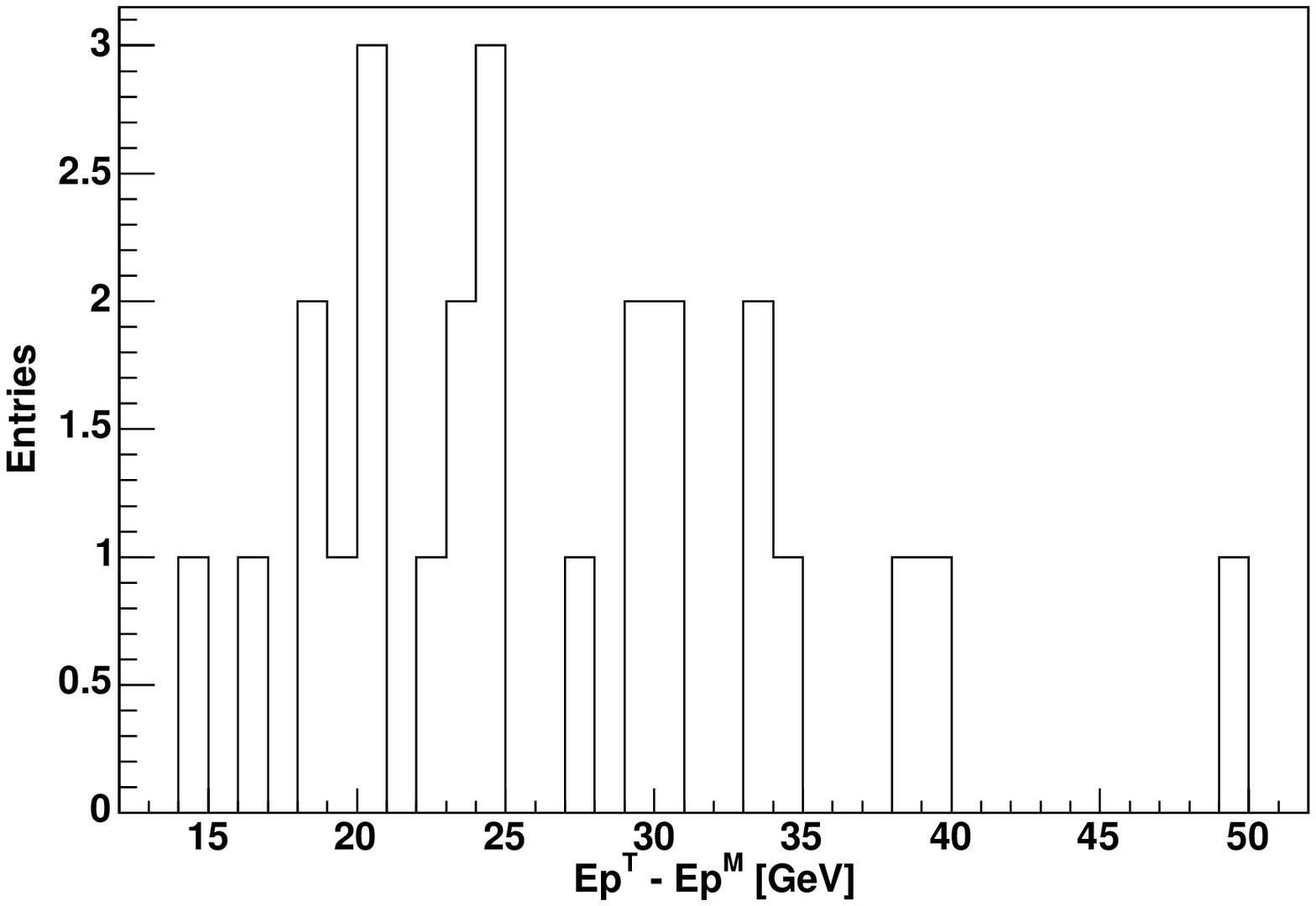} \caption{The difference of the
  theoretical and the measured $M_{\rho^{\pm}_{2}q}$ endpoint obtained
  with the linear fit in$\gev$ for all 25 points. It shows a shift of
  $26.8\gev$ with a root-mean-square deviation of $8.0\gev$.} 
  \label{Ml2q_Linear_TEP-MEP} \end{center}
\end{figure}

\begin{figure}[H]
  \begin{center} \includegraphics[height=\textheight/2 - 4cm]
  {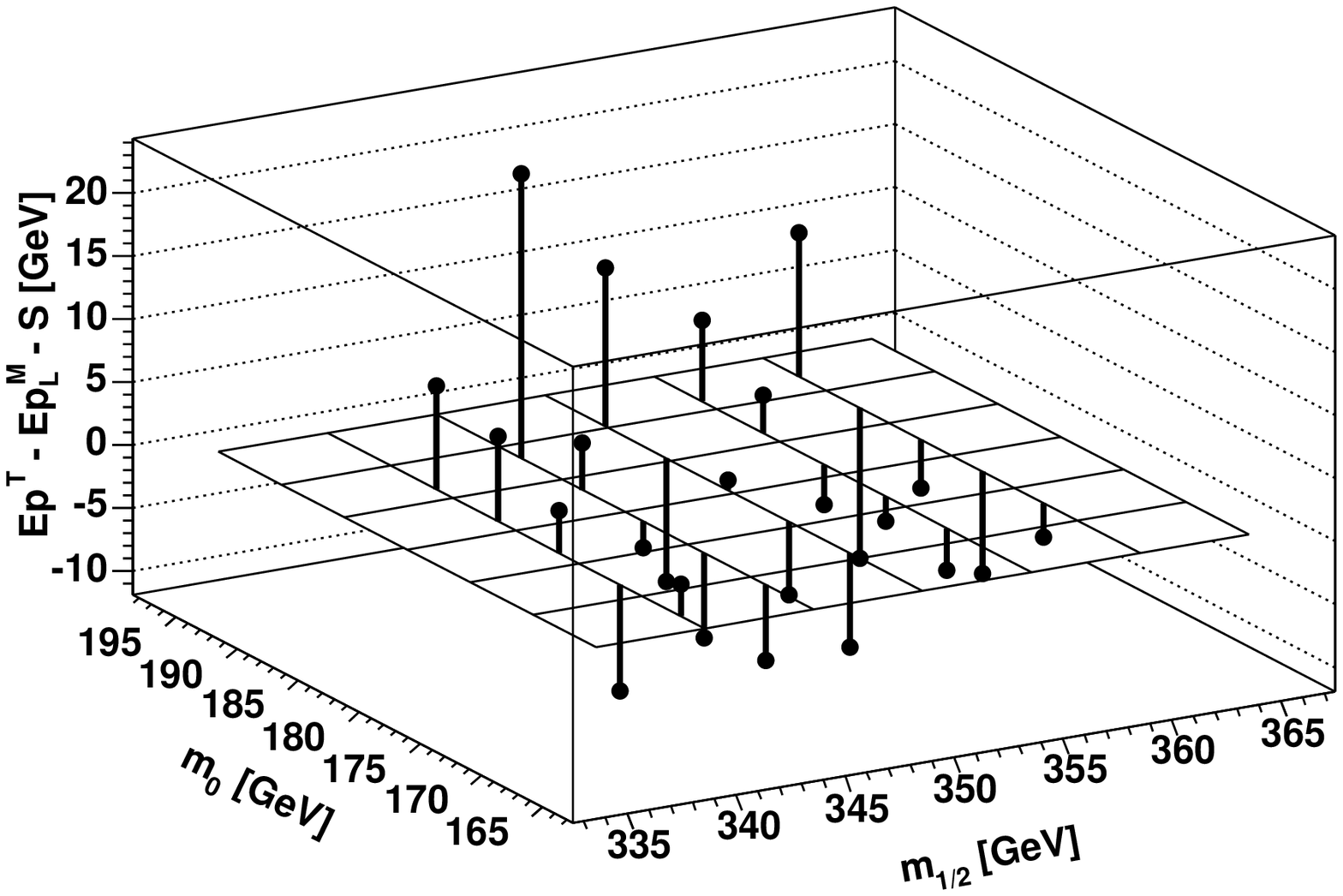} \caption{The theoretical
  $M_{\rho^{\pm}_{2}q}$ endpoint minus the measured endpoint obtained
  with the linear fit in$\gev$ after applying the constant correction
  of $26.8\gev$. The values decrease significantly for decreasing
  values of $m_{0}$.} \label{Ml2q_Linear_TEP-MEP-Cor} \end{center}
\end{figure}

For the measurements of the smaller second endpoint (given in
figure~\ref{Ml2q_TEP}) the first rho gives a background which has to
be avoided. In figure~\ref{R1R2_vs_R1R2Jet} the invariant mass
distribution of the single rho with one quark versus the invariant
mass of both opposite-sign rhos is shown. On the left upper quadrant
the events with the second rho dominate. Therefore, by ignoring the
events with an invariant mass of both rho mesons being less than or
equal to 55$\%$ of its upper limit the background can be well
reduced. This cut value has been uniformly optimised for all 25
points. The ratio of the signal to the first lepton background between
zero and the theoretical value for the second kinematic endpoint can
thus be increased by a factor which is between 1.4 and 1.9, depending
on the values for $m_{0}$ and $m_{1/2}$. This improvement can be seen
by comparing figure~\ref{Ml2q_Ex_181_350_NC} with
figure~\ref{Ml2q_Ex_181_350}. For each individual point in the
parameter space the signal to background ratio could even be improved
by changing the cut value. For the following analysis data samples of
between 5096 and 6806 events after the cut have been used.
The linear fit method has to be adapted to this case. In order to find
the endpoint on the background two linear fits are necessary. The
first fit with the fit function $f(M_{INV})$ is defined the same way
as in section~\ref{sec:methods} except that the chosen region is from
95$\%$ to 45$\%$ with respect to the maximum bin content. The second
fit with the fit function $g(M_{INV})$ iteratively searches the area
right of the second lepton peak where the slope is bigger than $- 0.2$
for a fit over seven bins. The minimal slope and the number of bins
have been optimised for a stable second fit. The aim of this method is
to find the second endpoint on the invariant mass distribution of the
first rho with the associated quark. Here, the endpoint $Ep^{M}_{i}$
is defined as the intercept point of both linear fits which are shown
in figure~\ref{Ml2q_Ex_181_350}. If the two linear fit functions are
defined by
\begin{eqnarray}
f(M_{INV}) &=& A \cdot M_{INV} + B\\ 
g(M_{INV}) &=& C \cdot M_{INV} + D
\end{eqnarray}
the measured endpoint $Ep^{M}_{i}$ is given by the intersection of $f$
and $g$:
\begin{eqnarray}
Ep^{M}_{i} &=& \frac{D_{i} - B_{i}}{A_{i} - B_{i}}
\end{eqnarray}
The associated statistical uncertainty is:
\begin{eqnarray}
(\delta_{stat} Ep^{M})^{2} &=& \frac{\delta_{DD}^{i} +
\delta_{BB}^{i}}{(A_{i} - C_{i})^{2}} + \frac{(D_{i} - B_{i})^{2}
\cdot (\delta_{AA}^{i} + \delta_{CC}^{i})}{(A_{i} - C_{i})^{4}}\\\nonumber 
&& + 2 \frac{(D_{i} - B_{i})^{2} \cdot
\delta_{CD}^{i}\delta_{AB}^{i}}{(A_{i} - C_{i})^{3}}
\end{eqnarray}
The values $\delta_{XY}^{i}$ are the entries of the error matrix which
contains the correlated statistical uncertainties on $A$, $B$, $C$ and
$D$. Since $g$ and $f$ are not correlated the elements
$\delta_{AC}^{i}$, $\delta_{AD}^{i}$, $\delta_{BC}^{i}$ and
$\delta_{BD}^{i}$ are zero. The result of this method is that the
measured endpoint is for every point in the parameter space too low
(Figure~\ref{Ml2q_Linear_TEP-MEP}). A correction $S$ of $26.8 \pm
8.0\gev$ has to be added. The actually needed correction here has the
same $m_{0}$ dependence as that in $M_{\rho^{\pm}\rho^{\mp}}$ which
can be seen in figure~\ref{Ml2q_Linear_TEP-MEP-Cor}.

\begin{figure}[H]
  \begin{center} \includegraphics[height=\textheight/2 - 4cm]
  {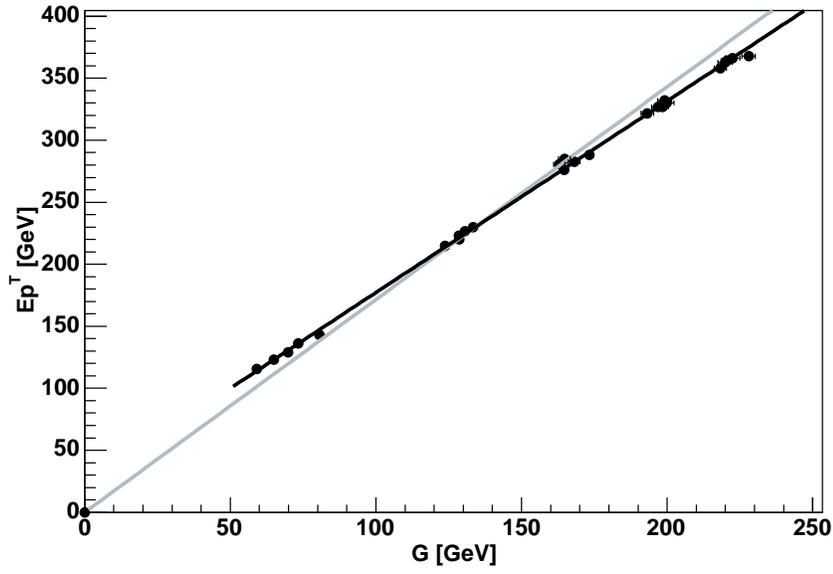} 
\caption{The theoretical kinematic endpoints of $M_{\rho^{\pm}_{2}q}$
as a function of the measured gaussian maximum for all 25 investigated
points in the $m_{0}$-$m_{1/2}$ plane. The gray line is forced to go
through the origin whereas the black has an optimised ordinate value
of $22.7\gev$.}  \label{Ml2q_Gauss_TE_vs_MM_Fit} \end{center}
\end{figure}

The gaussian method here remains the same except that the symmetric
area for the fit changes from 25 $\%$ to 45 $\%$ of the centre bin
value in order to achieve better results. The reason for this change
is that the peak here has a smaller width. As can be seen in
figure~\ref{Ml2q_Gauss_TE_vs_MM_Fit} it is reasonable to assume a
linear dependence of $Ep^{T}_{i}$ on $G_{i} \pm \delta G_{i}$. If the
origin is chosen to be fixed the slope $C$ of the fit is $1.7$. If
the ordinate value is a free parameter the slope $C$ decreases to
$1.5$ with $D = 22.7$. The ratio between both slopes is 1.1.

The calculated uncertainties and the systematic shifts for the linear
and both gaussian methods are given in table~\ref{sum:Ml2q} .

\begin{figure}[H]
  \begin{center} \includegraphics[height=\textheight/2 - 4cm]
  {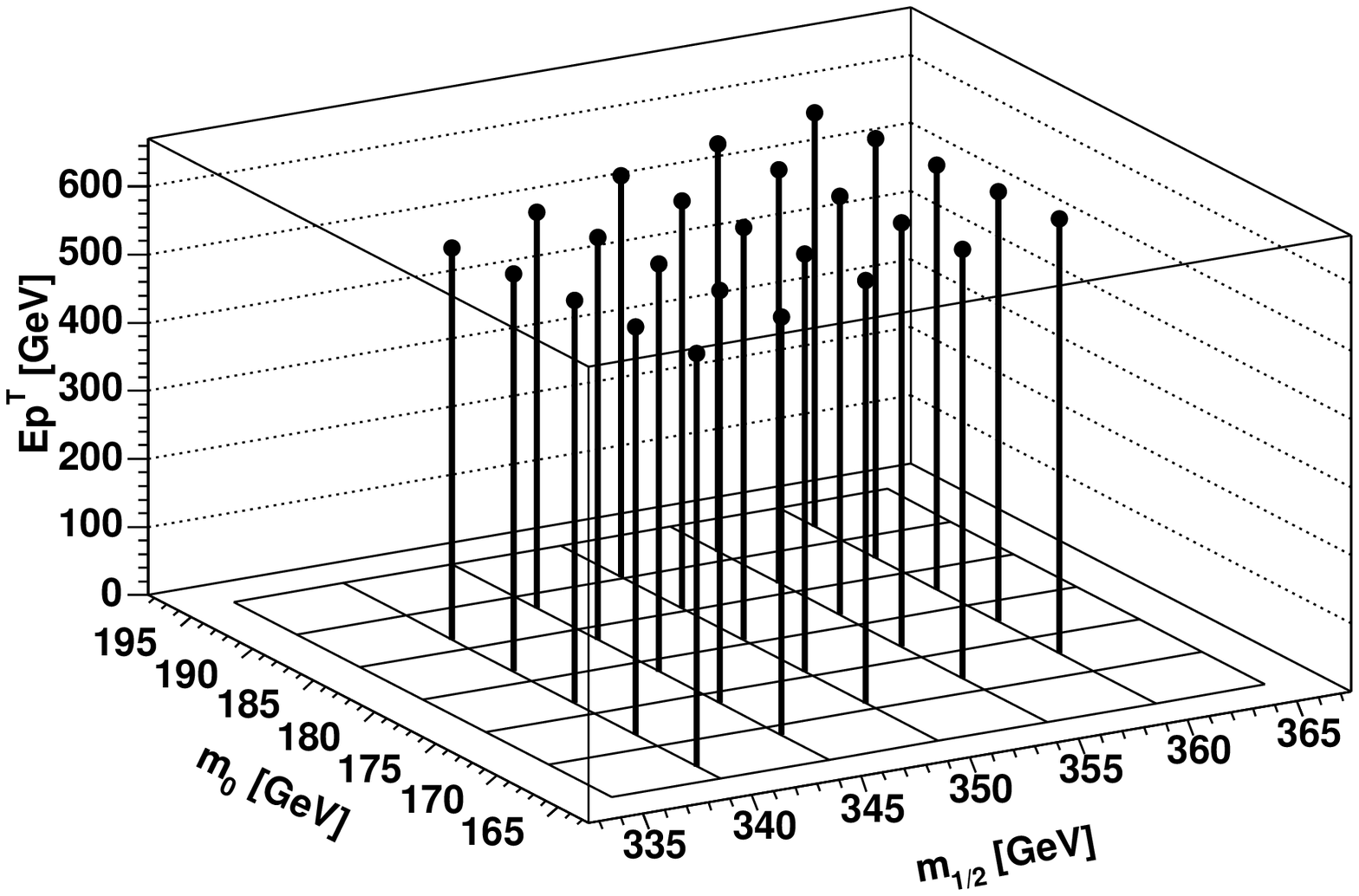} \caption{Theoretical kinematic endpoints in$\gev$
  for the invariant mass of first rho coming from the $\chinonn$ and
  the associated light quark coming from the squark in the
  $\tilde{s}_1,\tilde{d}_1 \ra s,d + \chinonn$ decay.}
  \label{Ml1q_TEP} \end{center}
\end{figure}

\begin{figure}[H]
  \begin{center} \includegraphics[height=\textheight/2 - 4cm]
  {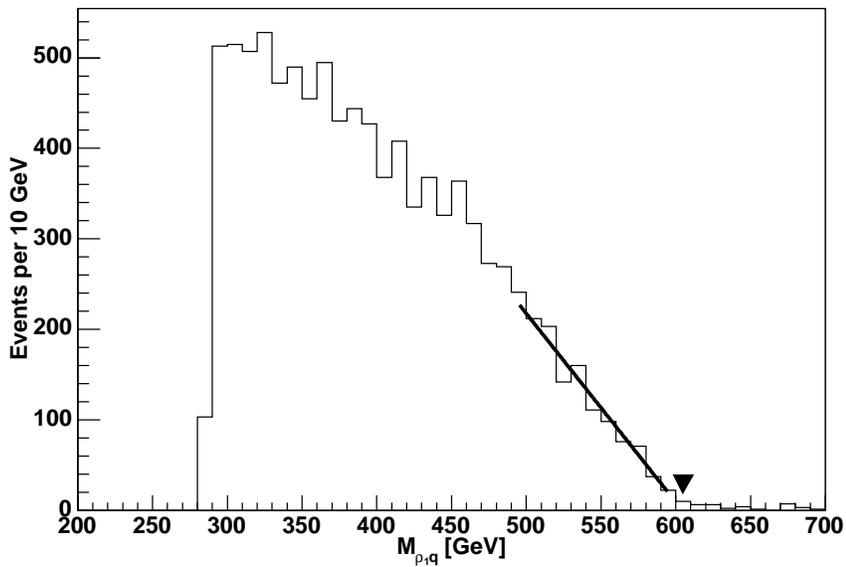} \caption{Example for the linear fit at
  $m_{0} = 181$ and $m_{1/2} = 350$ after applying a cut on the
  invariant mass $M_{\rho^{\pm}q}$ at the theoretical second endpoint
  value. In this example 9855 events are represented. The triangle
  shows the theoretical kinematic endpoint of $M_{\rho^{\pm}_{1}q}$ at
  $606.0\gev$.} \label{Ml1q_Ex_181_350} \end{center}
\end{figure}

\begin{figure}[H]
  \begin{center} \includegraphics[height=\textheight/2 - 4cm]
  {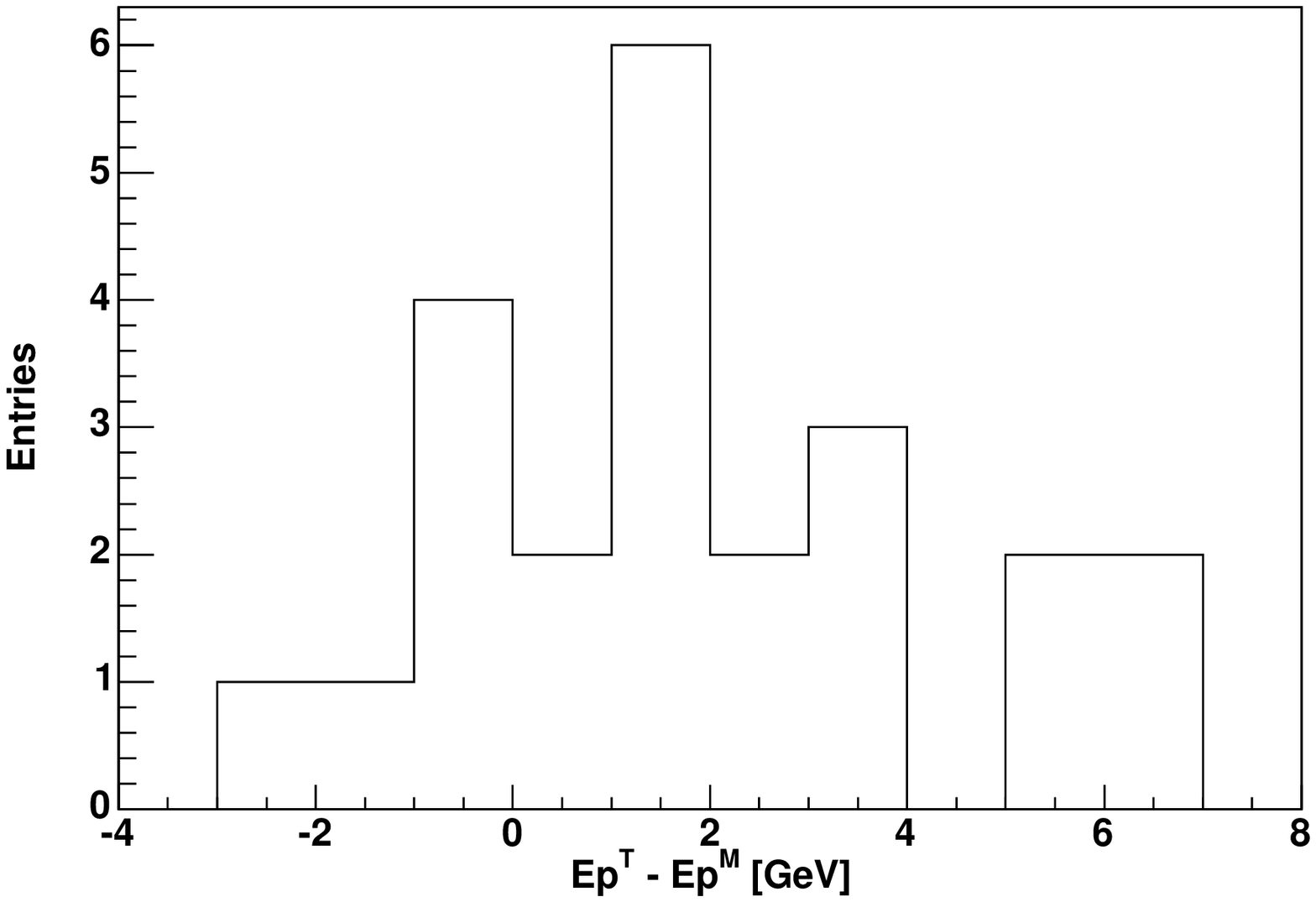} \caption{The difference of the
  theoretical and the measured $M_{\rho^{\pm}_{1}q}$ endpoint in$\gev$
  obtained with a linear fit for all 25 points. It shows a shift of
  $1.8\gev$ with a root-mean-square deviation of $2.4\gev$.} 
  \label{Ml1q_Linear_TEP-MEP} \end{center}
\end{figure}

\begin{figure}[H]
  \begin{center} \includegraphics[height=\textheight/2 - 4cm]
  {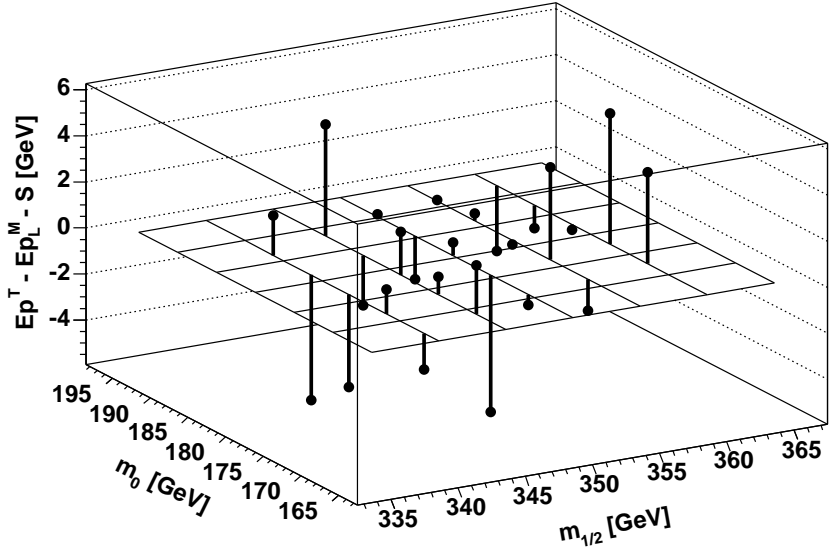} \caption{The theoretical
  $M_{\rho^{\pm}_{1}q}$ endpoint minus the measured endpoint in$\gev$
  obtained with the linear fit after applying the constant correction
  of $1.8\gev$. The uncertainties show no significant $m_{0}$ or
  $m_{1/2}$ dependence. Thus, a constant correction is sensible in
  this case.} \label{Ml1q_Linear_TEP-MEP-Cor} \end{center}
\end{figure}

\begin{table}[H]
\begin{center}
\begin{tabular}{|l|rrr|} \hline
\emph{Value} & Linear & Gaussian FO & Gaussian NFO\\\hline\hline
$\delta_{stat} Ep^{M}$ & 4.6 & 3.0 & 2.7\\
$S$ & 26.8 & -3.9 & 0.1\\
$\delta S$ & 8.0 & 9.9 & 2.9\\\hline
\end{tabular}
\caption{A summary of the calculated mean uncertainties and systematic
shifts in$\gev$ for the second $M_{\rho^{\pm}q}$ endpoint. FO is with
fixed origin and NFO with non-fixed origin as described in
section~\ref{sec:methods}.}\label{sum:Ml2q}
\end{center}
\end{table}
When the second endpoint is determined it can be used to make a cut on
the measured invariant mass in order to select solely first rho
events. The uncertainty of that cut value has no effect on the
following results since the tail of the distribution is not dependent
on the cut. The only important aspect is that the fit algorithm for
the linear fit takes the maximum of the first rho distribution and not
the maximum of the second rho distribution. For the following analysis
data samples of between 5142 and 16963 events after the cut have been
used. The large difference of the events in the data samples is caused
by the very different ratio between both endpoint masses. However, the
relevant number in the fitted region is in all data samples of the
same magnitude. The theoretical first rho endpoints are shown in
figure~\ref{Ml1q_TEP}. In the investigated area the maximum of the
first rho plus associated quark distribution is very close to the
endpoint concerning the second rho. Therefore, it is only possible to
make a linear fit at the tail of the distribution. An example is given
in figure~\ref{Ml1q_Ex_181_350}. The systematic shift here is compared
to the previous cases with $1.8 \pm 2.4\gev$
(Figure~\ref{Ml1q_Linear_TEP-MEP}) relatively small. Furthermore there
is no significant $m_{0}$ and $m_{1/2}$ dependence on that shift which
can be seen in figure~\ref{Ml1q_Linear_TEP-MEP-Cor}.

\begin{table}[H]
\begin{center}
\begin{tabular}{|l|r|} \hline
\emph{Value} & Linear \\\hline\hline
$\delta_{stat} Ep^{M}$ & 1.7\\
$S$ & 1.8\\
$\delta S$ & 2.4\\\hline
\end{tabular}
\caption{A summary of the calculated correction and the mean uncertainties
in$\gev$ for the first endpoint in $M_{\rho^{\pm}q}$.}\label{sum:Ml1q}
\end{center}
\end{table}
%-------------
\begin{figure}[H]
  \begin{center} \includegraphics[height=\textheight/2 - 4cm]
  {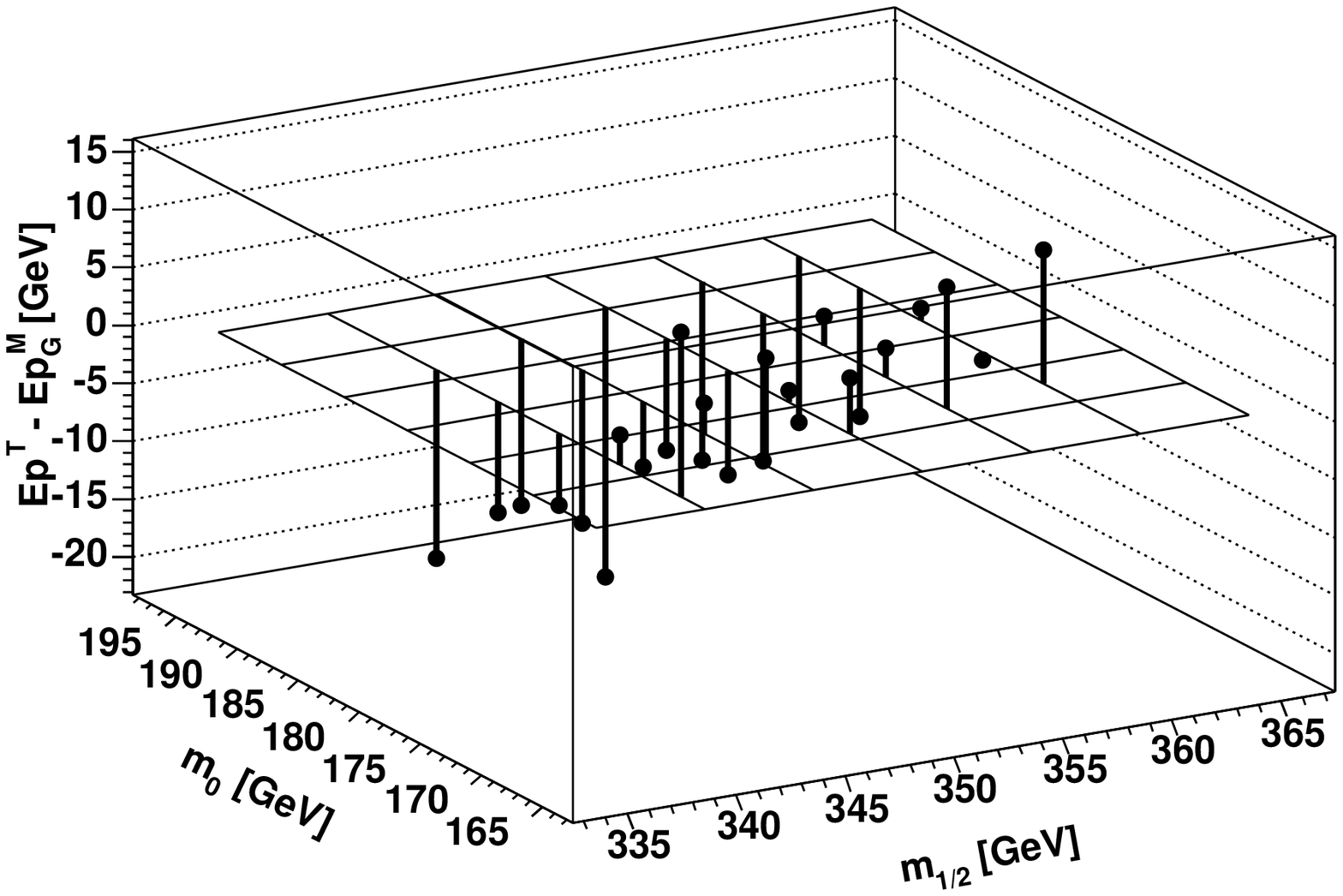} \caption{The difference of the
  theoretical and the measured $M_{\rho^{\pm}_{2}q}$ endpoint in$\gev$
  obtained with the gaussian fixed origin method. The measured values
  are on average too large by $9.1\gev$} \label{Ml2q_GaussFZ_TEP-MEP}
  \end{center}
\end{figure}
\begin{figure}[H]
  \begin{center} \includegraphics[height=\textheight/2 - 4cm]
  {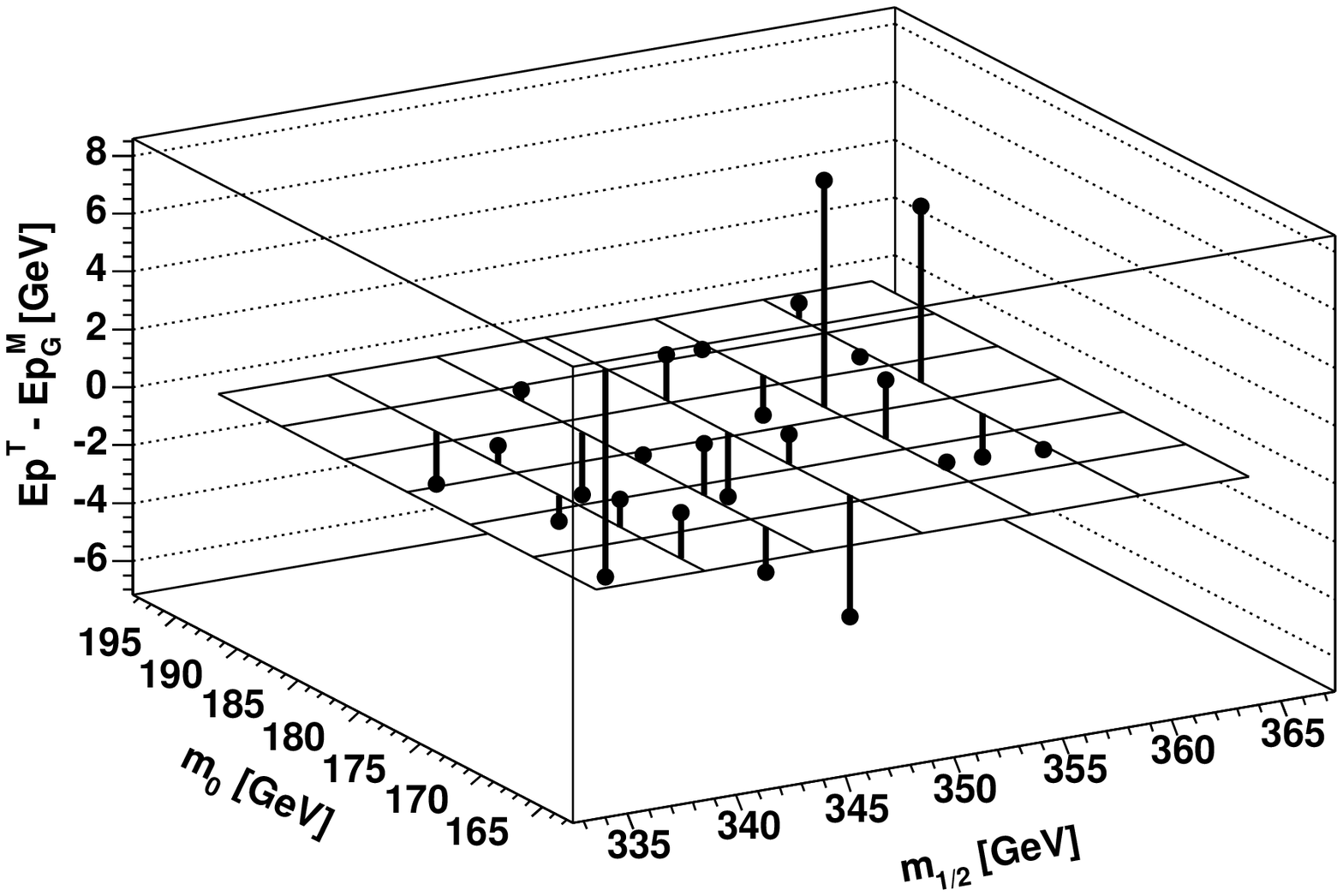} \caption{The difference of the
  theoretical and the measured $M_{\rho^{\pm}_{2}q}$ endpoint in$\gev$
  obtained with the gaussian non-fixed origin method. } 
  \label{Ml2q_GaussNFZ_TEP-MEP} \end{center}
\end{figure}
%-------------
%--------------------------------------------------------------------
\subsection{\texorpdfstring{The kinematic limit of $M_{\rho_{1}^{\pm}q} +
M_{\rho_{2}^{\mp}q}$ for light quarks}{The kinematic limit of M(r1q) +
M(r2q) for light quarks}}\label{sec:Ml1q_Ml2q} 
\begin{figure}[H]
  \begin{center}
    \shadowbox{\includegraphics[width=\textwidth/2]{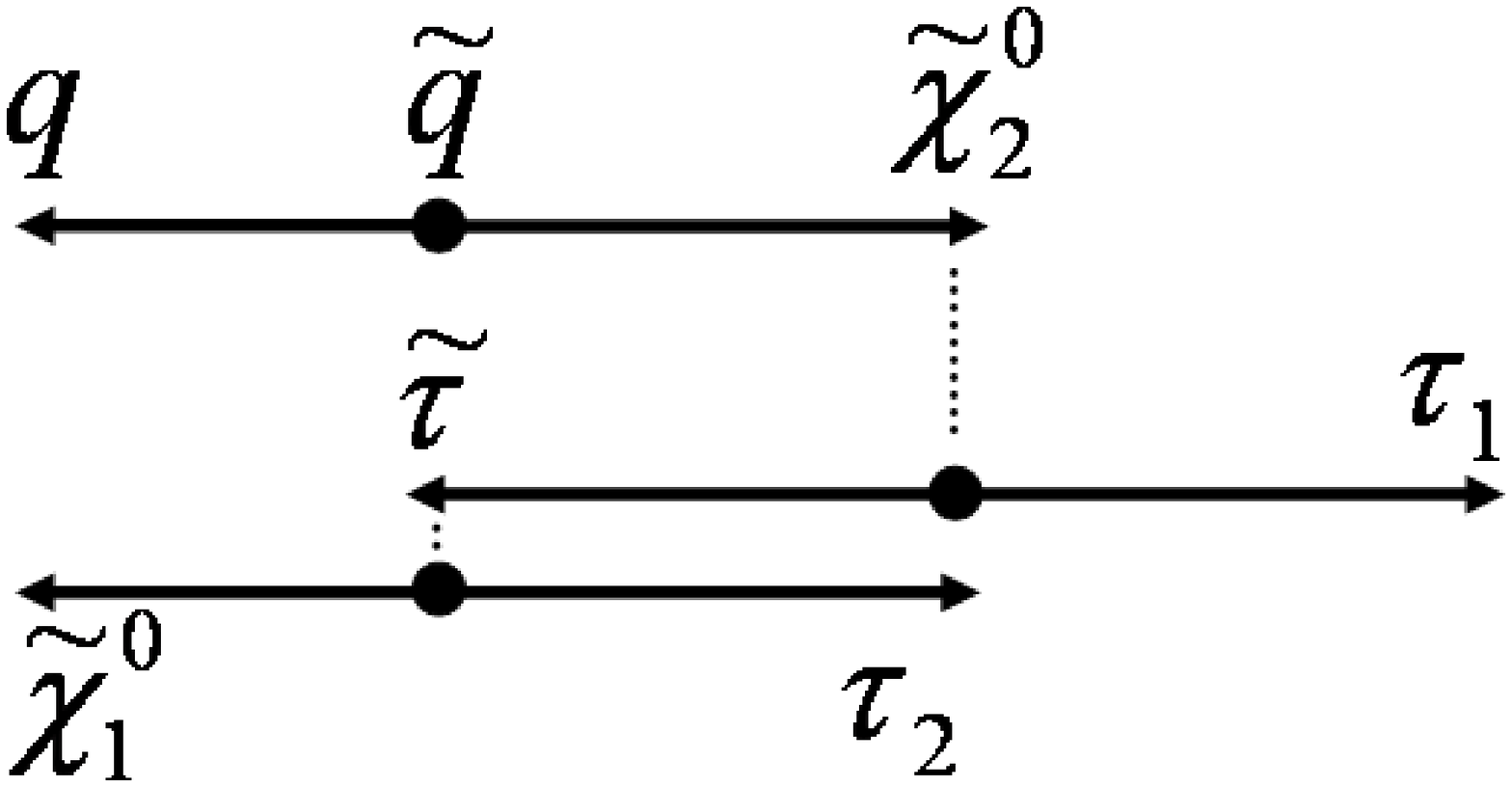}}
  \end{center}
\end{figure}
\begin{figure}[H]
  \begin{center} \includegraphics[height=\textheight/2 - 4cm]
  {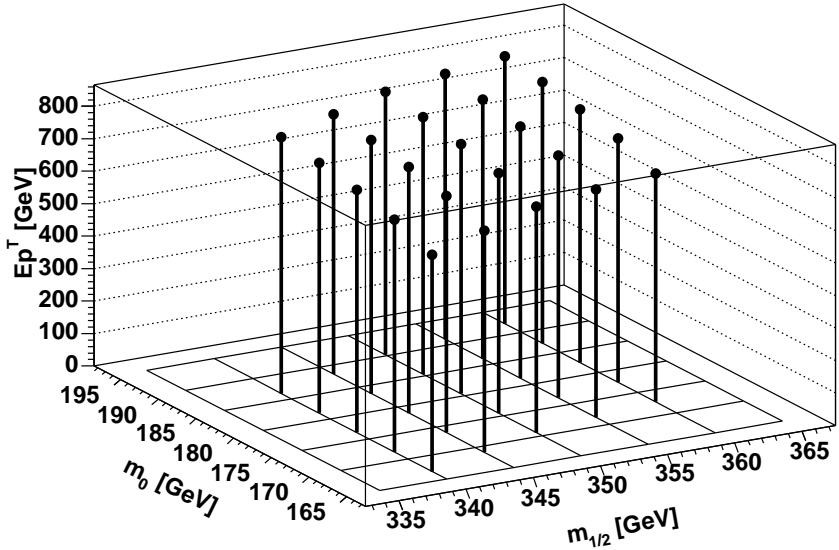} \caption{Theoretical kinematic endpoints for
  $M_{\rho_{1}^{\pm}q} + M_{\rho_{2}^{\mp}q}$ in$\gev$: The
  $\rho_{2}^{\mp}$ coming from the $\tilde{\tau}$ with the light quark
  and the $\rho_{1}^{\pm}$ coming from the $\chinonn$ and the same
  light quark from $\tilde{s}_1,\tilde{d}_1 \ra s,d + \chinonn$.}
  \label{Ml1q_Ml2q_TEP} \end{center}
\end{figure}

If all endpoints could be measured very precisely four measured
kinematic limits would be enough to reconstruct the four sparticle
masses $M_{\chinon}$, $M_{\chinonn}$, $M_{\tilde{\tau}}$ and
$M_{\tilde{d},\tilde{s}}$. But mainly due to the neutrinos even at
parton level there are large statistical and systematic uncertainties
for the measured endpoints. Although this fifth endpoint
(Figure~\ref{Ml1q_Ml2q_TEP}) depends on the previous four endpoints,
its measurement improves significantly the uncertainties for the mass
reconstruction which is performed in section~\ref{sec:30fb}. For the
following analysis the same data sample as in section~\ref{sec:Mllq}
is chosen.

\begin{figure}[H]
  \begin{center} \includegraphics[height=\textheight/2 - 4cm]
  {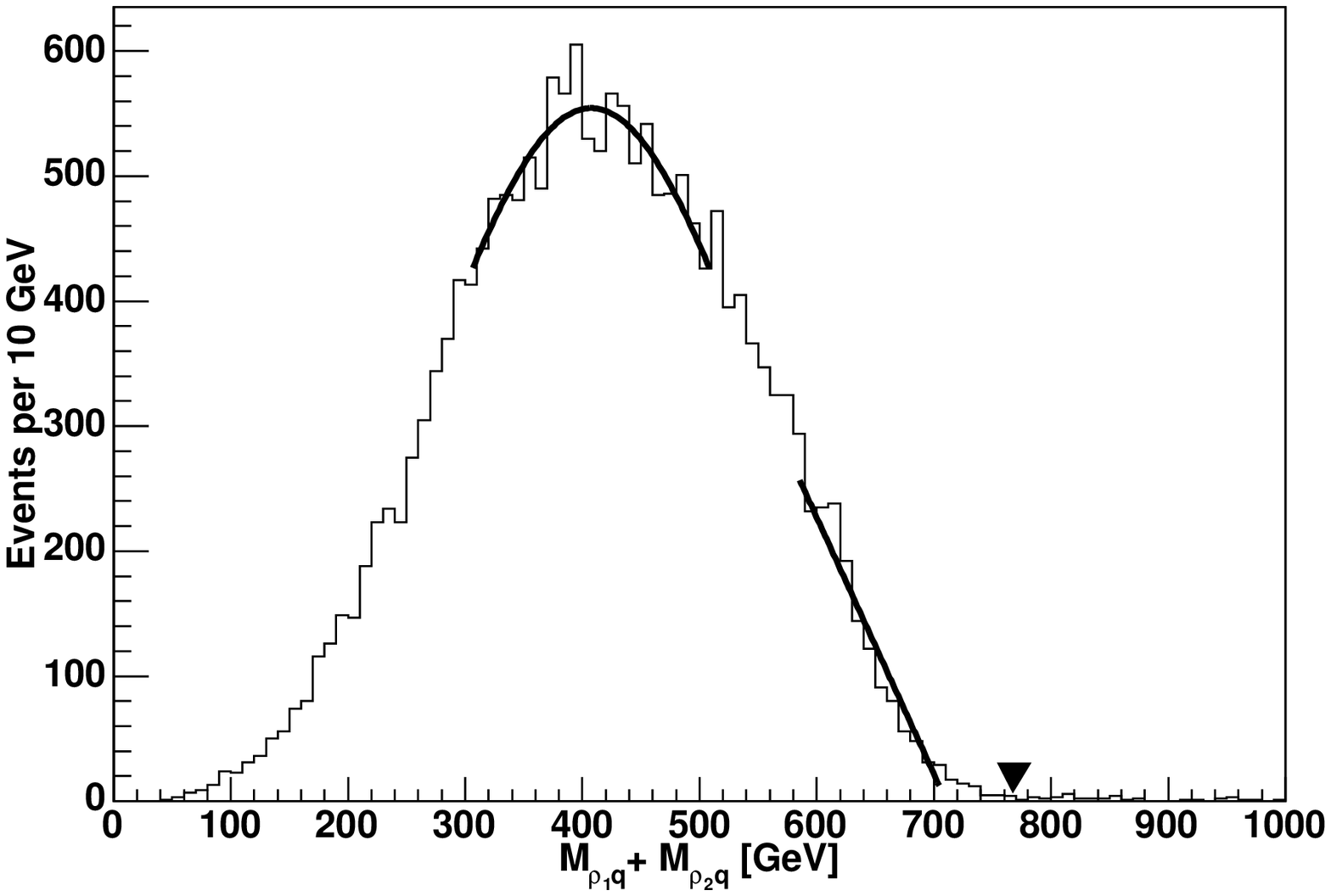} \caption{Example for the linear and
  the gaussian fit to the $M_{\rho_{1}^{\pm}q} + M_{\rho_{2}^{\mp}q}$
  distribution at $m_{0} = 181$ and $m_{1/2} = 350$. In this
  distribution 18714 events are represented. The triangle shows the
  theoretical kinematic endpoint at $768.8\gev$.} 
  \label{Ml1q_Ml2q_Ex_181_350} \end{center}
\end{figure}

\begin{figure}[H]
  \begin{center} \includegraphics[height=\textheight/2 - 4cm]
  {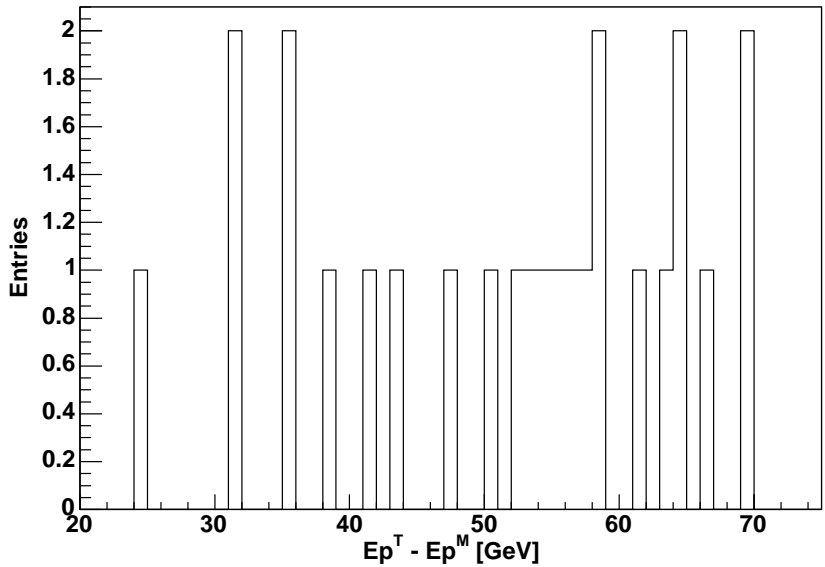} \caption{The difference of the
  theoretical and the measured $M_{\rho_{1}^{\pm}q} +
  M_{\rho_{2}^{\mp}q}$ endpoint in$\gev$ obtained with a linear fit
  for all 25 points. It shows a large shift of $51.4\gev$ with a
  root-mean-square deviation of $12.8\gev$.} 
  \label{Ml1q_Ml2q_Linear_TEP-MEP} \end{center}
\end{figure}

\begin{figure}[H]
  \begin{center} \includegraphics[height=\textheight/2 - 4cm]
  {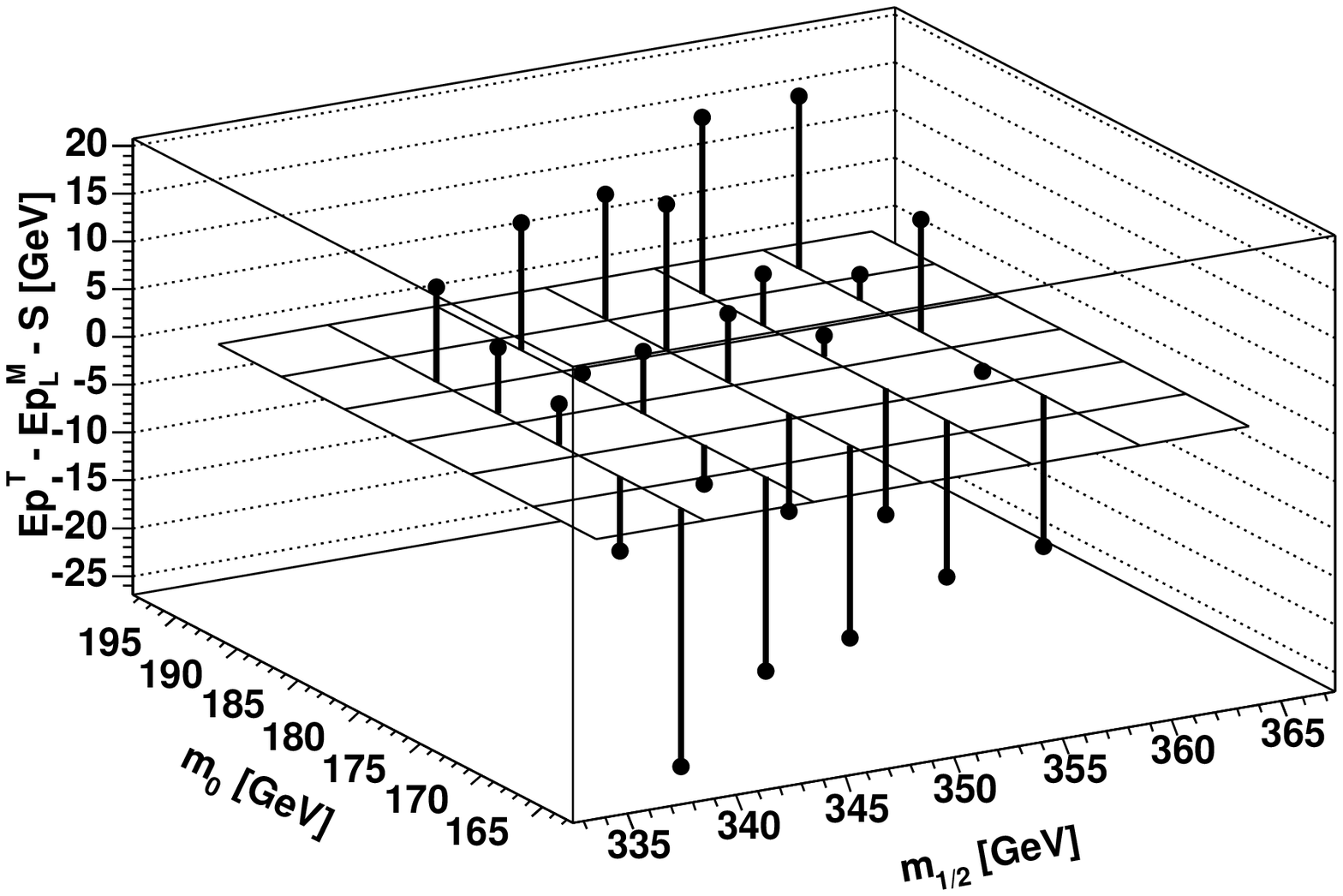} \caption{The theoretical
  $M_{\rho_{1}^{\pm}q} + M_{\rho_{2}^{\mp}q}$ endpoint minus the
  measured endpoint in$\gev$ obtained with the linear fit after
  applying the constant correction of $51.4\gev$. The values decrease
  significantly for decreasing values of $m_{0}$.} 
  \label{Ml1q_Ml2q_Linear_TEP-MEP-Cor} \end{center}
\end{figure}

The linear fit (Figure~\ref{Ml1q_Ml2q_Ex_181_350}) measures the
endpoints systematically too low by $51.4 \pm 12.8\gev$ as can be seen
in figure~\ref{Ml1q_Ml2q_Linear_TEP-MEP}. The same $m_{0}$ dependence
on the actually needed correction as has been determined for
$M_{\rho^{\pm}\rho^{\mp}}$ or $M_{\rho^{\pm}\rho^{\mp}q}$ is shown in
figure~\ref{Ml1q_Ml2q_Linear_TEP-MEP-Cor}.

\begin{figure}[H]
  \begin{center} \includegraphics[height=\textheight/2 - 4cm]
  {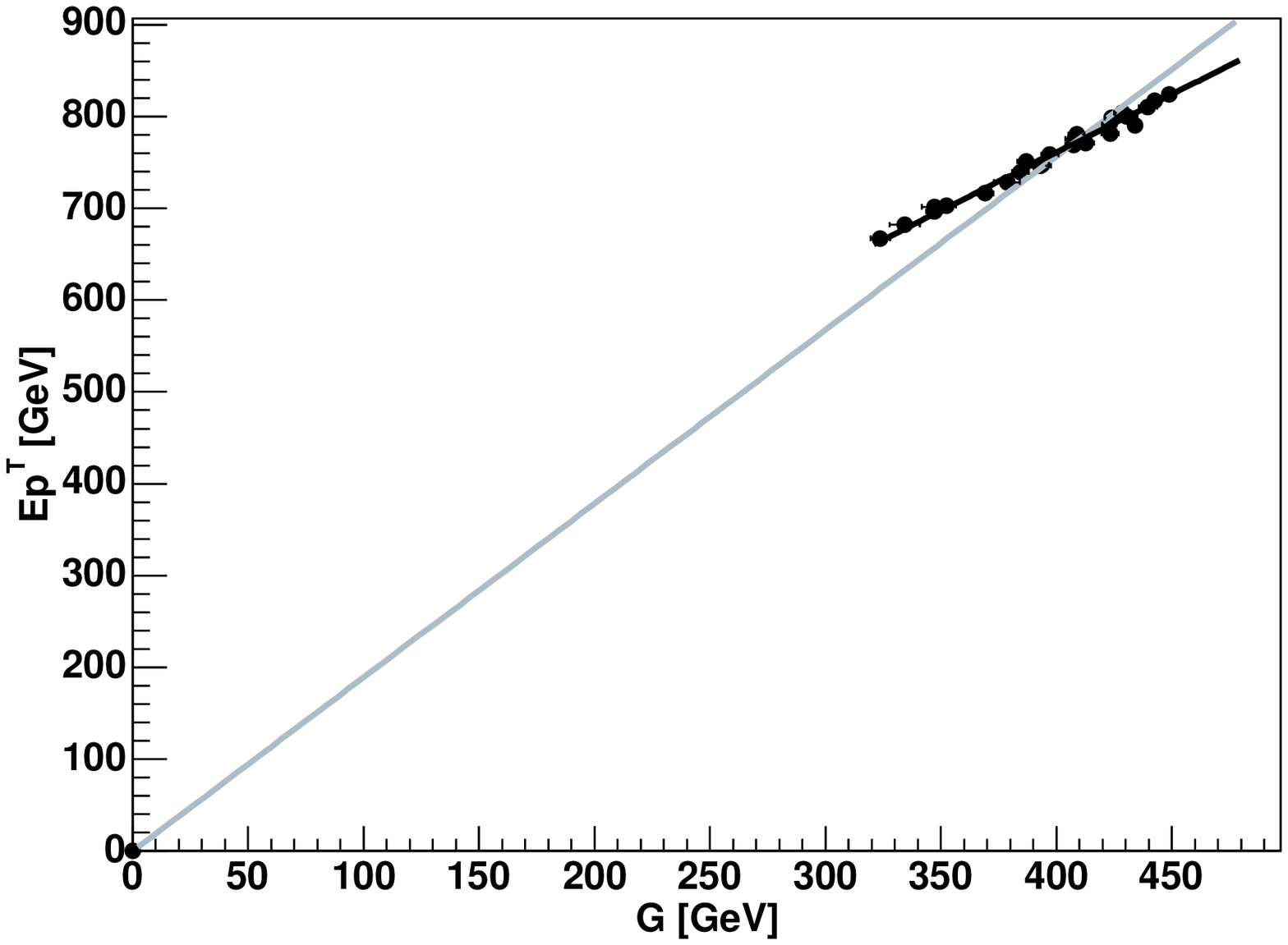} 
\caption{The theoretical kinematic $M_{\rho_{1}^{\pm}q} +
M_{\rho_{2}^{\mp}q}$ endpoints as a function of the measured gaussian
maximum for all 25 investigated points in the $m_{0}$-$m_{1/2}$
plane. The gray fit with $C = 1.9$ is forced to go through the origin
whereas the black with $C = 1.3$ has an optimised ordinate value of
$254.3\gev$.}
\label{Ml1q_Ml2q_Gauss_TE_vs_MM_Fit} \end{center}
\end{figure}

%--gaussian--
Figure~\ref{Ml1q_Ml2q_Gauss_TE_vs_MM_Fit} illustrates that the
gaussian method is suitable to determine the endpoint. Even if the fit
is forced to go through the origin, which leads to a slope value $C =
1.9$, good results can be achieved. If the fit can optimise the
ordinate value the slope $C$ decreases to $1.3$ with $D =
254.3\gev$. The calculated values for the uncertainties and the
systematic shifts are given in table~\ref{sum:Ml1q_Ml2q}.
%--------------------------------------------------------------------
\begin{table}[H]
\begin{center}
\begin{tabular}{|l|rrr|} \hline
\emph{Value} & Linear & Gaussian FO & Gaussian NFO\\\hline\hline
$\delta_{stat} Ep^{M}$ & 1.9 & 7.4 & 5.0\\
$S$ & 51.4 & -2.3 & 0.4\\
$\delta S$ & 12.8 & 13.2 & 5.6\\\hline
\end{tabular}
\caption{A summary of the calculated mean uncertainties and systematic
shifts in$\gev$ for the $M_{\rho_{1}^{\pm}q} + M_{\rho_{2}^{\mp}q}$
endpoint analysis. FO denotes the results with fixed origin and NFO
with non-fixed origin.}\label{sum:Ml1q_Ml2q}
\end{center}
\end{table}
\begin{figure}[H]
  \begin{center} \includegraphics[height=\textheight/2 - 4cm]
  {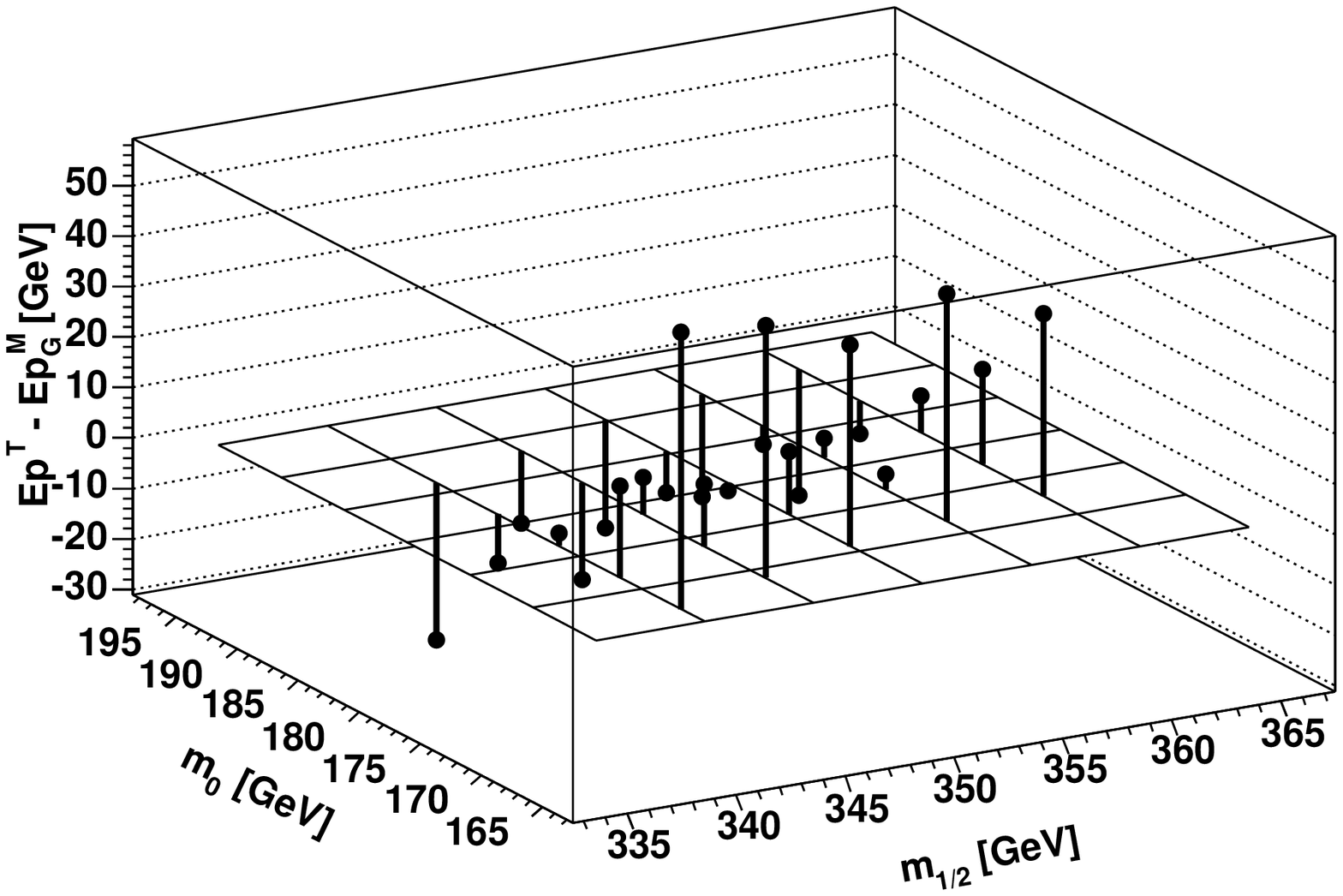} \caption{The difference of the
  theoretical and the measured $M_{\rho_{1}^{\pm}q} +
  M_{\rho_{2}^{\mp}q}$ endpoint in$\gev$ obtained with the gaussian
  fixed origin method. The measured values are too high with a
  systematic shift of $2.3\gev$} \label{Ml1q_Ml2q_GaussFZ_TEP-MEP}
  \end{center}
\end{figure}
\begin{figure}[H]
  \begin{center} \includegraphics[height=\textheight/2 - 4cm]
  {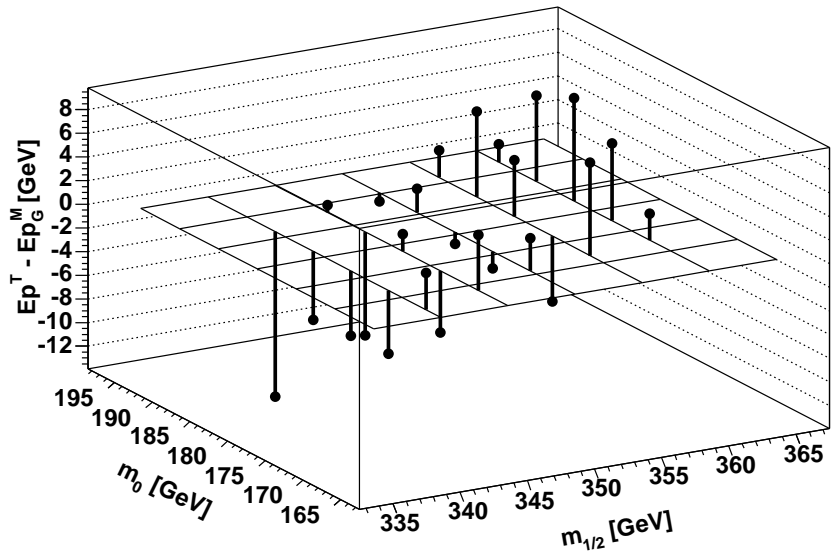} \caption{The difference of the
  theoretical $M_{\rho_{1}^{\pm}q} + M_{\rho_{2}^{\mp}q}$ endpoint and
  the measured endpoint in$\gev$ obtained with the gaussian non-fixed
  origin method. }\label{Ml1q_Ml2q_GaussNFZ_TEP-MEP} \end{center}
\end{figure}
%--------------------------------------------------------------------

 \subsection{\texorpdfstring{The kinematic limit of
$M_{\rho^{\pm}\rho^{\mp}q}$ for bottom quarks}{The kinematic limit of
M(rrq) for bottom quarks}}\label{sec:Mllb} 
\begin{figure}[H]
  \begin{center}
    \shadowbox{\includegraphics[width=\textwidth/2]{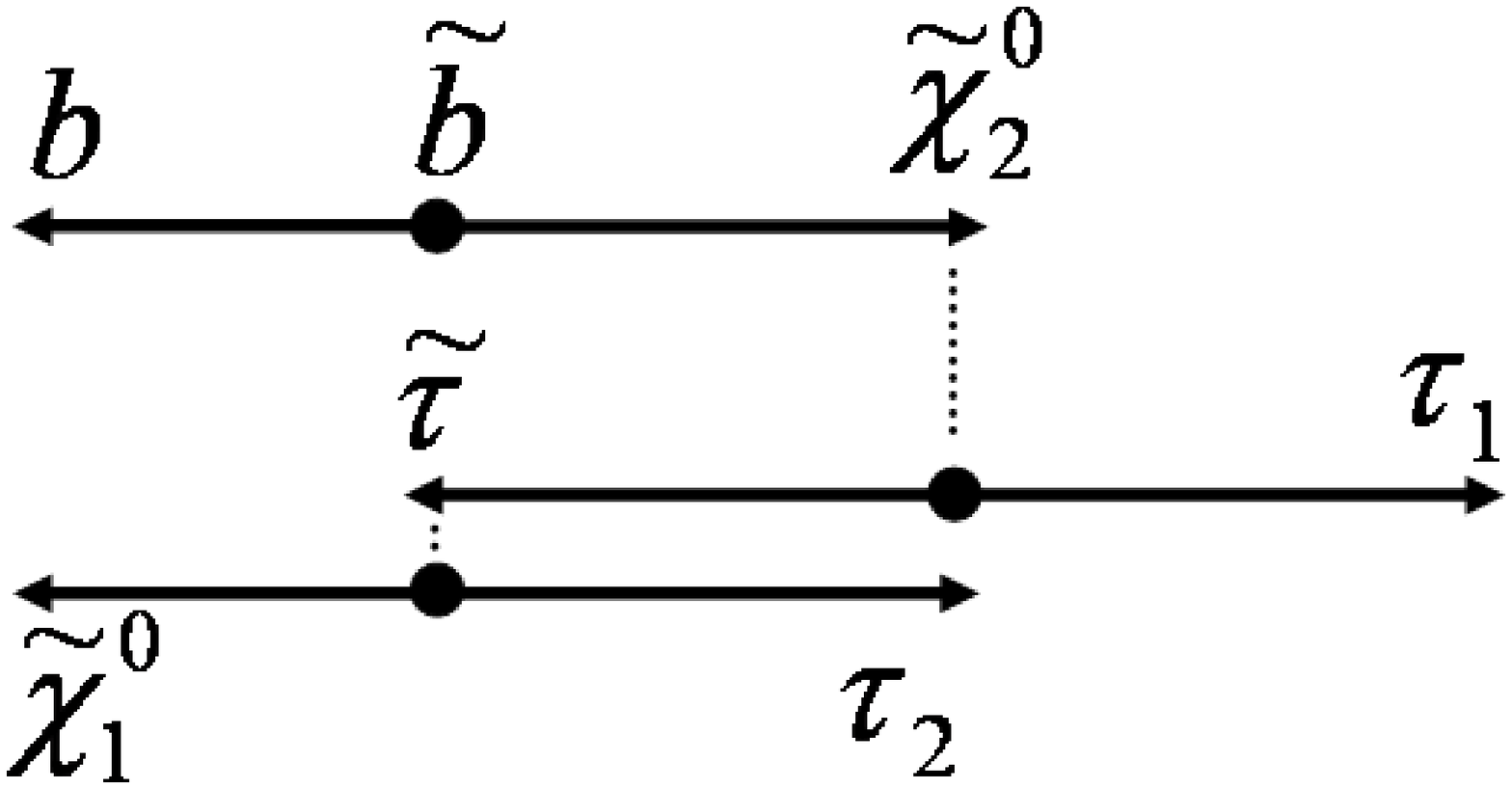}}
  \end{center}
\end{figure}
\begin{figure}[H]
  \begin{center} \includegraphics[height=\textheight/2 - 2.2cm]
  {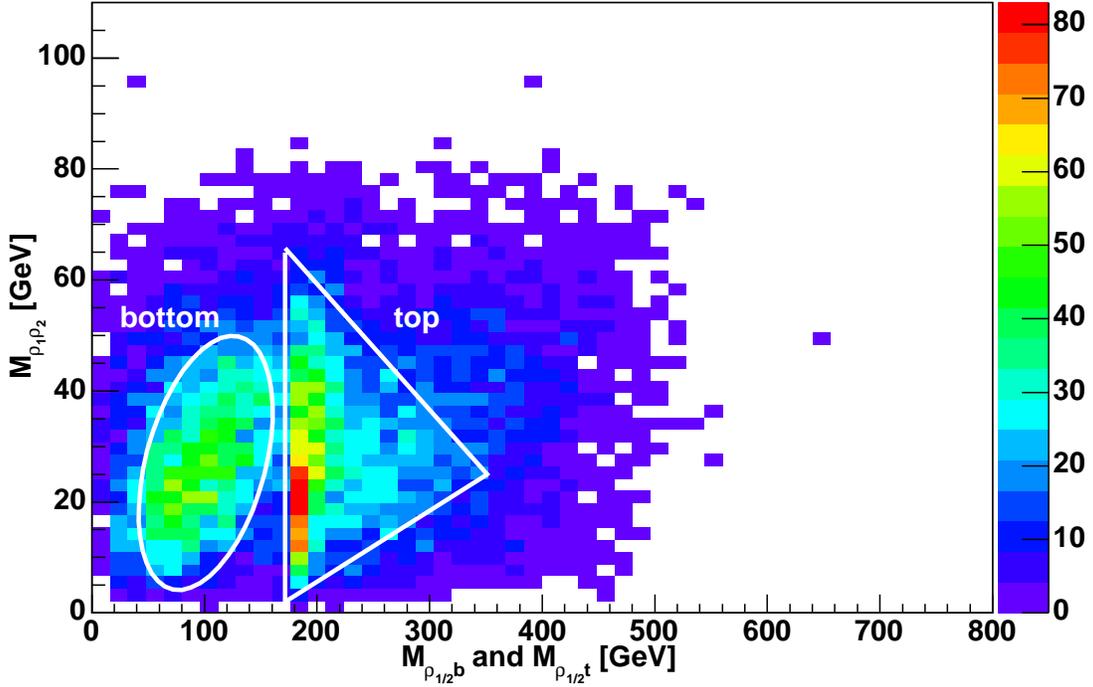} \caption{The invariant mass of
  both opposite-sign rho mesons versus the invariant mass of a single
  rho with the associated top or bottom quark in$\gev$. It is an
  example for the point I'. The left ellipse denotes the events with
  the bottom quarks whereas the top quark events are mainly located in
  the triangle on the right side.}  \label{R1R2_vs_R1R2Jet_Top_Bottom}
  \end{center}
\end{figure}

Indeed, with the b-tagging it is possible to select events with bottom
quarks. But the top quark which is used in the sections
\ref{ssec:llqtop} to \ref{sec:Ml1t_Ml2t} decays almost entirely
into bottom quarks such that the b-tagging is also sensitive to top
events. Nevertheless, there is a possibility to distinguish both kinds
of events, as shown in figure~\ref{R1R2_vs_R1R2Jet_Top_Bottom}. In the
invariant mass distribution of both opposite-sign rho mesons versus
the invariant mass of a single rho with the associated top or bottom
quark, both kinds of events are well separated. Both processes are
treated separately in the following. The analysis in this section is
based on 25 data samples with between 4224 and 4901 bottom quark
events.

\begin{figure}[H]
  \begin{center} \includegraphics[height=\textheight/2 - 4cm]
  {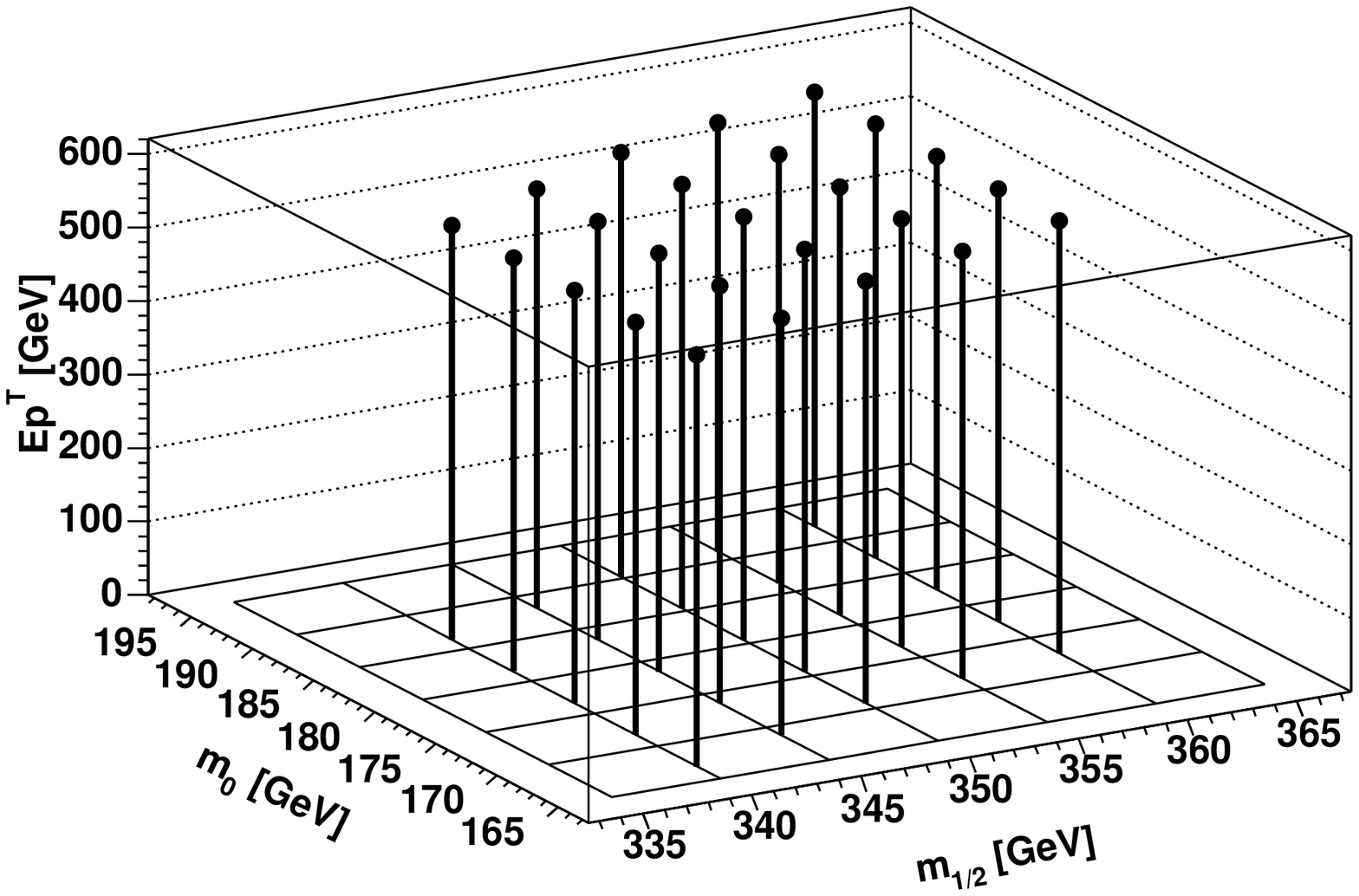} \caption{Theoretical kinematic endpoints in$\gev$
  of the invariant mass of both opposite-sign rhos, coming from the
  $\chinonn$ and the $\tilde{\tau}$ respectively, and the
  associated quark from $\tilde{b}_2 \ra b + \chinonn$.}
  \label{Mllb_TEP} \end{center}
\end{figure}

\begin{figure}[H]
  \begin{center} \includegraphics[height=\textheight/2 - 4cm]
  {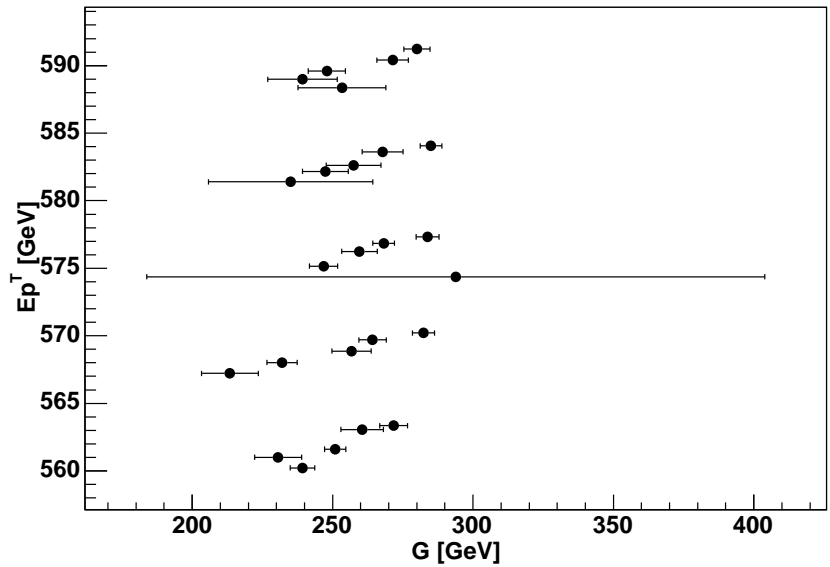} 
\caption{The theoretical kinematic $M_{\rho^{\pm}\rho^{\mp}b}$
  endpoints versus the measured gaussian maximum for all 25
  investigated points in the $m_{0}$-$m_{1/2}$ plane. Due to the flat
  shape near the maximum of one distribution, one point shows a large
  uncertainty. A unique mapping of the measured maximum to the
  theoretical endpoint can not be established since several
  theoretical endpoint values are associated with the same measured
  maximum.}  \label{Mllb_Gauss_TE_vs_MM_Fit} \end{center}
\end{figure}

For the kinematic limit of $M_{\rho^{\pm}\rho^{\mp}b}$, which is shown
in figure~\ref{Mllb_TEP} for all 25 points, the gaussian method cannot
be applied for two reasons. First, the shape of the distribution
changes significantly within the investigated region. In some cases
the region around the maximum of the distribution is flat, resulting
in a large $\chi^{2}$ for the gaussian fit.  Second, when applying the
gaussian fit on the distribution the maximum value shows no linear
dependence on the theoretical endpoint. This can be seen in
figure~\ref{Mllb_Gauss_TE_vs_MM_Fit}.

\begin{figure}[H]
  \begin{center} \includegraphics[height=\textheight/2 - 4cm]
  {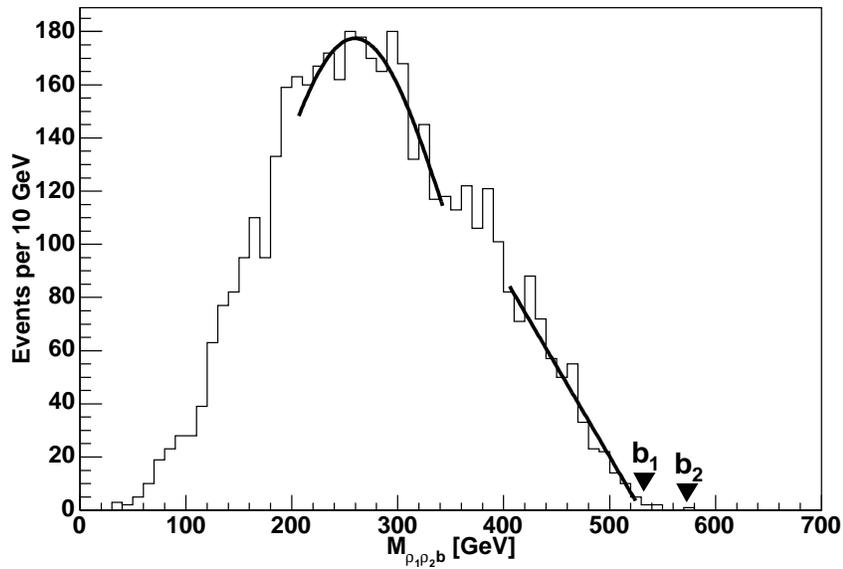} \caption{Example for the linear and the
  gaussian fit to the $M_{\rho^{\pm}\rho^{\mp}b}$ distribution at
  $m_{0} = 181$ and $m_{1/2} = 350$. The invariant mass distribution
  is based on 4498 bottom quark events. The two triangles show the
  theoretical endpoint values for events with $\tilde{b}_{1}$
  ($532.9\gev$) and $\tilde{b}_{2}$ ($576.3\gev$), respectively.} 
  \label{Mllb_Ex_181_350} \end{center}
\end{figure}

\begin{figure}[H]
  \begin{center} \includegraphics[height=\textheight/2 - 4cm]
  {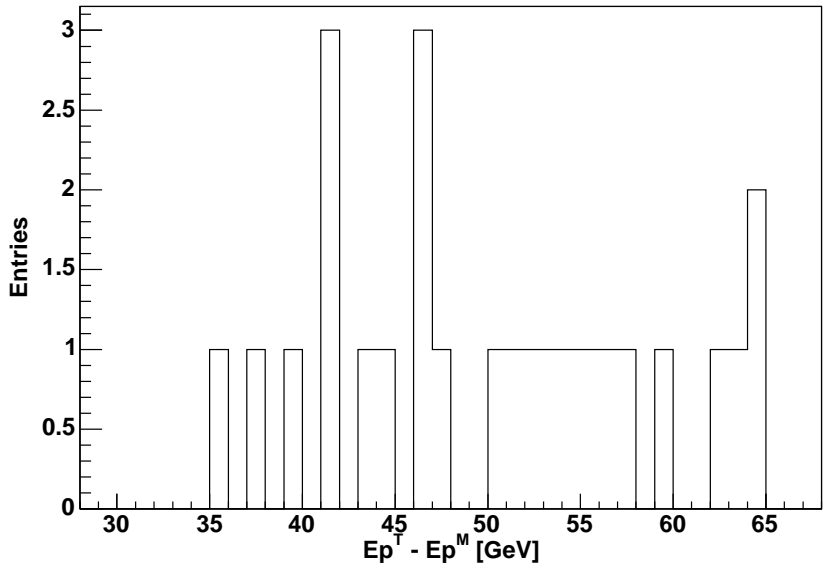} \caption{The theoretical
  $M_{\rho^{\pm}\rho^{\mp}b}$ endpoint minus the measured endpoint
  in$\gev$ for all 25 points. The mean value $S$ is $50.3\gev$ with a
  root-mean-square deviation of $8.5\gev$. If the $\tilde{b}_{1}$
  mass instead of the $\tilde{b}_{2}$ mass is taken for the
  theoretical endpoint the value $S$ changes to $7.2 \pm 8.6\gev$}
  \label{Mllb_Linear_TEP-MEP} \end{center}
\end{figure}

\begin{figure}[H]
  \begin{center} \includegraphics[height=\textheight/2 - 4cm]
  {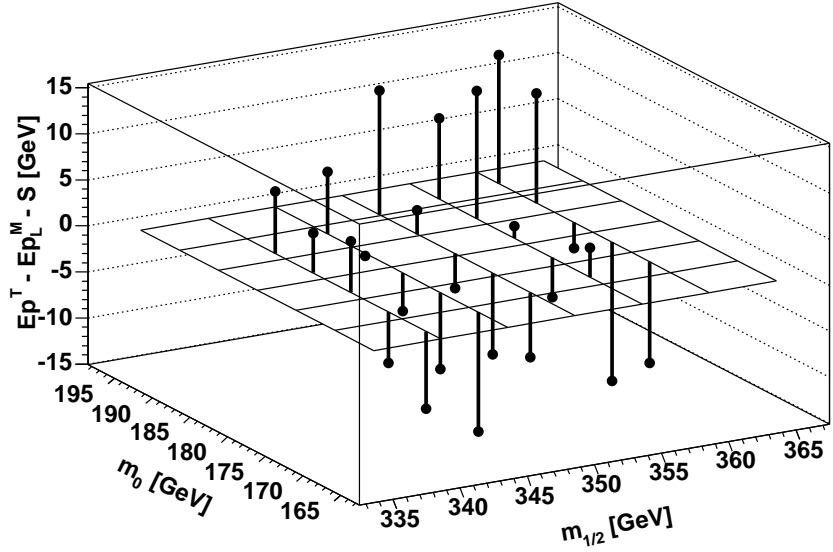} \caption{The theoretical
  $M_{\rho^{\pm}\rho^{\mp}b}$ endpoint minus the measured endpoint
  in$\gev$ after applying a constant correction of $50.3\gev$. The
  values decrease with decreasing values of $m_{0}$.} 
  \label{Mllb_Linear_TEP-MEP-Cor} \end{center}
\end{figure}

\begin{figure}[H]
  \begin{center} \includegraphics[width=\textwidth]
  {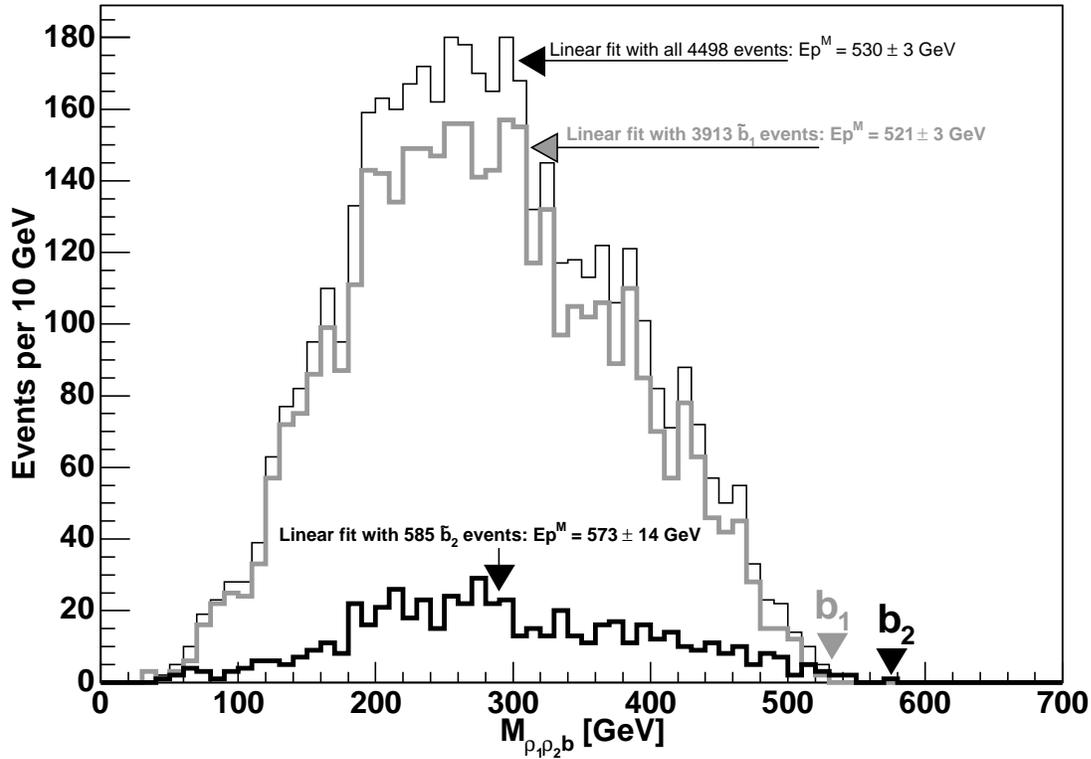} 
\caption{The invariant mass distribution of both rho with the
associated bottom quark. The bold black line shows 585 events coming
from a $\tilde{b}_{2}$, the bold gray line 3913 events coming form a
$\tilde{b}_{1}$ and the light black line shows the superposition of
both lines. Although the $\tilde{b}_{1}$ events clearly dominate, the
$\tilde{b}_{2}$ events have a significant influence on the measurement
of $Ep^{M}$. The triangles show the theoretical endpoints at $533\gev
(\tilde{b}_{1})$ and $576\gev (\tilde{b}_{2})$.}
\label{B_Mixing} \end{center}
\end{figure}

On the other hand, the linear fit (Figure~\ref{Mllb_Ex_181_350})
provides a good possibility to measure the $M_{\rho^{\pm}\rho^{\mp}b}$
endpoint. However, a large systematic correction
(Figure~\ref{Mllb_Linear_TEP-MEP}) is needed because there are about
six to seven times more events with the lighter $\tilde{b}_{1}$ than
with the heavier $\tilde{b}_{2}$ being responsible for the real
kinematic endpoint. This can easily be understood by comparing
table~\ref{tab:masses} with table~\ref{tab:sqtochi2}. The effect of
this mixing is shown and explained in figure~\ref{B_Mixing}. The same
$m_{0}$ dependence as in previous cases for the actually needed
correction can be seen in figure~\ref{Mllb_Linear_TEP-MEP-Cor}.
%-------
\begin{table}[H]
\begin{center}
\begin{tabular}{|l|r|} \hline
\emph{Value} & Linear \\\hline\hline
$\delta_{stat} Ep^{M}$ & 2.9 (2.9) \\
$S$ & 50.3 (7.2) \\
$\delta S$ & 8.5 (8.6) \\\hline
\end{tabular}
\caption{A summary of the calculated correction and mean uncertainties
in$\gev$. The values in brackets correspond to the case of
$\tilde{b}_{1}$ being the input for the kinematic
limit.}\label{sum:Mllb}
\end{center}
\end{table}
\vspace{-2cm}
%----------------------------------------------------------------------
\subsection{\texorpdfstring{Two kinematic limits in $M_{\rho^{\pm}q}$
for bottom quarks}{Two kinematic limits in M(rq) for bottom
quarks}}\label{sec:Mlb} 
\begin{figure}[H]
  \begin{center}
    \shadowbox{\begin{tabular}{c}
	\includegraphics[width=\textwidth/2]{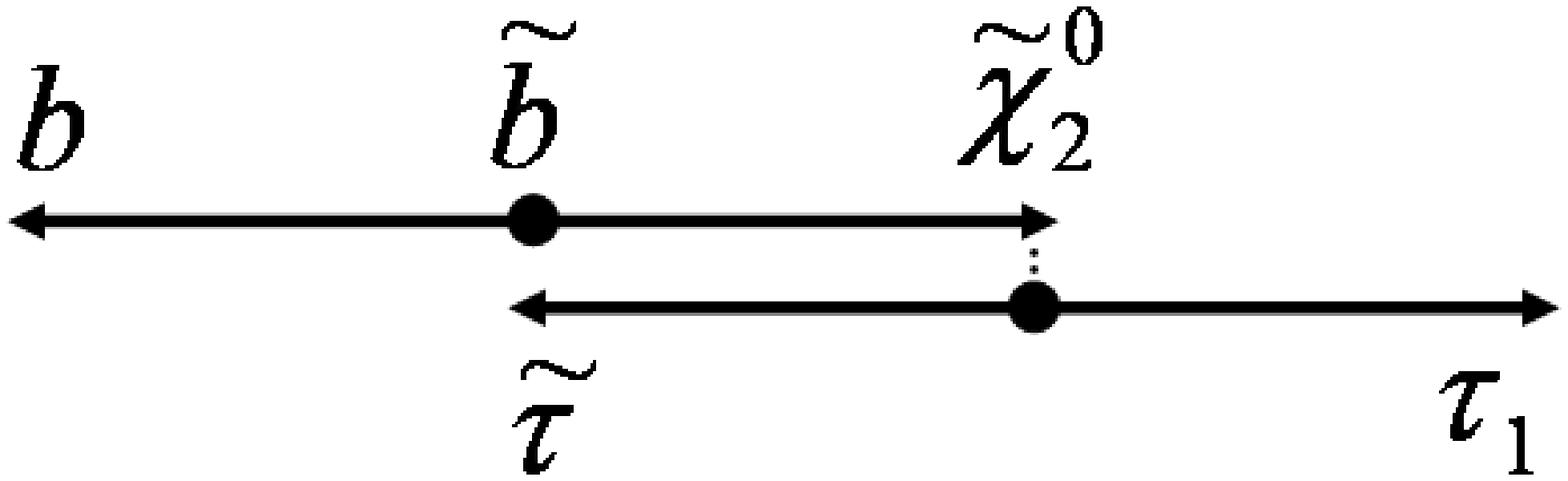}\\
    \includegraphics[width=\textwidth/2]{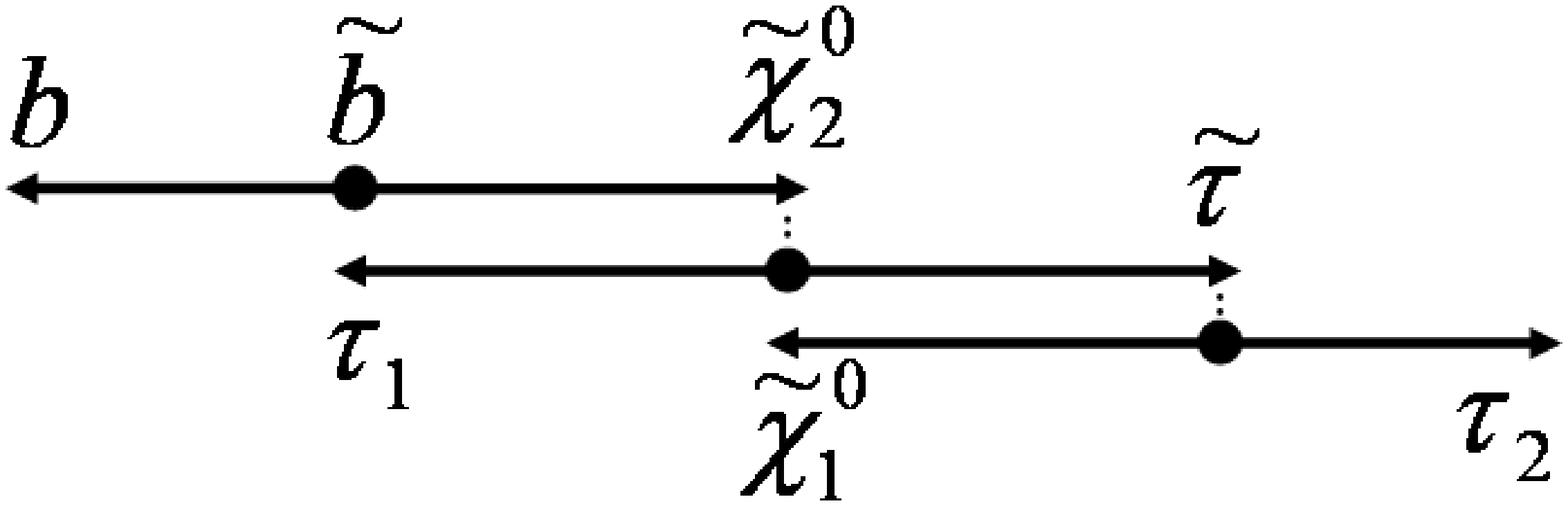}
	\end{tabular}}
  \end{center}
\end{figure}
\begin{figure}[H]
  \begin{center} \includegraphics[height=\textheight/2 - 4cm]
  {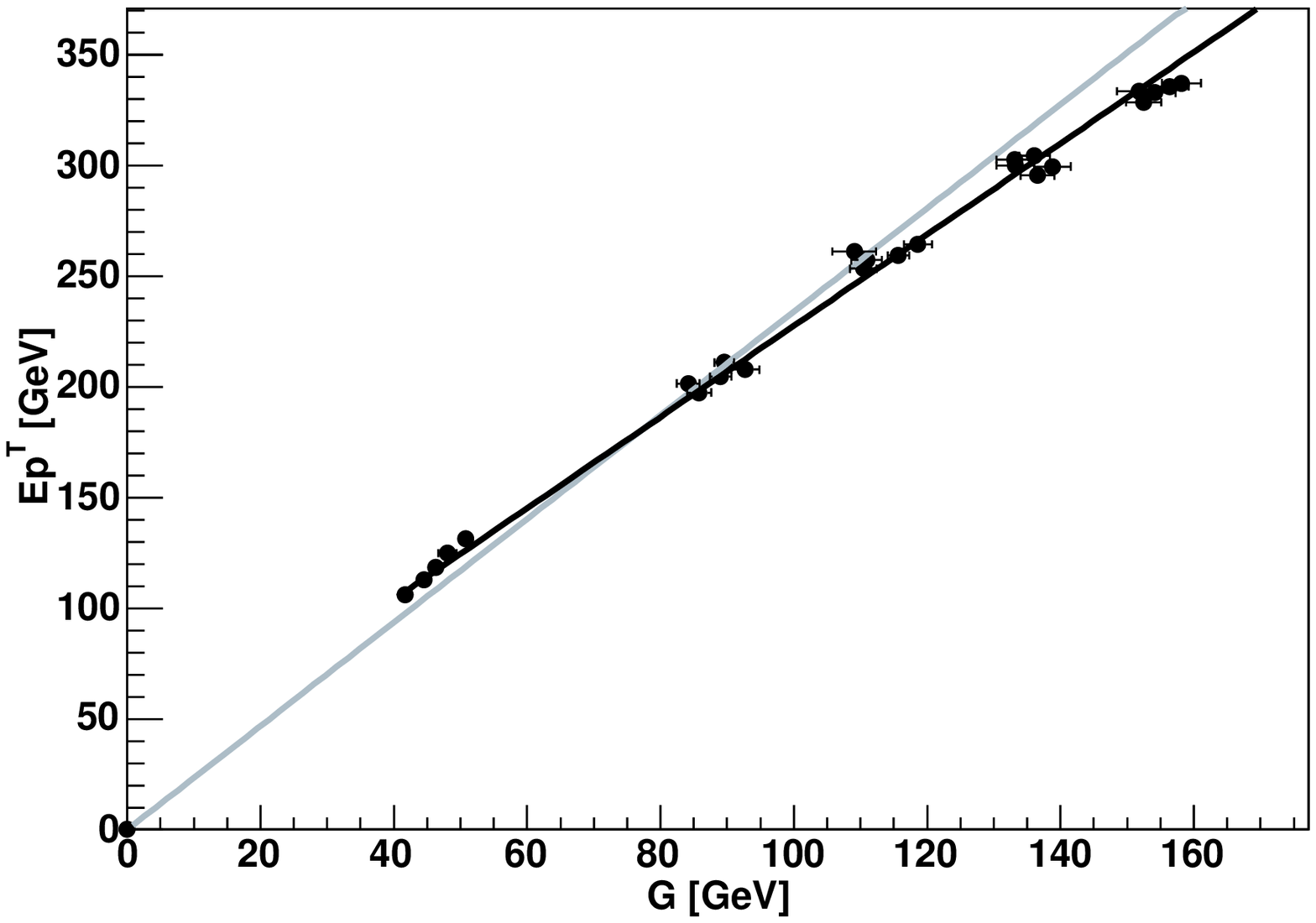} 
\caption{The theoretical kinematic $M_{\rho^{\pm}_{2}b}$
  endpoints as a function of the measured gaussian maximum for all 25
  investigated points in the $m_{0}$-$m_{1/2}$ plane. The gray line is
  forced to go through the origin whereas the black has an optimised
  ordinate value of $21.6\gev$.}  \label{Ml2b_Gauss_TE_vs_MM_Fit}
  \end{center}
\end{figure}
\begin{figure}[H]
  \begin{center} \includegraphics[height=\textheight/2 - 4cm]
  {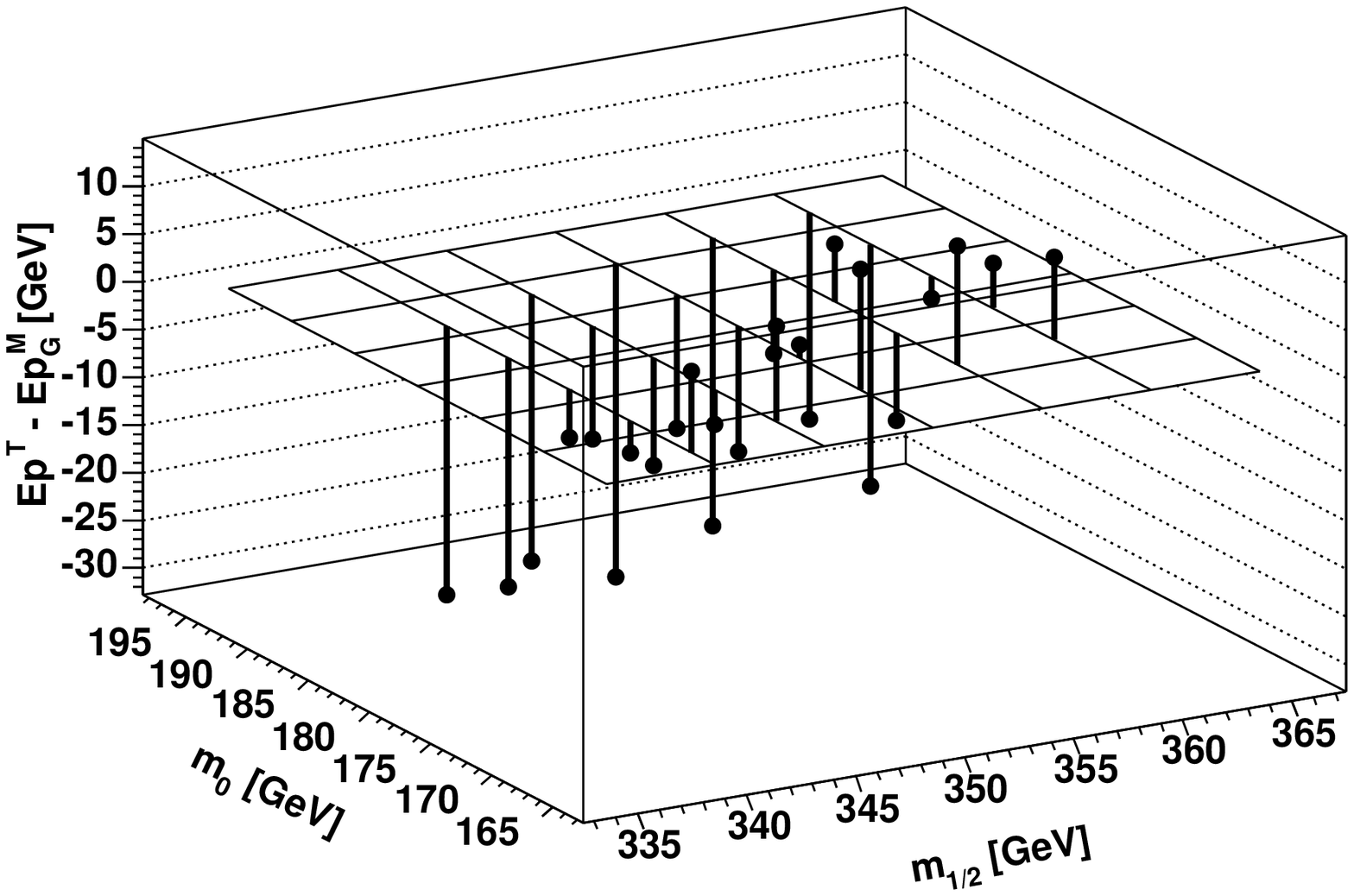} \caption{The difference of the
  theoretical and the measured $M_{\rho^{\pm}_{2}b}$ endpoint in$\gev$
  obtained with the gaussian fixed origin method. The measured values
  are too low with a systematic shift of $-8.3\gev$}
  \label{Ml2b_GaussFZ_TEP-MEP} \end{center}
\end{figure}

The principles described in section
\ref{sec:Mlq} remain the same for this invariant mass distribution. In
the following only the necessary changes and the results are
described. The region for the gaussian fit is changed from a symmetric
area of $25 \%$ of the bin value which is used for the fitting centre
to a symmetric area of $45\%$ similar to section~\ref{sec:Mlq}. The
gaussian method for this distribution can be applied since the
gaussian fit is stable for all 25 points and the assumed linear
dependence is sensible (see figure~\ref{Ml2b_Gauss_TE_vs_MM_Fit}). If
the linear fit of $Ep^{T}_{i}$ as a function of the fitted gaussian
maximum $G_{i} \pm
\delta G_{i}$ is forced to go to the origin the resulting slope $C$ is
2.3. The statistical uncertainty $\delta_{stat} Ep^{M}$ on the
endpoint measurement has then a mean value of $4.9\gev$. However, in
this case a systematic shift $S$ of $-8.3 \pm 14.1\gev$ is needed in
the whole region to improve the endpoint measurement (see
figure~\ref{Ml2b_GaussFZ_TEP-MEP}). If the fit is not restricted to
intersect the origin the value $C$ changes to 2.1, nearly no
systematic shift $S$ is needed and $\delta S$ decreases to
$6.0\gev$. The fit here intersects the ordinate at $D = 21.6\gev$.
The mean statistical uncertainty $\delta_{stat} Ep^{M}$ on the
endpoint measurement decreases slightly to $4.3\gev$. The ratio
between both slopes is 1.1. 

For the double linear fit method the second fit iteratively searches
the area right of the second lepton peak where the slope is bigger
than $- 0.25$, instead of $- 0.2$ like in section~\ref{sec:Mlq}. This
value leads to a stable fit over the whole investigated parameter
space. This method here again has the problem that a large shift of
$46.9 \pm 10.3\gev$ is needed. However, it can be used to estimate the
endpoint. Table~\ref{sum:Ml2b} summarises the results.

\begin{table}[H]
\begin{center}
\begin{tabular}{|l|rrr|} \hline
\emph{Value} & Linear & Gaussian FO & Gaussian NFO\\\hline\hline
$\delta_{stat} Ep^{M}$ & 5.8 (5.8) & 4.9 & 4.3\\
$S$ & 46.9 (28.7) & -8.3 & 0.2\\
$\delta S$ & 10.3 (7.7) & 14.1 & 6.0\\\hline
\end{tabular}
\caption{A summary of the calculated mean uncertainties and systematic
shifts in$\gev$ for the second endpoint in $M_{\rho^{\pm}b}$. FO is
with fixed origin and NFO with non-fixed origin.}\label{sum:Ml2b}
\end{center}
\end{table}

\begin{figure}[H]
  \begin{center} \includegraphics[height=\textheight/2 - 4cm]
  {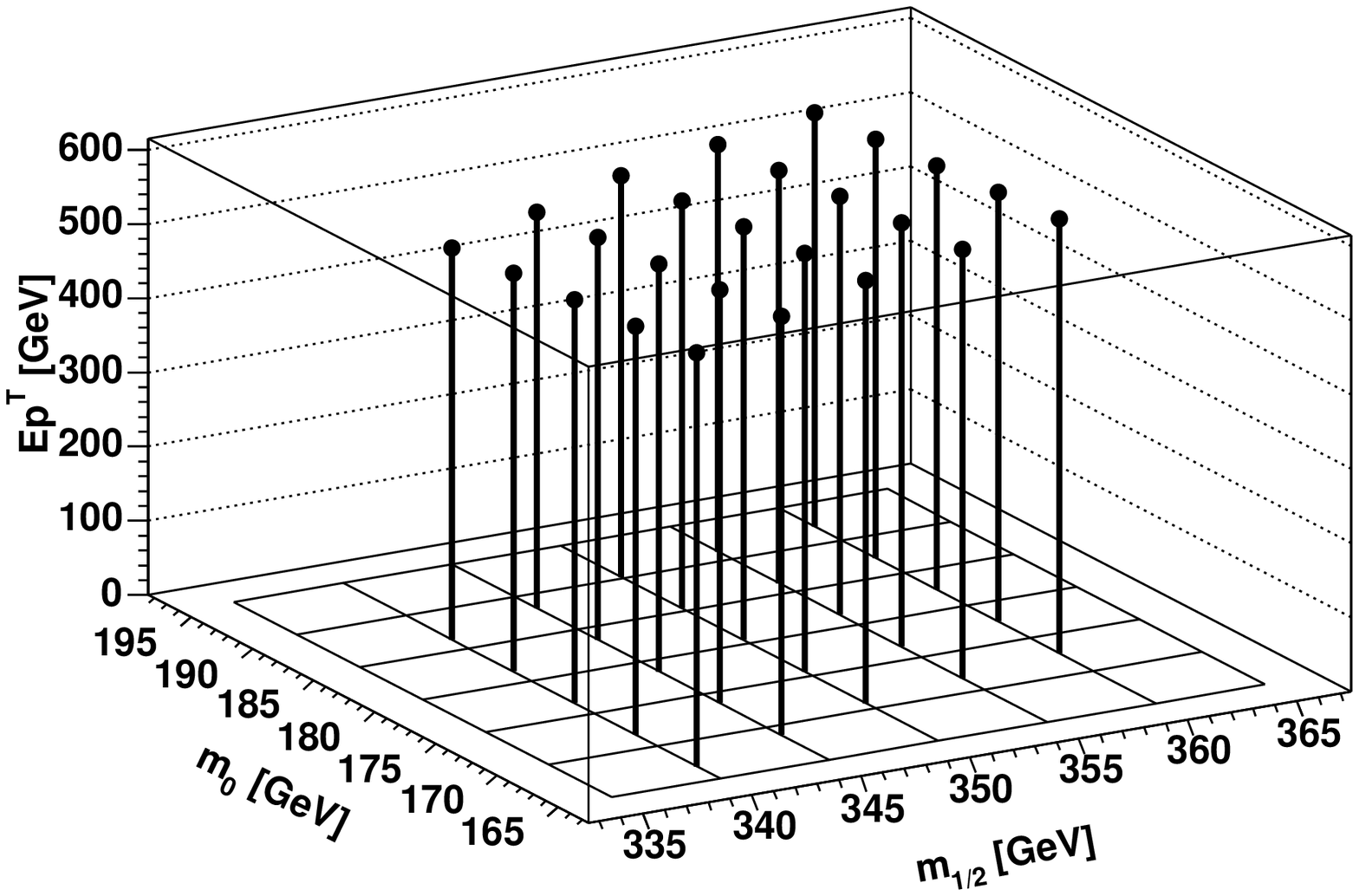} \caption{Theoretical kinematic endpoints in$\gev$
  for the invariant mass of first rho coming from the $\chinonn$ and
  the associated bottom quark coming from the sbottom in the
  $\tilde{b}_2 \ra b + \chinonn$ decay.}  \label{Ml1b_TEP}
  \end{center}
\end{figure}

\begin{figure}[H]
  \begin{center} \includegraphics[height=\textheight/2 - 4cm]
  {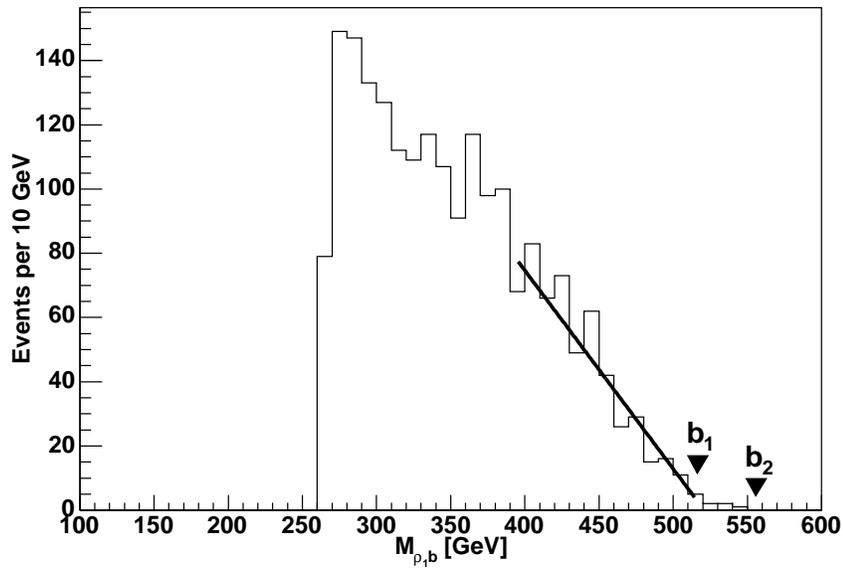} \caption{Example for the linear fit at
  $m_{0} = 181$ and $m_{1/2} = 350$ after applying a cut on the
  invariant mass $M_{\rho^{\pm}b}$ at the second endpoint. In this
  distribution 2036 events are represented. The two triangles show the
  theoretical endpoint values for events with $\tilde{b}_{1}$
  ($514.7\gev$) and $\tilde{b}_{2}$ ($556.4\gev$), respectively.} 
  \label{Ml1b_Ex_181_350} \end{center}
\end{figure}

\begin{figure}[H]
  \begin{center} \includegraphics[height=\textheight/2 - 4cm]
  {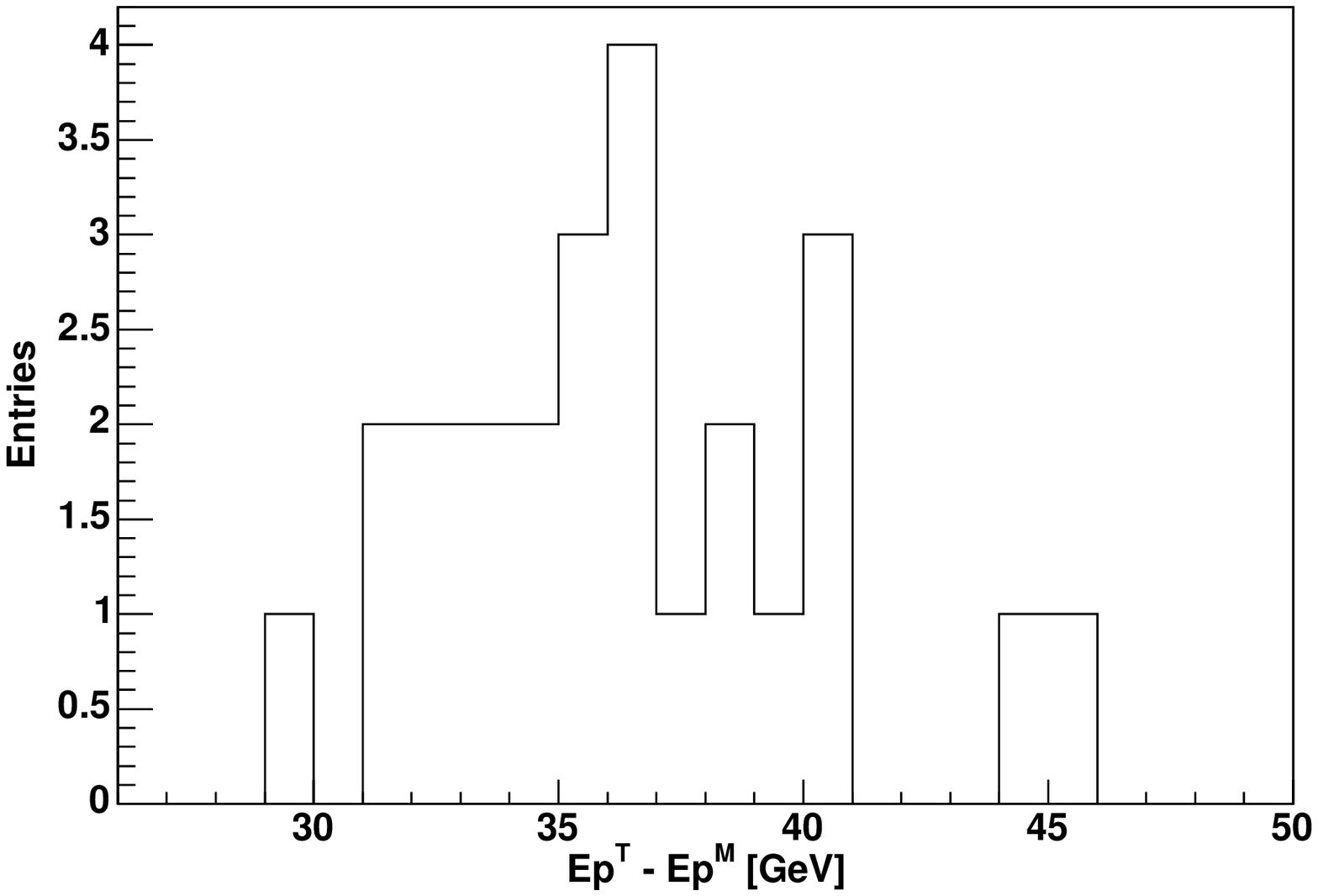} \caption{The difference of the
  theoretical and the measured $M_{\rho^{\pm}_{1}b}$ endpoint in$\gev$
  obtained with the linear fit. It shows a shift of $36.5\gev$ with a
  root-mean-square deviation of $3.9\gev$.} 
  \label{Ml1b_Linear_TEP-MEP} \end{center}
\end{figure}

\begin{figure}[H]
  \begin{center}
 \includegraphics[height=\textheight/2 - 4cm]
{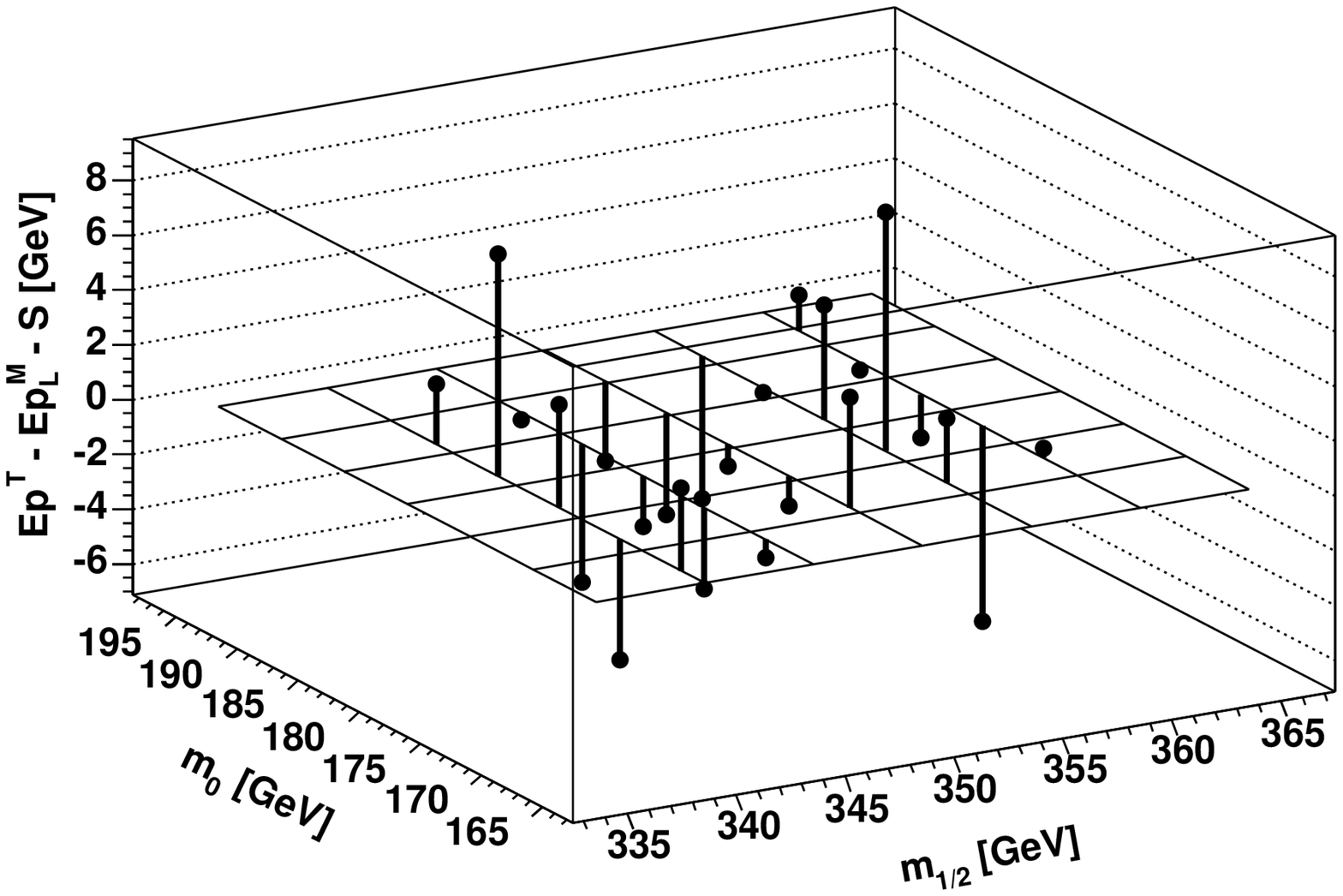} 
\caption{The theoretical $M_{\rho^{\pm}_{1}b}$ endpoint
  minus the measured endpoint in$\gev$ obtained with the linear fit
  after applying the constant correction of $36.5\gev$.} 
  \label{Ml1b_Linear_TEP-MEP-Cor} \end{center}
\end{figure}

The endpoint for the first rho with the associated bottom quark is
determined with a linear fit (Figure~\ref{Ml1b_Ex_181_350}) the same
way as it is described in section \ref{sec:Mlq}. In
figure~\ref{Ml1b_TEP} the theoretical value is shown which corresponds
to events coming from the $\tilde{b}_{2}$. In
figure~\ref{Ml1b_Linear_TEP-MEP} the large systematic shift $S =
36.5\pm 3.7\gev$ can be seen. There is no significant $m_{0}$ and
$m_{1/2}$ dependence on that shift as shown in
figure~\ref{Ml1b_Linear_TEP-MEP-Cor}. This is similar to the first
$M_{\rho^{\pm}q}$ endpoint for the light quarks
(Figure~\ref{Ml1q_Linear_TEP-MEP-Cor}).

\begin{table}[H]
\begin{center}
\begin{tabular}{|l|r|} \hline
\emph{Value} & Linear \\\hline\hline
$\delta_{stat} Ep^{M}$ & 2.6 (2.6) \\ $S$ & 36.5 (-4.8) \\ $\delta
S$ & 3.9 (3.7) \\\hline
\end{tabular}
\caption{A summary of the calculated correction and mean uncertainties
in$\gev$ for the first endpoint in the $M_{\rho^{\pm}q}$
distribution.}\label{sum:Ml1b}
\end{center}
\end{table}
%--------------------
\begin{figure}[H]
  \begin{center} \includegraphics[height=\textheight/2 - 4cm]
  {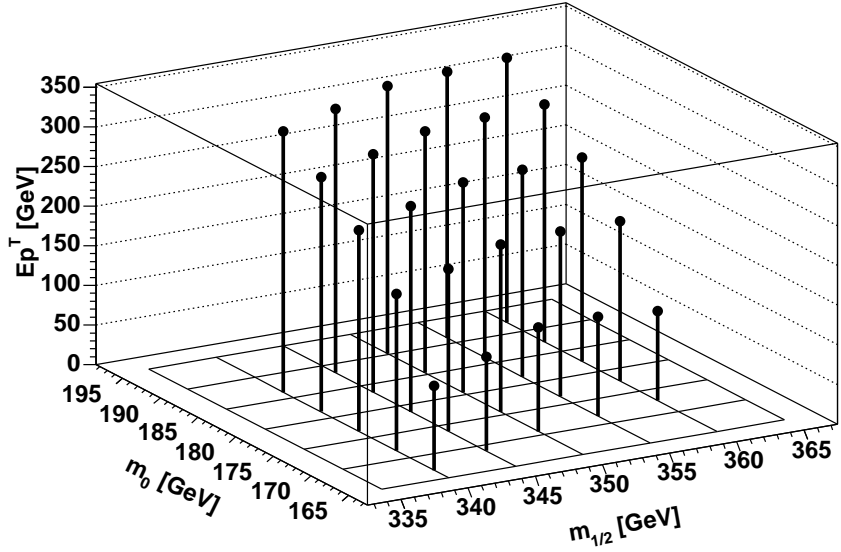} \caption{Theoretical kinematic endpoints in$\gev$
  for the invariant mass of the second rho coming from the
  $\tilde{\tau}^{\pm}$ and the associated bottom quark coming from
  the sbottom in the $\tilde{b}_2 \ra b + \chinonn$ decay. The
  endpoints decrease significantly for decreasing values of $m_{0}$.}
  \label{Ml2b_TEP} \end{center}
\end{figure}
\begin{figure}[H]
  \begin{center} \includegraphics[height=\textheight/2 - 4cm]
  {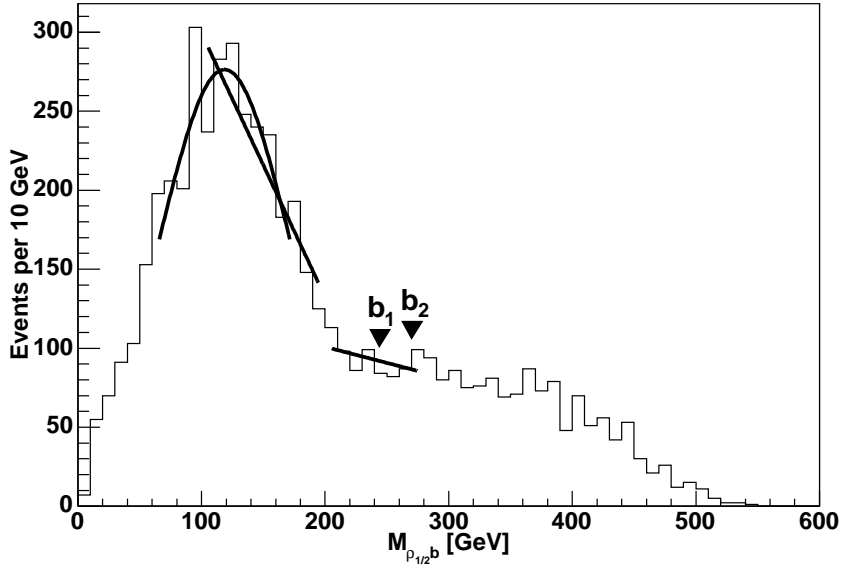} \caption{Example for the two linear fits
  and the gaussian fit to the $M_{\rho^{\pm}b}$ distribution at $m_{0}
  = 181$ and $m_{1/2} = 350$. In this distribution 5636 events after
  the cut at $30\%$ of the maximum invariant mass of both rhos (see
  section~\ref{sec:Mlq}) are represented. The two triangles show the
  theoretical endpoint values for events with $\tilde{b}_{1}$
  ($244.7\gev$) and $\tilde{b}_{2}$ ($264.5\gev$), respectively.} 
  \label{Ml2b_Ex_181_350} \end{center}
\end{figure}
\begin{figure}[H]
  \begin{center} \includegraphics[height=\textheight/2 - 4cm]
  {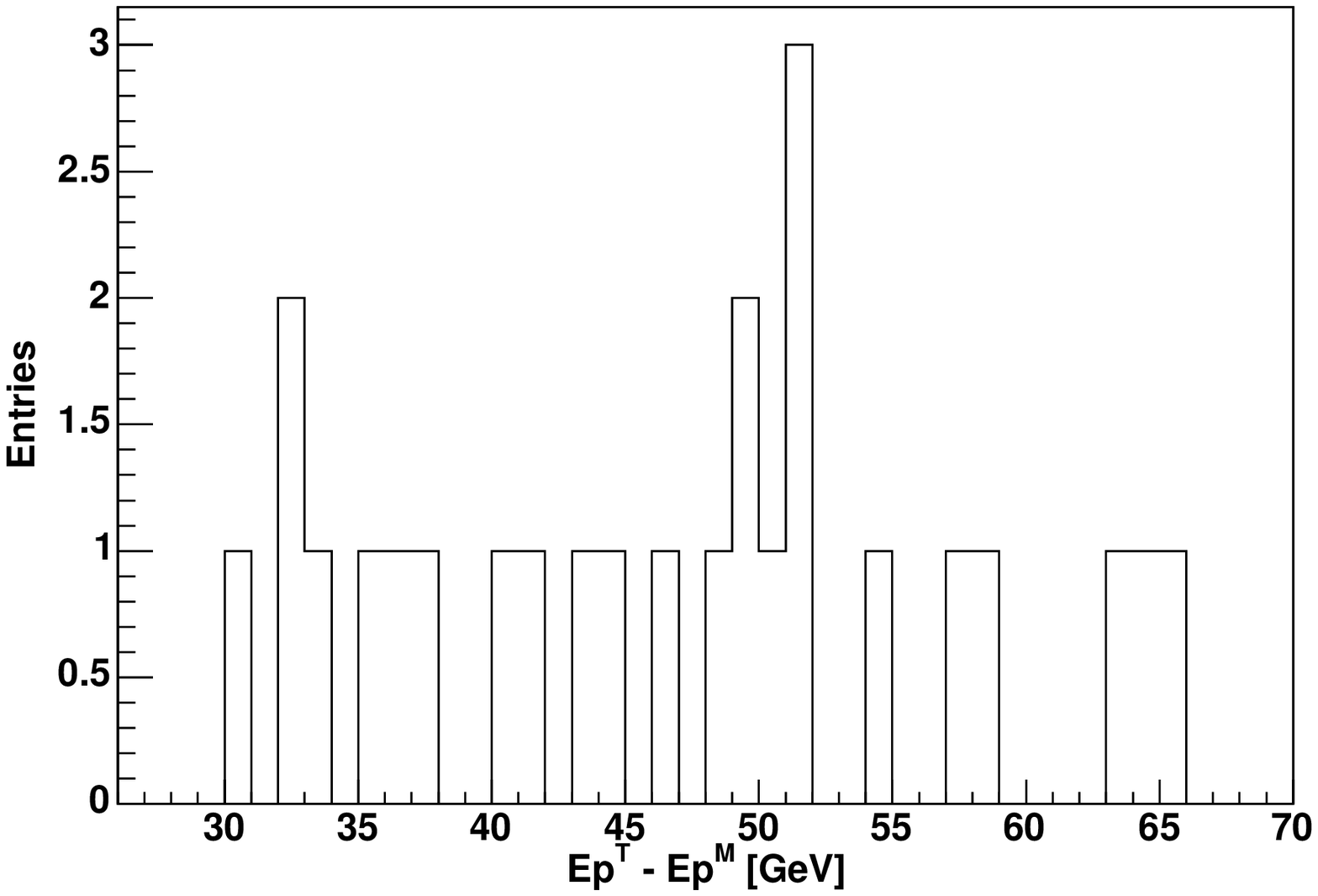} \caption{The difference of the
  theoretical and the measured $M_{\rho^{\pm}_{2}b}$ endpoint in$\gev$
  obtained with the linear fit for all 25 points. It shows a shift of
  $46.9\gev$ with a root-mean-square deviation of $10.3\gev$.} 
  \label{Ml2b_Linear_TEP-MEP} \end{center}
\end{figure}
\begin{figure}[H]
  \begin{center} \includegraphics[height=\textheight/2 - 4cm]
  {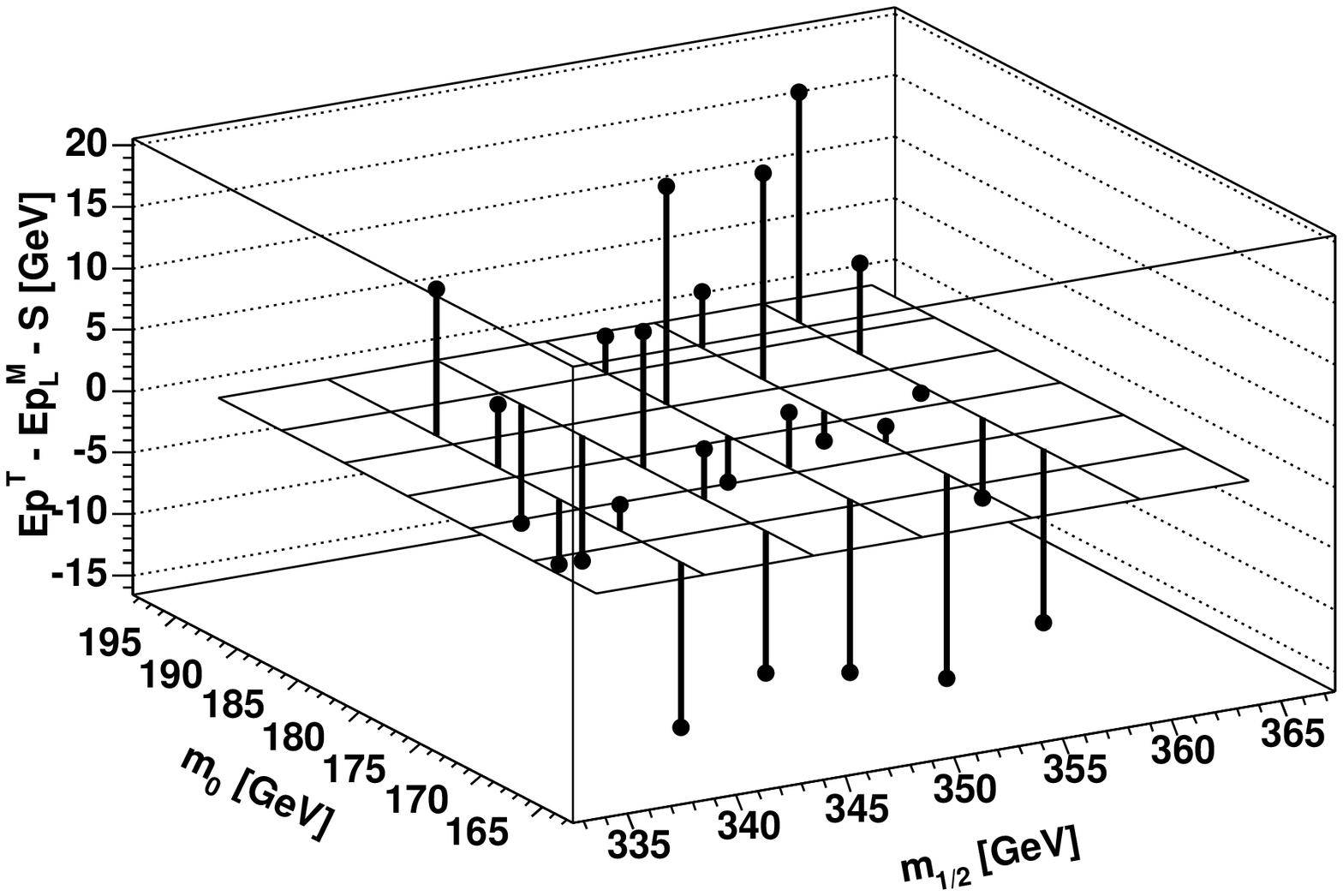} \caption{The theoretical
  $M_{\rho^{\pm}_{2}b}$ endpoint minus the measured endpoint in$\gev$
  obtained with the linear fit after applying the constant correction
  of $46.9\gev$.} \label{Ml2b_Linear_TEP-MEP-Cor} \end{center}
\end{figure}
\begin{figure}[H]
  \begin{center} \includegraphics[height=\textheight/2 - 4cm]
  {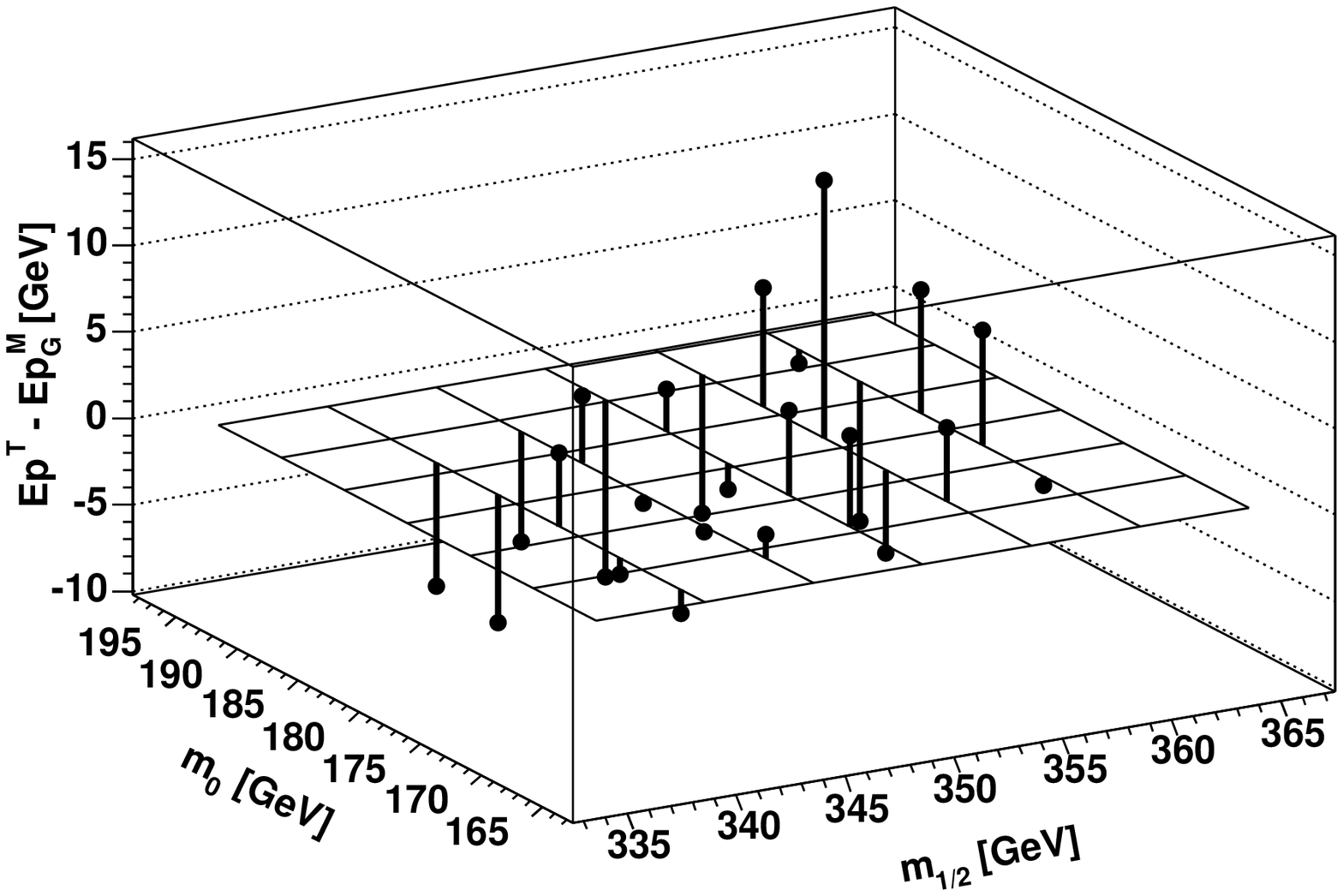} \caption{The difference of the
  theoretical and the measured $M_{\rho^{\pm}_{2}b}$ endpoint in$\gev$
  obtained with the gaussian non-fixed origin method. } 
  \label{Ml2b_GaussNFZ_TEP-MEP} \end{center}
\end{figure}
%-------------
%----------------------------------------------------------------------
\subsection{\texorpdfstring{The kinematic limit of $M_{\rho_{1}^{\pm}q} +
M_{\rho_{2}^{\mp}q}$ for bottom quarks}{The kinematic limit of M(r1q)
+ M(r2q) for bottom quarks}}\label{sec:Ml1b_Ml2b} 
\begin{figure}[H]
  \begin{center}
    \shadowbox{\includegraphics[width=\textwidth/2]{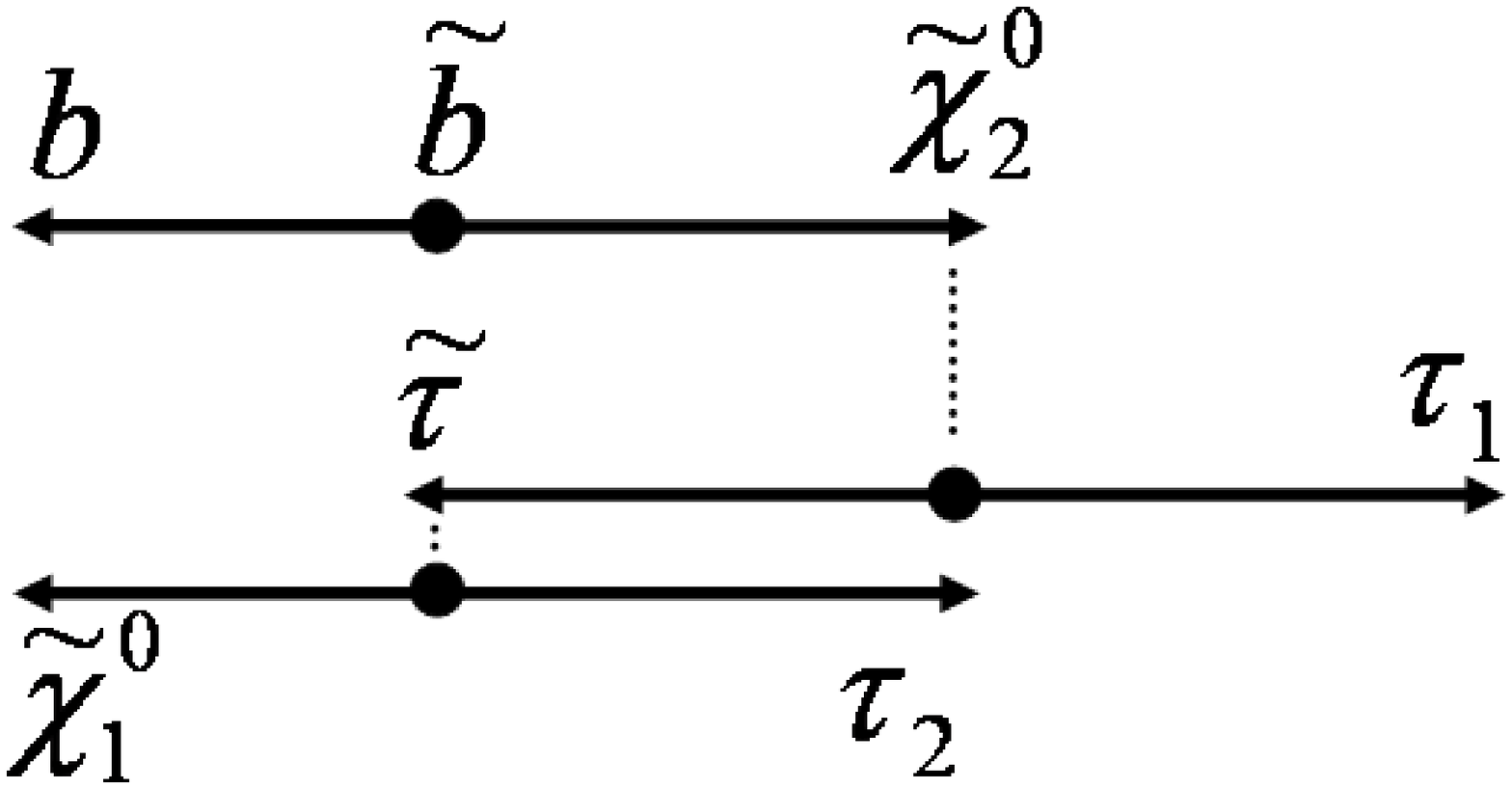}}
  \end{center}
\end{figure}
\begin{figure}[H]
  \begin{center} \includegraphics[height=\textheight/2 - 4cm]
  {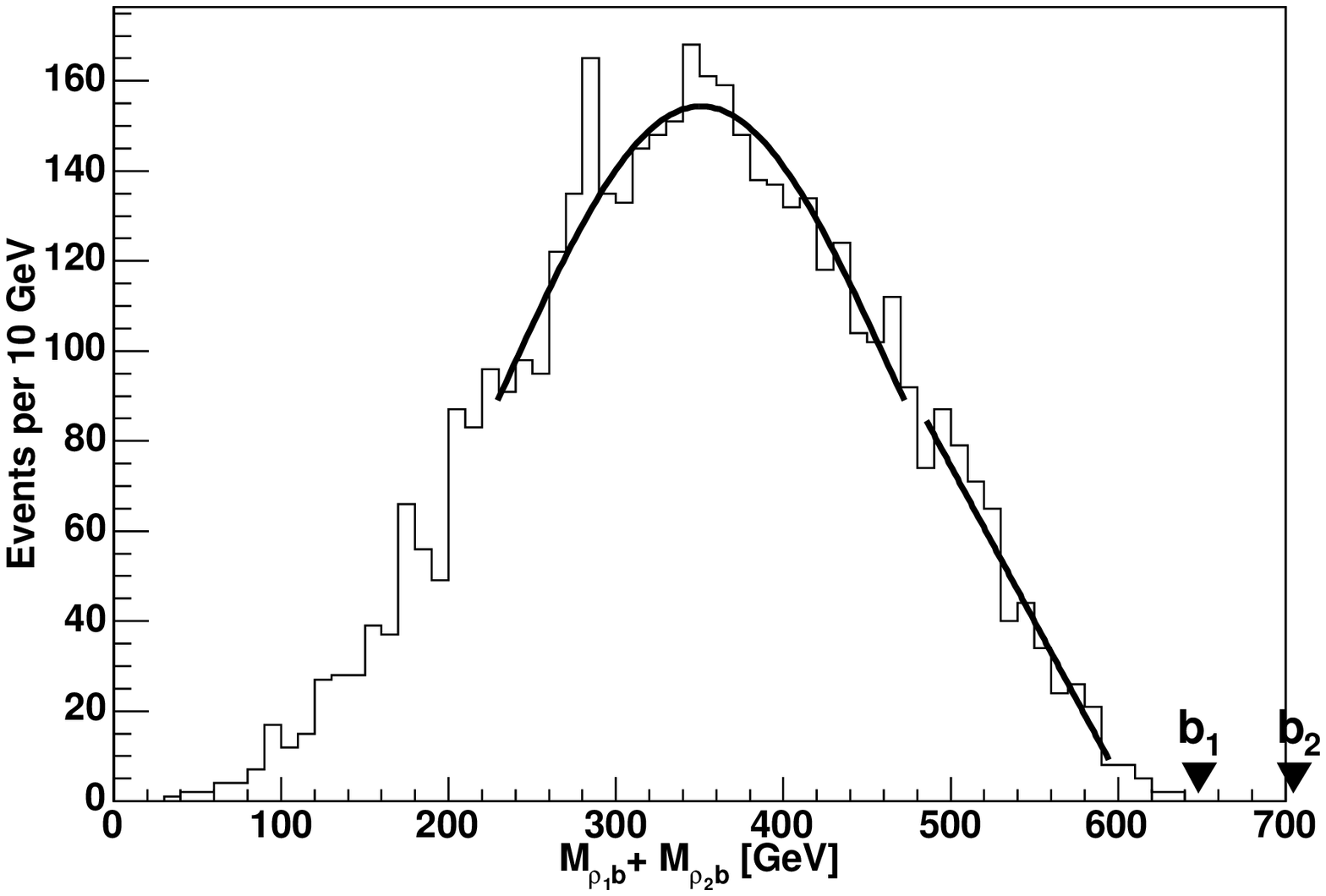} \caption{Example of a linear and a
  gaussian fit to the $M_{\rho_{1}^{\pm}b} + M_{\rho_{2}^{\mp}b}$
  distribution at $m_{0} = 181$ and $m_{1/2} = 350$. In this
  distribution 4498 events are represented. The two triangles show the
  theoretical values for events with $\tilde{b}_{1}$ ($653.1\gev$) and
  $\tilde{b}_{2}$ ($706.1\gev$), respectively.} 
  \label{Ml1b_Ml2b_Ex_181_350} \end{center}
\end{figure}

For this kinematic endpoint the linear and the gaussian method can
give a good estimate on its value. An example of a
$M_{\rho_{1}^{\pm}b} + M_{\rho_{2}^{\mp}b}$ distribution is given in
figure~\ref{Ml1b_Ml2b_Ex_181_350}. The data sample is the same as in
section~\ref{sec:Mllb}.

\begin{figure}[H]
  \begin{center} \includegraphics[height=\textheight/2 - 4cm]
  {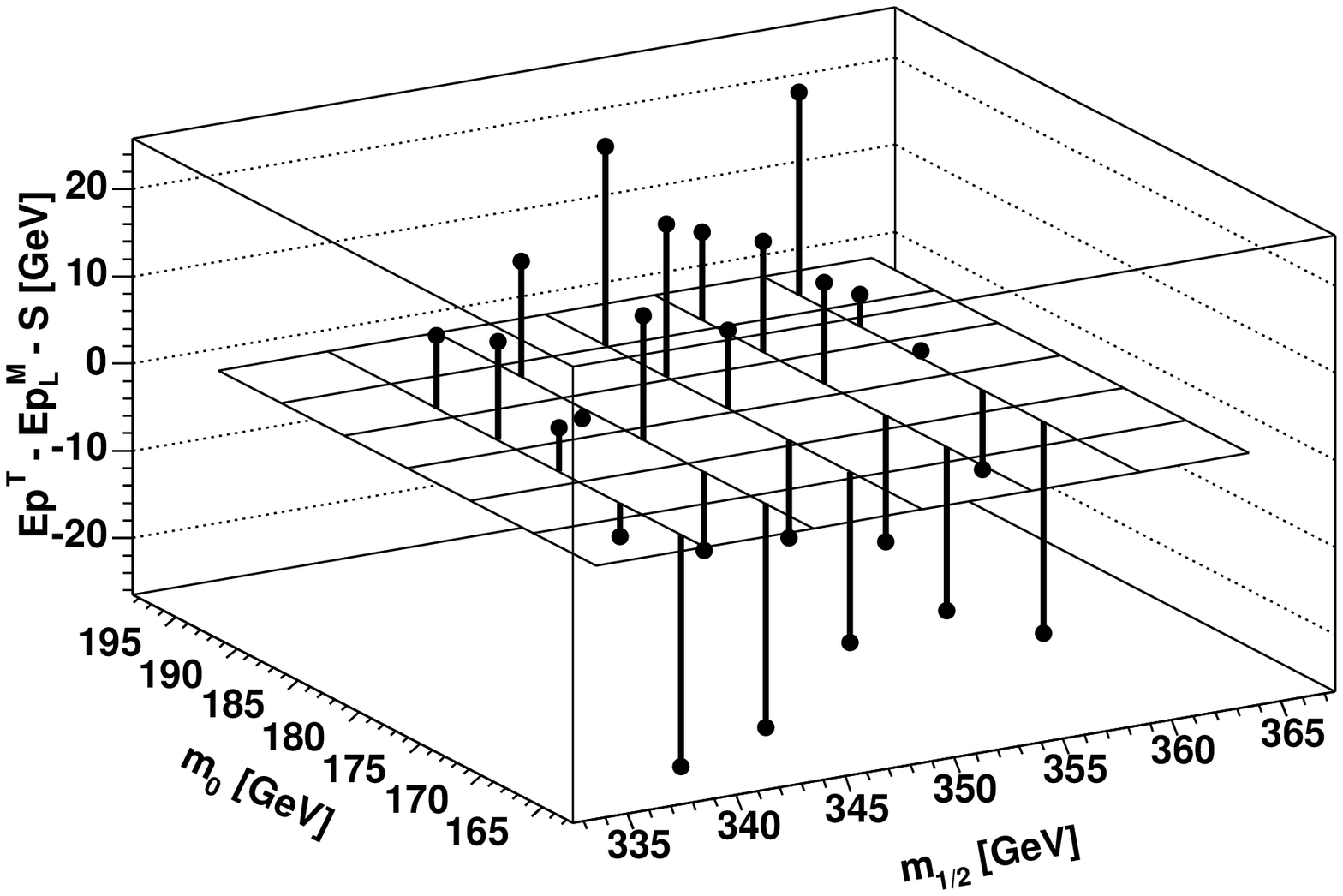} \caption{The theoretical
  $M_{\rho_{1}^{\pm}b} + M_{\rho_{2}^{\mp}b}$ endpoint minus the
  measured endpoint in$\gev$ obtained with the linear fit after
  applying the constant correction of $89.3\gev$. The values decrease
  significantly for decreasing values of $m_{0}$.} 
  \label{Ml1b_Ml2b_Linear_TEP-MEP-Cor} \end{center}
\end{figure}

\begin{figure}[H]
  \begin{center} \includegraphics[height=\textheight/2 - 4cm]
  {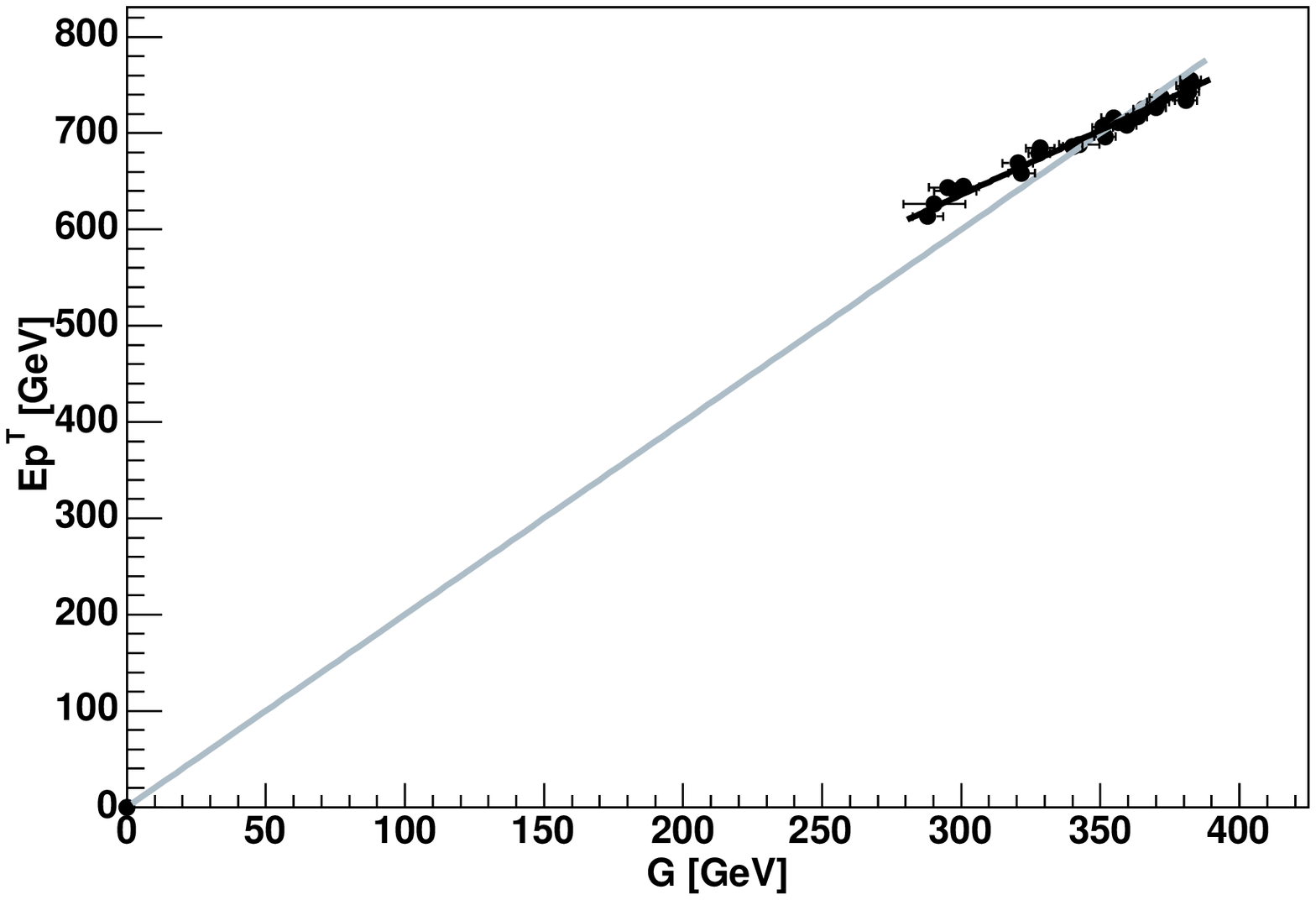} 
\caption{The theoretical kinematic $M_{\rho_{1}^{\pm}b} +
M_{\rho_{2}^{\mp}b}$ endpoints as a function of the measured gaussian
maximum for all 25 investigated points in the $m_{0}$-$m_{1/2}$
plane. The gray line with $C = 2.0$ is forced to go through the origin
whereas the black with $C = 1.3$ has an optimised ordinate value of
$234.5\gev$.}
\label{Ml1b_Ml2b_Gauss_TE_vs_MM_Fit} \end{center}
\end{figure}

For the linear fit the large correction $S = 89.3 \pm 15.1\gev$ is
again caused by the origin of the bottom quark. A clear $m_{0}$
dependence can be seen in figure~\ref{Ml1b_Ml2b_Linear_TEP-MEP-Cor}
which should be included into the correction. Similar to the light
quark case the gaussian method can be applied as well. In order to
achieve good results for the gaussian fit the symmetric area is chosen
to be $35\%$ of the maximum bin value. By comparing the two linear
fits for the light quark gaussian case in
figure~\ref{Ml1q_Ml2q_Gauss_TE_vs_MM_Fit} with the bottom quark
gaussian case in figure~\ref{Ml1b_Ml2b_Gauss_TE_vs_MM_Fit} an
interesting similarity can be found. In both cases the ratio between
the slope in the fixed origin fit and the slope in the non-fixed
origin fit has an universal value of $1.5$. A summary of the measured
values is given in table~\ref{sum:Ml1b_Ml2b}.

%---------------------------
\begin{table}[H]
\begin{center}
\begin{tabular}{|l|rrr|} \hline
\emph{Value} & Linear & Gaussian FO & Gaussian NFO\\\hline\hline
$\delta_{stat} Ep^{M}$ & 3.2 (3.2) & 9.7 & 6.5 \\
$S$ & 89.3 (37.2) & 6.2 & 1.0 \\
$\delta S$ & 15.1 (12.6) & 21.3 & 6.4 \\\hline
\end{tabular}
\caption{A summary of the calculated mean uncertainties and systematic
shifts in$\gev$ for the $M_{\rho_{1}^{\pm}b} + M_{\rho_{2}^{\mp}b}$
distribution. FO is with fixed origin and NFO with non-fixed
origin.}\label{sum:Ml1b_Ml2b}
\end{center}
\end{table}

\begin{figure}[H]
  \begin{center} \includegraphics[height=\textheight/2 - 4cm]
  {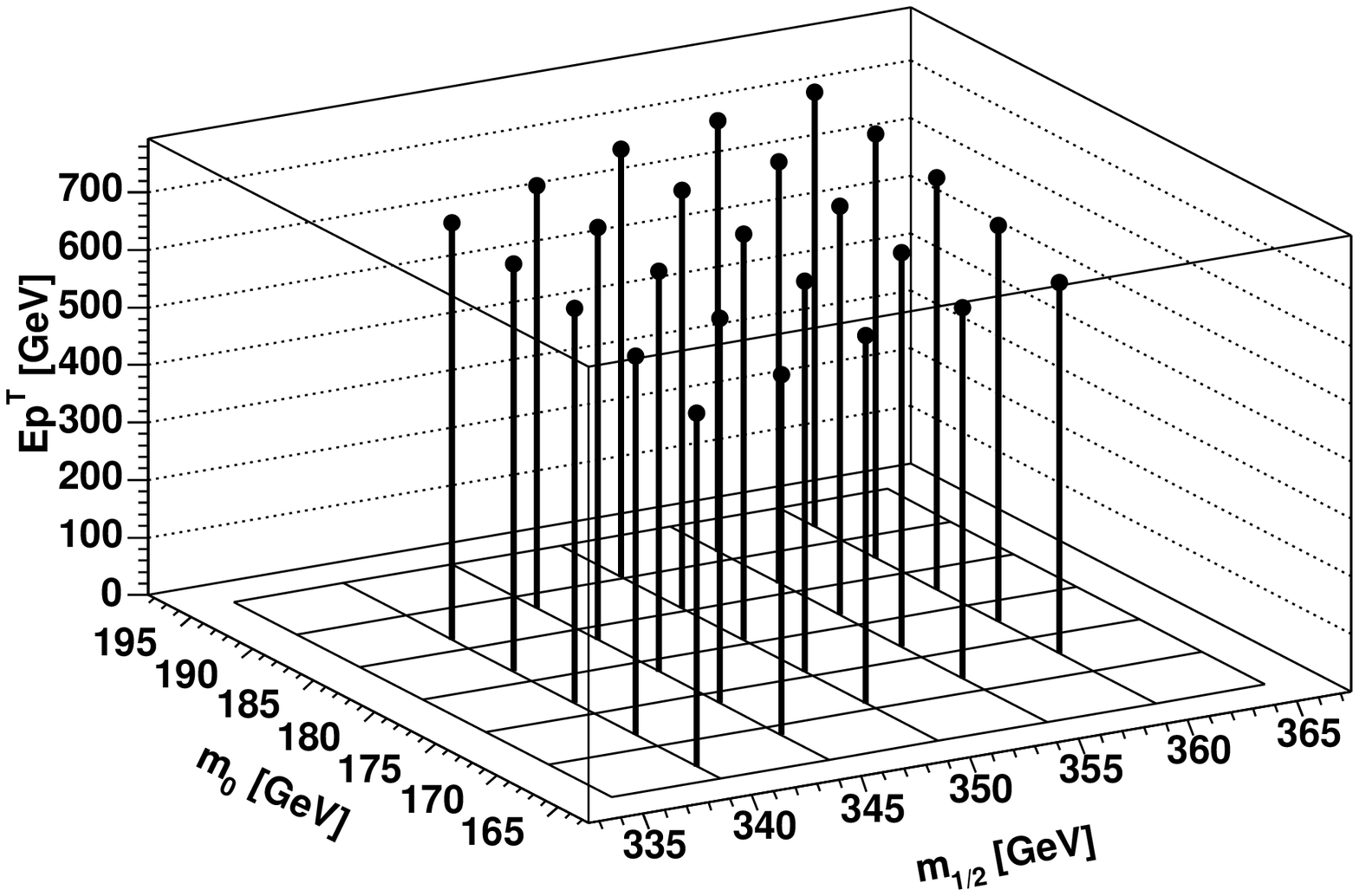} \caption{Theoretical kinematic endpoints for
  $M_{\rho_{1}^{\pm}b} + M_{\rho_{2}^{\mp}b}$ in$\gev$: The
  $\rho_{2}^{\mp}$ coming from the $\tilde{\tau}$ with the bottom
  quark and the $\rho_{1}^{\pm}$ coming from the $\chinonn$ and the
  same bottom quark from the $\tilde{b}_2 \ra b +
\chinonn$ decay.}
  \label{Ml1b_Ml2b_TEP} \end{center}
\end{figure}
\begin{figure}[H]
  \begin{center} \includegraphics[height=\textheight/2 - 4cm]
  {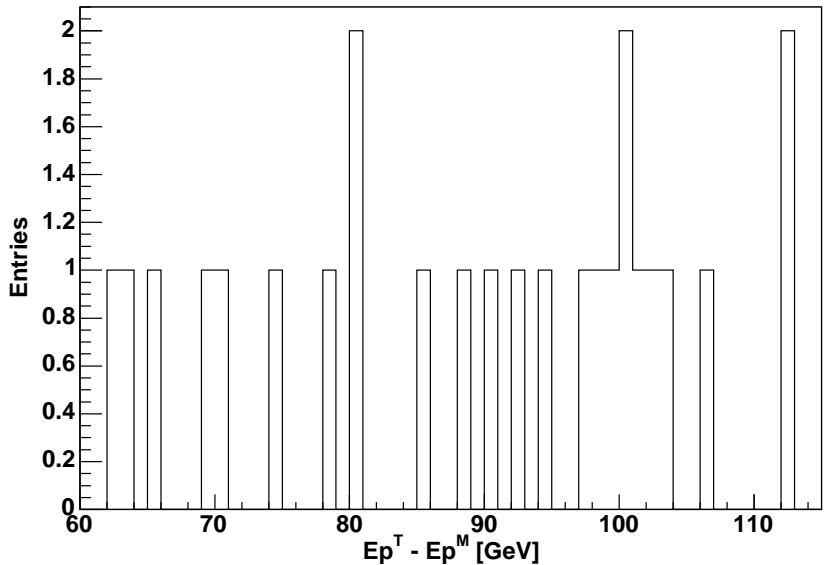} \caption{The difference of the
  theoretical and the measured $M_{\rho_{1}^{\pm}b} +
  M_{\rho_{2}^{\mp}b}$ endpoint in$\gev$ obtained with the linear fit
  for all 25 points. It shows a large mean shift of $89.3\gev$ with a
  root-mean-square deviation of $15.1\gev$.} 
  \label{Ml1b_Ml2b_Linear_TEP-MEP} \end{center}
\end{figure}

\begin{figure}[H]
  \begin{center} \includegraphics[height=\textheight/2 - 4cm]
  {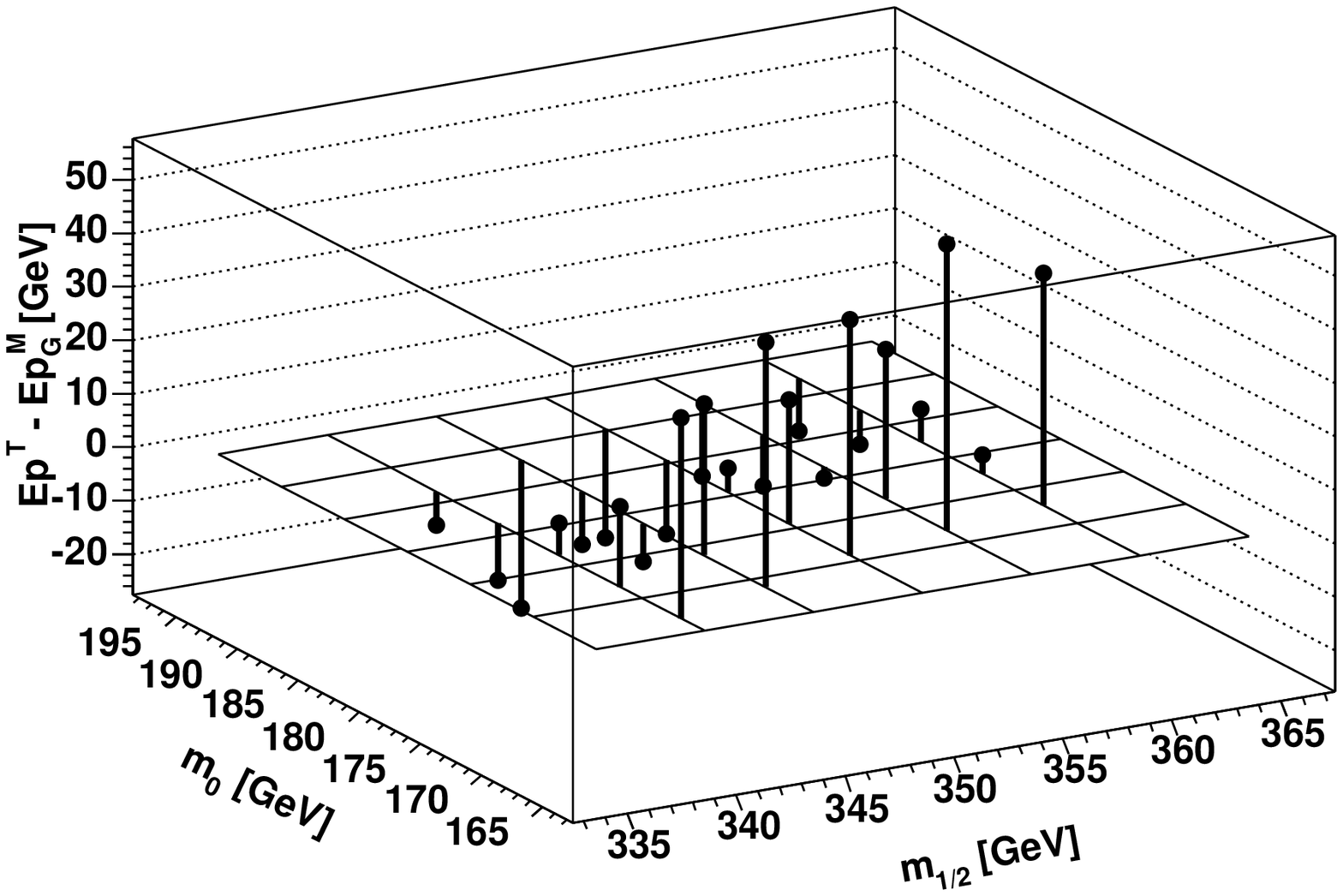} \caption{The difference of the
  theoretical and the measured $M_{\rho_{1}^{\pm}b} +
  M_{\rho_{2}^{\mp}b}$ endpoint in$\gev$ obtained with the gaussian
  fixed origin method. The measured values are too low with a mean
  systematic shift of $6.2\gev$} \label{Ml1b_Ml2b_GaussFZ_TEP-MEP}
  \end{center}
\end{figure}
\begin{figure}[H]
  \begin{center} \includegraphics[height=\textheight/2 - 4cm]
  {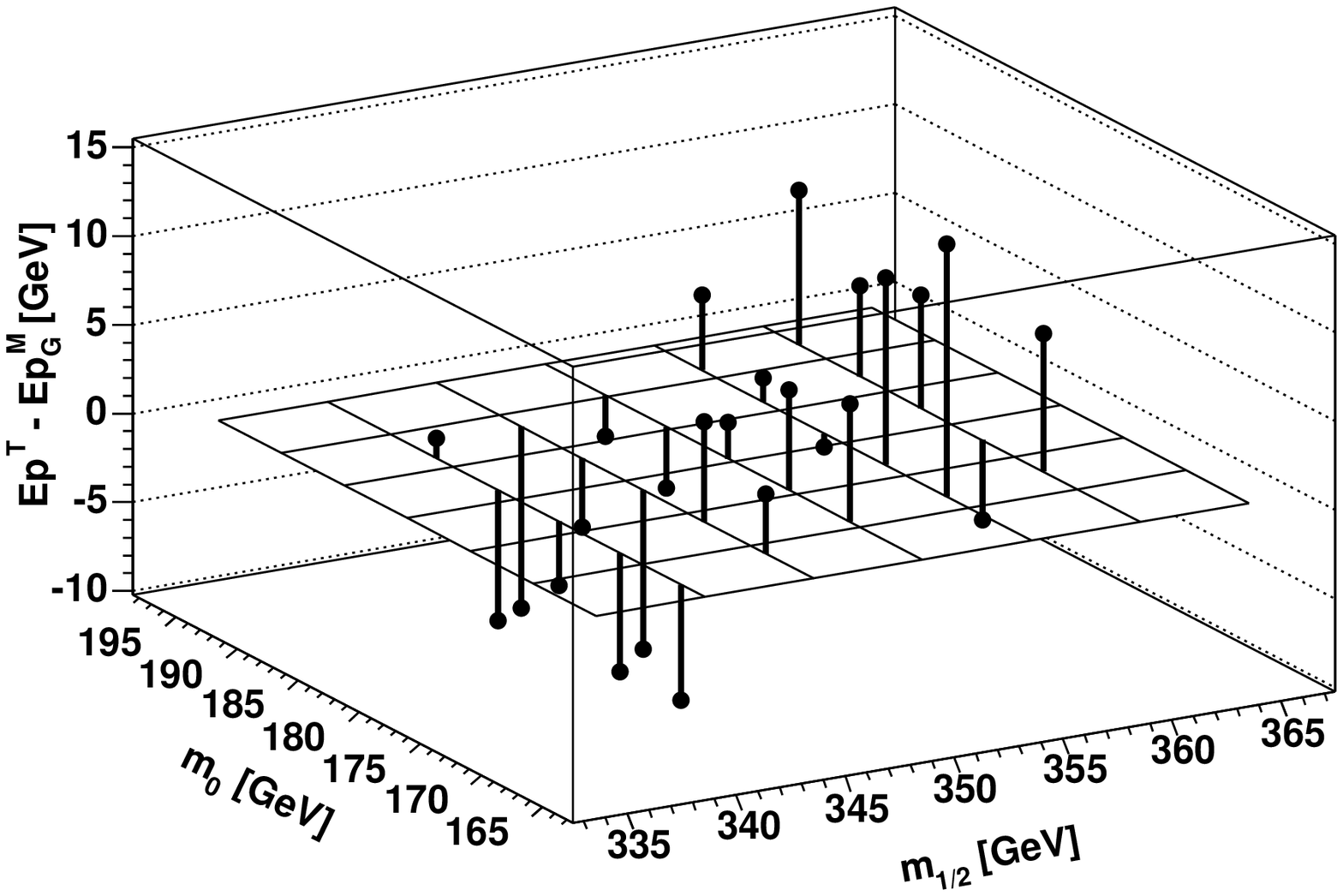} \caption{The difference of the
  theoretical and the measured $M_{\rho_{1}^{\pm}b} +
  M_{\rho_{2}^{\mp}b}$ endpoint in$\gev$ obtained with the gaussian
  non-fixed origin method. }\label{Ml1b_Ml2b_GaussNFZ_TEP-MEP}
  \end{center}
\end{figure}
%----------------------------------------------------------------------
%----------------------------------------------------------------------
\subsection{\texorpdfstring{The kinematic limit of $M_{\rho^{\pm}\rho^{\mp}q}$ for
top quarks}{The kinematic limit of M(rrq) for top
quarks}}\label{ssec:llqtop} 
\begin{figure}[H]
  \begin{center}
    \shadowbox{\includegraphics[width=\textwidth/2]{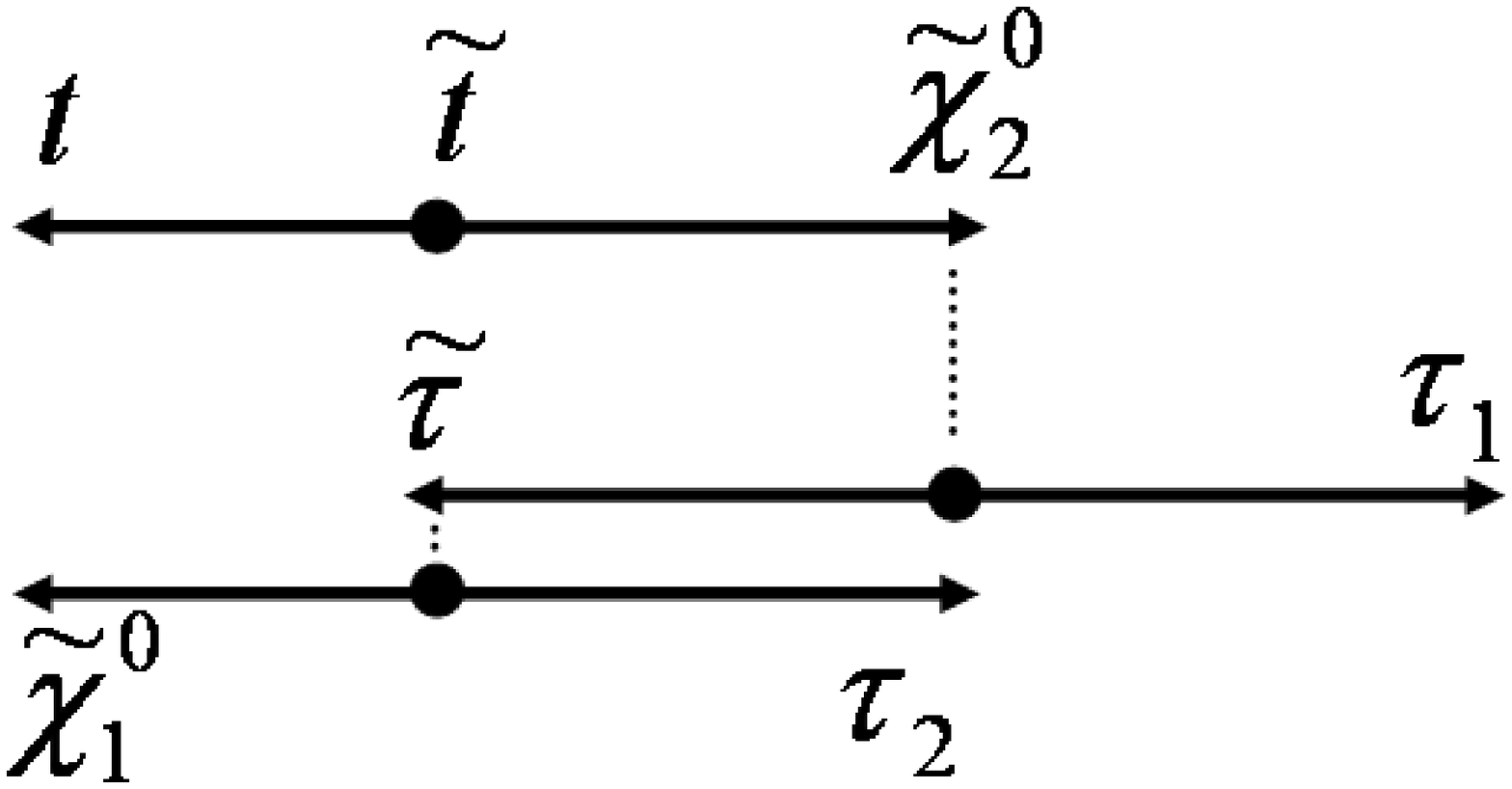}}
  \end{center}
\end{figure}
\begin{figure}[H]
  \begin{center} \includegraphics[height=\textheight/2 - 4cm]
  {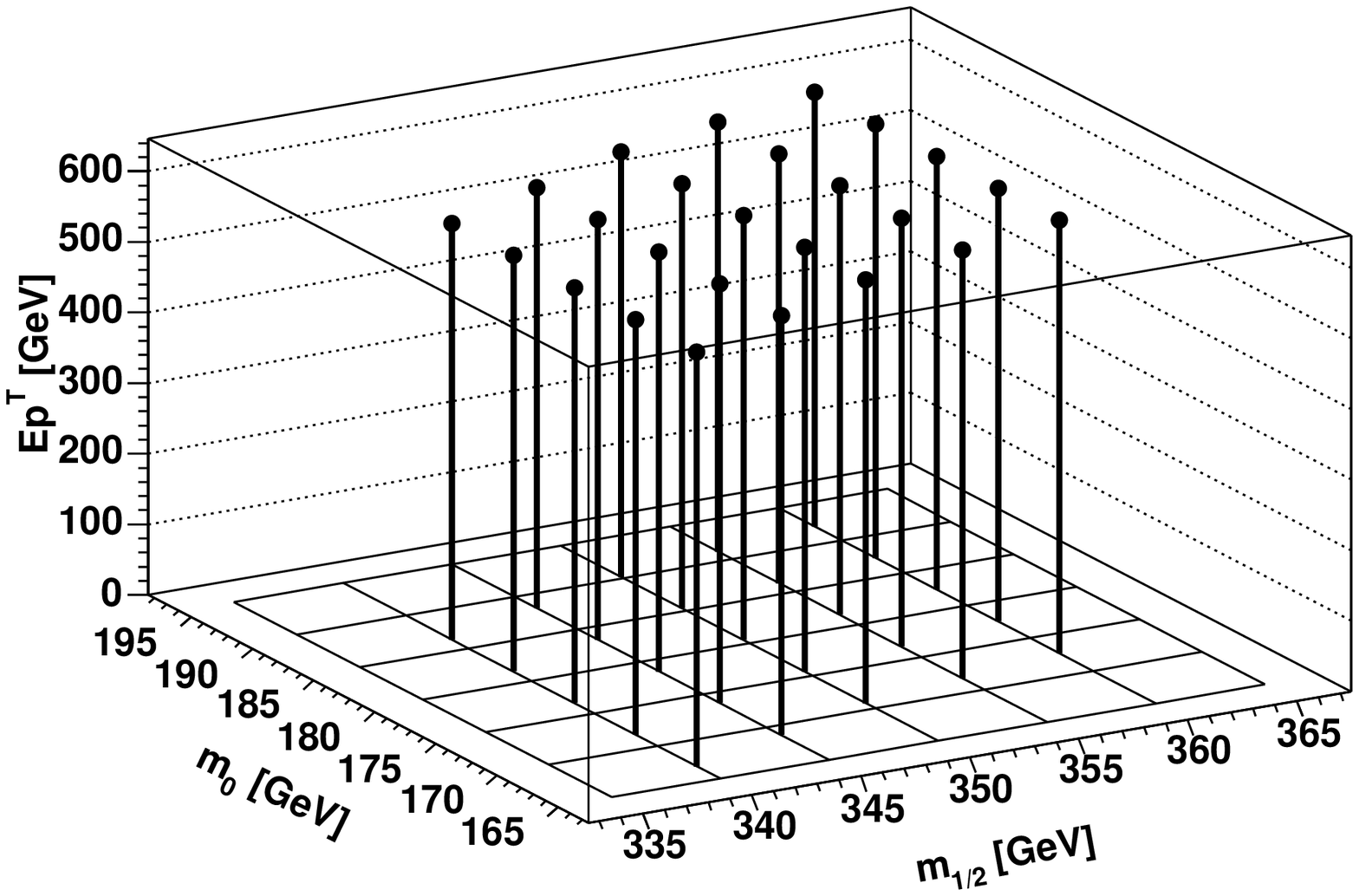} \caption{Theoretical kinematic endpoints in$\gev$
  for the invariant mass of both opposite-sign rhos, coming from the
  $\chinonn$ and the $\tilde{\tau}$ respectively, and the
  associated top quark from $\tilde{t}_2 \ra t + \chinonn$.}
  \label{Mllt_TEP} \end{center}
\end{figure}
\begin{figure}[H]
  \begin{center} \includegraphics[height=\textheight/2 - 4cm]
  {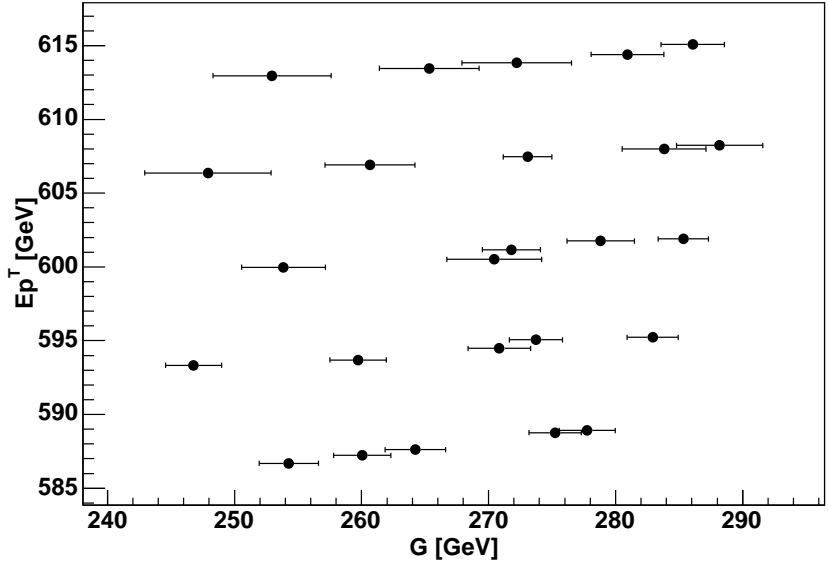} 
\caption{The theoretical kinematic $M_{\rho^{\pm}\rho^{\mp}t}$
  endpoints versus the measured gaussian maximum for all 25
  investigated points in the $m_{0}$-$m_{1/2}$ plane. A unique mapping
  of the measured maximum to the theoretical endpoint can not be
  established since several theoretical endpoint values are associated
  with the same measured maximum.}  \label{Mllt_Gauss_TE_vs_MM_Fit}
  \end{center}
\end{figure}

For the analysis of events with top quarks it is necessary to use the
endpoint formulae given in section~\ref{sec:top}. The input for the
upper limit is the mass of the $\tilde{t}_{2}$ and for the lower limit
the lighter $\tilde{t}_{1}$. For the following analysis 25 data
samples with about 1750 events have been used. Similar to the previous
cases, the gaussian method cannot be applied here which can be seen in
figure~\ref{Mllt_Gauss_TE_vs_MM_Fit}.  The lower kinematic limit which
is given by formula~(\ref{Mlltmin}) does not provide any information
which can be used for the mass reconstruction since the theoretical
value hardly depends on $m_{0}$ and $m_{1/2}$. It is $193.5 \pm
0.1\gev$ over the whole investigated area. The upper limit given by
formula~(\ref{Mlltmax}) varies from $586.7\gev$ at $m_{0} = 167$ and
$m_{1/2} = 340$ to $615.1\gev$ at $m_{0} = 193$ and $m_{1/2} = 360$
which can be seen in figure~\ref{Mllt_TEP}. The linear fit gives the
endpoint which arises from the $\tilde{t}_{1}$ mass with adequate
precision. However, the real kinematic endpoint is given by the
$\tilde{t}_{2}$ mass. Thus, a quite large correction $S$ of $155.5
\pm 5.7\gev$ is necessary. The reason for this is the stop
mixing. The principle is described in section~\ref{sec:Mllb} for the
sbottom mixing.
%-------
\begin{table}[H]
\begin{center}
\begin{tabular}{|l|r|} \hline
\emph{Value} & Linear \\\hline\hline
$\delta_{stat} Ep^{M}$ & 3.1 (3.1)\\
$S$ & 155.5 (0.4)  \\
$\delta S$ & 5.7 (5.6) \\\hline
\end{tabular}
\caption{A summary of the calculated correction and mean uncertainties
in$\gev$.  The values in brackets correspond to the case of
$\tilde{t}_{1}$ being the input for the upper kinematic
limit.}\label{sum:Mllt}
\end{center}
\end{table}

\begin{figure}[H]
  \begin{center} \includegraphics[height=\textheight/2 - 4cm]
  {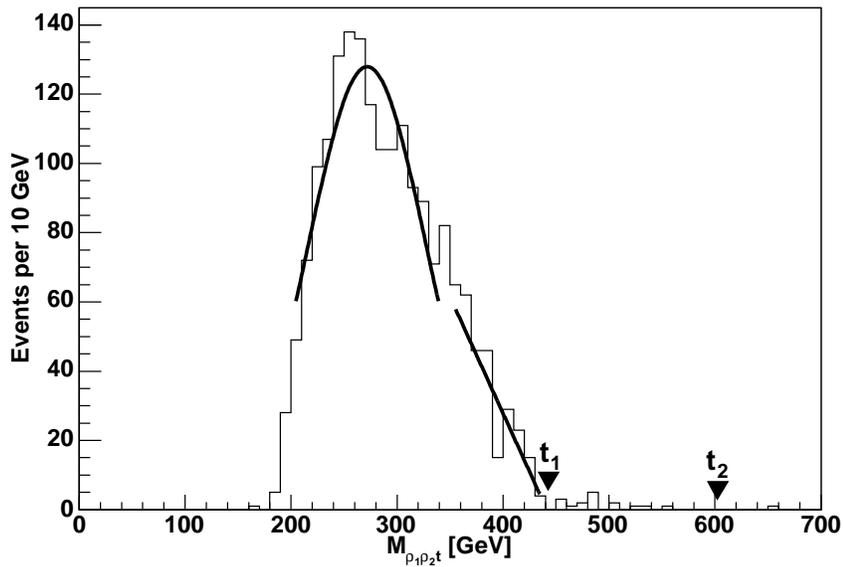} \caption{Example of a linear and a gaussian
  fit to the $M_{\rho^{\pm}\rho^{\mp}t}$ distribution at $m_{0} = 181$
  and $m_{1/2} = 350$. The invariant mass distribution is based on
  1859 top quark events. The two triangles show the theoretical values
  for events with $\tilde{t}_{1}$ ($446.1\gev$) and $\tilde{t}_{2}$
  ($601.2\gev$), respectively.} \label{Mllt_Ex_181_350} \end{center}
\end{figure}
\begin{figure}[H]
  \begin{center} \includegraphics[height=\textheight/2 - 4cm]
  {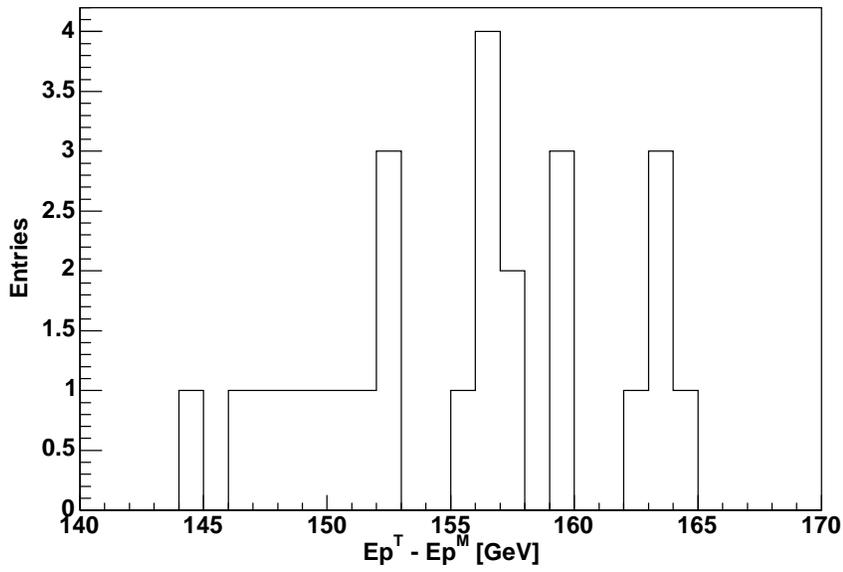} \caption{The theoretical
  $M_{\rho^{\pm}\rho^{\mp}t}$ endpoint minus the measured endpoint
  in$\gev$. The mean value $S$ is $155.5\gev$ with a root-mean-square
  deviation of $5.7\gev$. If the $\tilde{t}_{1}$ mass is taken for
  the theoretical endpoint the value $S$ changes to $0.6 \pm
  5.6\gev$} \label{Mllt_Linear_TEP-MEP} \end{center}
\end{figure}
\begin{figure}[H]
  \begin{center} \includegraphics[height=\textheight/2 - 4cm]
  {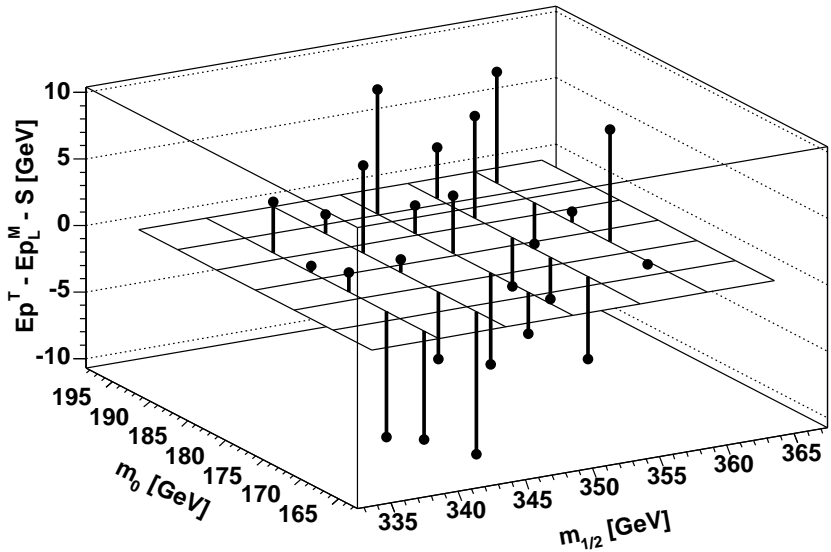} \caption{The theoretical
  $M_{\rho^{\pm}\rho^{\mp}t}$ endpoint minus the measured endpoint
  in$\gev$ obtained with the linear fit after applying a constant
  correction of $155.5\gev$.} \label{Mllt_Linear_TEP-MEP-Cor}
  \end{center}
\end{figure}
%----------------------------------------------------------------------
\subsection{\texorpdfstring{Two kinematic limits in $M_{\rho^{\pm}q}$
for top quarks}{Two kinematic limits in M(rq) for top
quarks}}\label{ssec:l12qtop} 
\begin{figure}[H]
  \begin{center}
    \shadowbox{\begin{tabular}{c}
	\includegraphics[width=\textwidth/2]{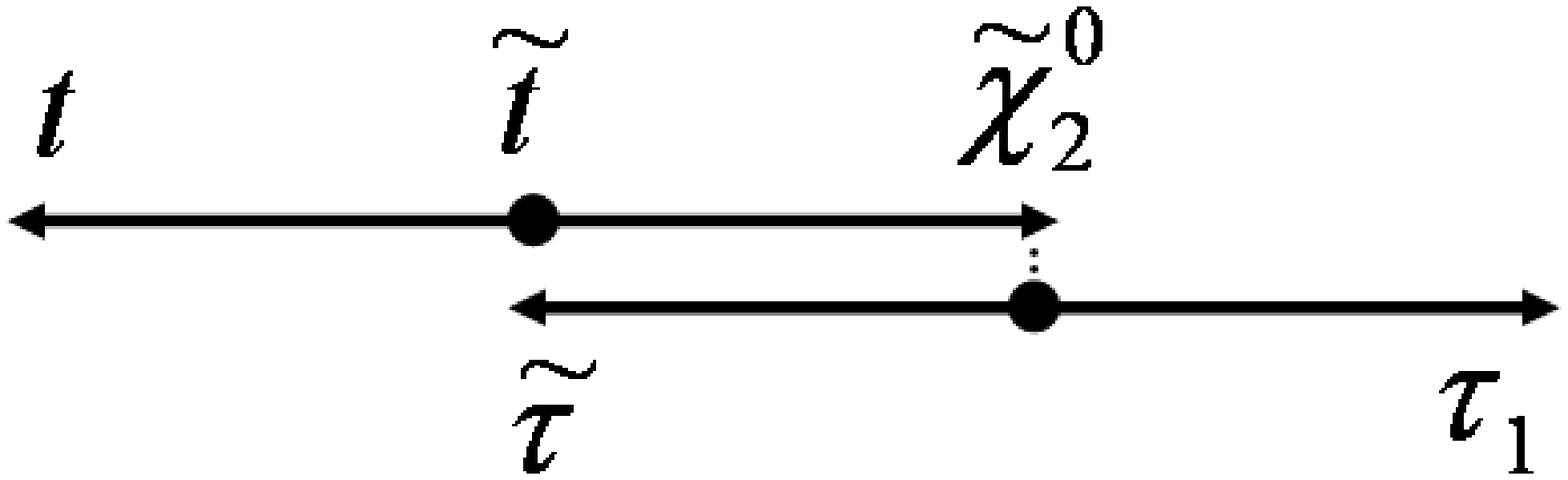}\\
    \includegraphics[width=\textwidth/2]{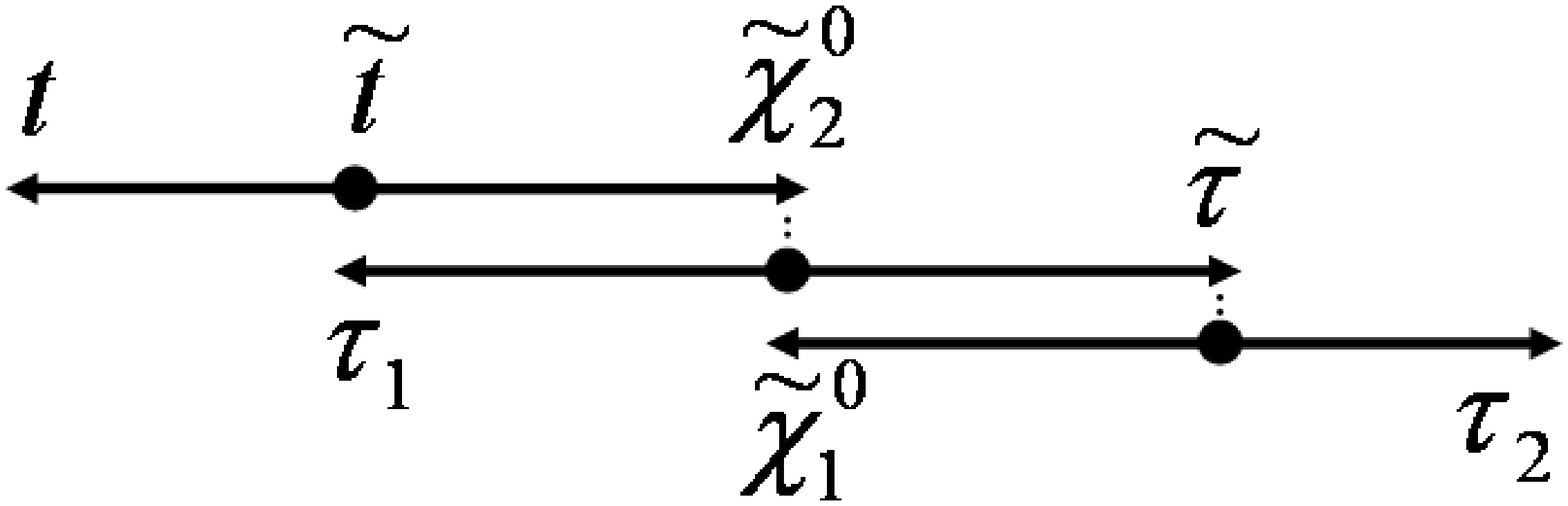}
	\end{tabular}}
  \end{center}
\end{figure}
\begin{figure}[H]
  \begin{center} \includegraphics[height=\textheight/2 - 4cm]
  {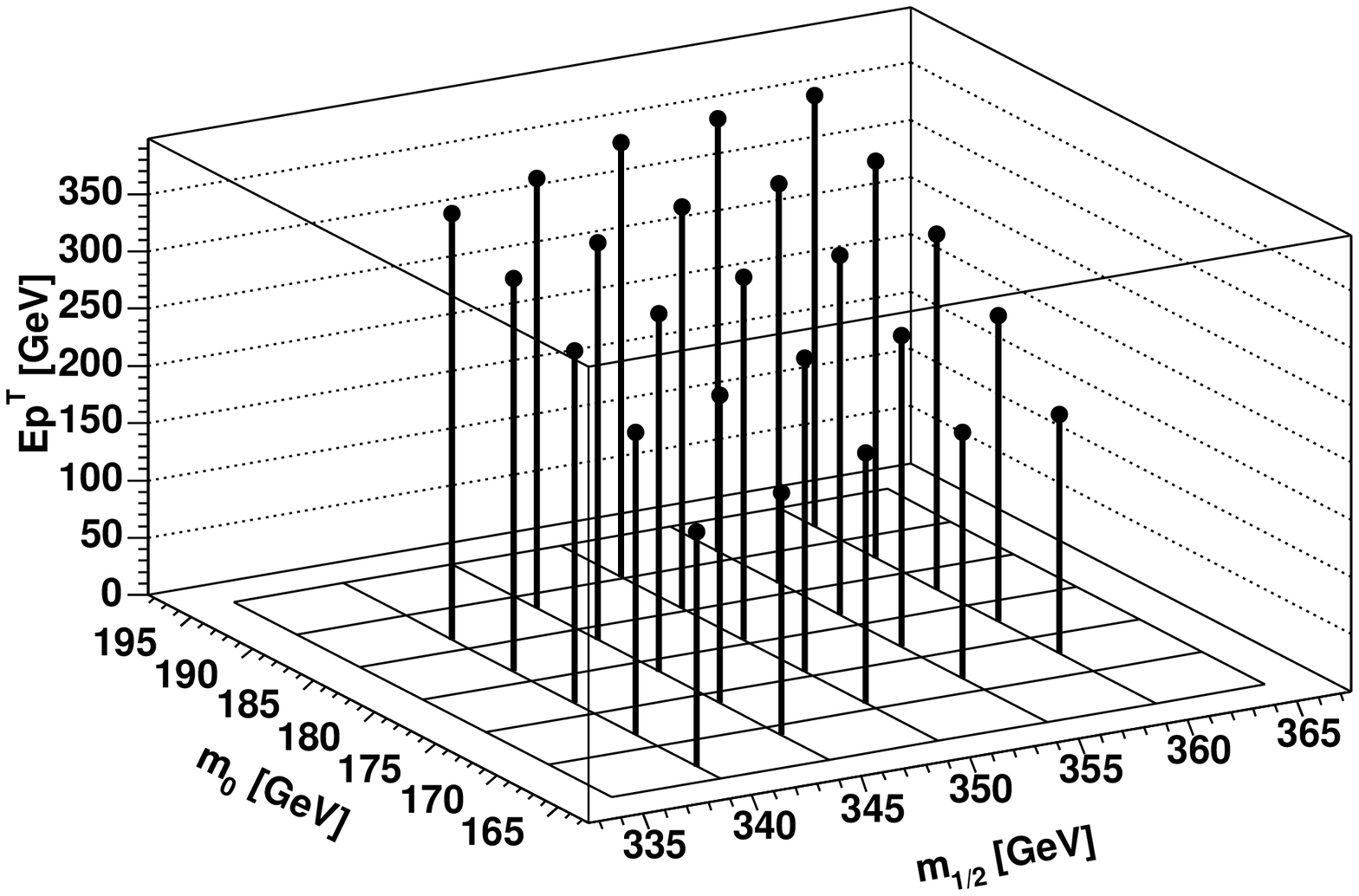} \caption{Theoretical kinematic endpoints in$\gev$
  for the invariant mass of the second rhos coming from the
  $\tilde{\tau}^{\pm}$ and the associated top quark coming from the
  stop in the $\tilde{t}_2 \ra t + \chinonn$ decay. The endpoints
  decrease for decreasing values of $m_{0}$.}  \label{Ml2t_TEP}
  \end{center}
\end{figure}
\begin{figure}[H]
  \begin{center} \includegraphics[height=\textheight/2 - 4cm]
  {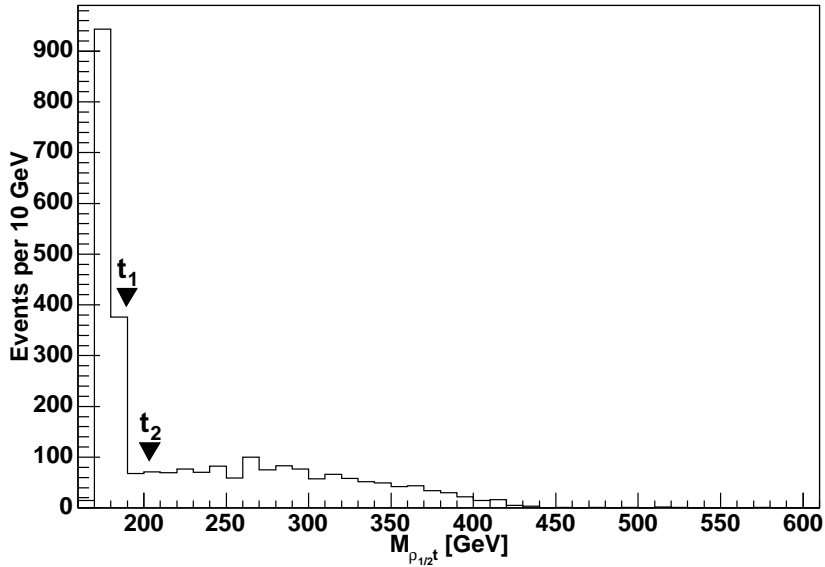} \caption{Example of $M_{\rho^{\pm}t}$ at
  $m_{0} = 167$ and $m_{1/2} = 340$ after the cut on
  $M_{\rho^{\pm}\rho^{\mp}}$ at $30\%$ of its maximum invariant mass
  (see section~\ref{sec:Mlq}). 2676 events are represented. The two
  triangles show the theoretical values for events with
  $\tilde{t}_{1}$ ($190.4\gev$) and $\tilde{t}_{2}$ ($204.7\gev$),
  respectively. In this case it is possible to measure the endpoint of
  the $M_{\rho^{\pm}_{2}t}$ distribution.} \label{Ml2t_Ex_167_340}
  \end{center}
\end{figure}
\begin{figure}[H]
  \begin{center} \includegraphics[height=\textheight/2 - 4cm]
  {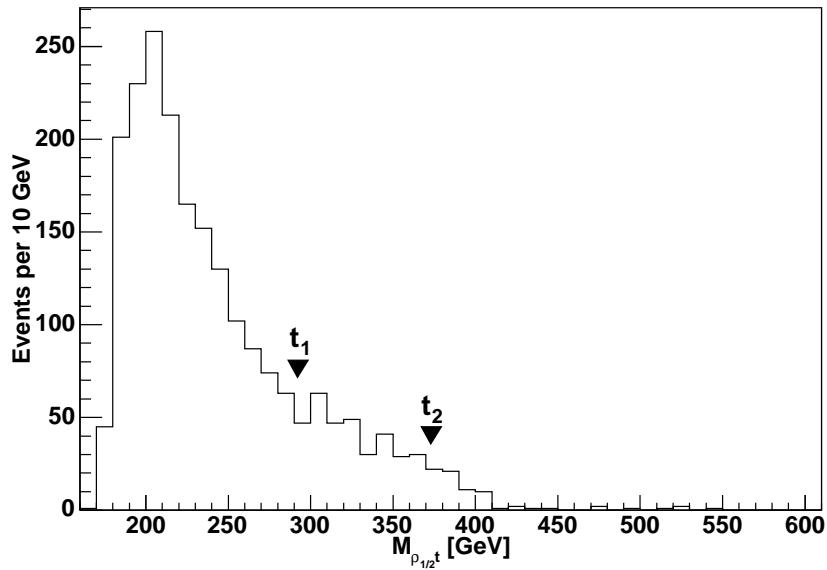} \caption{Example of $M_{\rho^{\pm}t}$ at
  $m_{0} = 187$ and $m_{1/2} = 340$ after the cut at $30\%$ of the
  maximum invariant mass of both rhos. 2140 events are
  represented. The two triangles show the theoretical values for
  events with $\tilde{t}_{1}$ ($291.0\gev$) and $\tilde{t}_{2}$
  ($371.8\gev$), respectively. In this case, even with the cut, it is
  not possible to measure the endpoint of the $M_{\rho^{\pm}_{2}t}$
  distribution.} \label{Ml2t_Ex_187_340} \end{center}
\end{figure}
\begin{figure}[H]
  \begin{center} \includegraphics[height=\textheight/2 - 4cm]
  {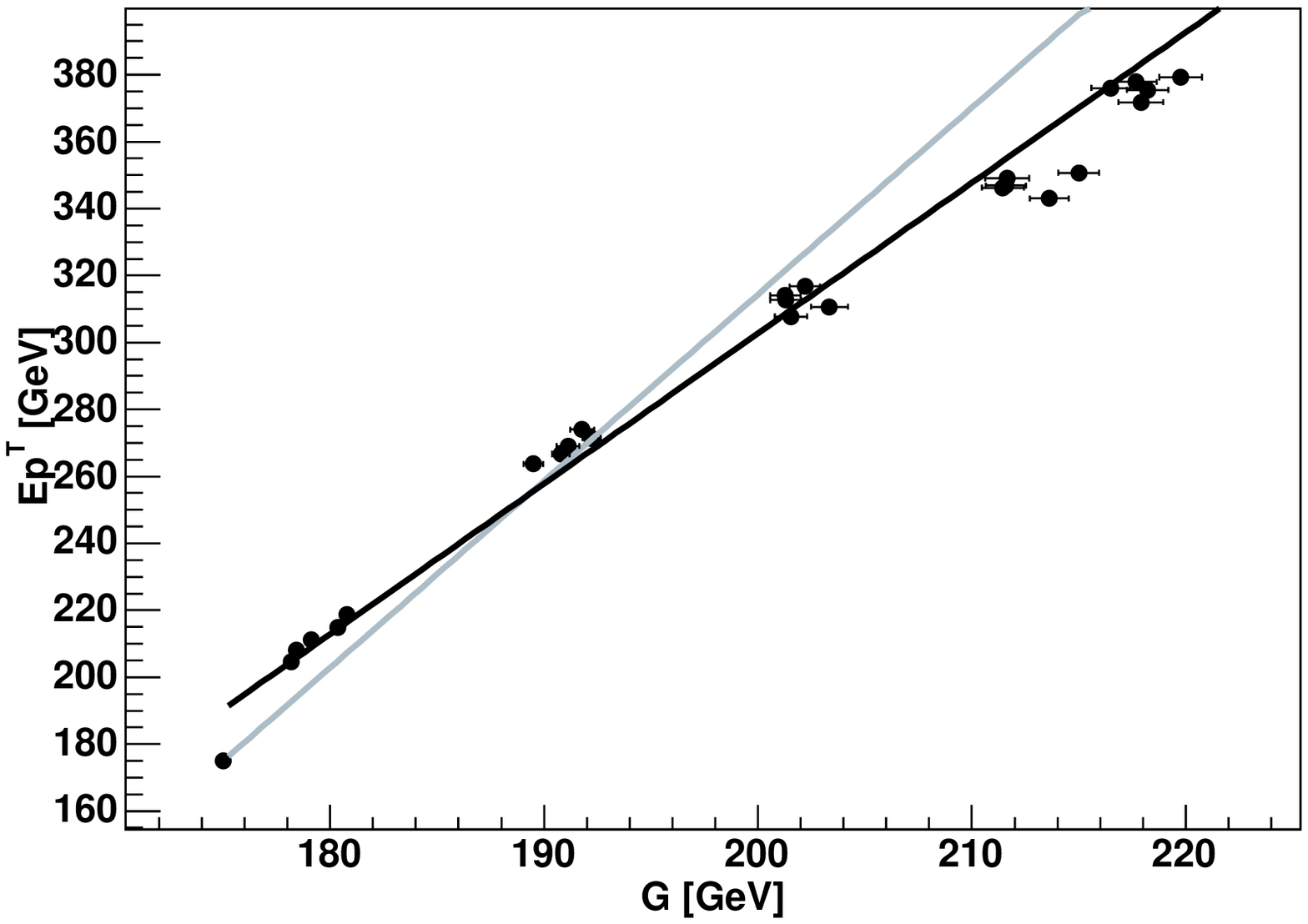} 
\caption{The theoretical kinematic $M_{\rho^{\pm}_{2}t}$
  endpoints as a function of the measured gaussian maximum for all 25
  investigated points in the $m_{0}$-$m_{1/2}$ plane. The gray line is
  forced to go through the adapted origin whereas the black has an
  optimised adapted ordinate value of $15.4\gev$.}
  \label{Ml2t_Gauss_TE_vs_MM_Fit} \end{center}
\end{figure}

The theoretical lower limit of the first rho with associated top quark
distribution is constant at a value 192.4 $\pm$ 0.8$\gev$ and for the
second it is 176.3 $\pm$ 0.7$\gev$. The upper limit of the second
distribution shown in figure~\ref{Ml2t_TEP} varies from 204.7$\gev$ at
$m_{0} = 167$ and $m_{1/2} = 340$ to 376.0$\gev$ at $m_{0} = 193$ and
$m_{1/2} = 360$. The linear fit method can in principal be applied in
the area with a $m_{0}$ value around $170\gev$. However, for $m_{0}$
values around $190\gev$ this is not possible, even if the cut at
$30\%$ of the invariant mass of both rhos is used. An example for the
first case at $m_{0} = 167$ and $m_{1/2} = 340$ is given in
figure~\ref{Ml2t_Ex_167_340} and figure~\ref{Ml2t_Ex_187_340} shows an
example of the latter at $m_{0} = 187$ and $m_{1/2} = 340$. The
gaussian method is applicable here as can be seen in
figure~\ref{Ml2t_Gauss_TE_vs_MM_Fit}. The only necessary change is
that due to the large top mass the origin in the $Ep^{T}$-$G$ plane
used in the previous sections has to be moved from (0,0) to
(175,175). This is called hence the ``adapted'' origin. The
formula~(\ref{eq:gaussfit}) has to be changed to
\begin{eqnarray}
Ep^{T}_{i} &=& C \cdot (G_{i} - 175) + D + 175.
\end{eqnarray}
If the adapted origin is fixed the slope $C$ is 5.6. If it is not, it
decreases to $C = 4.5$ with $D = 15.4$.
\begin{table}[H]
\begin{center}
\begin{tabular}{|l|rr|} \hline
\emph{Value} & Gaussian FO & Gaussian NFO\\\hline\hline
$\delta_{stat} Ep^{M}$ & 3.7 & 2.9 \\
$S$ & -14.6 & -1.2 \\
$\delta S$ & 21.6 & 6.8 \\\hline
\end{tabular}
\caption{A summary of the calculated mean uncertainties and systematic
shifts in$\gev$ for $M_{\rho^{\pm}t}$ endpoint measurements. FO
denotes the values with fixed adapted origin and NFO with non-fixed
adapted origin.}\label{sum:Mlt}
\end{center}
\end{table}

\begin{figure}[H]
  \begin{center} \includegraphics[height=\textheight/2 - 4cm]
  {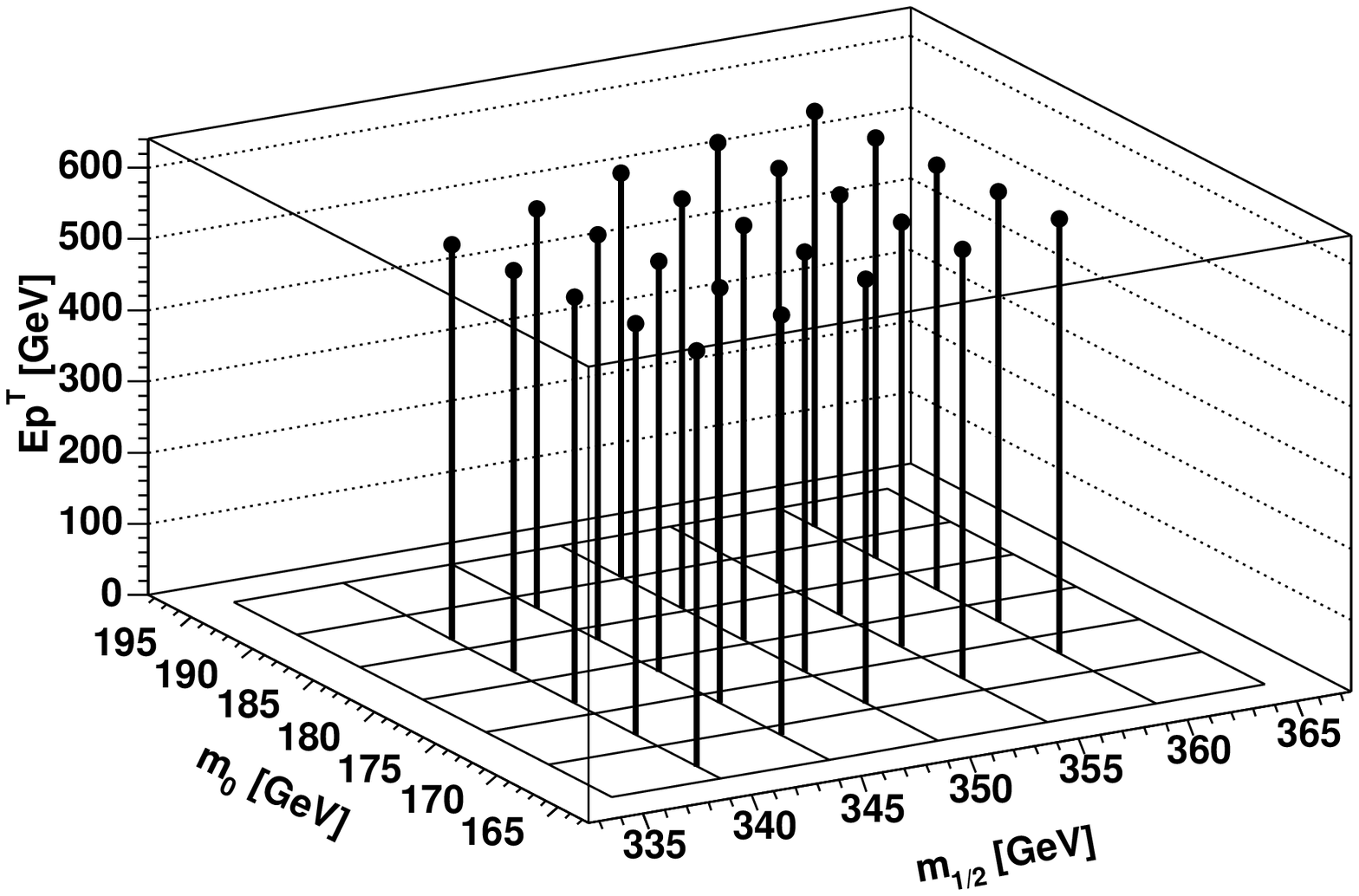} \caption{Theoretical kinematic endpoints in$\gev$
  for the invariant mass of first rho coming from the $\chinonn$ and
  the associated top quark coming from the stop in the
  $\tilde{t}_2 \ra t + \chinonn$ decay.}
  \label{Ml1t_TEP} \end{center}
\end{figure}

\begin{figure}[H]
  \begin{center} \includegraphics[height=\textheight/2 - 4cm]
  {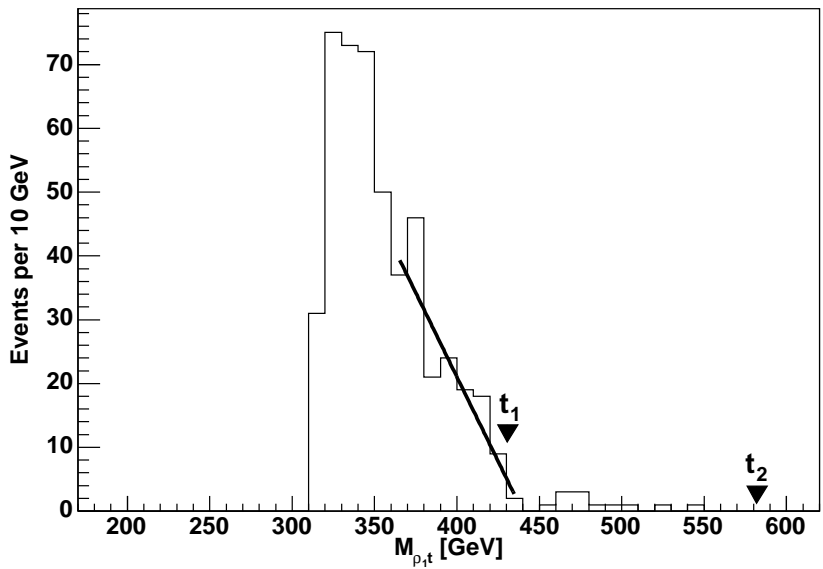} \caption{Example for the linear fit to the
  $M_{\rho^{\pm}t}$ distribution at $m_{0} = 181$ and $m_{1/2} = 350$
  after applying a cut on $M_{\rho^{\pm}t}$ at the theoretical second
  endpoint value. In this example 490 events are represented. The two
  triangles show the theoretical values for events with
  $\tilde{t}_{1}$ ($429.2\gev$) and $\tilde{t}_{2}$ ($582.3\gev$),
  respectively.} \label{Ml1t_Ex_181_350} \end{center}
\end{figure}

\begin{figure}[H]
  \begin{center} \includegraphics[height=\textheight/2 - 4cm]
  {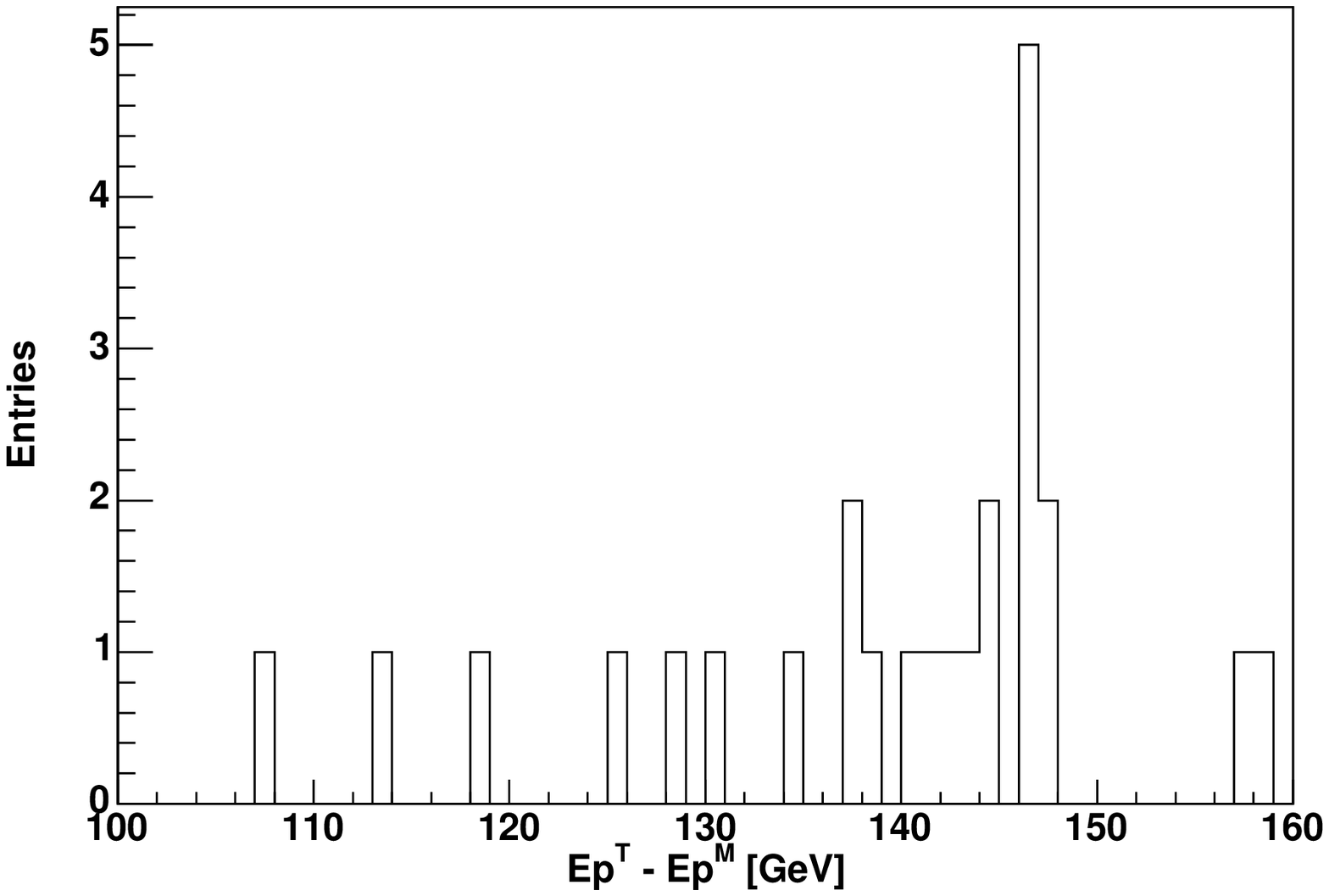} \caption{The difference of the
  theoretical and the measured $M_{\rho^{\pm}_{1}t}$ endpoint in$\gev$
  obtained with the linear fit for all 25 points. It shows a shift of
  $138.8\gev$ with a root-mean-square deviation of $12.2\gev$.} 
  \label{Ml1t_Linear_TEP-MEP} \end{center}
\end{figure}

\begin{figure}[H]
  \begin{center}
 \includegraphics[height=\textheight/2 - 4cm]
{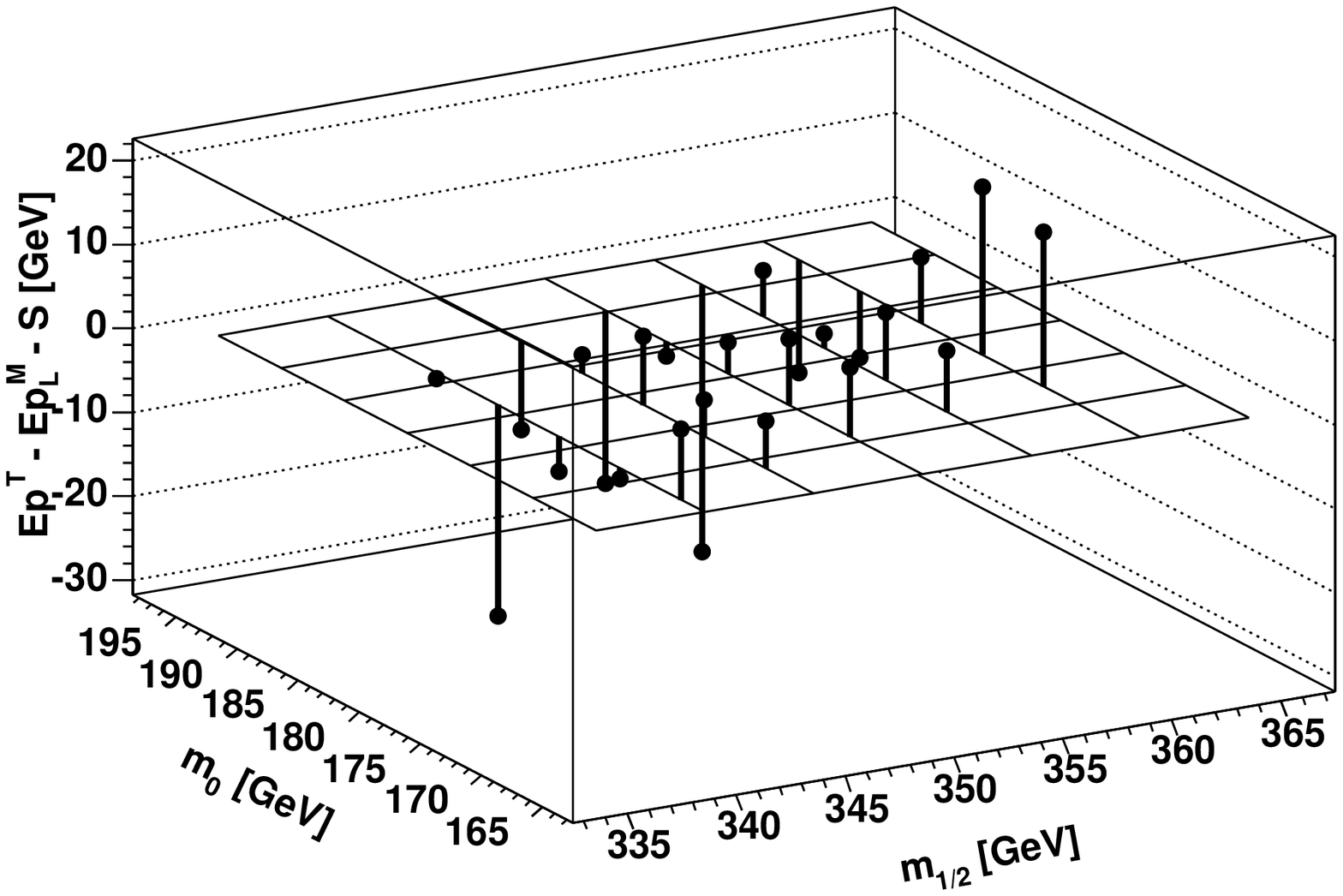} 
\caption{The theoretical $M_{\rho^{\pm}_{1}t}$ endpoint
  minus the measured endpoint in$\gev$ obtained with the linear fit
  after applying the constant correction of $138.8\gev$.} 
  \label{Ml1t_Linear_TEP-MEP-Cor} \end{center}
\end{figure}

The upper limit of the first distribution shown in
figure~\ref{Ml1t_TEP} varies from $555.0\gev$ at $m_{0} = 187$ and
$m_{1/2} = 340$ to $610.1\gev$ at $m_{0} = 173$ and $m_{1/2} =
360$. In order to achieve stable results the linear fit area has to
start at the bin with $60\%$ and ending at the bin with $5\%$ of the
maximum bin content. This change is necessary, because - due to the
much lower statistics - for some points in the parameter space
MINUIT~\cite{root} tries to fit a single bin. As can be seen in
figure~\ref{Ml1t_Ex_181_350} the linear fit actually measures the
endpoint given by the $\tilde{t}_1$. Therefore also the shift $S$ and
its uncertainty for measuring this endpoint is given in
table~\ref{sum:Ml1t}. Figure~\ref{Ml1t_Linear_TEP-MEP} and
figure~\ref{Ml1t_Linear_TEP-MEP-Cor}, however, show only the results
for the real kinematic endpoint which is determined by $\tilde{t}_2$.

\begin{table}[H]
\begin{center}
\begin{tabular}{|l|r|} \hline
\emph{Value} & Linear \\\hline\hline
$\delta_{stat} Ep^{M}$ & 3.2 (3.2) \\ $S$ & 138.8 (-5.7) \\
$\delta S$ & 12.2 (4.2)\\\hline
\end{tabular}
\caption{A summary of the calculated mean uncertainties
in$\gev$ for the $\rho^{\pm}_{1}$ endpoint in $M_{\rho^{\pm}t}$. The
values in brackets correspond to the case of $\tilde{t}_{1}$ being the
input for the kinematic limit.}\label{sum:Ml1t}
\end{center}
\end{table}

\begin{figure}[H]
  \begin{center} \includegraphics[height=\textheight/2 - 4cm]
  {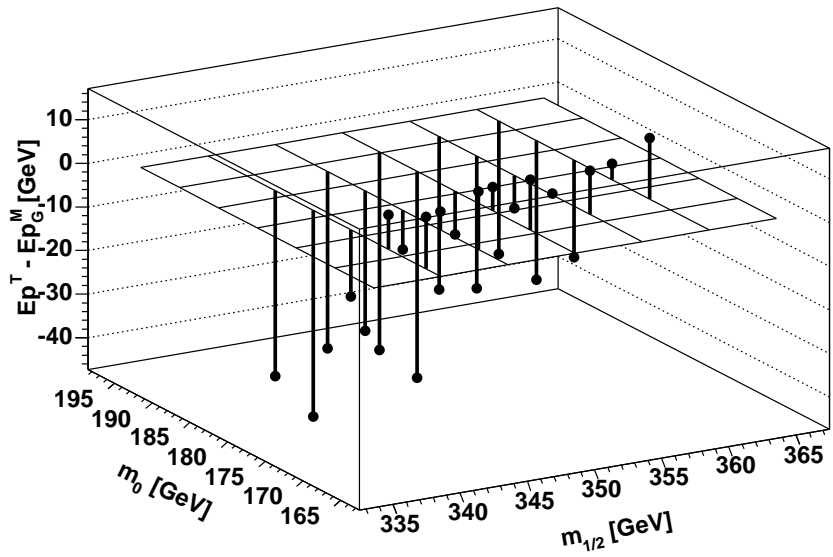} \caption{The difference of the
  theoretical and the measured $M_{\rho^{\pm}_{2}t}$ endpoint in$\gev$
  obtained with the gaussian fixed adapted origin method for all 25
  points. The value is systematically too low on average by
  $14.6\gev$.} \label{Ml2t_GaussFZ_TEP-MEP} \end{center}
\end{figure}
\begin{figure}[H]
  \begin{center} \includegraphics[height=\textheight/2 - 4cm]
  {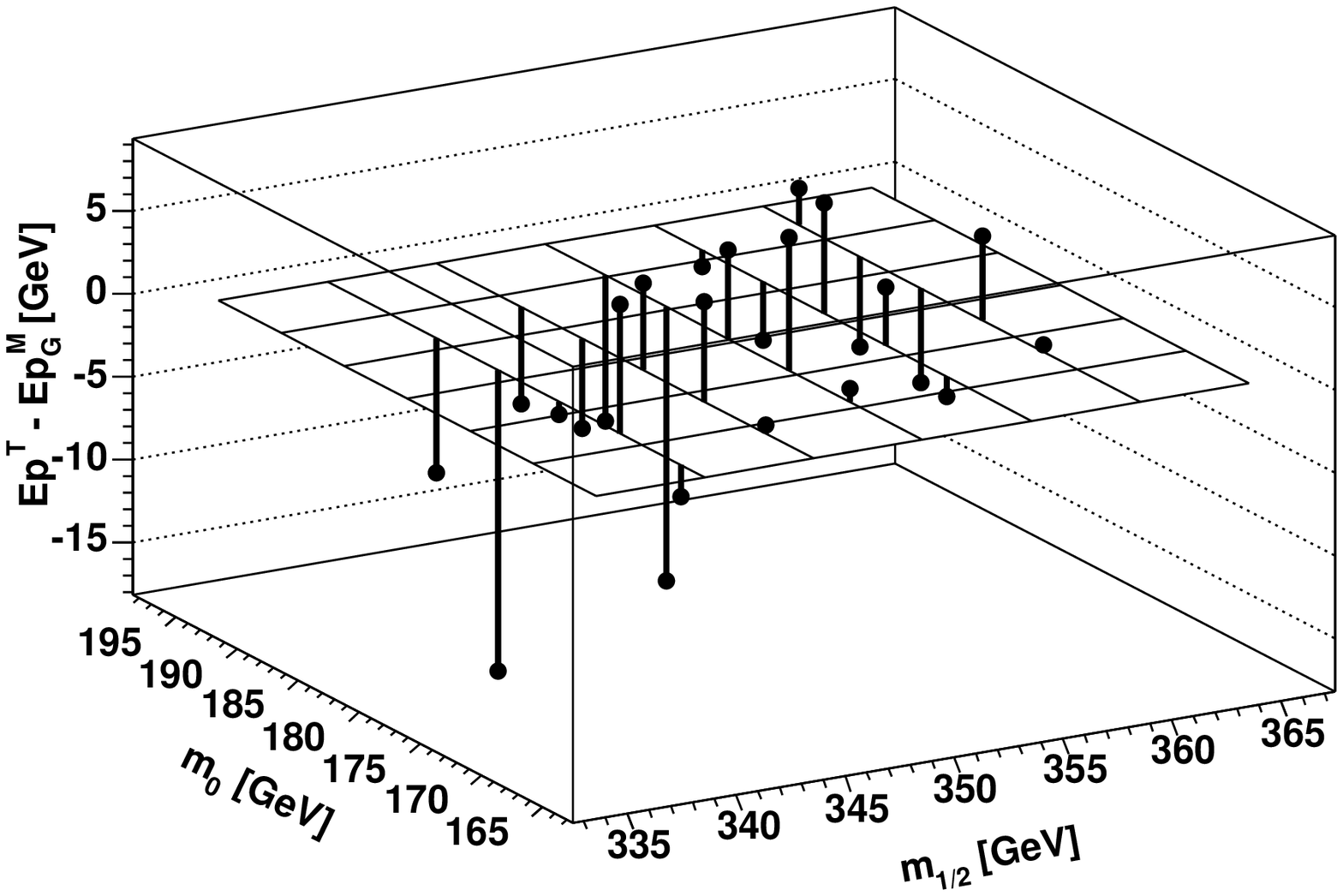} \caption{The difference of the
  theoretical and the measured $M_{\rho^{\pm}_{2}t}$ endpoint in$\gev$
  obtained with the gaussian non-fixed adapted origin method.} 
  \label{Ml2t_GaussNFZ_TEP-MEP} \end{center}
\end{figure}
%-------------

%----------------------------------------------------------------------
\subsection{\texorpdfstring{The kinematic limit of $M_{\rho_{1}^{\pm}q} +
M_{\rho_{2}^{\mp}q}$ for top quarks}{The kinematic limit of M(r1q) +
M(r2q) for top quarks}}\label{sec:Ml1t_Ml2t} 
\begin{figure}[H]
  \begin{center}
    \shadowbox{\includegraphics[width=\textwidth/2]{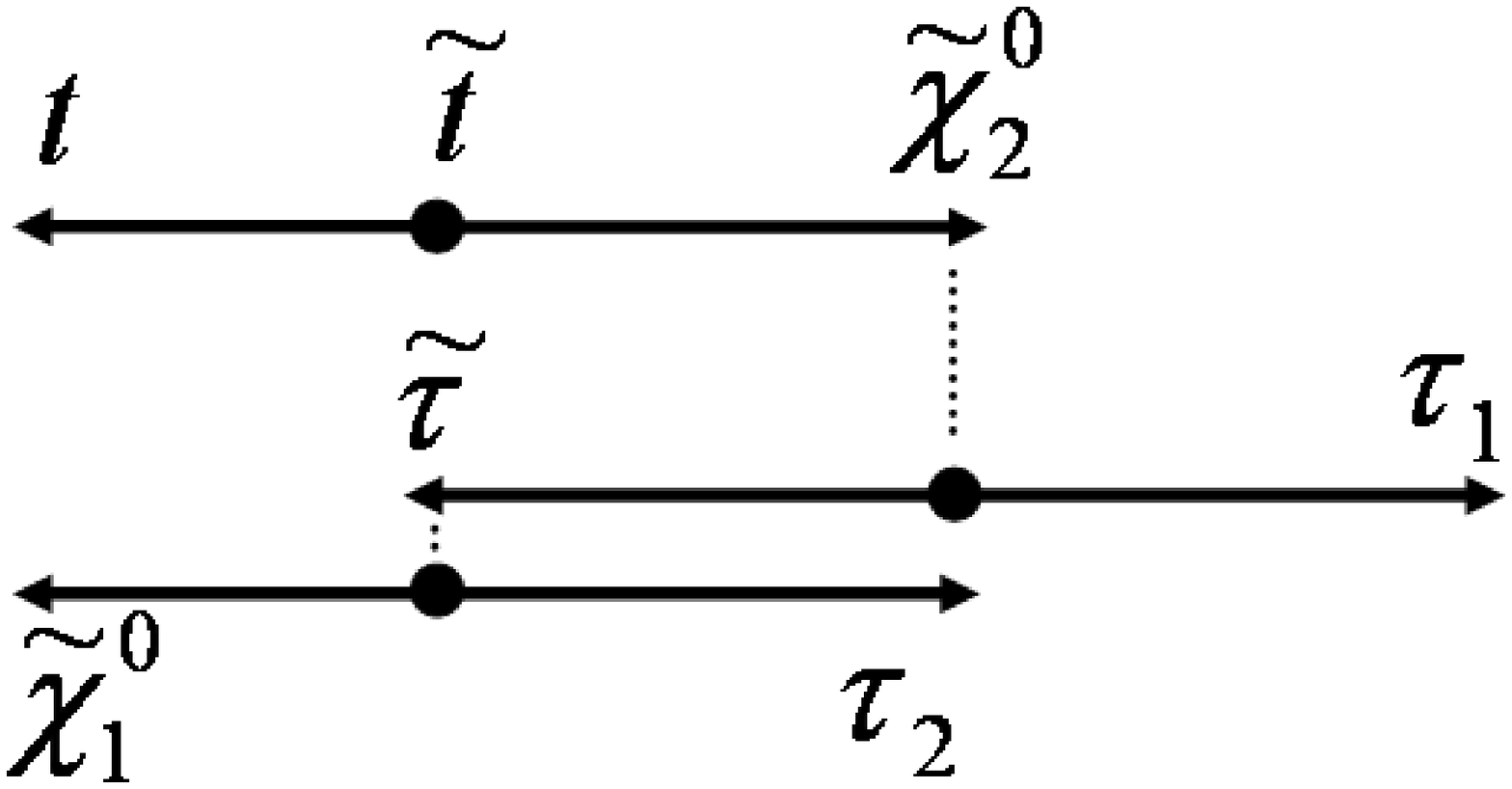}}
  \end{center}
\end{figure}
\begin{figure}[H]
  \begin{center} \includegraphics[height=\textheight/2 - 4cm]
  {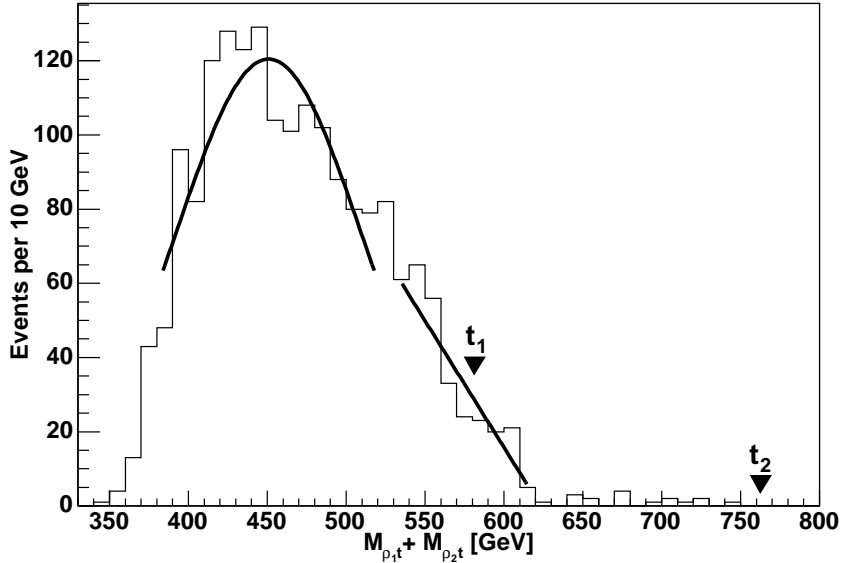} \caption{Example of a linear and a
  gaussian fit to $M_{\rho_{1}^{\pm}t} + M_{\rho_{2}^{\mp}t}$ at
  $m_{0} = 181$ and $m_{1/2} = 350$. In this distribution 1859 events
  are represented. The two triangles show the theoretical values for
  events with $\tilde{t}_{1}$ ($578.6\gev$) and $\tilde{t}_{2}$
  ($761.4\gev$), respectively.} \label{Ml1t_Ml2t_Ex_181_350}
  \end{center}
\end{figure}
\begin{figure}[H]
  \begin{center} \includegraphics[height=\textheight/2 - 4cm]
  {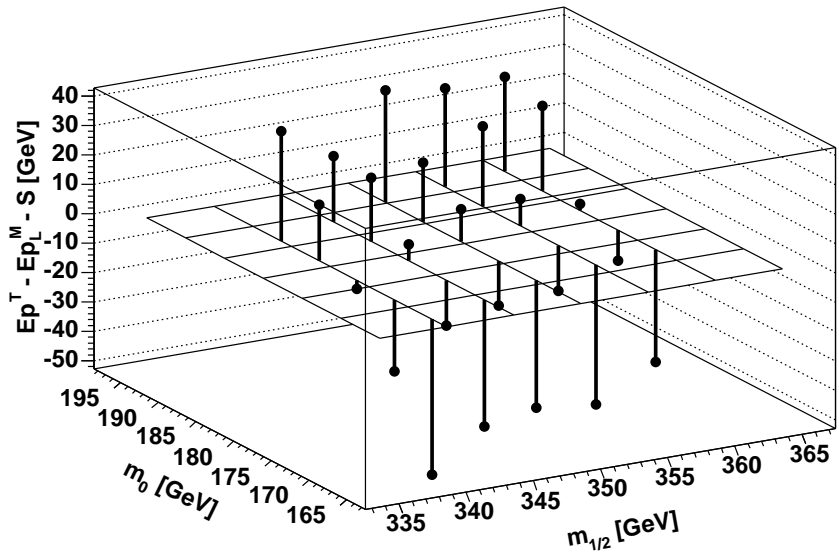} \caption{The theoretical
  endpoint minus the measured $M_{\rho_{1}^{\pm}t} +
  M_{\rho_{2}^{\mp}t}$ endpoint in$\gev$ obtained with the linear fit
  after applying the constant correction of $127.3\gev$. The values
  decrease significantly for decreasing values of $m_{0}$.} 
  \label{Ml1t_Ml2t_Linear_TEP-MEP-Cor} \end{center}
\end{figure}
\begin{figure}[H]
  \begin{center} \includegraphics[height=\textheight/2 - 4cm]
  {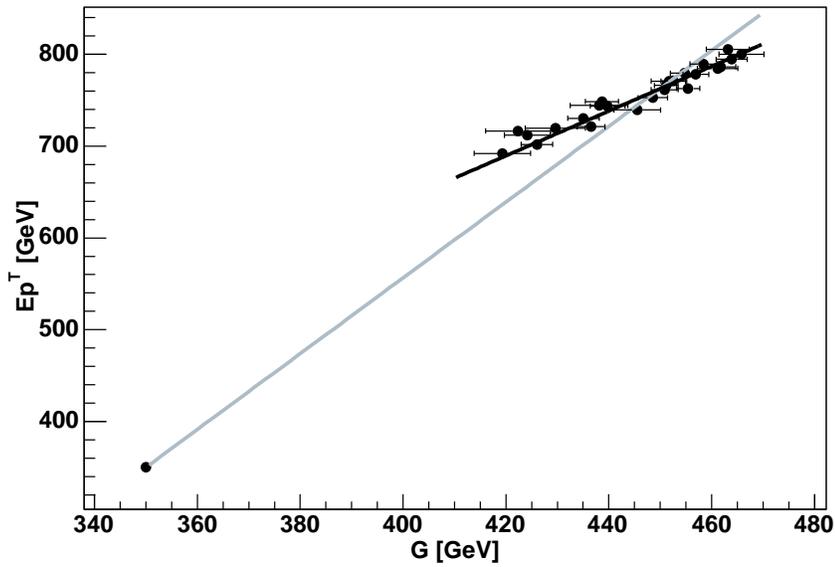} 
\caption{The theoretical kinematic endpoints as a function of the
measured gaussian maximum for all 25 investigated points in the
$m_{0}$-$m_{1/2}$ plane. The gray line with $C = 4.3$ is forced to go
through the adapted origin whereas the black line with $C = 2.4$ has an optimised
adapted ordinate value of $169.0\gev$ associated to the real
ordinate value of $-335.0\gev$.}
\label{Ml1t_Ml2t_Gauss_TE_vs_MM_Fit} \end{center}
\end{figure}

Both methods can give an adequate estimate on the kinematic endpoint
of the $M_{\rho_{1}^{\pm}t} + M_{\rho_{2}^{\mp}t}$ distribution. An
example of such a distribution is shown in
figure~\ref{Ml1t_Ml2t_Ex_181_350}. The data sample is the same as used
in section~\ref{ssec:llqtop}. The large correction $S = 127.3 \pm
28.5\gev$ for the linear fit is mainly caused by the stop mixing. A
significant $m_{0}$ dependence which should be included into the
correction can be seen in
figure~\ref{Ml1t_Ml2t_Linear_TEP-MEP-Cor}. Similar to the light and
bottom quark case the gaussian method can be applied, which can be
seen in figure~\ref{Ml1t_Ml2t_Gauss_TE_vs_MM_Fit}. However, the
gaussian function does not fit well in every case. In order to achieve
adequate results for the gaussian fit the symmetric area is chosen to
be $15\%$ of the maximum bin value. The origin in the $Ep^{T}$-$G$
plane has to be moved from (0,0) to (350,350) since two top quarks are
involved. This is hence again called the ``adapted'' origin. Thus, the
formula~(\ref{eq:gaussfit}) changes to
\begin{eqnarray}
Ep^{T}_{i} &=& C \cdot (G_{i} - 350) + D + 350.
\end{eqnarray}
The ratio between the slope of the fixed adapted origin fit and the
slope in the non-fixed adapted origin fit has a value of $1.7$. A
summary of the measured values is given in table~\ref{sum:Ml1t_Ml2t}
which show that the $\tilde{t}_2$ definitely has an important
influence on the endpoint measurement.
%---------------------------
\begin{table}[H]
\begin{center}
\begin{tabular}{|l|rrr|} \hline
\emph{Value} & Linear & Gaussian FO & Gaussian NFO\\\hline\hline
$\delta_{stat} Ep^{M}$ & 3.0 (3.0) & 14.9 & 8.8 \\
$S$ & 127.3 (-52.6) & 1.3 & 1.9 \\
$\delta S$ & 28.5 (18.2) & 21.6 & 8.1 \\\hline
\end{tabular}
\caption{A summary of the calculated mean uncertainties and systematic
shifts in$\gev$ for endpoint measurements in $M_{\rho_{1}^{\pm}t} +
M_{\rho_{2}^{\mp}t}$. FO is with fixed origin and NFO with non-fixed
origin. The values in brackets correspond to the case of
$\tilde{t}_{1}$ being the input for the kinematic
limit.}\label{sum:Ml1t_Ml2t}
\end{center}
\end{table}

\begin{figure}[H]
  \begin{center} \includegraphics[height=\textheight/2 - 4cm]
  {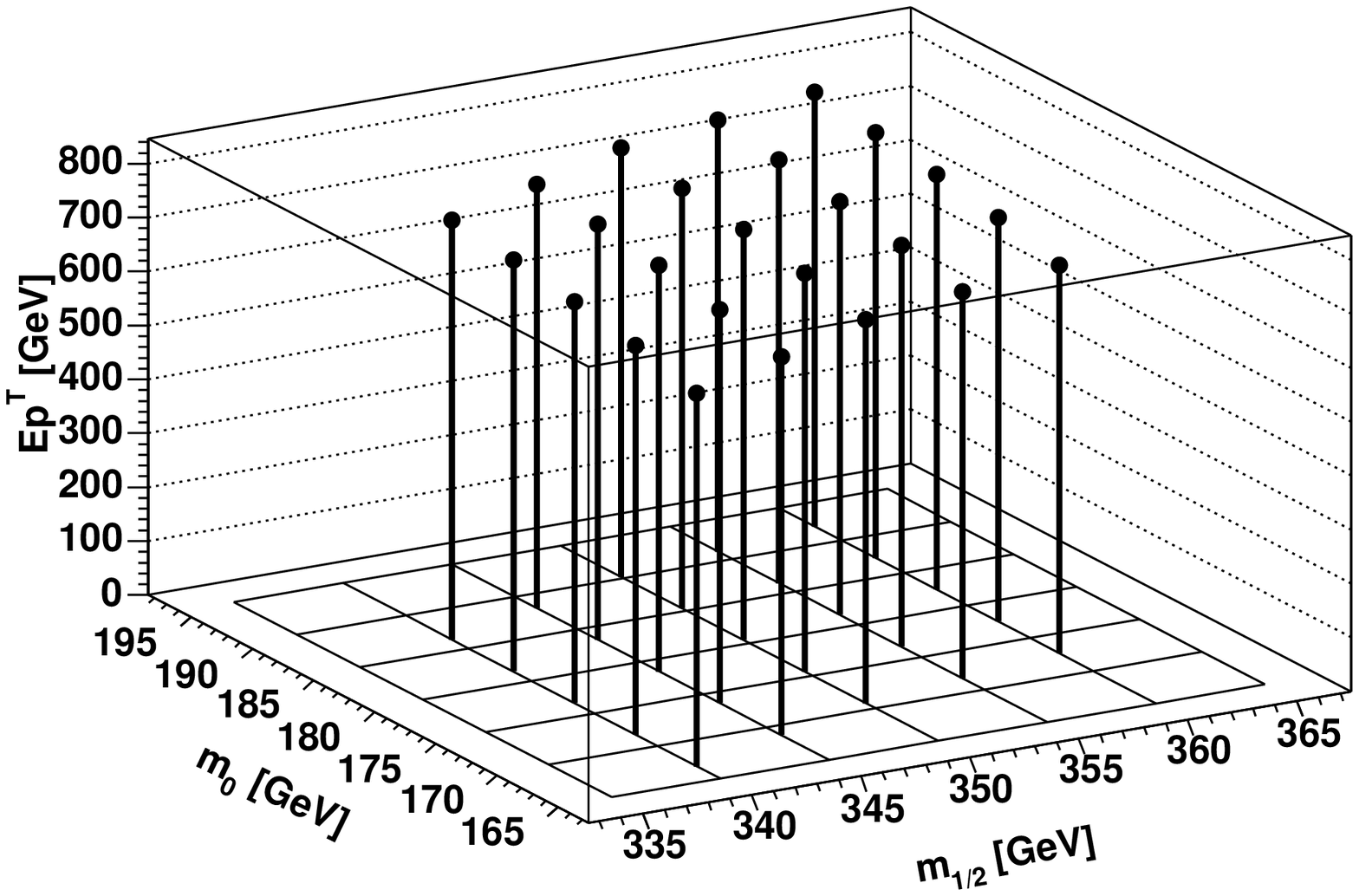} \caption{Theoretical kinematic endpoints for
  $M_{\rho_{1}^{\pm}t} + M_{\rho_{2}^{\mp}t}$ in$\gev$: The
  $\rho_{2}^{\mp}$ coming from the $\tilde{\tau}$ with the top quark
  and the $\rho_{1}^{\pm}$ coming from the $\chinonn$ and the same top
  quark from the $\tilde{t}_2 \ra t + \chinonn$ decay.}
  \label{Ml1t_Ml2t_TEP} \end{center}
\end{figure}
\begin{figure}[H]
  \begin{center} \includegraphics[height=\textheight/2 - 4cm]
  {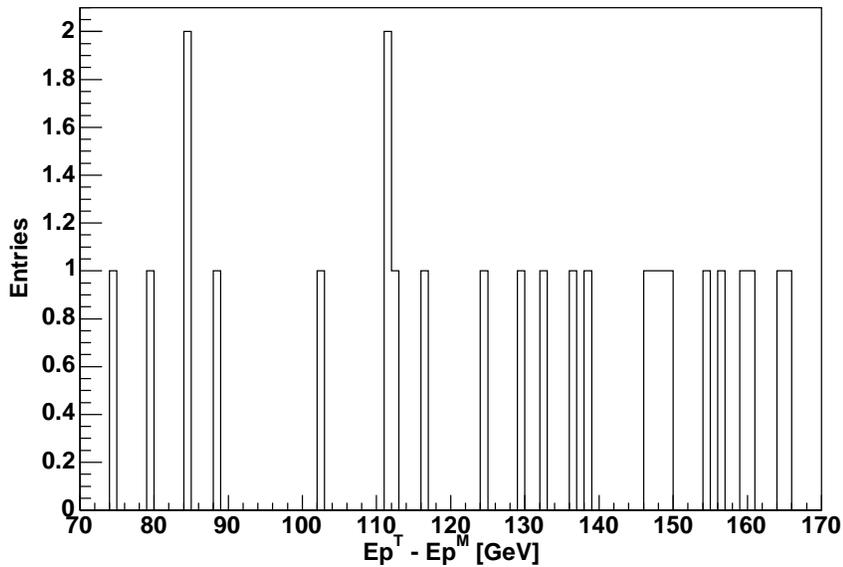} \caption{The difference of the
  theoretical and the measured $M_{\rho_{1}^{\pm}t} +
  M_{\rho_{2}^{\mp}t}$ endpoint in$\gev$ obtained with the linear
  fit. It shows a large mean shift of $127.3\gev$ with a
  root-mean-square deviation of $28.5\gev$.} 
  \label{Mltb_Ml2t_Linear_TEP-MEP} \end{center}
\end{figure}

\begin{figure}[H]
  \begin{center} \includegraphics[height=\textheight/2 - 4cm]
  {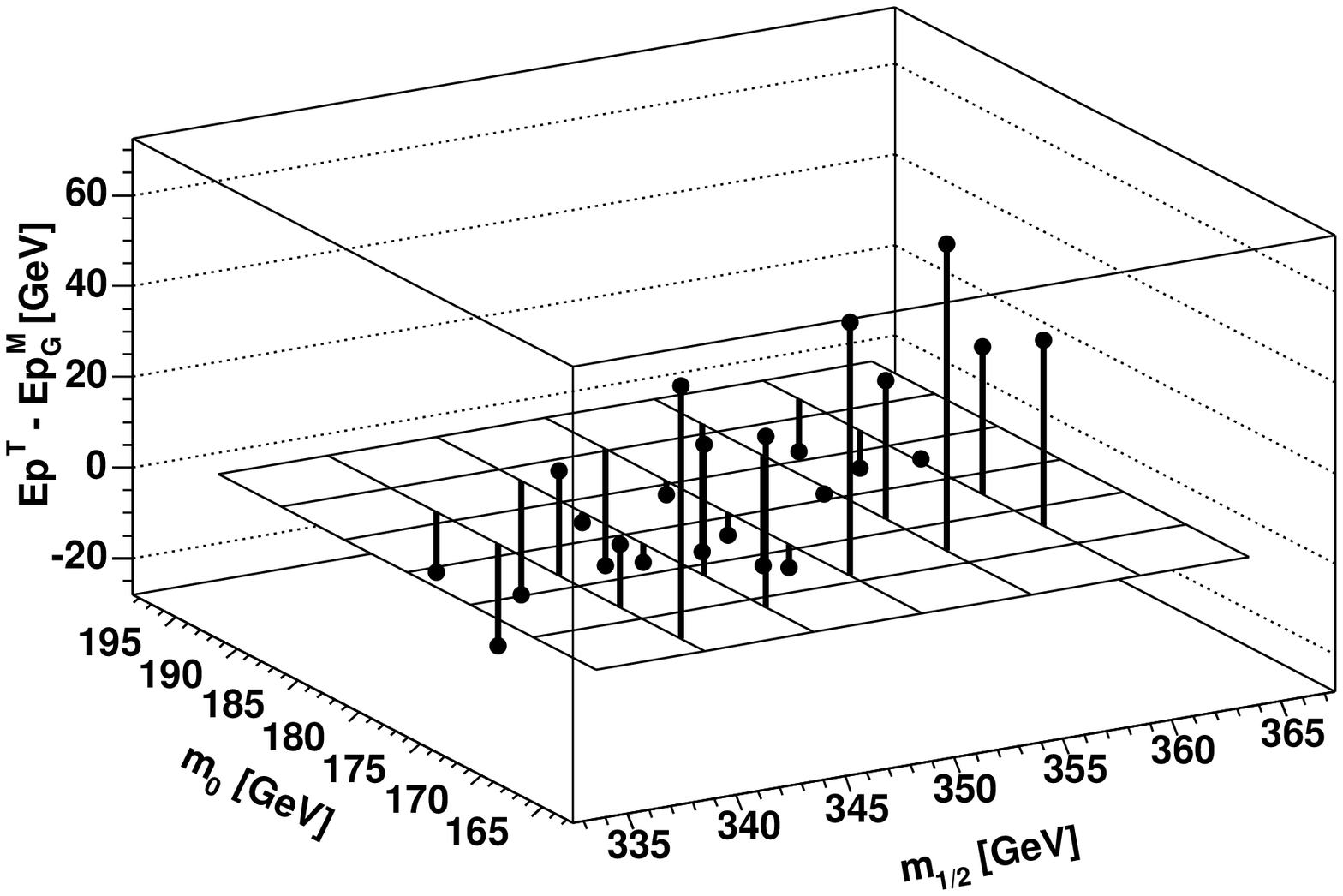} \caption{The difference of the
  theoretical and the measured $M_{\rho_{1}^{\pm}t} +
  M_{\rho_{2}^{\mp}t}$ endpoint in$\gev$ obtained with the gaussian
  fixed origin method.} \label{Ml1t_Ml2t_GaussFZ_TEP-MEP} \end{center}
\end{figure}
\begin{figure}[H]
  \begin{center} \includegraphics[height=\textheight/2 - 4cm]
  {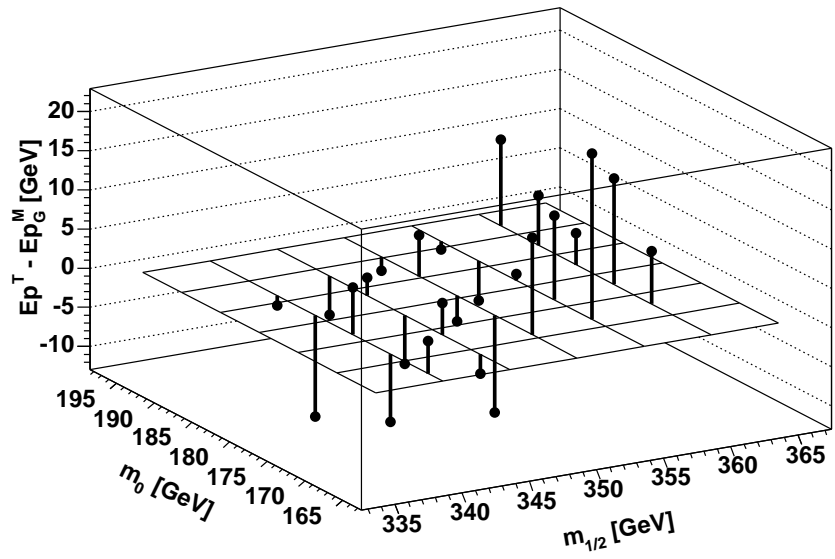} \caption{The difference of the
  theoretical and the measured $M_{\rho_{1}^{\pm}t} +
  M_{\rho_{2}^{\mp}t}$ endpoint in$\gev$ obtained with the gaussian
  non-fixed adapted origin method.}\label{Ml1t_Ml2t_GaussNFZ_TEP-MEP}
  \end{center}
\end{figure}
\pagebreak
%----------------------------------------------------------------------
\subsection{Examples of measurements with 30 $fb^{-1}$}\label{sec:30fb}
\begin{figure}[H]
\begin{minipage}[t]{7.5cm}
 \begin{center} \includegraphics[width=\textwidth]
 {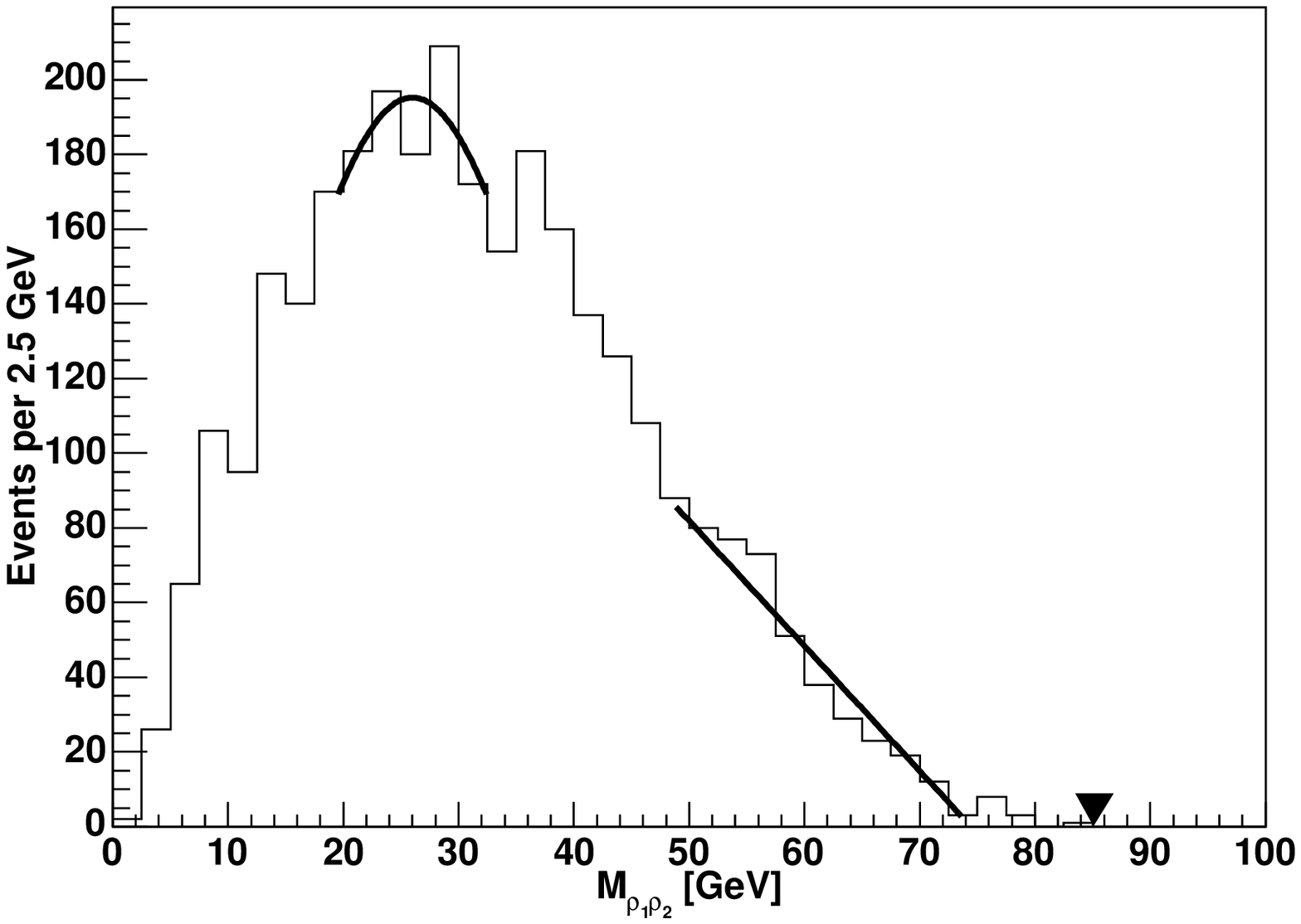} \caption{The invariant mass distribution
 of both opposite-sign rhos is based on 3062 events. The triangle
 shows the theoretical endpoint value at $85.8\gev$.} 
 \label{Mll_Ex_181_350_30fb}
\end{center}
\end{minipage}
\hfill
%\end{figure}
%---------------------
%\begin{figure}[H]
 \begin{minipage}[t]{7.5cm}
\begin{center} 
  \includegraphics[width=\textwidth] {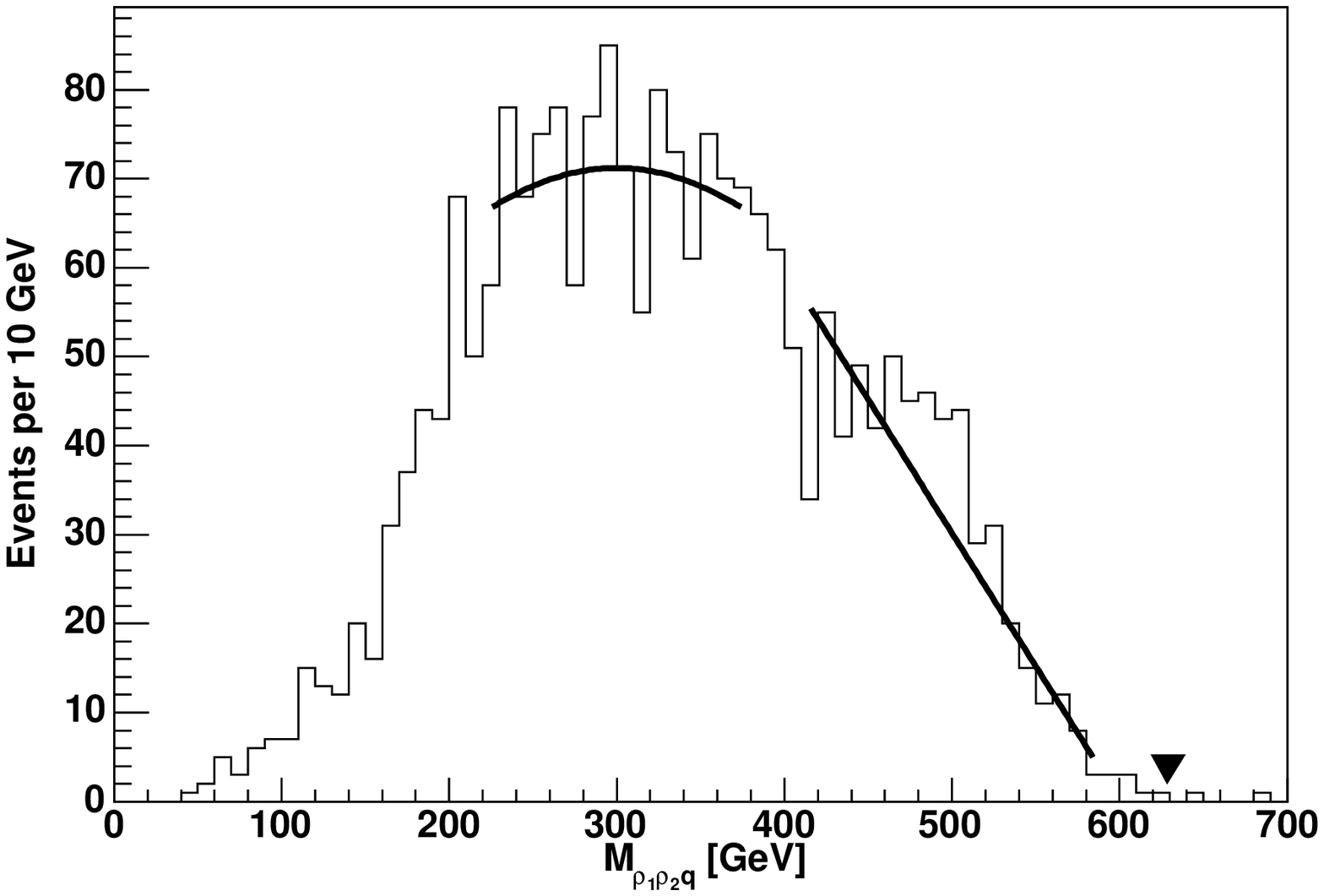}
  \caption{The invariant mass distribution of both rhos and the
  associated light quark is based on 2282 events. The triangle shows
  the theoretical kinematic endpoint of $M_{\rho^{\pm}\rho^{\mp}q}$ at
  $627.8\gev$.} \label{Mllq_Ex_181_350_30fb}
\end{center}
\end{minipage}
\hfill
\end{figure}

\begin{figure}[H]
\begin{minipage}[t]{7.5cm}
  \begin{center} 
\includegraphics[width=\textwidth]{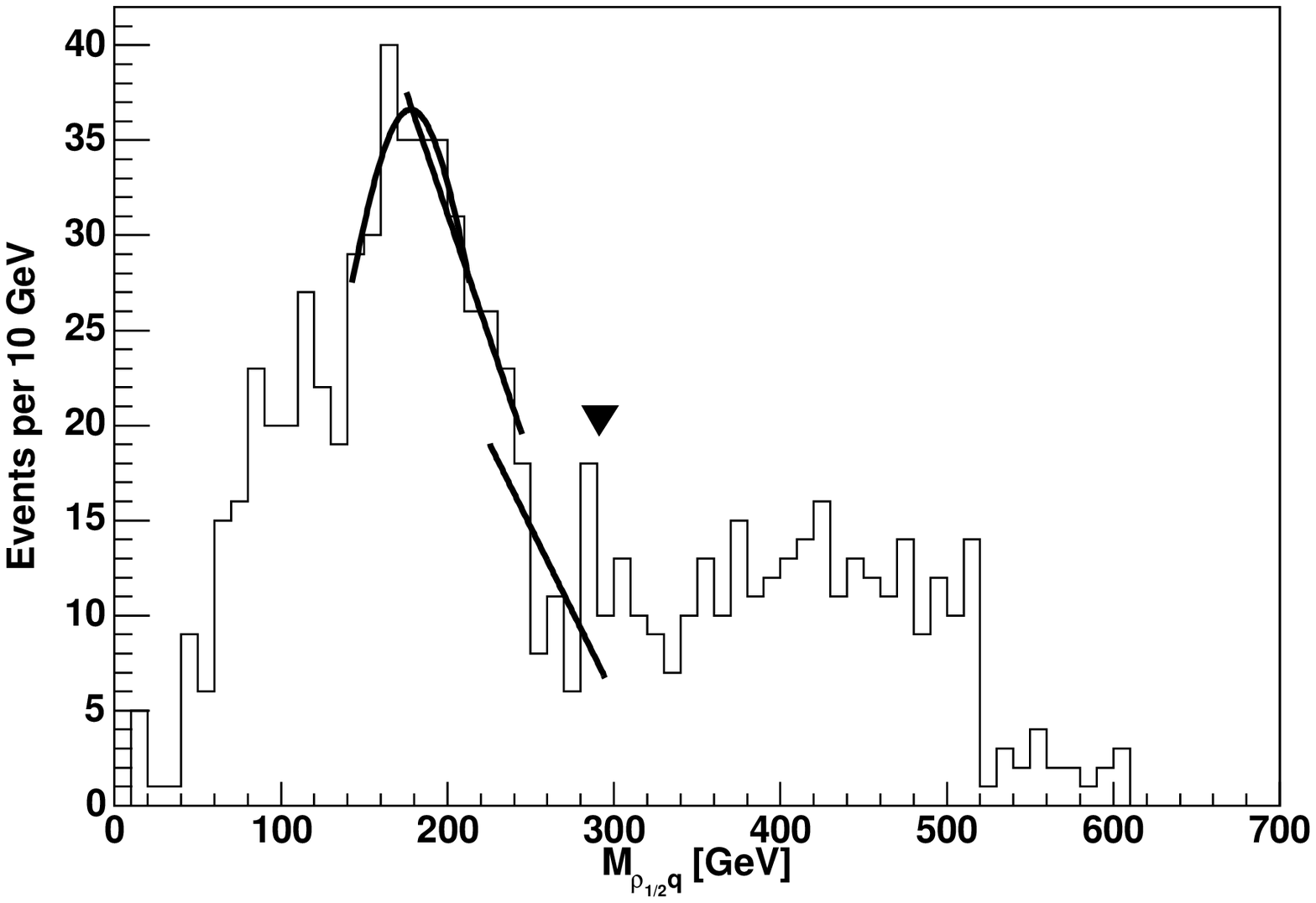} \caption{The invariant mass
distribution of the second rho and the associated light quark with the
first rho background is after the cut (55\%) based on 844 events. The
triangle shows the theoretical endpoint value at $288.0\gev$.}
\label{Ml2q_Ex_181_350_30fb}
\end{center}
\end{minipage}
\hfill
%\end{figure}
%\begin{figure}[H]
 \begin{minipage}[t]{7.5cm}
  \begin{center} 
\includegraphics[width=\textwidth]
  {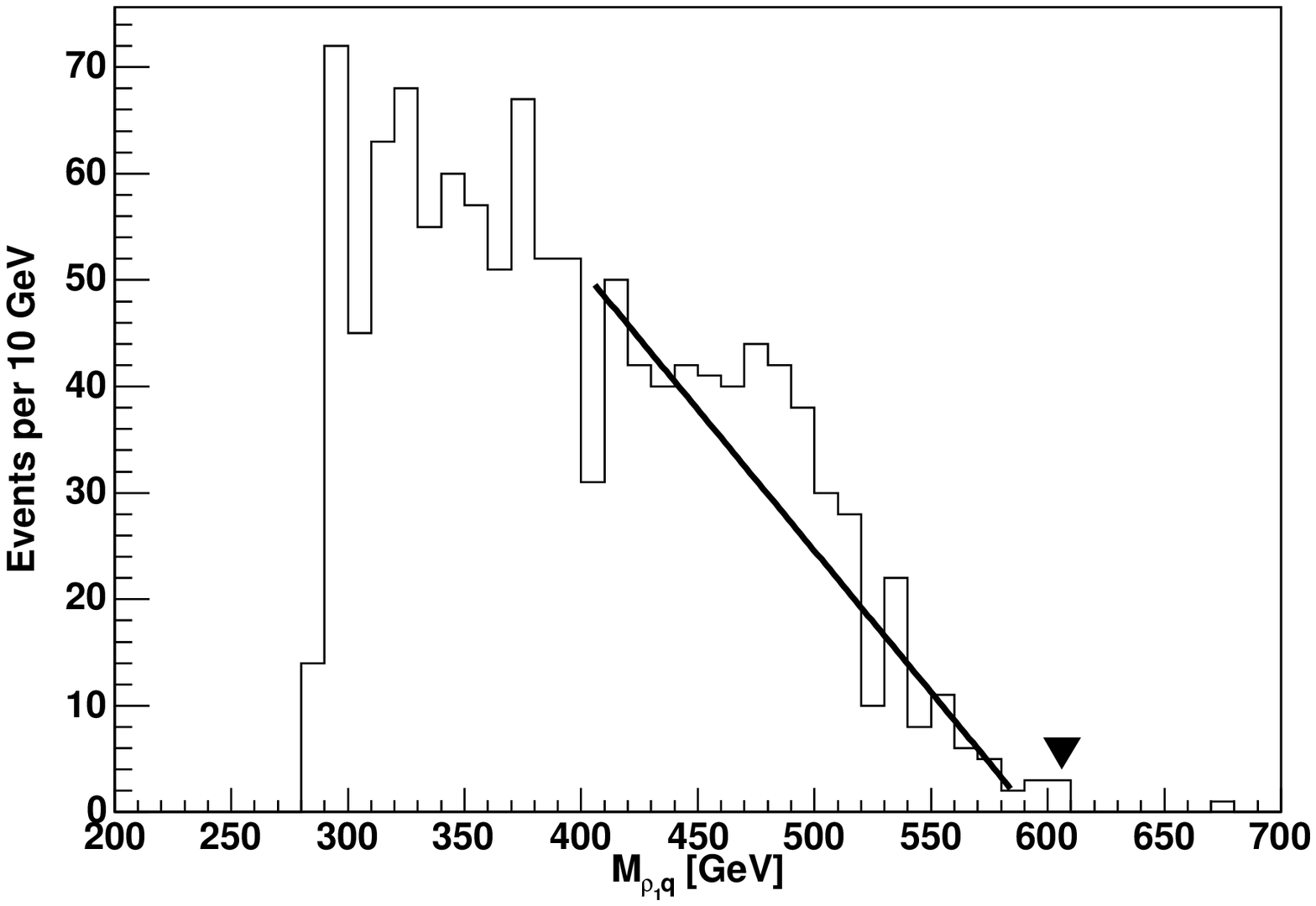} \caption{The invariant mass
  distribution of the first rho and the associated light quark is
  after the cut on the invariant mass $M_{\rho^{\pm}q}$ at the second
  endpoint based on 1199 events. The triangle shows the theoretical
  endpoint value at $606.0\gev$.} \label{Ml1q_Ex_181_350_30fb}
\end{center}
\end{minipage}
\hfill
\end{figure}

\begin{figure}[H]
\begin{minipage}[t]{7.5cm}
  \begin{center} \includegraphics[width=\textwidth]
  {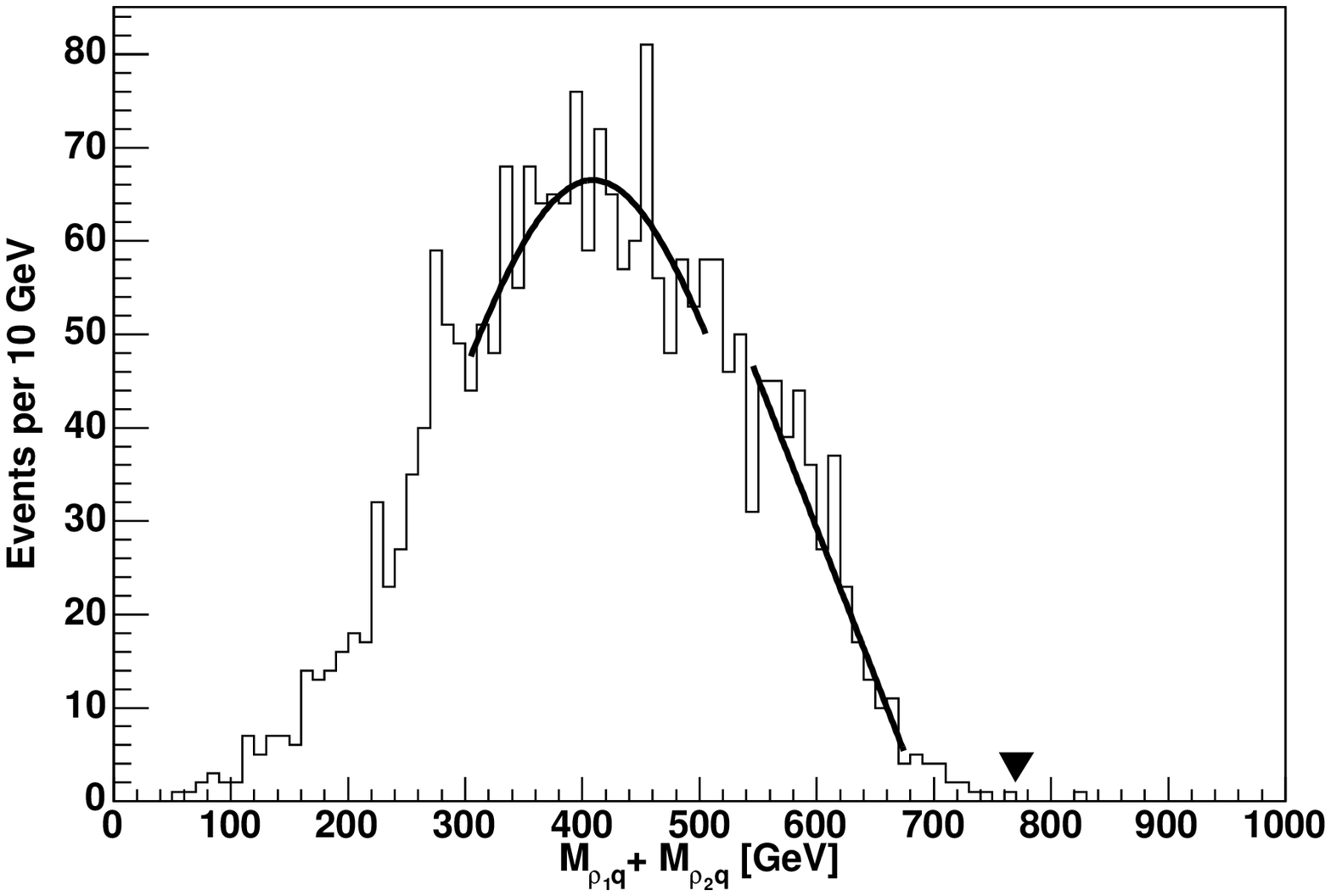} \caption{Example for the linear
  and the gaussian fit to the $M_{\rho_{1}^{\pm}q} +
  M_{\rho_{2}^{\mp}q}$ distribution at $m_{0} = 181$ and $m_{1/2} =
  350$. In this distribution 2282 events are represented. The triangle
  shows the theoretical kinematic endpoint at $768.8\gev$.}
  \label{Ml1q_Ml2q_Ex_181_350_30fb} \end{center}
\end{minipage}
\hfill
%\end{figure}
%---------------------
%\begin{figure}[H]
\begin{minipage}[t]{7.5cm}
  \begin{center} \includegraphics[width=\textwidth]
  {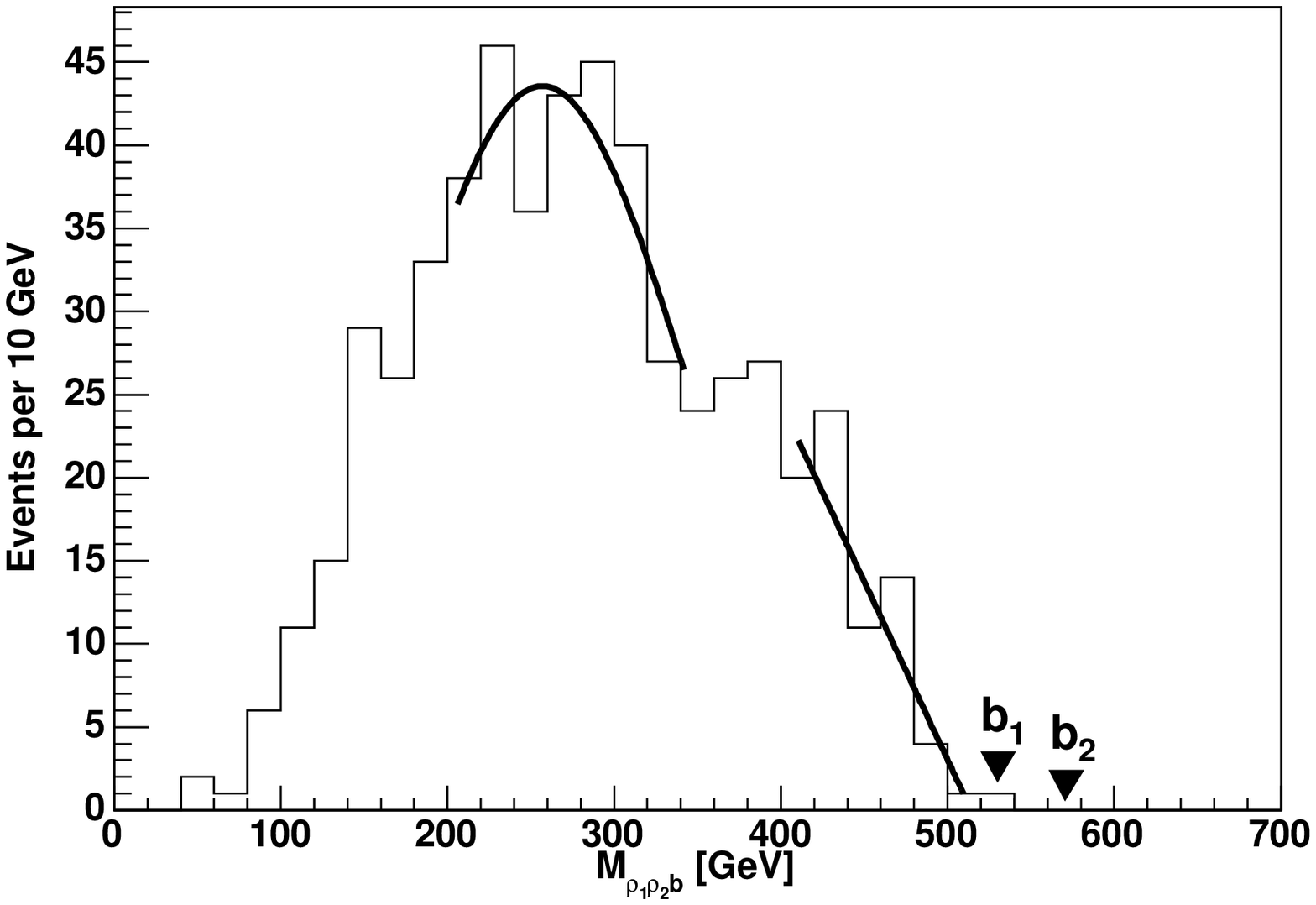} \caption{The invariant mass
  distribution of both rhos and the associated bottom quark is based
  on 550 events. The two triangles show the theoretical endpoint
  values for events with $\tilde{b}_{1}$ ($532.9\gev$) and
  $\tilde{b}_{2}$ ($576.3\gev$), respectively.} 
  \label{Mllb_Ex_181_350_30fb} \end{center}
\end{minipage}
\hfill
\end{figure}

\begin{figure}[H]
\begin{minipage}[t]{7.5cm}
  \begin{center} \includegraphics[width=\textwidth]
  {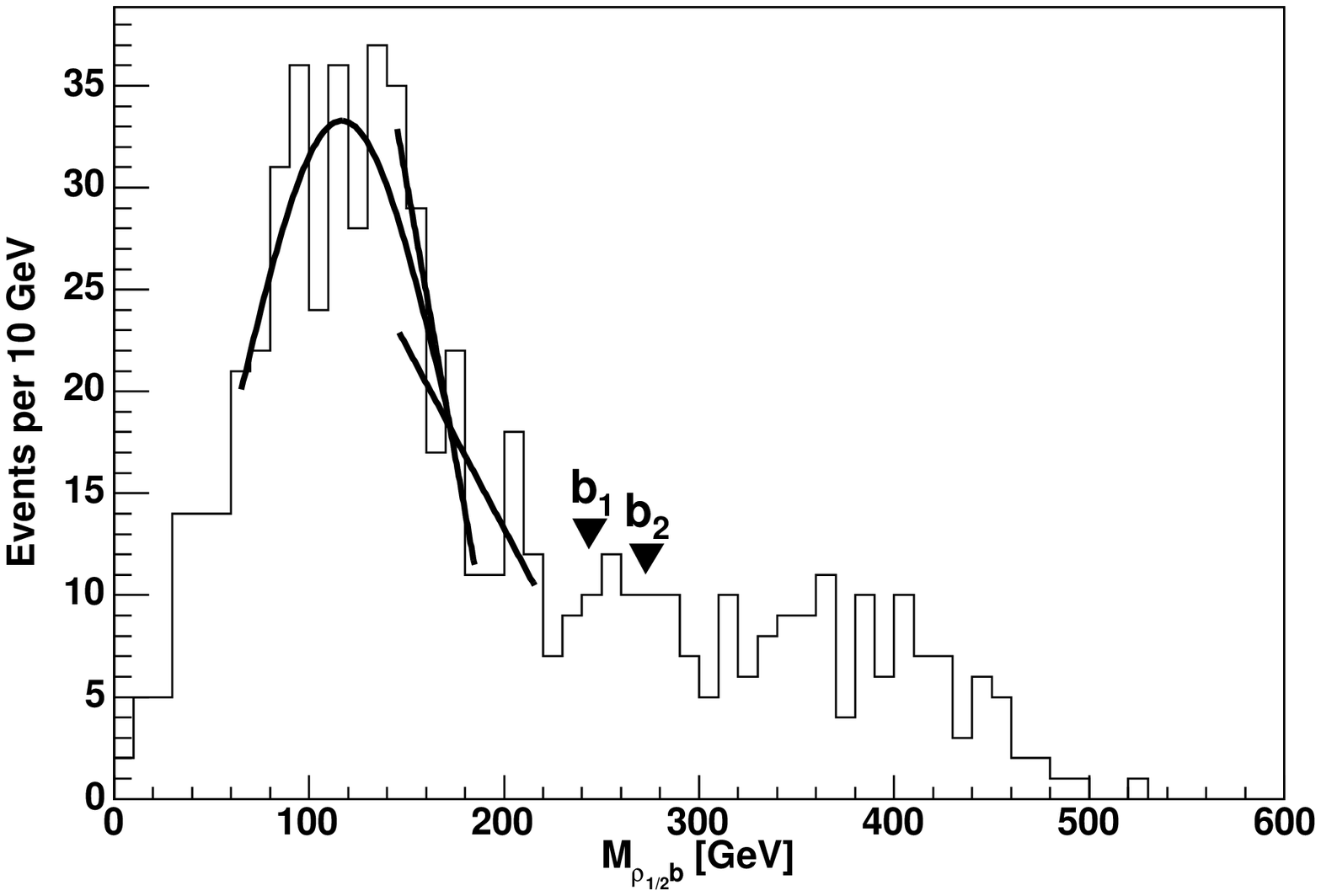} \caption{The invariant mass
  distribution of the second rho and the associated bottom quark with
  the first rho background is after the cut ($30\%$) based on 642
  events. The two triangles show the theoretical endpoint values for
  events with $\tilde{b}_{1}$ ($244.7\gev$) and $\tilde{b}_{2}$
  ($264.5\gev$), respectively.} \label{Ml2b_Ex_181_350_30fb}
  \end{center}
\end{minipage}
\hfill
%\end{figure}
%\begin{figure}[H]
\begin{minipage}[t]{7.5cm}
  \begin{center} \includegraphics[width=\textwidth]
  {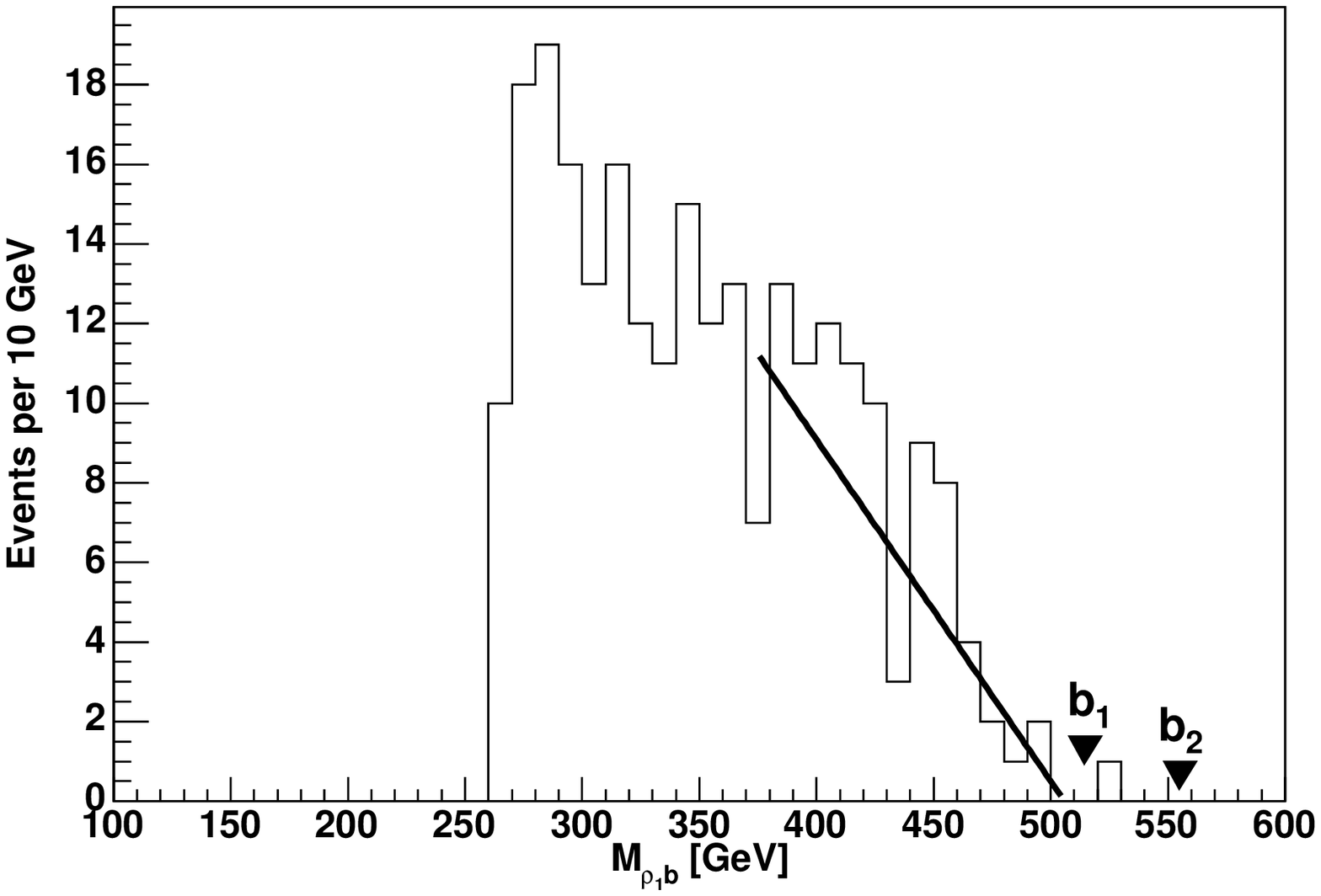} \caption{The invariant mass
  distribution of the first rho and the associated bottom quark is
  after the cut on the invariant mass $M_{\rho^{\pm}b}$ at the second
  endpoint based on 249 events. The two triangles show the theoretical
  endpoint values for events with $\tilde{b}_{1}$ ($514.7\gev$) and
  $\tilde{b}_{2}$ ($556.4\gev$), respectively.} 
  \label{Ml1b_Ex_181_350_30fb} \end{center}
\end{minipage}
\hfill
\end{figure}

\begin{figure}[H]
\begin{minipage}[t]{7.5cm}
  \begin{center} \includegraphics[width=\textwidth]
  {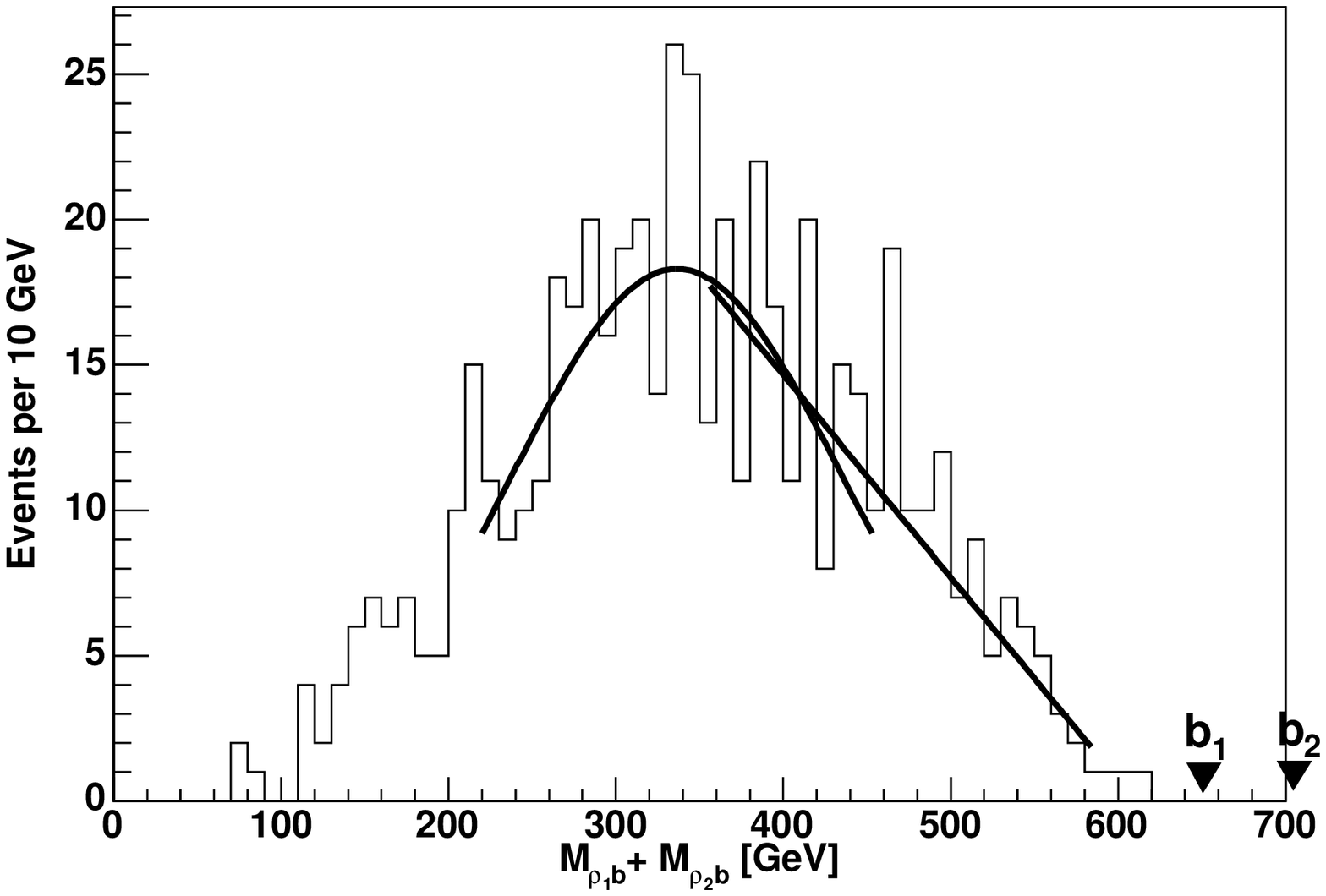} \caption{Example of a linear and
  a gaussian fit to the $M_{\rho_{1}^{\pm}b} + M_{\rho_{2}^{\mp}b}$
  distribution at $m_{0} = 181$ and $m_{1/2} = 350$. In this
  distribution 550 events are represented. The two triangles show the
  theoretical values for events with $\tilde{b}_{1}$ ($653.1\gev$) and
  $\tilde{b}_{2}$ ($706.1\gev$), respectively.}
  \label{Ml1b_Ml2b_Ex_181_350_30fb} \end{center}
\end{minipage}
\hfill
%\end{figure}
%---------------------
%\begin{figure}[H]
\begin{minipage}[t]{7.5cm}
  \begin{center} \includegraphics[width=\textwidth]
  {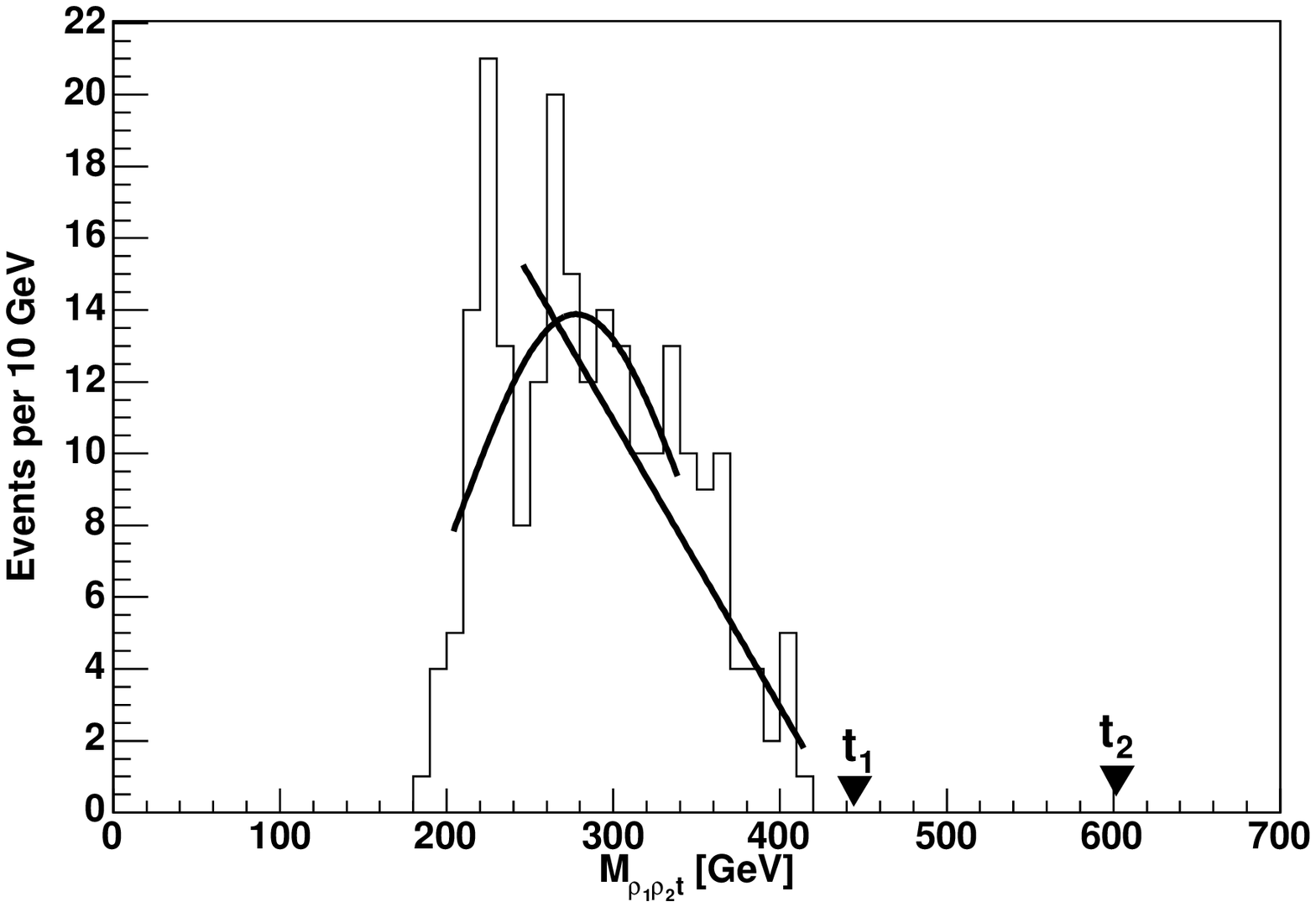} \caption{The invariant mass
  distribution of both rhos and the associated top quark is based on
  230 events. The two triangles show the theoretical values for events
  with $\tilde{t}_{1}$ ($446.1\gev$) and $\tilde{t}_{2}$
  ($601.2\gev$), respectively.} \label{Mllt_Ex_181_350_30fb}
  \end{center}
\end{minipage}
\hfill
\end{figure}

\begin{figure}[H]
\begin{minipage}[t]{7.5cm}
  \begin{center} \includegraphics[width=\textwidth]
  {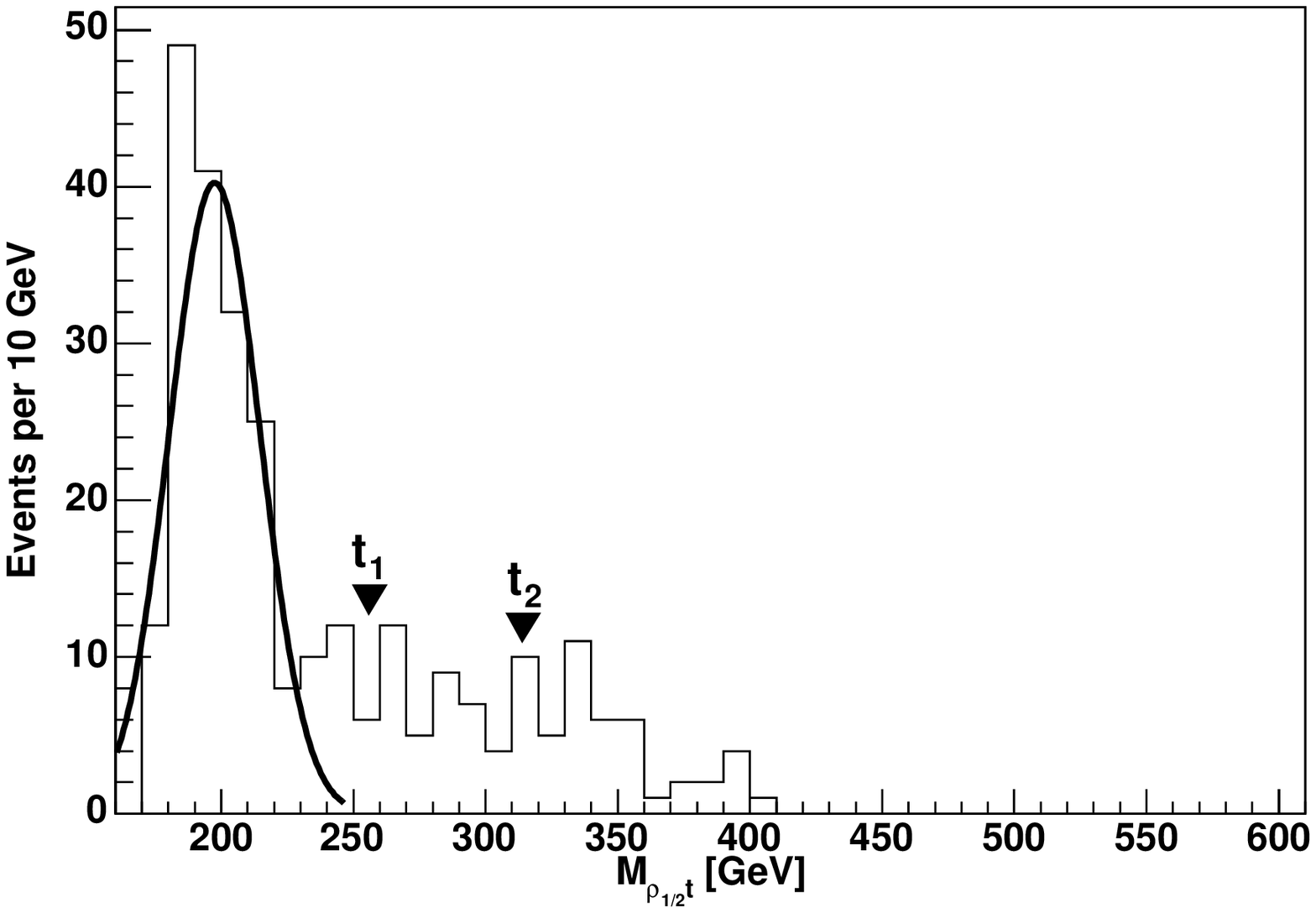} \caption{The invariant mass
  distribution of the second rho and the associated top quark with the
  first rho background is after the cut ($30\%$) based on 280
  events. The two triangles show the theoretical values for events
  with $\tilde{t}_{1}$ ($257.1\gev$) and $\tilde{t}_{2}$
  ($316.7\gev$), respectively.} \label{Ml2t_Ex_181_350_30fb}
  \end{center}
\end{minipage}
\hfill
%\end{figure}
%\begin{figure}[H]
\begin{minipage}[t]{7.5cm}
  \begin{center} \includegraphics[width=\textwidth]
  {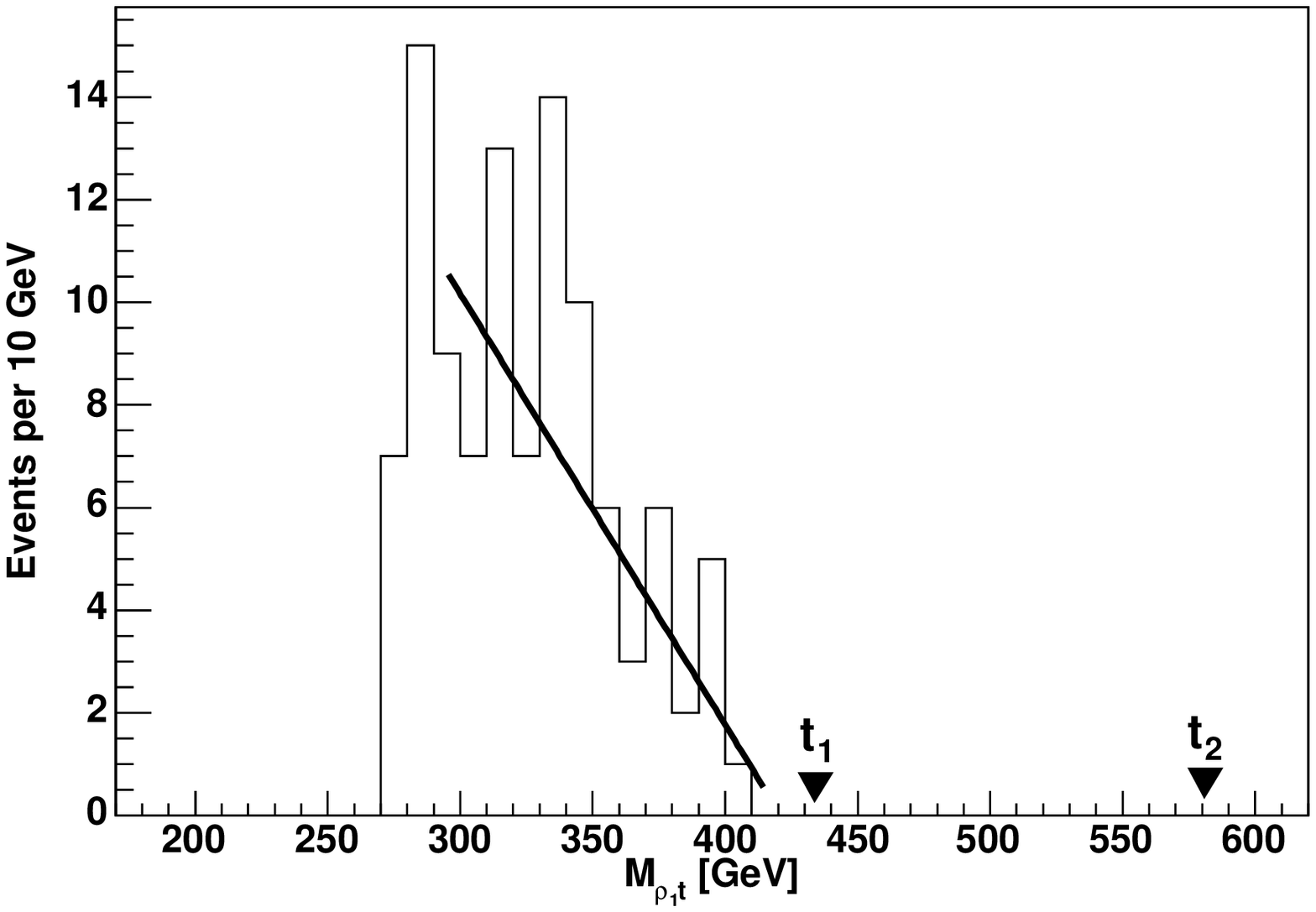} \caption{The invariant mass
  distribution of the first rho and the associated top quark is after
  the cut on $M_{\rho^{\pm}t}$ at the theoretical second endpoint
  value based on 112 events. The two triangles show the theoretical
  values for events with $\tilde{t}_{1}$ ($429.2\gev$) and
  $\tilde{t}_{2}$ ($582.3\gev$), respectively.} 
  \label{Ml1t_Ex_181_350_30fb} \end{center}
\end{minipage}
\hfill
\end{figure}

\begin{figure}[H]
\begin{minipage}[t]{7.2cm}
  \begin{center} \includegraphics[width=\textwidth]
  {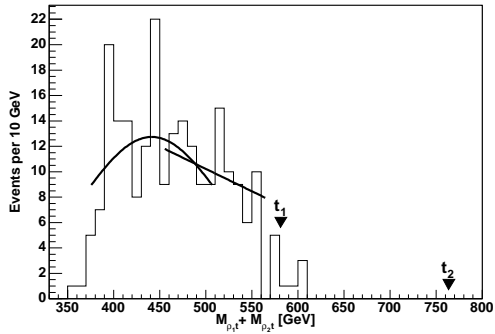} \caption{Example of a linear and
  a gaussian fit to $M_{\rho_{1}^{\pm}t} + M_{\rho_{2}^{\mp}t}$ at
  $m_{0} = 181$ and $m_{1/2} = 350$. In this distribution 230 events
  are represented. The two triangles show the theoretical values for
  events with $\tilde{t}_{1}$ ($578.6\gev$) and $\tilde{t}_{2}$
  ($761.4\gev$), respectively. Especially this figure shows the
  necessity to further optimise the fitting methods for low statistic
  samples.}  \label{Ml1t_Ml2t_Ex_181_350_30fb} \end{center}
\end{minipage}
\hfill
\end{figure}

In the following a Monte Carlo sample at I' is used which corresponds
to an integrated luminosity of 30 $fb^{-1}$ if all efficiencies and
acceptances were $100\%$. The aim is to test the methods introduced in
the previous sections for a low statistics sample. Table~\ref{30fb}
shows the results of the endpoint measurements and
figure~\ref{Mll_Ex_181_350_30fb} to \ref{Ml1t_Ml2t_Ex_181_350_30fb}
show all associated distributions.

\begin{table}[H]
\begin{center}
\begin{tabular}{|l|ccc|r|} \hline
\emph{Distribution} & Linear & Gaussian FO & Gaussian NFO & $Ep^{T}$\\\hline\hline
$M_{\rho\rho}$ & 
$83.5^{\pm 0.5}_{\pm 1.8}$ & 
$81.0 ^{\pm 4.3}_{\pm 2.3}$ & 
$81.2^{\pm 4.7}_{\pm 1.9}$ & 
$85.8$\\\hline	
$M_{\rho\rho{q}}$ & 
$614.8^{\pm 5.0}_{\pm 7.8}$ & 
$n.a$ & 
$n.a.$ & 
$627.8 $\\
$M_{\rho_{2}q}$ & 
$320.0^{\pm 83.7}_{\pm 8.0}$ & 
$300.2^{\pm 8.6}_{\pm 9.9}$ & 
$296.7^{\pm 7.7}_{\pm 2.9}$ & 
$288.0$\\
$M_{\rho_{1}q}$ & 
$594.3^{\pm 4.2}_{\pm 2.4}$ & 
$n.a.$ & 
$n.a.$ & 
$606.0 $\\
$\sum M_{\rho_{i}q}$ & 
$742.7^{\pm 5.7}_{\pm 12.8}$ & 
$770.6^{\pm 15.3}_{\pm 13.19}$ & 
$774.1^{\pm 10.3}_{\pm 5.61}$ & 
$768.8$\\\hline
$M_{\rho\rho{b}}$ & 
$560.4^{\pm 11.7}_{\pm 8.5}$ & 
$n.a.$ & 
$n.a.$ & 
$576.3$\\
$M_{\rho_{2}b}$ & 
$219.2^{\pm 25.4}_{\pm 10.3}$ & 
$265.2^{\pm 12.0}_{\pm 14.1}$ & 
$262.4^{\pm 10.6}_{\pm 6.0}$ & 
$264.5$\\
$M_{\rho_{1}b}$ & 
$542.3^{\pm 10.4}_{\pm 3.9}$ & 
$n.a.$ & 
$n.a$ & 
$556.4 $\\
$\sum M_{\rho_{i}b}$ & 
$700.0^{\pm 11.7}_{\pm 15.1}$ & 
$679.0^{\pm 17.9}_{\pm 21.3}$ & 
$686.3^{\pm 12.0}_{\pm 6.42}$ & 
$706.1$\\\hline
$M_{\rho\rho{t}}$ & 
$592.2^{\pm 11.3}_{\pm 5.7}$ & 
$n.a.$ & 
$n.a$ & 
$601.2$\\
$M_{\rho_{2}t}$ & 
$n.a.$ & 
$285.9^{\pm 8.4}_{\pm 21.6}$ & 
$290.2^{\pm 6.8}_{\pm 6.8}$ & 
$316.7$\\
$M_{\rho_{1}t}$ & 
$559.9^{\pm 11.9}_{\pm 12.2}$ & 
$n.a.$ & 
$n.a.$ & 
$582.3$\\
$\sum M_{\rho_{i}t}$ & 
$917.5^{\pm 236.3}_{\pm 28.5}$ & 
$744.7^{\pm 61.2}_{\pm 21.6}$ & 
$743.6^{\pm 34.6}_{\pm 8.1}$ & 
$761.4$\\\hline\hline
\end{tabular}
\caption{The measured values $Ep^{M}$ ${}^{\pm stat}_{\pm syst}$ for 30
$fb^{-1}$. If endpoints are not available they are denoted with
``n.a.''. All values are in$\gev$.}\label{30fb}
\end{center}
\end{table}
With the 13 measured endpoints it is now possible to reconstruct the
masses of the involved sparticles. For the mass reconstruction the
minimisation package MINUIT within the ROOT framework~\cite{root} is
chosen. The input to the minimisation are the measured endpoint values
with their associated formulae from
section~\ref{sec.seqmassless.summary} and \ref{sec:top} and the
measurement uncertainties $\delta = \sqrt{stat^{2} + syst^{2}}$.

First, the measurements of the dilepton mass and the masses with the
light quark with the smallest $\delta$ have been chosen. Thus, five
endpoints are used to measure four masses. The result (with the masses
calculated by ISASUGRA in brackets) is:
\begin{eqnarray}
M_{\tilde{d}/\tilde{s}_{1}} &=& 787.5 \pm 35.5\gev\ (782.6\gev)\label{30fb5EP}\\\nonumber
M_{\chinonn} &=& 261.6 \pm 25.5\gev\ (265.5\gev)\\\nonumber
M_{\tilde{\tau}_{1}} &=& 157.4 \pm 26.5\gev\ (150.1\gev)\\\nonumber
M_{\chinon} &=& 144.3\pm 25.7\gev\ (138.1\gev)
\end{eqnarray}
The uncertainties given in \ref{30fb5EP} are noticeably larger if only
the first four endpoints are used.  There are two reasons for these,
nevertheless, large uncertainties. First, the endpoints cannot be
measured as precisely as it is possible for low $\tan \beta$ models
due to the presence of neutrinos in the decay chains. Second, the
equations for the endpoints are strongly correlated and therefore do
not contain sufficiently complementary information which could
compensate the endpoint uncertainties. By adding the endpoint
measurements which involve the bottom quark the uncertainties can be
improved and the sbottom mass can be measured. Here nine measured
values are taken to determine five unknown masses.
\begin{eqnarray}
M_{\tilde{d}/\tilde{s}_{1}} &=& 778.4 \pm 28.2\gev\ (782.6\gev)\\\nonumber
M_{\tilde{b}_{2}} &=& 710.8 \pm 25.7\gev\ (724.9\gev)\\\nonumber
M_{\chinonn} &=& 256.9 \pm 20.4\gev\ (265.5\gev)\\\nonumber
M_{\tilde{\tau}_{1}} &=& 150.9 \pm 21.2\gev\ (150.1\gev)\\\nonumber
M_{\chinon} &=& 138.2 \pm  20.5\gev\ (138.1\gev)
\end{eqnarray}
With adding the endpoint measurements which involve the top quark the
uncertainties on the reconstructed masses slightly increase due to the
large endpoint uncertainties. However, the stop mass can be
estimated. Here 13 measured values are taken for the determination of
six unknown masses.
\begin{eqnarray}
M_{\tilde{d}/\tilde{s}_{1}} &=& 795.2 \pm 29.9\gev\
(782.6\gev)\\\nonumber
M_{\tilde{t}_{2}} &=& 743.1 \pm 28.8\gev\ (747.9\gev)\\\nonumber
M_{\tilde{b}_{2}} &=& 727.2 \pm 27.2\gev\ (724.9\gev)\\\nonumber
M_{\chinonn} &=& 272.4 \pm 21.7\gev\ (265.5\gev)\\\nonumber
M_{\tilde{\tau}_{1}} &=& 164.5 \pm 22.6\gev\ (150.1\gev)\\\nonumber
M_{\chinon} &=& 151.9 \pm  21.9\gev\ (138.1\gev)
\end{eqnarray}
The uncertainties on the masses are larger than the mass differences
within the investigated area of the mSUGRA parameter space. However,
it is at this level possible to give good estimates on these six
sparticle masses. 

\pagebreak
\vspace*{1cm}

\vspace*{1cm}
\pagebreak
\vspace*{1cm}

\vspace*{1cm}
\pagebreak
\vspace*{1cm}
\vspace*{1.5cm}
%----------------------------------------------------------------------
%----------------------------------------------------------------------

 \newpage
 \cleardoublepage

 \chapter{Conclusions}\label{chap:Conclusions} 
 If nature is described by mSUGRA in the investigated region around I'
there will be some difficulties to reconstruct the investigated
sparticle masses in the dileptonic decay of the $\chinonn$. However,
with the Large Hadron Collider the CMS and the ATLAS experiments would
be able to find signals for Supersymmetry and to give an estimate of
the sparticle masses. The results of this study can be summarised as
follows:
\begin{enumerate}
\item  Since the cascade decay of the $\chinonn$ is dominated by tau
leptons at high $\tan \beta$ the dileptonic invariant mass distribution
has no sharp edge at the kinematic limit.
\item The $\tau^{\pm} \ra \rho^{\pm} \nu_{\tau}$ channel offers a good
opportunity to search for endpoints.
\item In the investigated parameter space it is possible - at the
Monte Carlo level - to reconstruct five to six sparticle masses in a
30 $fb^{-1}$ data sample, with an uncertainty of about 20 to 30$\gev$,
by using the methods introduced in this study.
\item There is a linear dependence between the gaussian maximum and
the theoretical endpoint for several mass distributions. Thus,
endpoint measurements with a linear fit at the tail of the
distribution can be well improved.
\item In principle it is possible to obtain two kinematic limits -
the limit of $M_{\rho^{\pm}_{1}q}$ and $M_{\rho^{\pm}_{2}q}$ - 
by measuring the $M_{\rho^{\pm}q}$ distribution with a cut on the
invariant mass of both opposite-sign rho mesons.
\item Top and bottom quark events can be well separated in the
$M_{\rho\rho} - M_{\rho{q}}$ plane. 
\item In the investigated parameter space the endpoints of the
distributions involving bottom and top quarks are determined by the
$\tilde{b}_{2}$ and $\tilde{t}_{2}$, respectively. However, the
measured endpoints with the linear fit method correspond more to the
$\tilde{b}_{1}$ mass and $\tilde{t}_{1}$ mass, respectively, because
$\tilde{b}_{1}$ and $\tilde{t}_{1}$ significantly dominate the
distributions. Thus, it is generally very important to take the mixing
of the involved particles into account.
\end{enumerate}

 \newpage
 \cleardoublepage

 \chapter{Outlook}\label{chap:Outlook}
 In this study many simplifications have been applied. In order to
obtain a more realistic prediction, at least the improvements given
below have to be included. Furthermore, even at the Monte Carlo level
additional studies like the investigation of the 3-prong channel may
give useful information. The next steps towards a more detailed
analysis are:
\begin{enumerate}
\item \emph{Estimate of background}\\
This study does not take Standard Model backgrounds into
account. Additionally, there is also a mSUGRA background coming from
taus in chargino and other decays. An estimate of these backgrounds is
necessary.
\item \emph{Estimate of detector effects}\\
Since this study is purely at the Monte Carlo level, no detector
effects have been taken into account. The CMS programs OSCAR for the
full simulation, ORCA for the reconstruction and FAMOS for a combined
fast detector simulation and reconstruction are necessary to give an
estimate of detector effects.
\item \emph{Investigation of the 3-prong channel}\\
Not only the $\tau^{\pm} \ra \rho^{\pm} \nu_{\tau}$ channel can be
used for the $\chinonn$ discovery at high $\tan \beta$. Due to the
nearly similar particles involved in the decay chain, the $\tau^{\pm}
\ra \pi^{\pm} \pi^{\mp} \pi^{\pm} \nu_{\tau}$ channel should have
related properties. Assuming the same efficiency and acceptance, its
use would increase the statistics by a factor 2.5.
\item \emph{Possible effect through spin correlations}\\
In the simulations no spin correlations are taken into
account. However, it is possible to make an estimate how the spin
correlations might have an effect on the invariant mass
distributions. Since the statistics in the region of an kinematic
endpoint depends on whether that special configuration is favoured or
not the investigation of that configuration with respect to the spin
gives a hint of the a possible effect.
\end{enumerate}

 \newpage
 \cleardoublepage
 
 \appendix
 \renewcommand{\chaptermark}[1]{\markboth{\sf\appendixname\ \thechapter:\ #1}{}}

 \chapter{Problems using PYTHIA 6.220}
 \section{Bug in the hadronic tau decay}\label{sec:bug} We found that
in PYTHIA 6.220 the invariant mass distribution of the visible
particles in the $\tau^{\pm} \ra \rho^{0} \pi^{\pm} \nu_{\tau}$ decay
has a $\tau$-energy dependence which is unphysical. Further
investigations showed that the energy distribution of the $\nu_{\tau}$
in the rest frame of the $\tau^{\pm}$ depends on the $\tau$-energy in
the laboratory frame of the program, which is a clear violation of
Lorentz Invariance. According to the author the intended neutrino
spectrum correction had not been obtained for taus with $E_{\tau} > 20
m_{\tau}$, which is the limit where taus are boosted to their rest
frame for better numerical precision
\cite{pythiaweb}. The bug has been fixed in version 6.222, which has
been released on the web~\cite{pythiaweb} on $21^{st}$ January
2004. However, the author remarks that the correction of the neutrino
spectrum is still crude.
\section{Particle listing} \label{sec:pylis}
In the particle listings of PYTHIA~6.220 interfaced with ISASUGRA~7.69
we found some problems depending on the value of MSTP(128). It is not
clear whether the problem is caused by PYTHIA in this case. In the
following the change of the particle listing due to different values
of MSTP(128) is briefly presented. Particles in the documentation
section have status three and particles in the rest of the particle
listing have status one or two.
\begin{enumerate}
\item \emph{MSTP(128) = 0:} For this value there is a huge overlap
between the documentation section and the rest of the particle
listing. All intermediate states are declared here, but it happens
that particles point to their children, but the children point to the
mother particle of their mother. This causes huge problems when
selecting special decay chains.
\item \emph{MSTP(128) = 1:} This is the value which is chosen in this
study. Even here there are difficulties in finding the right decay
chain, but it is possible to make unique assignments.
\item \emph{MSTP(128) = 2:} In this case the problem is that particles
in the particle listing with status one or two do not always point to
the right mother particle in the documentation section with status
three. It happens that intermediate states are totally missing
sometimes. Since the particles in the documentation section carry no
information about their children, it is therefore difficult to
distinguish between different decay chains.
\end{enumerate}

\section{PYTHIA 6.220 and TAUOLA}
We tried to use TAUOLA to solve the tau problem described in section
\ref{sec:bug} but we did not achieve to run TAUOLA with
PYTHIA and ISASUGRA. For Standard Model processes we found no
problem. C. Biscarat mentions this problem on her homepage
\cite{CBiscarat} and we assume that the mixed order of
particles in PYTHIA (see \ref{sec:pylis}) is responsible for that.

\section{Calculated cross sections}\label{sec:cs}
It seems that there is a dependence of the cross section values on the
selected production channels. We found changes in the production cross
section of about 20 $\%$.

 \newpage
 \cleardoublepage

%\chapter{Tables}
% \input{DTtables.tex}
% \newpage
% \cleardoublepage

  \bibliographystyle{unsrt}
 \bibliography{DTmybib}

\begin{thebibliography}{10}

\bibitem{SM}
S.~L. Glashow, \NP {\bf 22} (1961) 579; S. Weinberg, \PRL {\bf 19} (1967) 1264;
  A. Salam, in {\em Elementary Particle Theory}, ed. N. Svartholm, Stockholm,
  Alm\-qvist and Wiksell (1968), 367.

\bibitem{lightHiggs}
M. Spira and P.M. Zerwas \href{http://arxiv.org/abs/hep-ph/9803257}
  {arXiv:hep-ph/9803257}.

\bibitem{mSUGRA}
L. Alvarez-Gaume, J. Polchinski and M.B. Wise, Nucl. Phys. B221, 495 (1983); L.
  Iba{\~n}ez, Phys. Lett. 118B, 73 (1982); J.Ellis, D.V. Nanopolous and K.
  Tamvakis, Phys. Lett. 121B, 123 (1983); K. Inoue et al. Prog. Theor. Phys.
  68, 927 (1982); A.H. Chamseddine, R. Arnowitt, and P. Nath, Phys. Rev. Lett.,
  49, 970 (1982).

\bibitem{FirstEP}
H. Baer, C.-H. Chen, F. Paige and X. Tata, Phys.Rev. D50 (1994) 4508 and
  references therein,
  \href{http://arXiv.org/abs/hep-ph/9404212}{arXiv:hep-ph/9404212}.

\bibitem{wmappara}
M. Battaglia et al.,
  \href{http://arXiv.org/abs/hep-ph/0306219}{arXiv:hep-ph/0306219} v1,
  $23^{rd}$ July 2003, $(9)-(12)$.

\bibitem{g2}
G.W. Bennett et al.
  \href{http://arxiv.org/abs/hep-ex/0401008}{arXiv:hep-ex/0401008}; H.N. Brown
  et al. 2001 \href{http://tinyurl.com/3ypax} \PRL {\bf 86} 2227.

\bibitem{axion}
S. J. Asztalos et al.
  \href{http://arxiv.org/abs/astro-ph/0104200}{arXiv:astro-ph/0104200}.

\bibitem{MSSMworks}
S. Dawson, \href{http://arXiv.org/abs/hep-ph/9712464}{arXiv:hep-ph/9712464} v1,
  $19^{th}$ December 1997.

\bibitem{SusyPrimer}
S.P. Martin, \href{http://arXiv.org/abs/hep-ph/9709356}{arXiv:hep-ph/9709356}
  v3, $7^{th}$ April 1999.

\bibitem{isajet}
H. Baer et al., ISAJET 7.69,
  \href{http://arxiv.org/abs/hep-ph/0312045}{arXiv:hep-ph/0312045}.

\bibitem{pythia}
T. Sj{\"o}strand, P. Ed\'en, C. Friberg, L. L{\"o}nnblad, G. Miu, S. Mrenna and
  E. Norrbin, \href{http://www.jinr.ru/programs/cpc_in10/ADNN.html}{Computer
  Physics Commun. 135 (2001) 238}.

\bibitem{topmass}
K. Hagiwara et al. (Particle Data Group), Phys. Rev. D 66, 010001.

\bibitem{lepsusy2}
LEP Higgs Working Group,
  \href{http://arxiv.org/abs/hep-ex/0107030}{arXiv:hep-ex/0107030}.

\bibitem{lepsusy}
LEPSUSYWG, ALEPH, DELPHI, L3 and OPAL experiments, note LEPSUSYWG/02-06.2,
  \href{http://lepsusy.web.cern.ch/lepsusy/}{LEP2 Susy Working Group}.

\bibitem{wmappic}
\href{http://map.gsfc.nasa.gov/}{http://map.gsfc.nasa.gov/} - image gallery.

\bibitem{CMB_PW}
Wilson, R.W.; Penzias, A.A. Bell Telephone Labs., Holmdel, N. J. Science, 156:
  1100-1 (May 26, 1967).

\bibitem{bnl}
M. Davier et al.,
  \href{http://arxiv.org/abs/hep-ph/0208177}{arXiv:hep-ph/0208177}.

\bibitem{lhcpic}
\href{http://hepweb.rl.ac.uk/ppUKpics/}{http://hepweb.rl.ac.uk/ppUKpics/} -
  archive.

\bibitem{CernCourier}
CERN Courier, February 1999, ``A cold start for LHC''.

\bibitem{pape}
Luc Pape, private communication.

\bibitem{N2Studies}
I. Iashvili and A. Kharchilava, CMS Note 1997/065; S. Abdullin et al., CMS Note
  1998/006; D. Denegri, W. Majerotto and L. Rurua, CMS Note 1997/094.

\bibitem{root}
Rene Brun and Fons Rademakers, ROOT - An Object Oriented Data Analysis
  Framework, Proceedings AIHENP'96 Workshop, Lausanne, Sep. 1996, Nucl.
  Inst.{\&} Meth. in Phys. Res. A 389 (1997) 81-86. See also
  \href{http://root.cern.ch/}{ROOT Webpage}.

\bibitem{pythiaweb}
\href{http://www.thep.lu.se/~torbjorn/pythia/pythia6222.update}
  {http://www.thep.lu.se/torbjorn/pythia/pythia6222.update}.

\bibitem{CBiscarat}
\href{http://cbiscara.home.cern.ch/cbiscara/tauola_valid.html}
  {http://cbiscara.home.cern.ch/cbiscara/tauola\_valid.html}.

\end{thebibliography}

 \addcontentsline{toc}{chapter}{\bibname}

 \chapter* {Acknowledgements}
\addcontentsline{toc}{chapter}{Acknowledgements}
 First of all, I want to thank Prof.\ Felicitas Pauss for giving me the
possibility to carry out my diploma thesis in her group on a highly
interesting topic, and for great support in various concerns. I am
very grateful for the excellent supervision by the two experts Dr.\
Andr\'e Holzner and Dr.\ Luc Pape. It was a great pleasure to work
with Giovanna Davatz, Prof.\ G\"unther \mbox{Dissertori}, Dr. Michael
Dittmar, PD Klaus Freudenreich, Gabriele Kogler, Anne-Sylvie
Nicollerat and \mbox{Radoslaw} Ofierzynski. My special thanks go to my
friends who exceedingly enriched my life during my studies at ETH
Zurich. I want to express my gratitude to my family who supported me
unconditionally during my life.

\markboth{}{}
 \newpage
 \cleardoublepage

\end{document}